\documentclass[amssymb,aps,showkeys,showpacs,twocolumn]{revtex4-1}

\usepackage{fancyvrb}
\usepackage{graphicx}

\usepackage{amssymb}

\begin{document}

\title{The Wigner $3n$-j Graphs up to 12 Vertices}

\author{Richard J. Mathar}
\homepage{http://www.strw.leidenuniv.nl/~mathar}
\affiliation{Leiden Observatory, Leiden University, P.O. Box 9513, 2300 RA Leiden, The Netherlands}


\date{\today}

\begin{abstract}
The
3-regular
graphs representing sums over products of
Wigner $3-jm$ symbols are drawn on up to 12 vertices (complete to $18j$-symbols),
and the
irreducible graphs
on up to 14 vertices (complete to $21j$-symbols).
The Lederer-Coxeter-Frucht notations of the Hamiltonian cycles in these graphs
are tabulated to support search operations.
\end{abstract}

\maketitle
\section{Wigner symbols and cubic graphs}
\subsection{Wigner Sums}
We consider sums
of the form
\begin{equation}
\sum_{m_{01},m_{02},\ldots}
\left(\begin{array}{ccc}
j_{01} & j_{02} & j_{03} \\
m_{01} & m_{02} & m_{03} \\
\end{array}\right)
\left(\begin{array}{ccc}
j_{01} & j_{..} & j_{..} \\
-m_{01} & m_{..} & m_{..} \\
\end{array}\right)
\cdots
\label{eq.wsum}
\end{equation}
over products of Wigner $3jm$-symbols
which are closed in the sense (i) that the sum is over all
tupels of magnetic quantum numbers $m_{..}$ admitted by the standard spectroscopic
multiplicity of the factors, (ii) that for each column designed by $j_{..}$ and $m_{..}$
another column with sign-reversed $m$ appears in another factor
\cite{StoneMPCR52,StonePPS70,Edmonds}.

Each term contains $n$ factors---each factor a $3jm$-symbol---and
$3n/2$ independent variables $j_{..}$ for which
a pair of distinct 
indices in the
interval 0 to $n-1$ will be used in this script.
The numerical value
of each factor, internal symmetries or selection rules
are basically irrelevant for most of this work.

\subsection{Yutsis Reduction}

The Yutsis method maps the product structure of a Wigner $3n$-j symbol
onto a labeled 3-regular (also known as cubic) digraph
\cite{Yutsis,MassotRMP39,AndersonJPCA113,KentPRA40}.
Each factor is represented by a vertex.
An edge is drawn between
each pair of vertices which share one of the $j_{..}$; an edge
\emph{is} a $j$-value associated with a ``bundle'' of $m$-values.
Since each factor
comprises three $j_{..}$, the graph becomes 3-regular, i.e.,
in-degree and out-degree are both 3\@. The graphs are
directed (i.e., digraphs) where head and tail of the edge denote 
which of the factors carries which of the two signs of the $m$-value.
We shall enumerate vertices from $0$ to $n-1$ further below;
the two indices of the $j_{..}$ and its associated $m_{..}$ are just
the two labels of the two vertices that are connected by the edge.

Once an undirected unlabeled
connected graph is set up, adding a sign label
and a direction to the edges (i.e., an order and sign of the
three quantum numbers in the Wigner symbol) adds no
information besides phase factors.

A related question is whether and which cuts through the edges exist that split
any of these graphs into vertex-induced binary trees. The two trees generated
by these means represent recoupling schemes 
\cite{DanosPR304,vanDyckCPC173,vanDyckDM307,FackCPC101,AldredDAM157,AquilantiTCA123,Louck}.
The association generalizes the relation between Clebsch-Gordan coefficients
(connection coefficients between sets of orthogonal polynomials
\cite{LievensJMP43, GranovskiiJPA26})
and the Wigner $3j$ symbols to higher numbers of coupled angular momenta.

\subsection{Connectivity}
The rules of splitting the sum (\ref{eq.wsum})
into sums of lower vertex count depend on the edge-connectivity
of the cubic graph, i.e., the minimum number of edges that must be removed
to cut the graph into at least two disconnected parts.
Cubic graphs are at most 3-connected because removal of the
three edges that run into any vertex turns that vertex into a singleton.

Wigner sums can be hierarchically decomposed
for 1-connected, 2-connected
and those 3-connected diagrams which are separated by
cutting 3 lines into subgraphs with more than 1 vertex left \cite{Yutsis,BalcarSTMP234}.
These will be plotted subsequently with
one to three
red edges to illustrate this property.
Focus is therefore shifted to the remaining, ``irreducible'' graphs. Every
cycle (closed path along a set of edges) in those consists of at least
4 edges, because a cycle of 3 edges can clearly
be disconnected cutting the external 3 edges.
All of their edges are kept black;
they define ``classes'' of $j$-symbols \cite{Yutsis,NewmanJPA9}.

\subsection{LCF notation}
In the majority of our cases, simple cubic graphs are Hamiltonian, which means
they support at least one Hamiltonian cycle, a closed path along the edges
which visits each vertex exactly once and uses each edge at most once \cite{BauAJC2}.
(See A001186 and A164919 in the Encyclopedia of Integer Sequences
for a statistics of this feature \cite{EIS}.)
The structure of the graph is in essence caught by arranging the
vertices of such a walk on a circle---which uses already two third of
the edges to complete the cycle---and then specifying which
chords
need to be drawn to account for the remaining one third of the edges.
The chords are potentially crossing. Whether the graph is planar
or not (i.e., whether it could be drawn on a flat sheet of paper without
crossing lines) is not an issue.

The Lederberg-Coxeter-Frucht (LCF) notation is an ASCII representation
of these chords (diagonals) in cubic Hamiltonian graphs \cite{FruchtJGT1,Coxeter}.
For each vertex visited, starting with the first, the distance
to the vertex is noted where the chord originating there re-joins the cycle.
The distance is an integer counting after how many additional steps along the
cycle that opposite vertex of the chord will be visited, positive for a forward
direction along the cycle, negative for a backward direction.
The direction is chosen to minimize the absolute value of this distance,
and to use the positive value if there is a draw.
This generates a comma-separated list of $n$
integers in the half-open interval
$(-n/2,n/2]$, where $n$ is the number of vertices.
The values $0$ or $\pm 1$ do not appear because
we are considering only simple graphs (loopless, without multiple edges).

Because the choice of the starting vertex of a Hamiltonian
cycle is arbitrary, and because one may reverse the
walking direction,
two LCF strings may be trivially equivalent in two ways:
(i) a cyclic permutation
or (ii) reverting the order while flipping all signs (unless the entry is $n/2$)
is an irrelevant modification.

There are two notational contractions that are accompanied by
some symmetry
of the graph:
\begin{itemize}
\item
If the vector of $n$ distances is a repeated block of numbers of the form $[a,b,c,\ldots x,a,b,c,\ldots x,\ldots]$,
the group is written down once with an exponent counting the frequency of occurrences, $[a,b,c,\ldots x]^f$.
\item
If the distance vector has an inverted palindromic symmetry of the form $[a,b,c,\ldots x,-x,\ldots -c,-b,-a]$,
the repeated part is replaced by a semicolon and dash $[a,b,c,\ldots x;-]$.
\end{itemize}

If more than one Hamiltonian cycle exist in the graph, non-trivial but
equivalent LCF notations appear. In the following chapters,
lines
\begin{verbatim}
LCF ... = ...
\end{verbatim}
with one or more equal signs signal graphs which support more than one cycle.

The structure of the graph may also be visualized as a carbon
or silicate molecule with some graphic viewers
if this information is encoded as a SMILES string
\cite{WeiningerJCIC28}.
The Hamiltonian cycle defines the backbone of a ring, and the cords are enumerated 
and serve as indices to the atoms to recover the missing bonds.

The Wiener index of the undirected graph (sum of the distances of unordered
pairs of vertices) will be reported as an integer number attached to a \texttt{W} \cite{GraovacJMC8}.
The diameter of the undirected graph (largest distance between any two vertices)
is written down attached to a \texttt{d},
and the girth of the undirected graph (length of the shortest cycle)
is attached to a \texttt{g}.
Finally, the Estrada index (sum of the exponentials of the eigenvalue spectrum
of the adjacency matrix) follows after a \texttt{EE} \cite{GutmanCPL436}.
(These numbers are rounded to $10^{-5}$, the minimum precision to
generate unique indices for the graphs on 14 nodes.)

\section{$4$ and $6$ vertices}
The main part of the manuscript shows the nonequivalent (up to a permutation
of the vertex labels) simple cubic graphs, sorted along increasing
number of vertices and increasing edge-connectivity.

The labels are
an indication of at least one Hamiltonian cycle through the graphs
where one was found.
In
the applications, the labels are replaced by the two sign labels of the
node's orientation, i.e., basically a phase label which relates to
the ordering of the $j$-symbols in the Wigner $3jm$-symbol
at that vertex \cite{BarCPC50,LimaCPC66}.

The directions of the edges are an almost arbitrary choice as well,
pointing from the vertex labeled with the lower number  to
the vertex labeled with the higher number.

\begin{figure}
\includegraphics[scale=0.3]{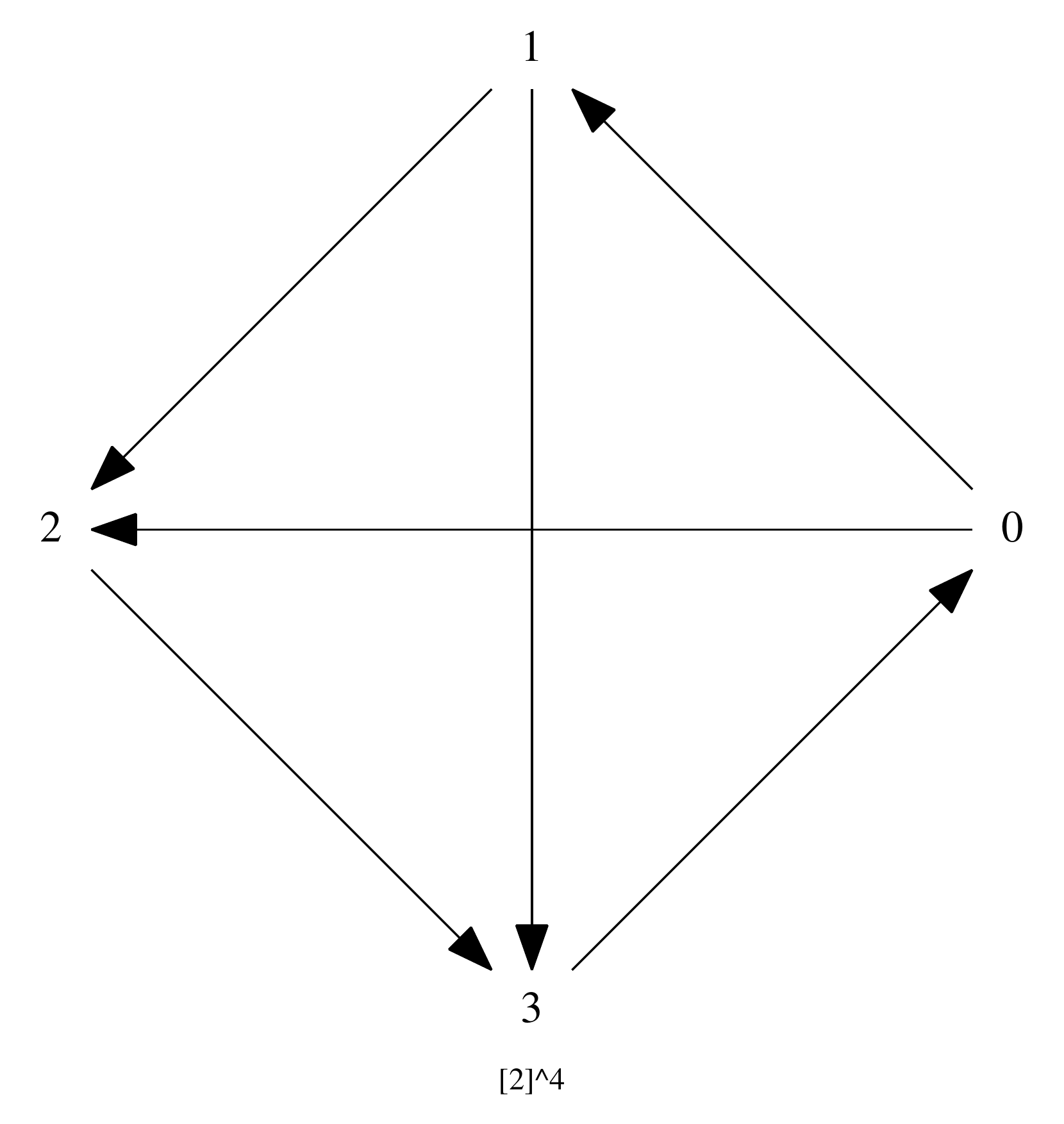}
\caption{The graph on $n=4$ vertices, defining
the $6j$-symbol.
}
\label{fig.4}
\end{figure}

On 4 vertices we find the planar version of a tetrahedron,
Figure \ref{fig.4}.

6 vertices support the two graphs in Figure \ref{fig.6}.
Their LCF notations are:
\begin{Verbatim}
   LCF [3,-2,2]^2  W21 d2 g3 EE25.07449   
   LCF [3]^6  W21 d2 g4 EE24.13532   
\end{Verbatim}

\begin{figure}
\includegraphics[scale=0.35]{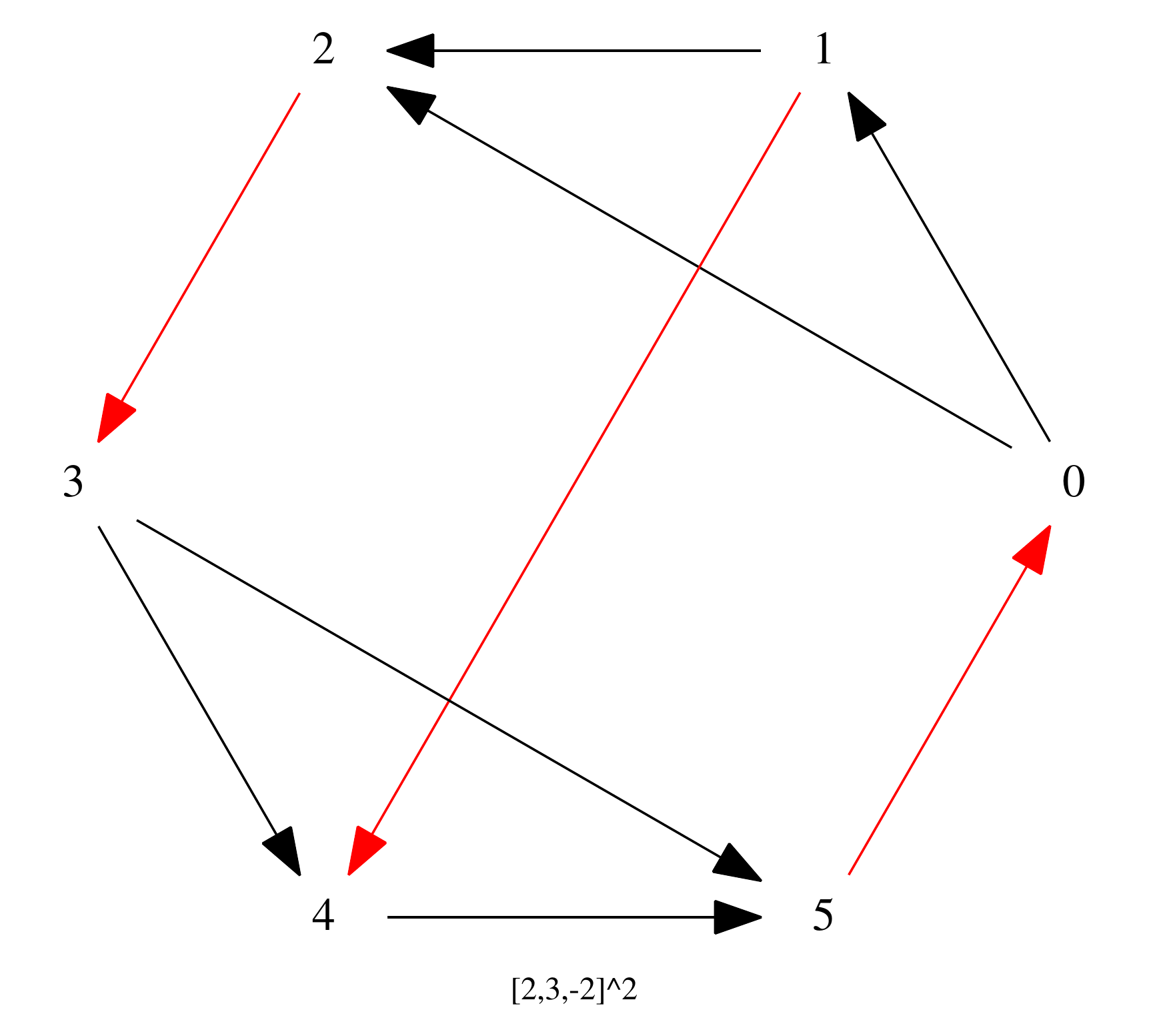}
\includegraphics[scale=0.35]{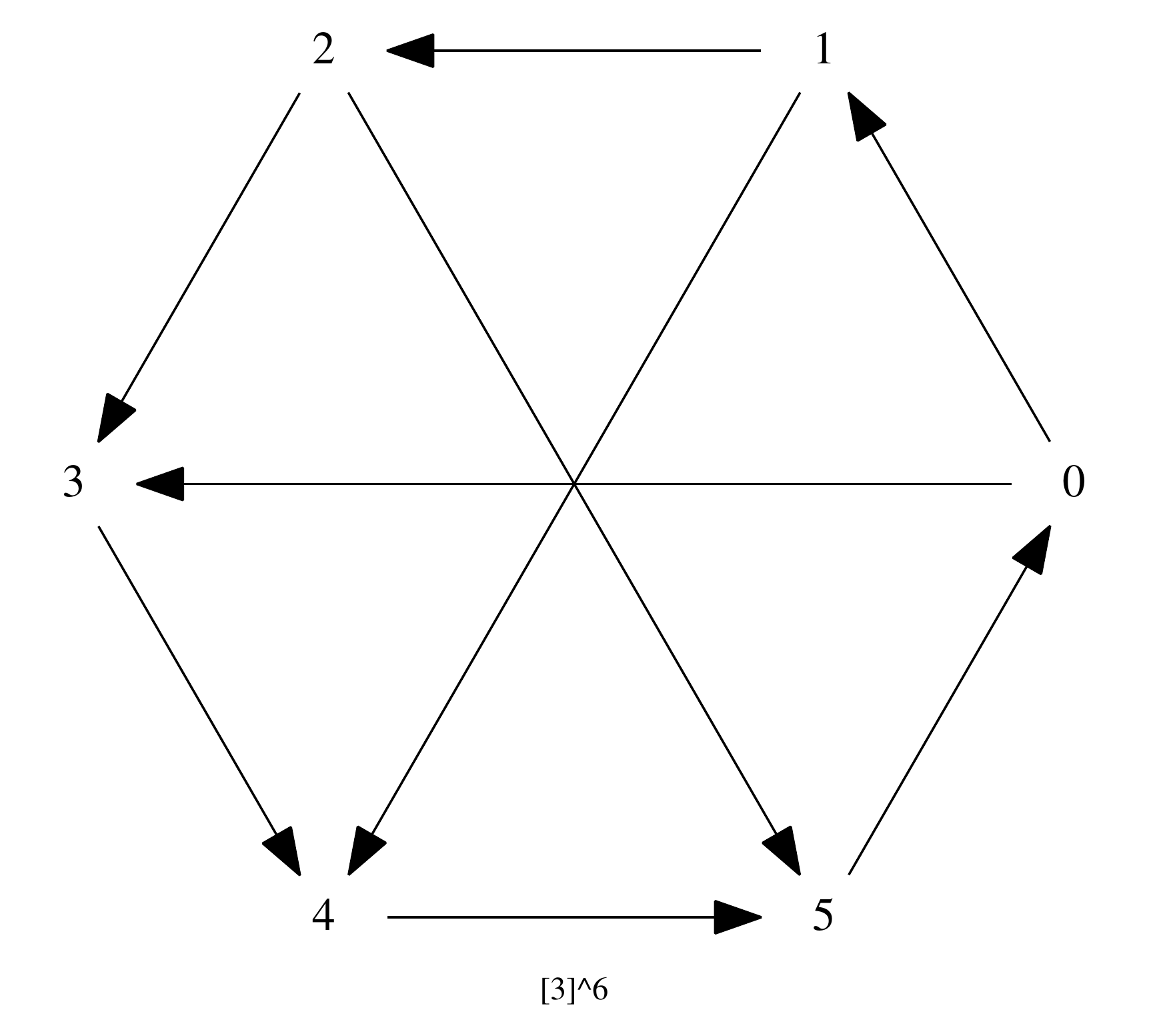}
\caption{The 2 graphs on $n=6$ vertices.
\texttt{[3]\char94 6}
(called the utility graph if undirected, unlabeled)
defines the $9j$-symbol \cite{JahnPR93}.
}
\label{fig.6}
\end{figure}

\begin{widetext}
\section{$8$ vertices}
All 5 cubic graphs on 8 vertices are shown in Figure \ref{fig.8}\@.
Their representations by LCF strings are:

\begin{Verbatim}
   LCF [2,-2,-2,2]^2  W50 d3 g3 EE33.73868   
   LCF [2,3,-2,3;-]  = [4,-2,4,2]^2 W46 d3 g3 EE30.97135   
   LCF [3,3,4,-3,-3,2,4,-2]  W44 d2 g3 EE30.03607   
   LCF [-3,3]^4  W48 d3 g4 EE29.39381   
   LCF [4]^8  = [4,-3,3,4]^2 W44 d2 g4 EE29.09522   
\end{Verbatim}

The two Hamiltonian cycles indicated by the first two LCF representations
for the graph \verb+[2,3,-2,3;-]+
in Figure \ref{fig.8} are: Walking along the vertices
labeled $0\to 1\to 2\to 3\to 4\to 5 \to 6 \to 7 \to 0$ generates the
LCF name \verb+[2,3,-2,3;-]+. The alternative
Hamiltonian cycle
$0\to 2\to 3\to 6\to 7\to 5\to 4\to 1\to 0$ is described by
the name \verb+[4,-2,4,2]^2+.

The last graph in Figure \ref{fig.8} is another example hosting two cycles,
equivalent to switching between Figures 19.1a and 19.1b in the
Yutsis--Levinson--Vanagas book \cite{Yutsis}:
The notation \verb+[4]^8+ describes a Hamiltonian Path along the
vertices $0\to 1\to 2\to 3\to 4\to 5\to 6 \to 7 \to 0$.
The alternative \verb+[4,-3,3,4]^2+ corresponds to the
path $7\to 0\to 1 \to 2 \to 6 \to 5 \to 4 \to 3 \to 7$.

\begin{figure}[hb]
\includegraphics[scale=0.4]{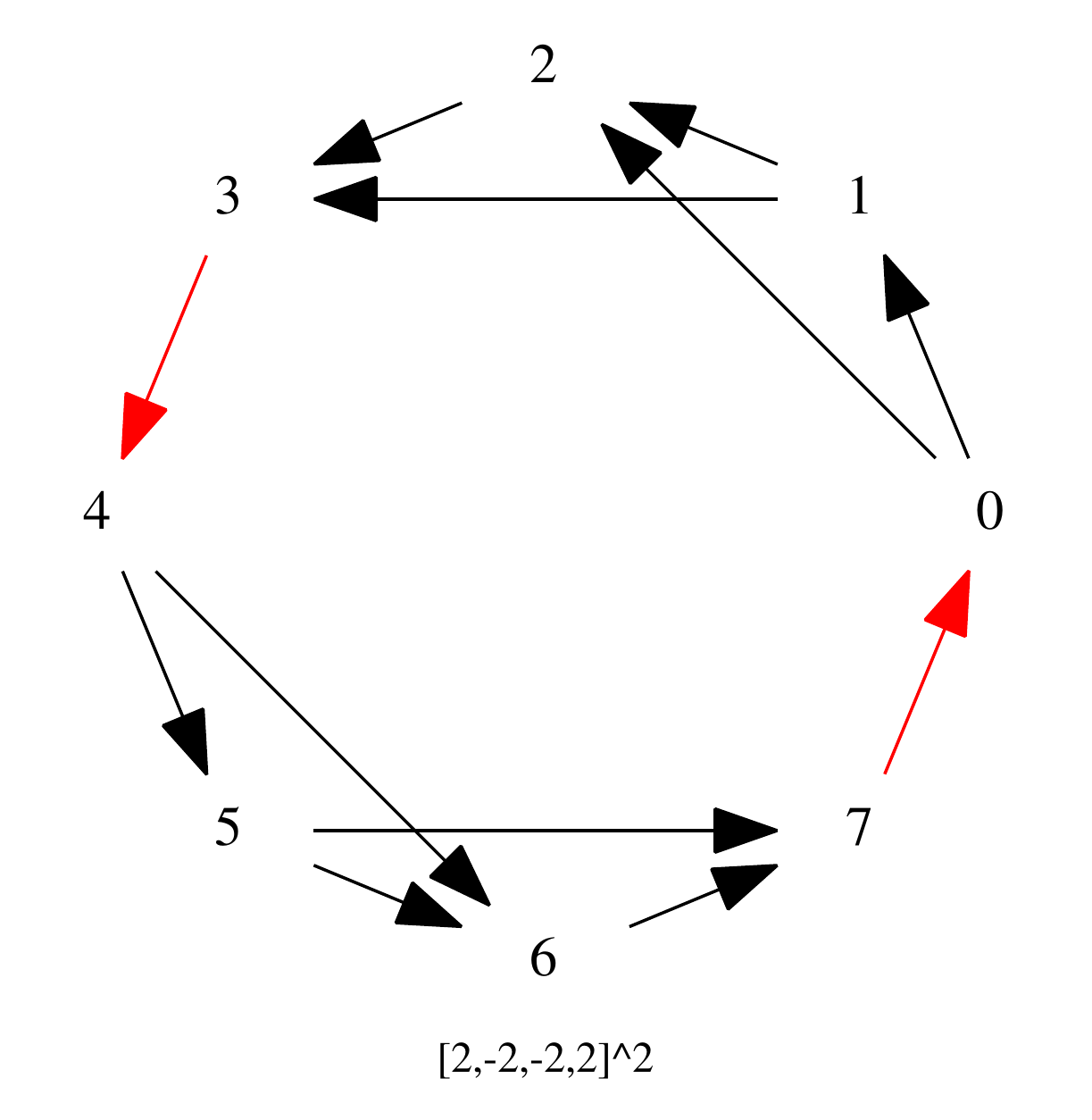}
\includegraphics[scale=0.4]{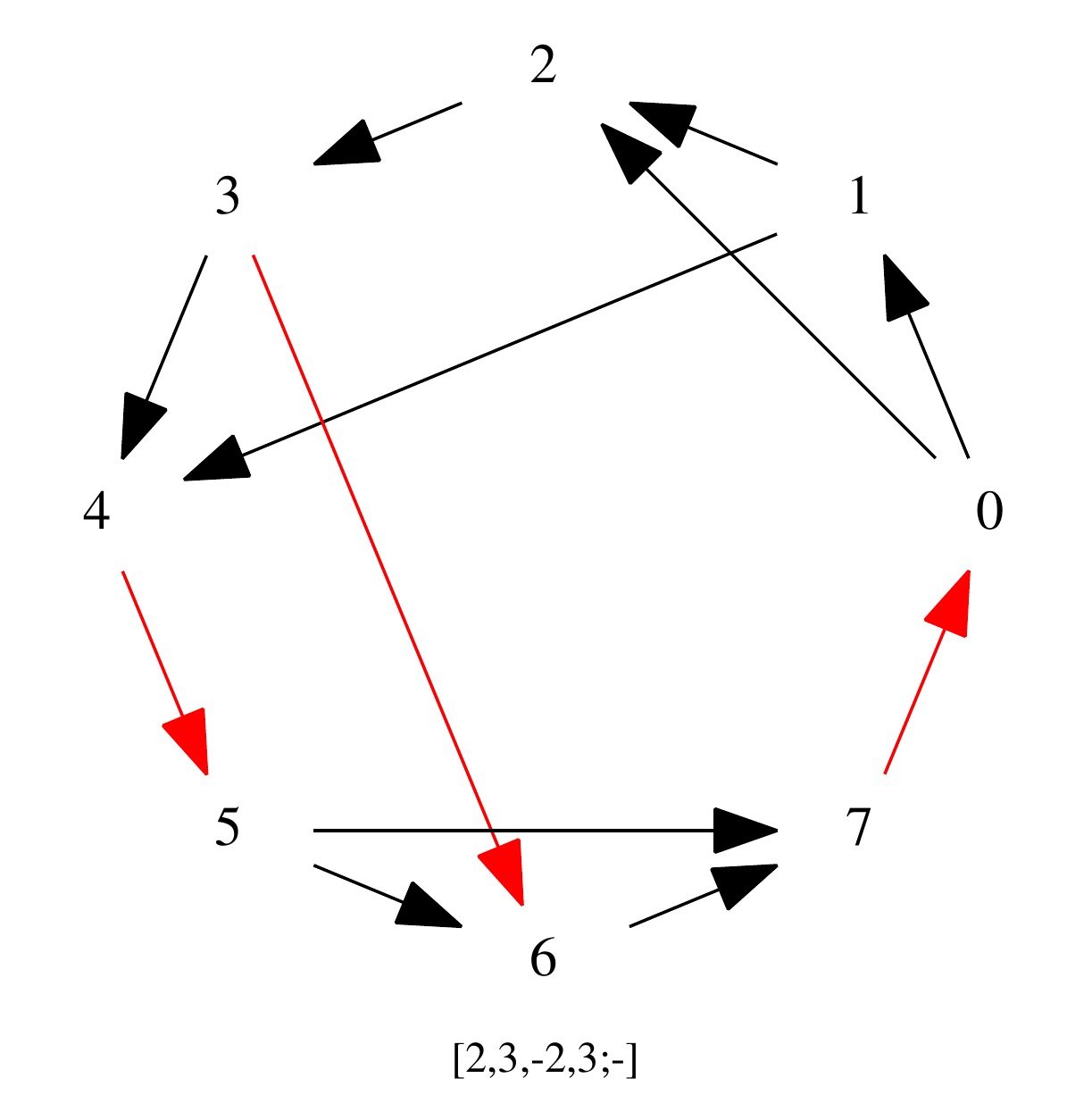}
\includegraphics[scale=0.4]{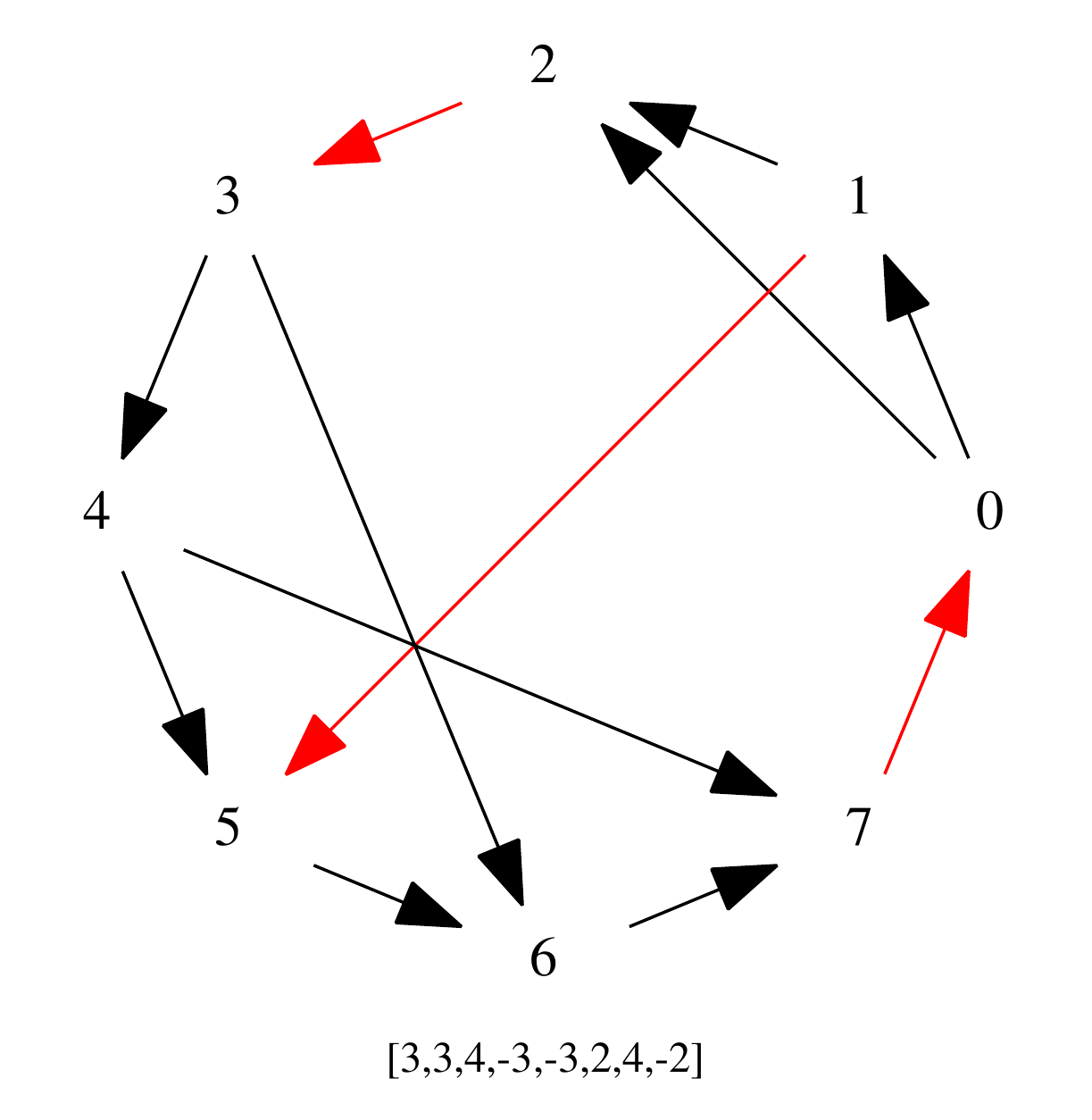}
\includegraphics[scale=0.4]{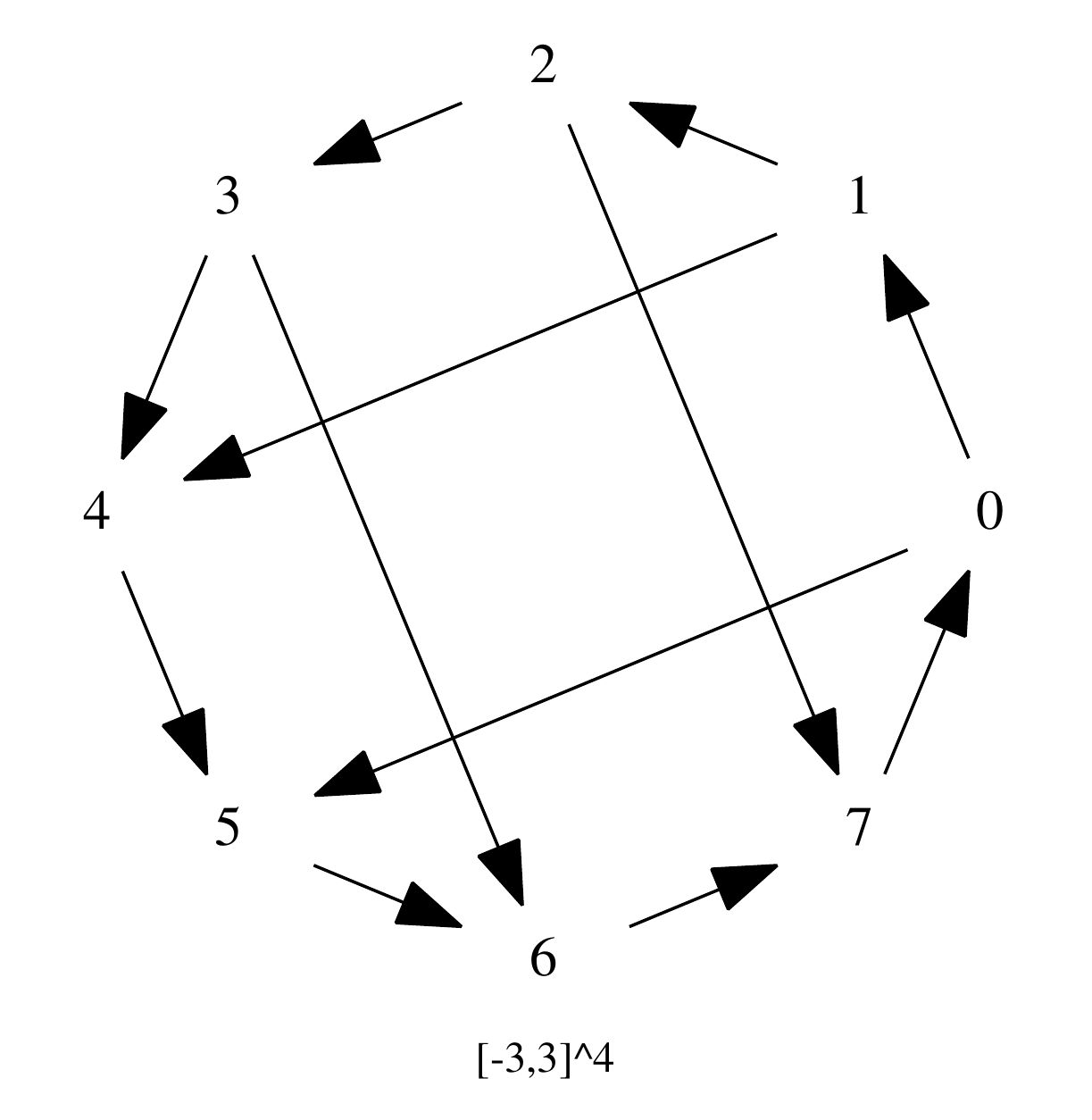}
\includegraphics[scale=0.4]{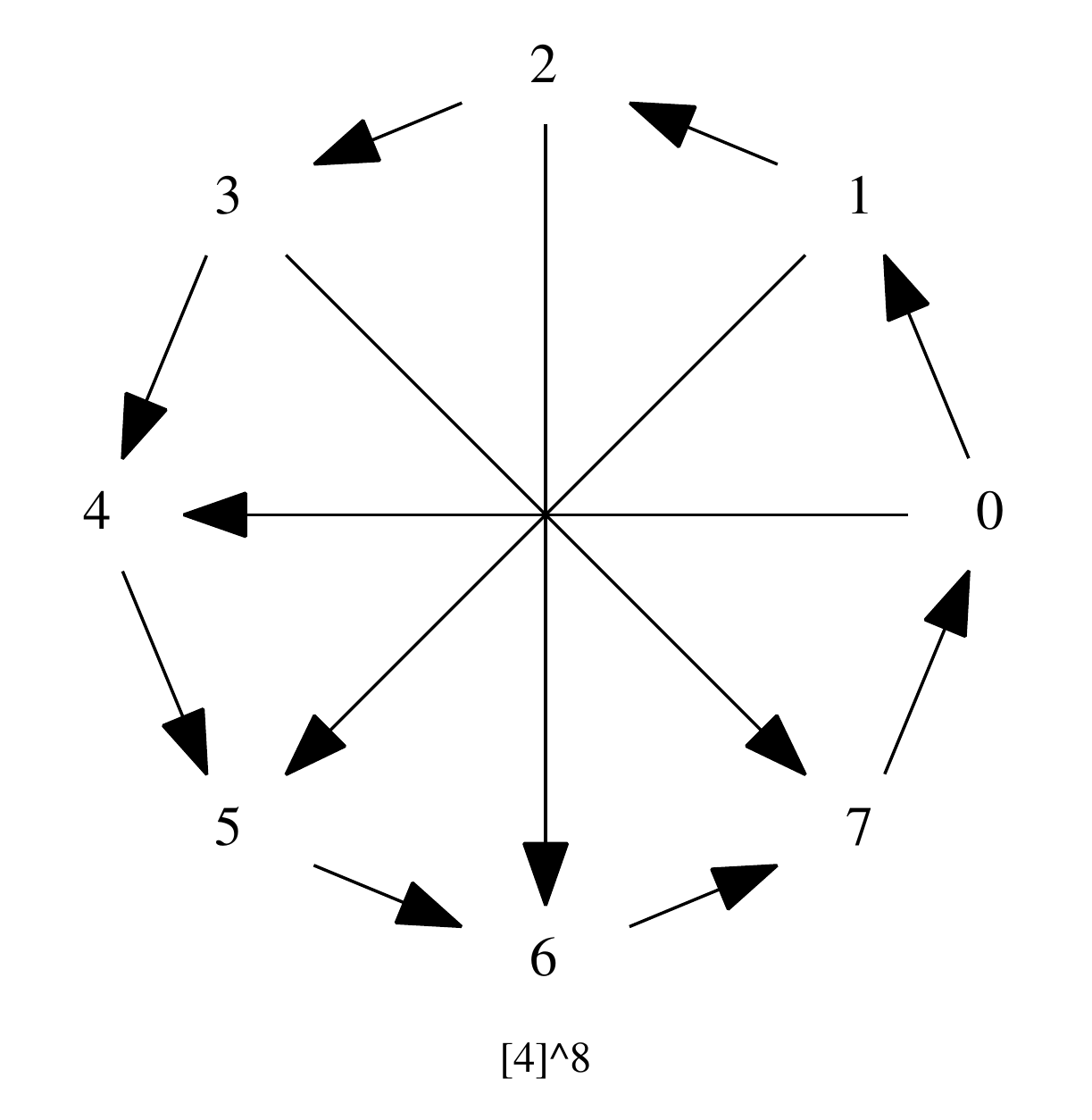}
\caption{Graphs on $n=8$ vertices. The two which are
cyclically 4-connected define the two $12j$-symbols \cite{OrdPR94,AlisauskasJMP43}.
The undirected, unlabeled version of
\texttt{[-3,3]\char94 4}
is
the cubical graph.
}
\label{fig.8}
\end{figure}
\clearpage

\section{$10$ vertices}
The 19 graphs with 10 vertices (15 edges)
are shown in Figure \ref{fig.10n1}  if 1- or 2-connected,
Figure \ref{fig.10n3} if 3-connected reducible,
and Figure \ref{fig.10n4} if irreducible \cite{ImrichAE6}.

Two of the 19 graphs, one in Figure \ref{fig.10n1} (\texttt{W111 d5 g3 EE42.60094})
and one in Figure \ref{fig.10n4} (\texttt{W75 d2 g5 EE34.21829}), are not Hamiltonian, so we are left
with 17 lines of
LCF strings of their Hamiltonian cycles:

Figure \ref{fig.10n1}:

\begin{Verbatim}
   LCF [3,-2,-4,-3,2,2,-2,-2,4,2]  W91 d4 g3 EE39.41746   
   LCF [-2,-2,3,3,3;-]  W90 d3 g3 EE38.90980   
   LCF [2,-3,-2,2,2;-]  W90 d3 g3 EE40.39508   
   LCF [-2,5,2,2,-2]^2  W93 d4 g3 EE40.69426   
\end{Verbatim}

Figure \ref{fig.10n3}:

\begin{Verbatim}
   LCF [3,-2,5,-3,2]^2  = [3,-2,4,-3,4,2,-4,-2,-4,2] W85 d3 g3 EE37.44960   
   LCF [-3,5,2,5,-2,4,5,3,5,-4]  = [-4,2,5,-2,4,4,4,5,-4,-4] = [-3,2,4,-2,4,4,-4,3,-4,-4] W82 d3 g3 EE36.00347   
   LCF [-4,3,3,5,-3,-3,4,2,5,-2]  = [3,-4,-3,-3,2,3,-2,4,-3,3] W85 d3 g3 EE36.68162   
   LCF [3,-3,5,-3,2,4,-2,5,3,-4]  W84 d3 g3 EE36.25442   
   LCF [-4,-2,5,2,4,-2,4,5,-4,2]  W81 d3 g3 EE36.77120   
   LCF [2,3,-2,3,-3;-]  = [-4,4,2,5,-2]^2 W87 d3 g3 EE37.69671   
   LCF [4,-2,5,2,-4,-2,2,5,-2,2]  W84 d3 g3 EE38.01880   
   LCF [2,4,-2,3,4;-]  = [2,5,-2,5,5]^2 W83 d3 g3 EE37.01785   
   LCF [-3,3,3,5,-3]^2  W85 d3 g4 EE35.83204   
\end{Verbatim}

Figure \ref{fig.10n4}:

\begin{Verbatim}
   LCF [5,-4,4,-4,4]^2  = [5,-4,-3,3,4,5,-3,4,-4,3] W79 d3 g4 EE34.72233   
   LCF [5,5,-4,4,5]^2  = [-3,4,-3,3,4;-] = [4,-3,4,4,-4;-] = [-4,3,5,5,-3,4,4,5,5,-4] W81 d3 g4 EE34.97449   
   LCF [5]^10  = [-3,3]^5 = [5,5,-3,5,3]^2 W85 d3 g4 EE35.40679   
   LCF [3,-4,4,-3,5]^2  W85 d3 g4 EE35.47908   
\end{Verbatim}

In terms of the standard
nomenclature
\begin{itemize}
\item
\verb+[5]^10+ is the $15j$-coefficient
of the first kind (the M\"obius ladder graph for that vertex count),
\item
\verb+[3,-4,4,-3,5]+ is the second kind,
\item
\verb+[5,-4,4,-4,4]^2+ the third,
\item
\verb+[5,5,-4,4,5]^2+ the fourth
\item
and
the Petersen Graph
(which has no Hamiltonian cycle \cite{ClarkPMH14,SchwenkJCTB47}) the fifth \cite{Yutsis}.
\end{itemize}

The number of classes of $3n-j$ symbols for even $n=4,6,8,\ldots$ grows as
1, 1, 2, 5, 18, 84, 607, 6100, 78824, 1195280, 20297600, 376940415,\ldots
\cite{BrouderJEESR86}.
(An apparently erroneous 576 is sometimes quoted instead of 607 \cite{DurrNC53,WormerAQC51}).

The volume of such lists grows with the number of vertices, which
leads to the main objective of this work. Starting from a Wigner product
of the form (\ref{eq.wsum}),
its cubic graph is quickly drawn, but whether the graph is the same as
(in our geometric mathematical framework isomorphic to) another one needs a kind
of signature or classification. One might build a frequency
statistics of the number of shortest cycles in the spirit of finding
faces of the the polytope of a 3-dimensional ball-and-stick model of the graph, or count the number of
cut sets and compare these.

Another approach is supported here: find at least one Hamiltonian cycle,
generate the LCF string, and use a reverse lookup
in the LCF table to see whether any two strings occur in the same line.
The common idea is to replace strenuous visual recognition of graphs
by a comparison of ASCII representations.

The ancillary files contain the source code of a small Java program 
which supports the detection of Hamiltonian cycles. Its input is an edge list
of a simple cubic Hamiltonian graph. The cycles are computed by walking
from the first node of the first edge in all three directions and generating a
tree of non-interfering walks recursively \cite{MartelloTOMS9}. The output is a LCF string and a vertex chain
along each cycle found, and optionally a representation in \texttt{dot}
format which can be plotted by the \texttt{graphviz} commands.

The ancilliary files contain also cage-type graphs detailed as sets of \texttt{gnuplot}
commands and molfiles \cite{DalbyJCIC32}. These graphs can be rotated interactively which helps to decipher the cycle
structure and symmetries.

\begin{figure}
\includegraphics[scale=0.45]{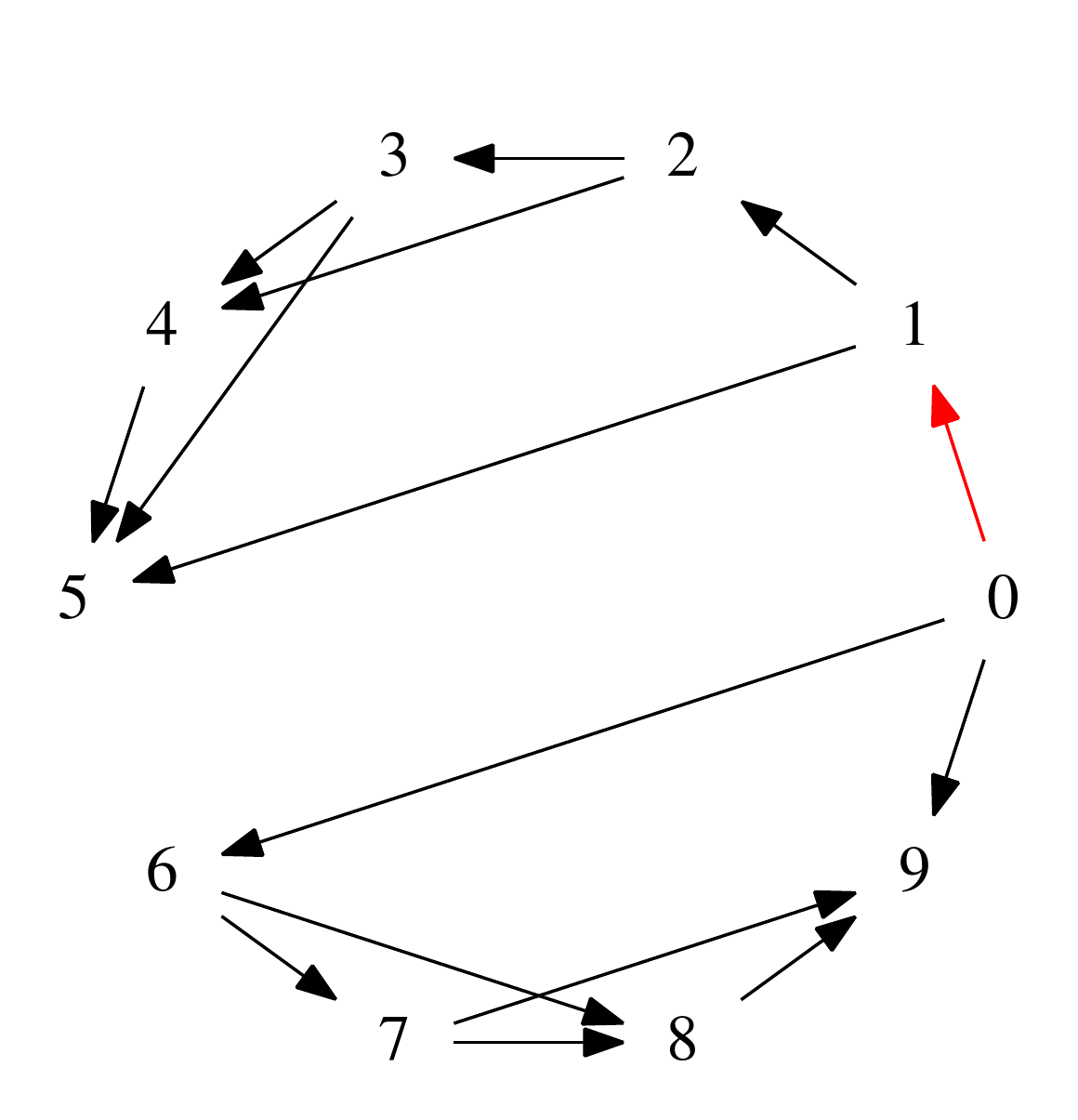}
\includegraphics[scale=0.45]{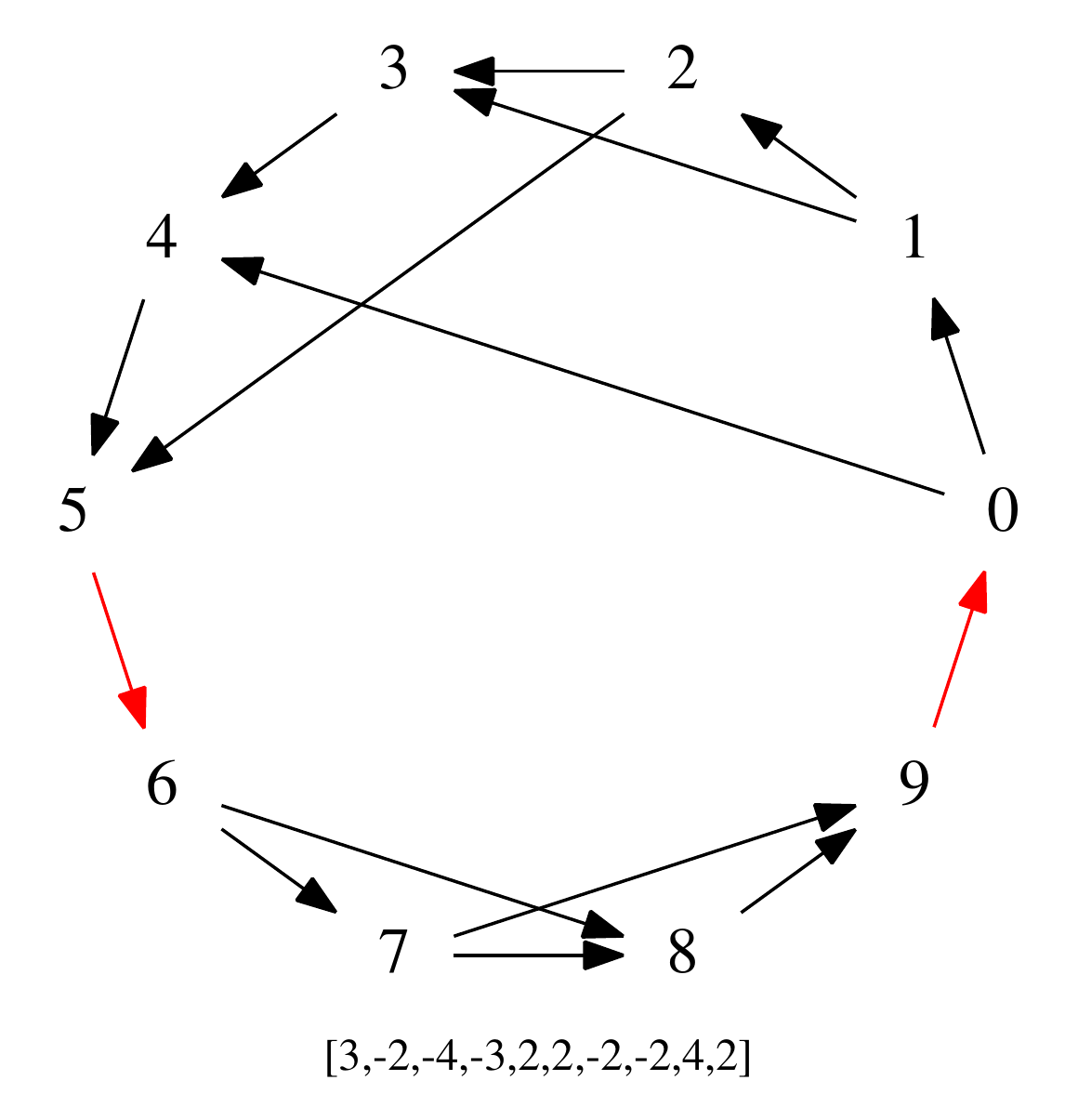}
\includegraphics[scale=0.45]{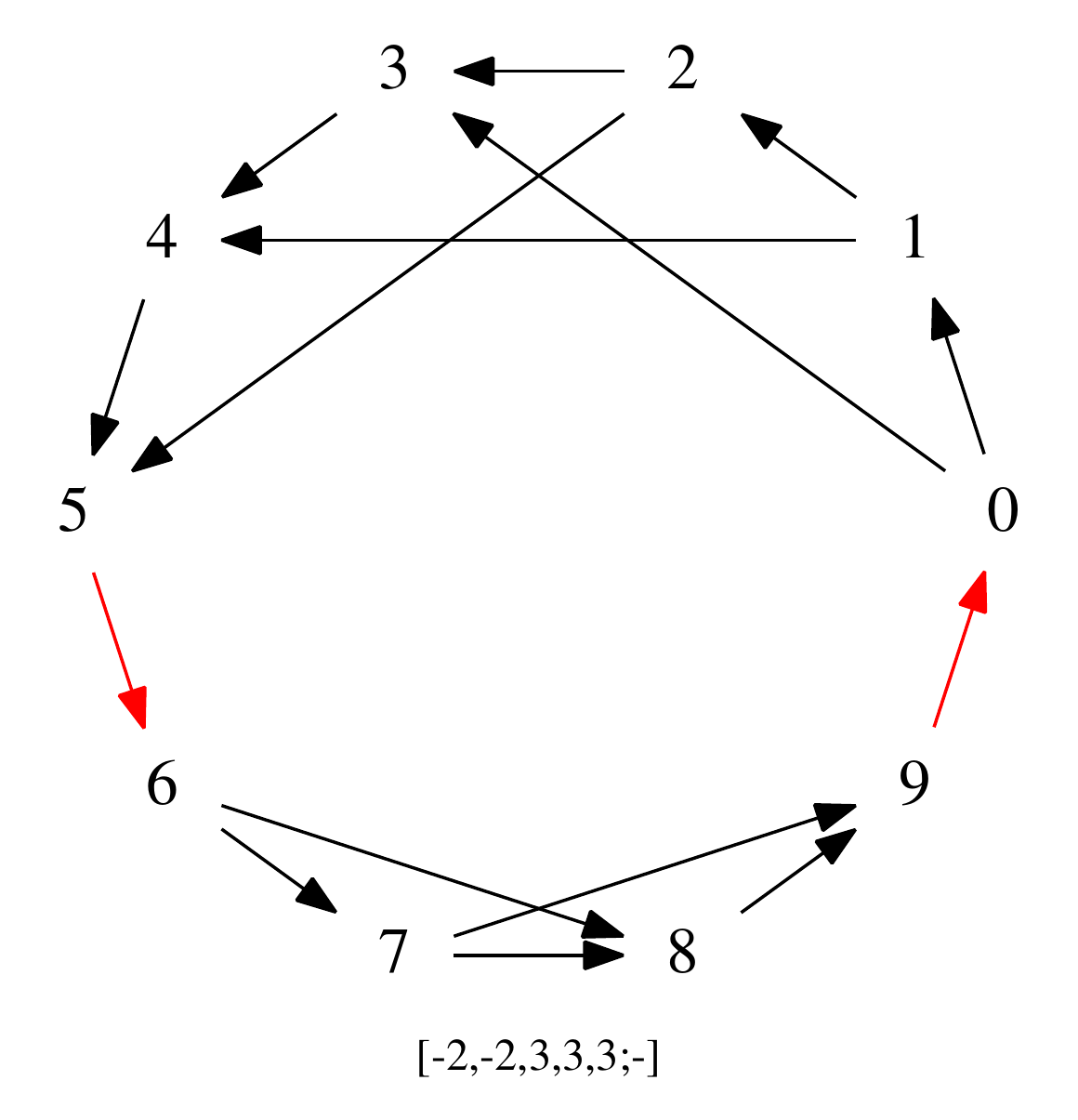}
\includegraphics[scale=0.45]{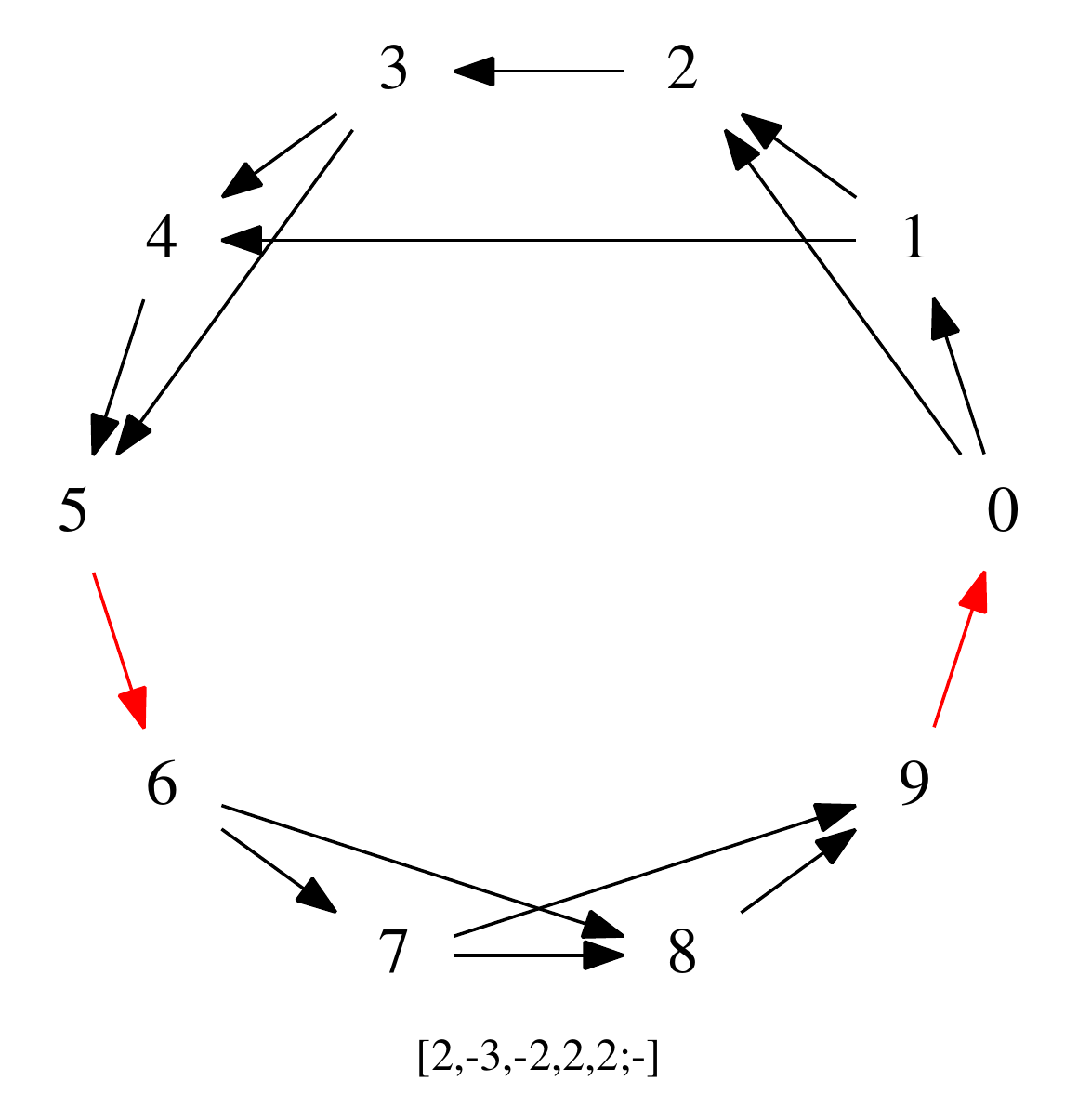}
\includegraphics[scale=0.45]{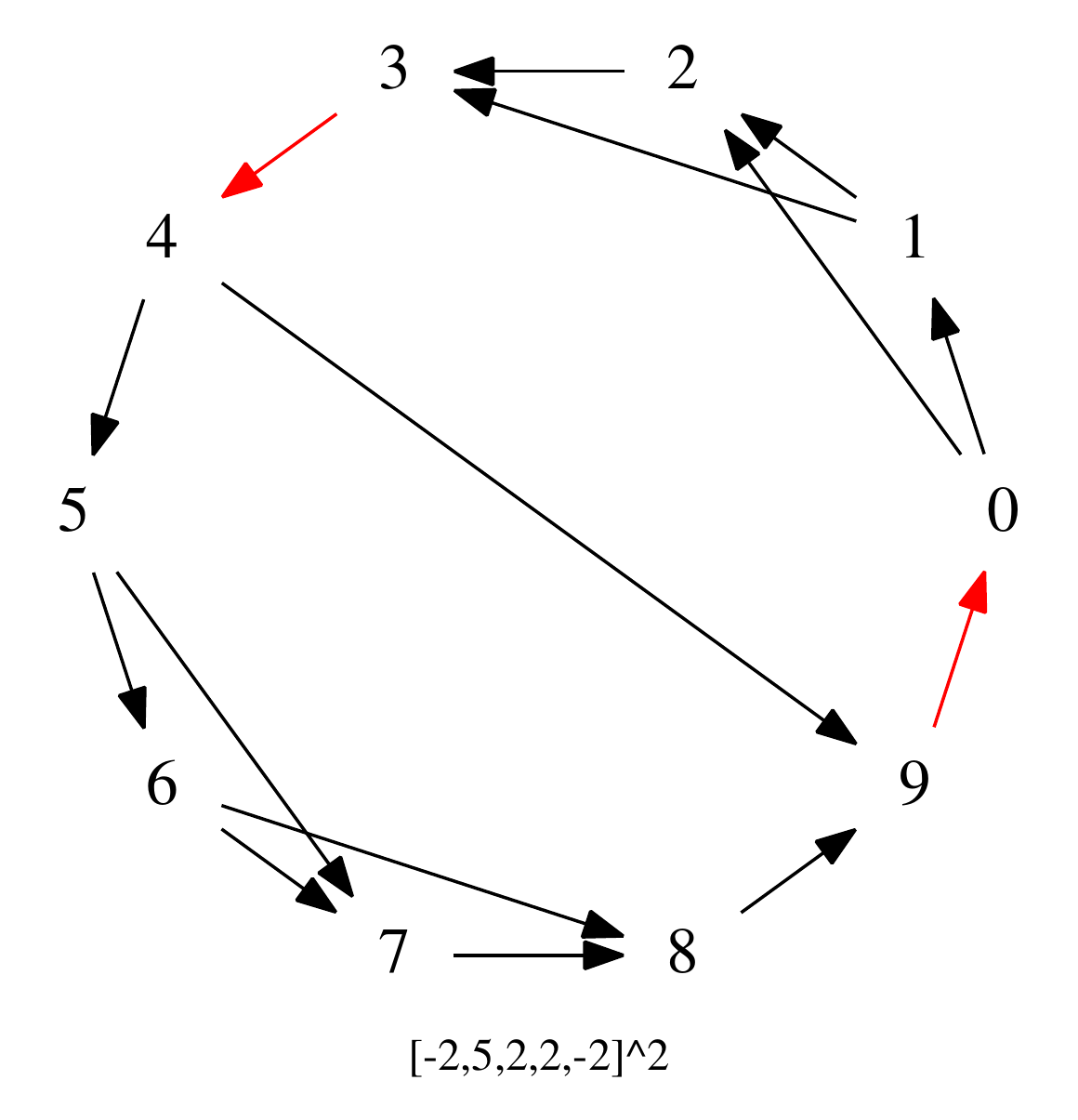}
\caption{Graphs on $n=10$ vertices which disconnect on 1 or 2 edges.
}
\label{fig.10n1}
\end{figure}

\begin{figure}
\includegraphics[scale=0.45]{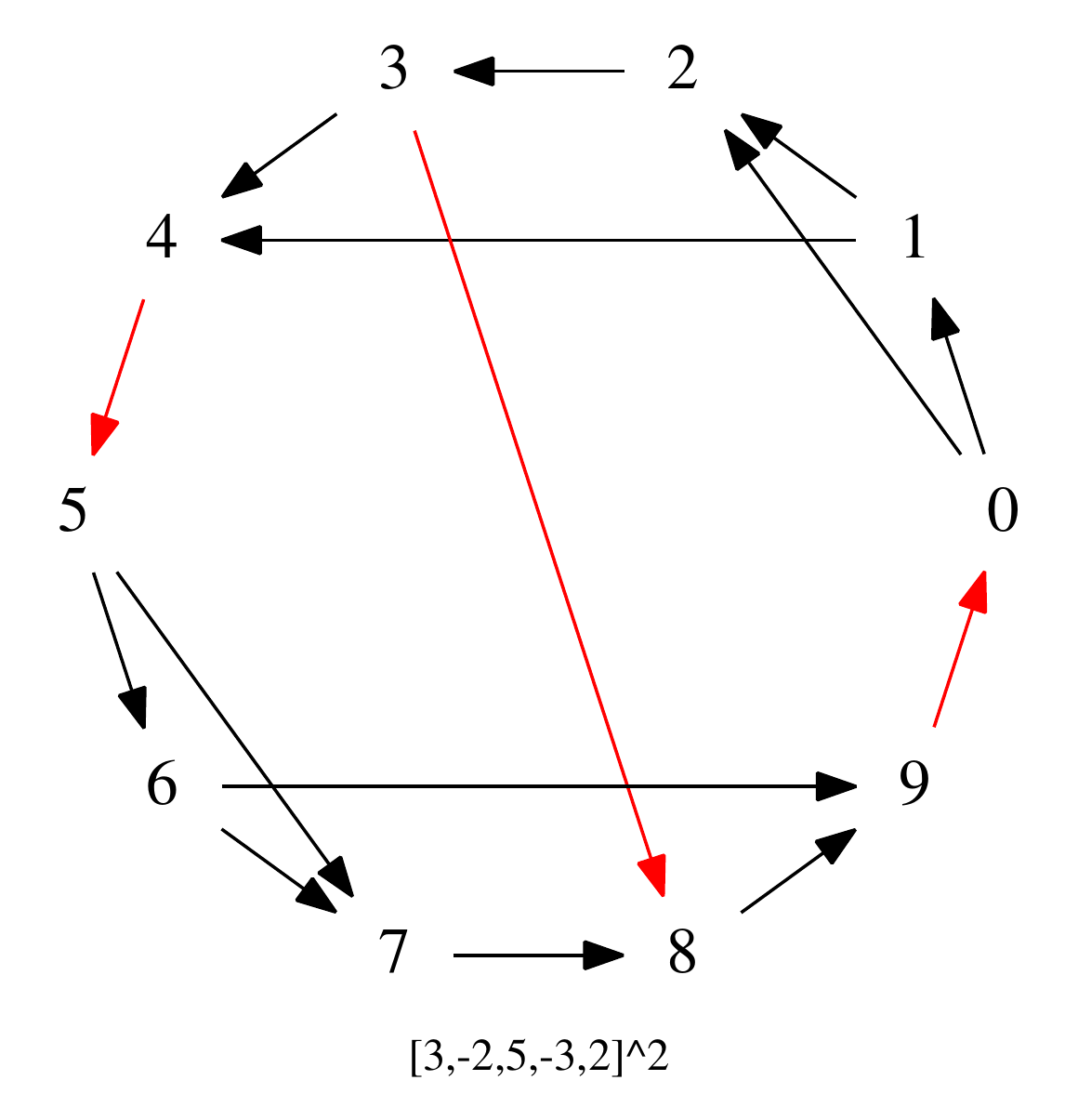}
\includegraphics[scale=0.45]{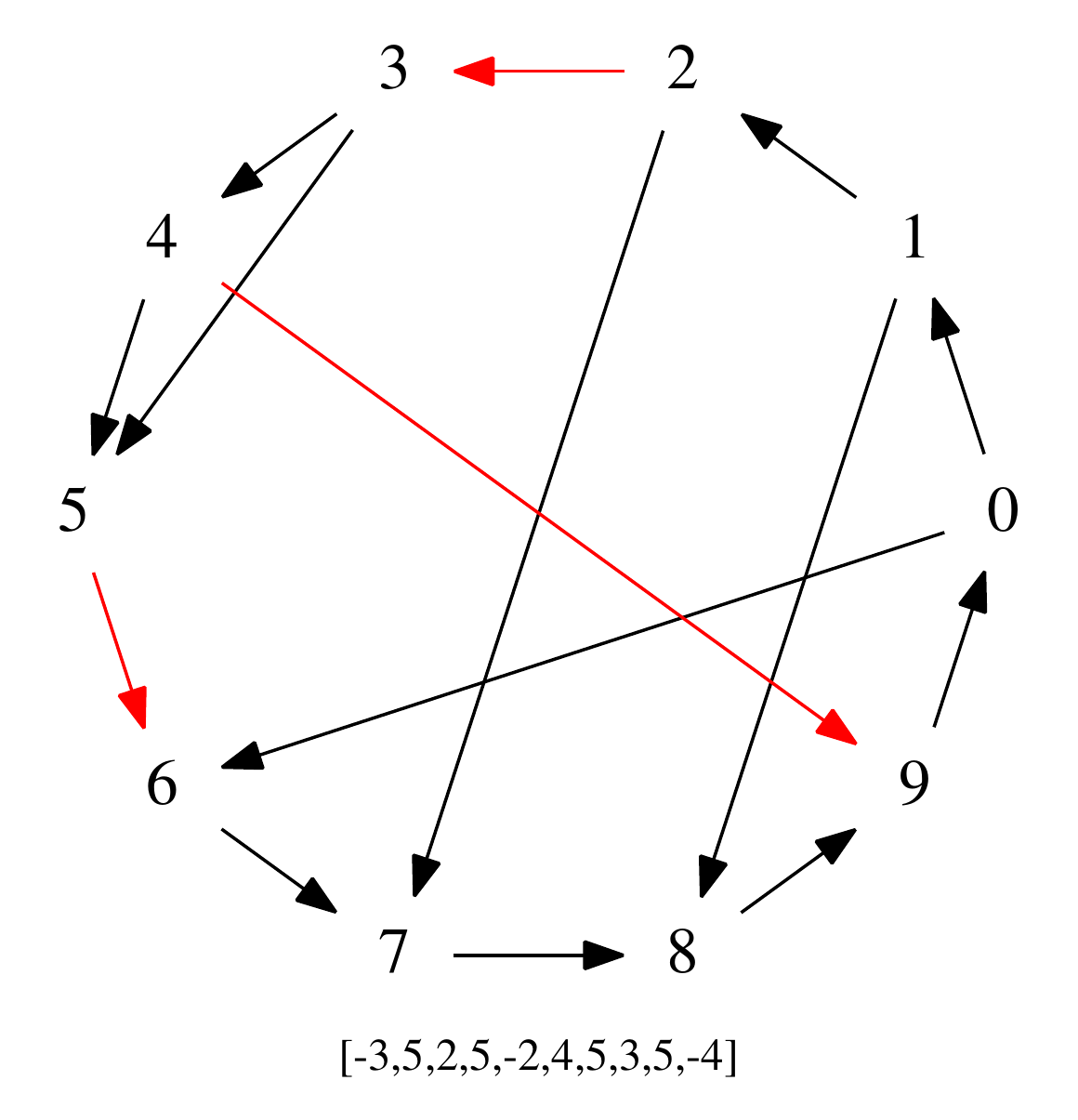}
\includegraphics[scale=0.45]{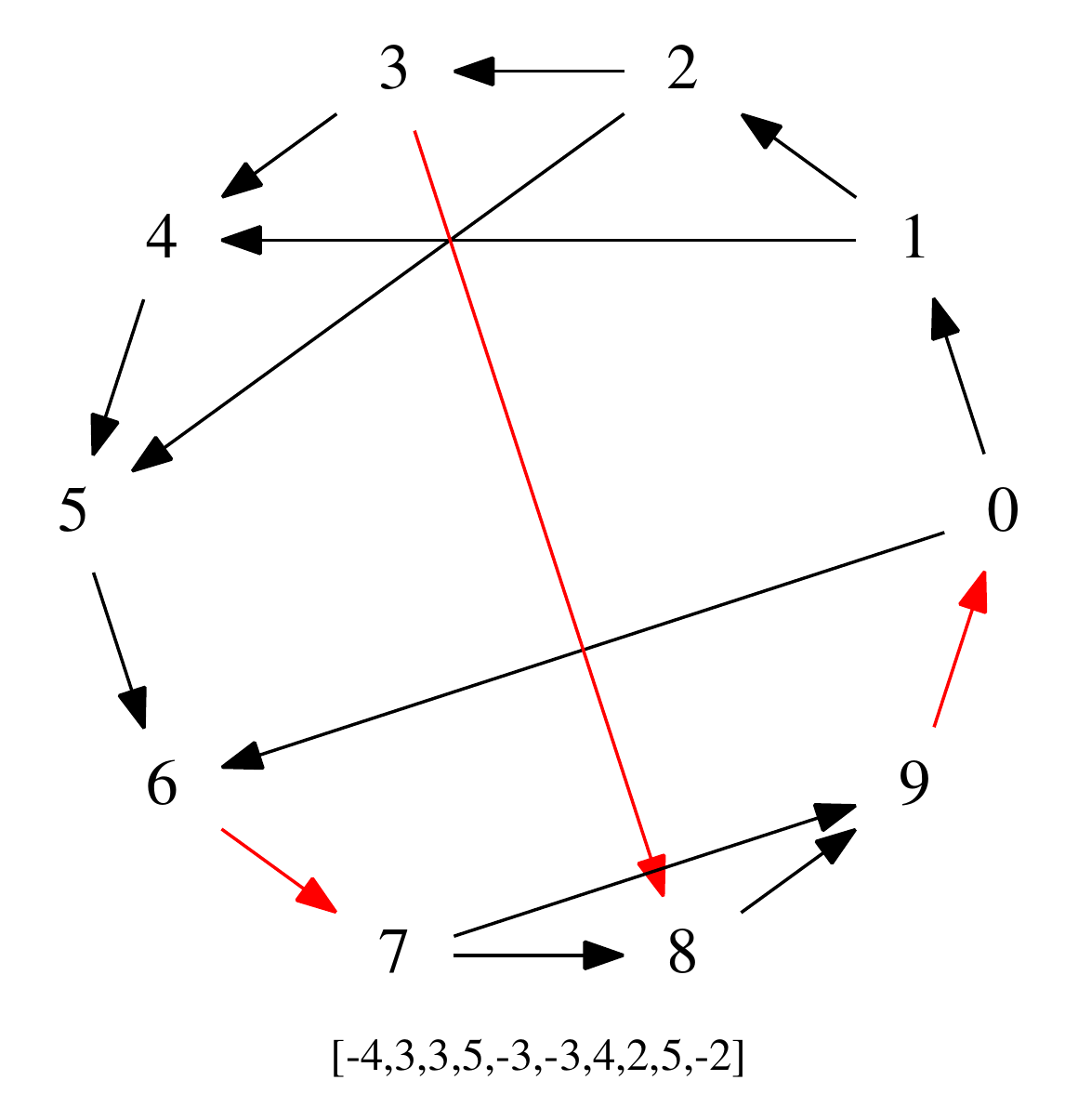}
\includegraphics[scale=0.45]{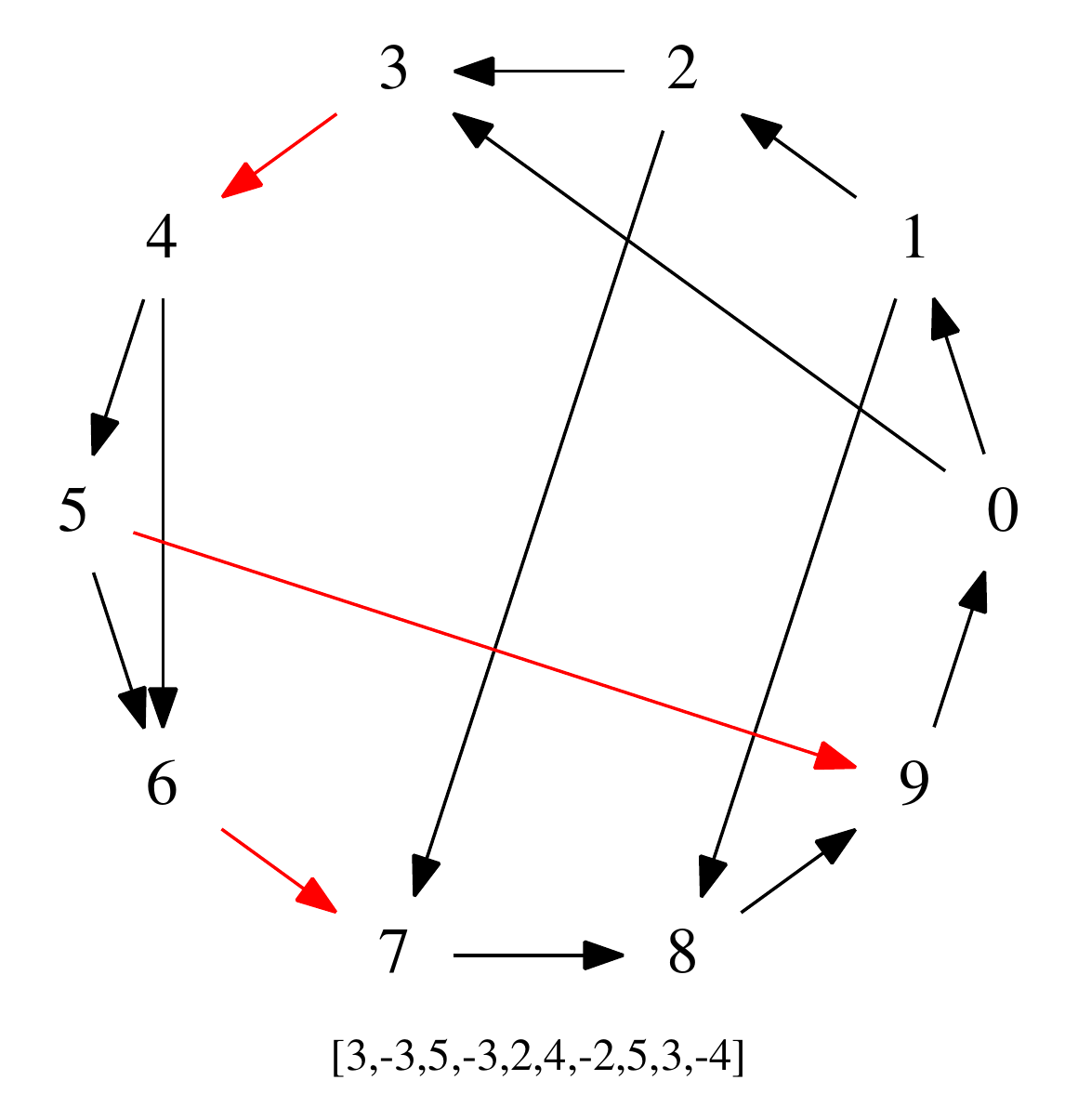}
\includegraphics[scale=0.45]{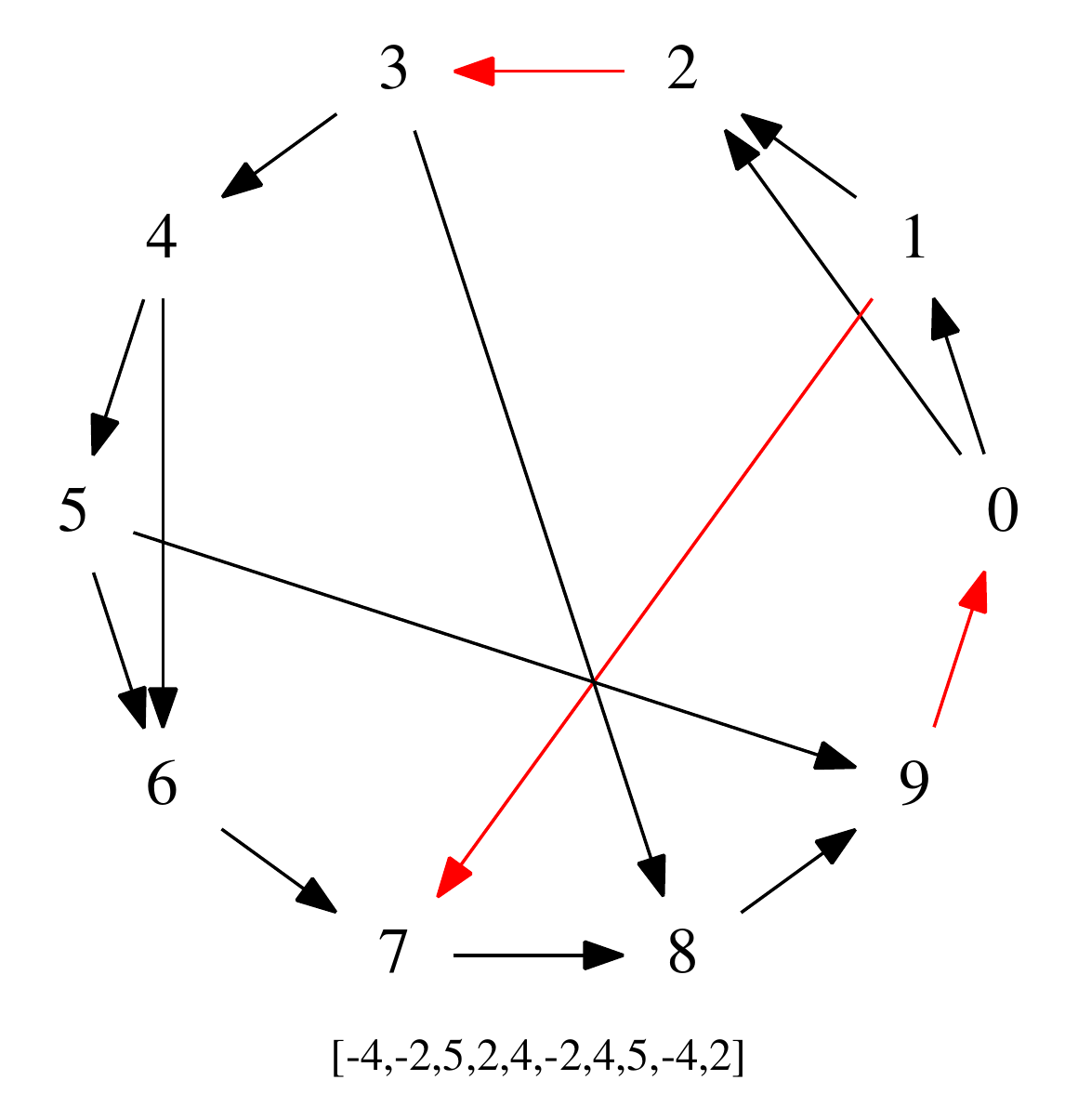}
\includegraphics[scale=0.45]{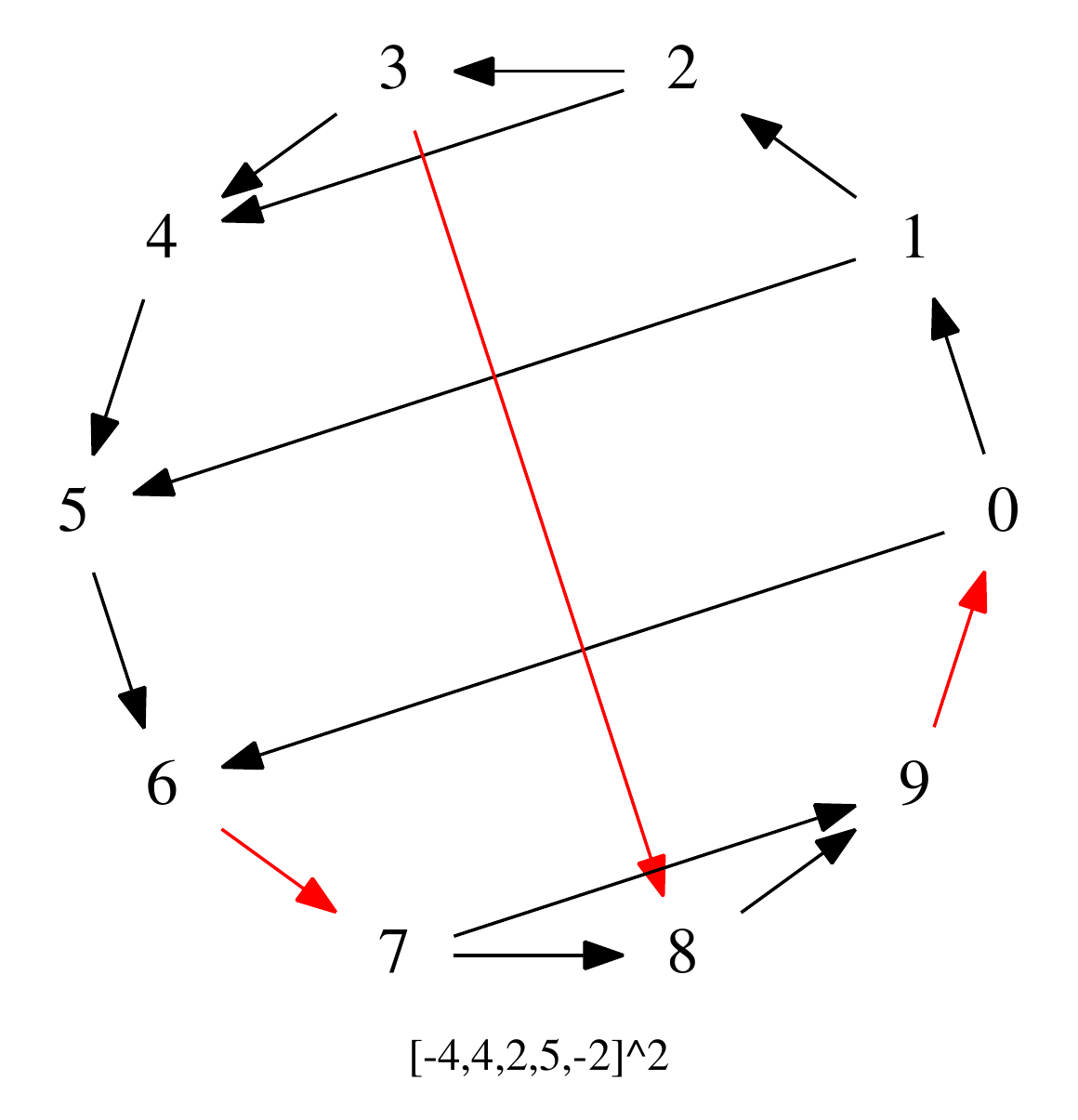}
\includegraphics[scale=0.45]{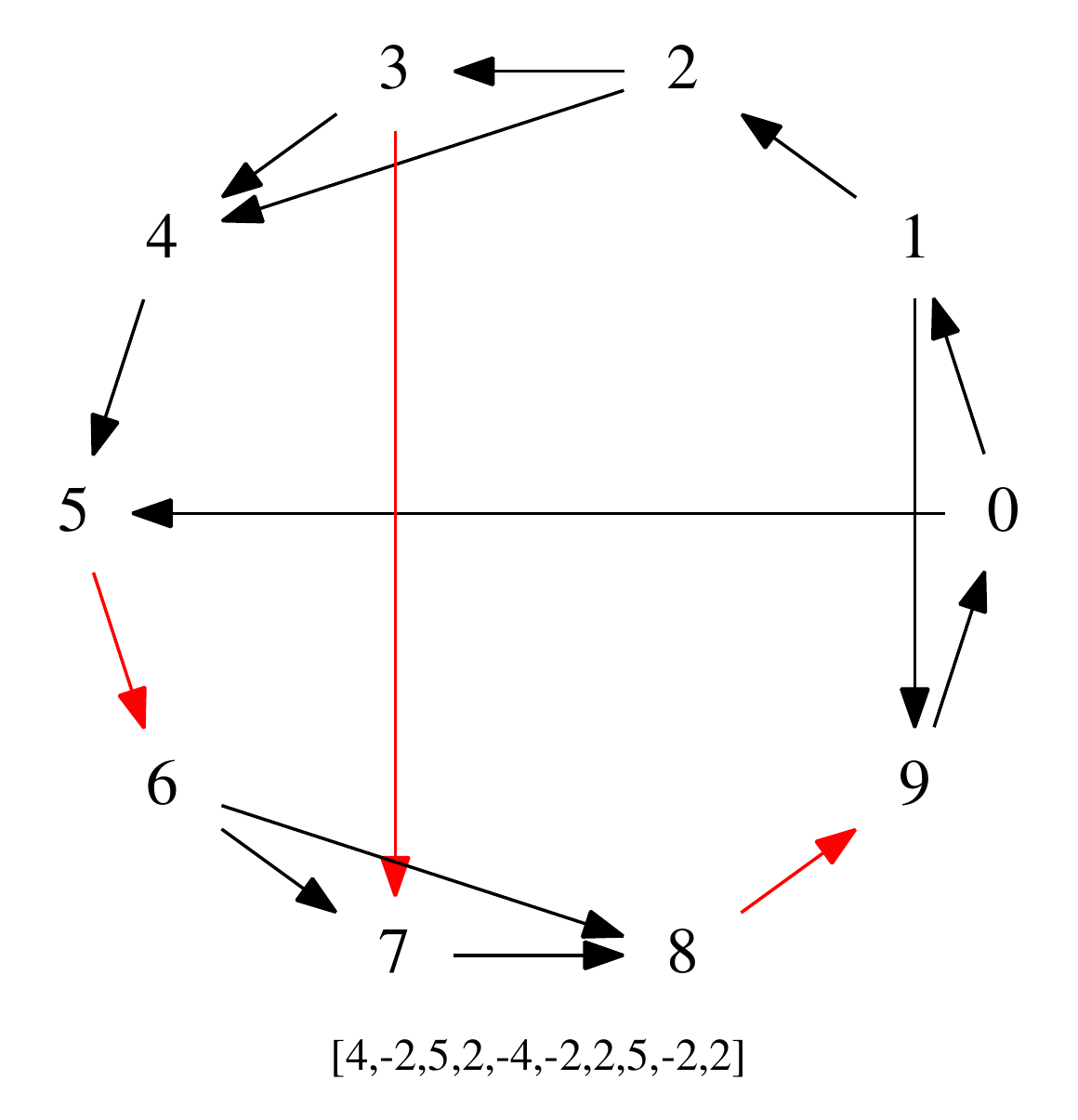}
\includegraphics[scale=0.45]{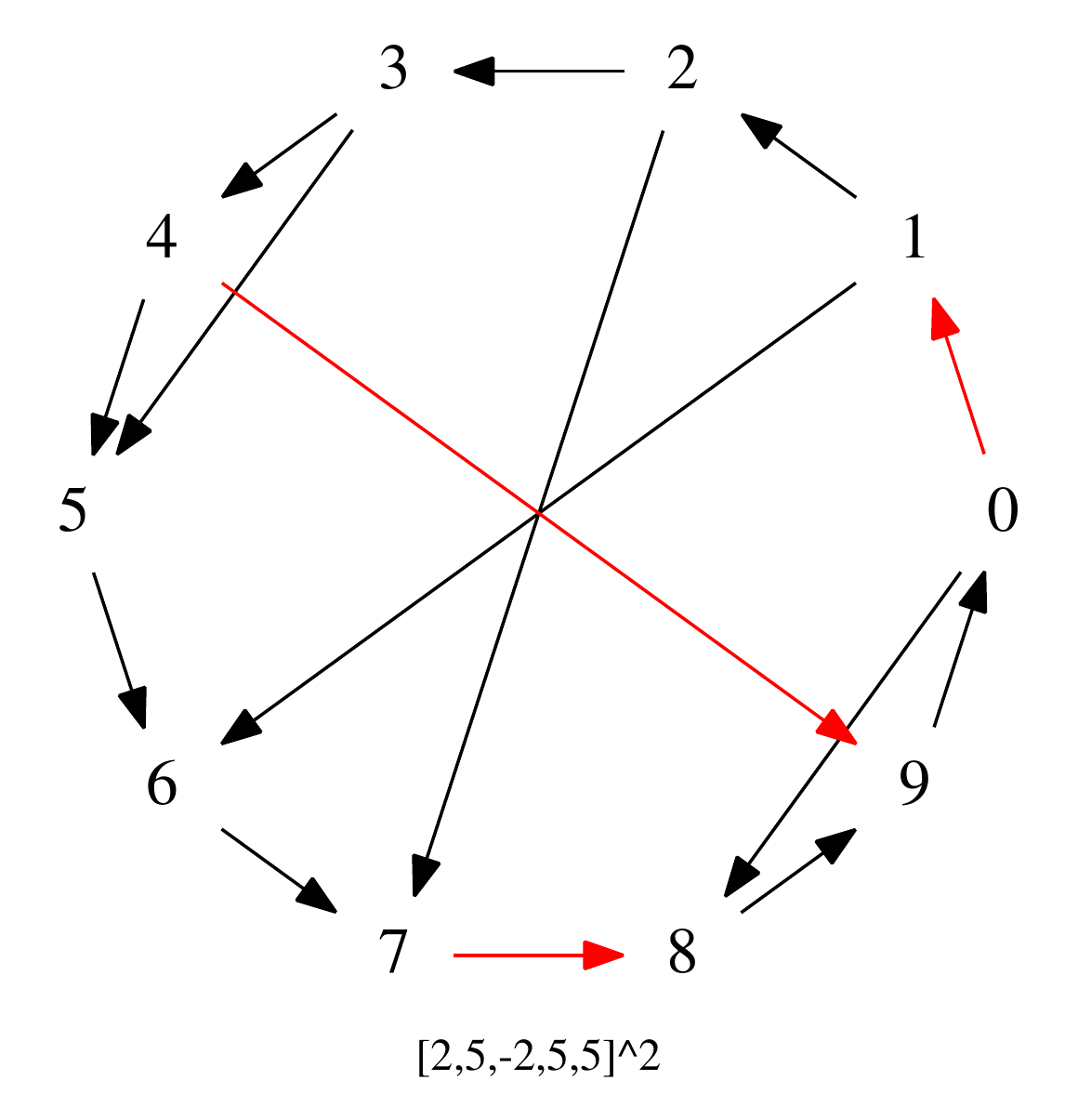}
\includegraphics[scale=0.45]{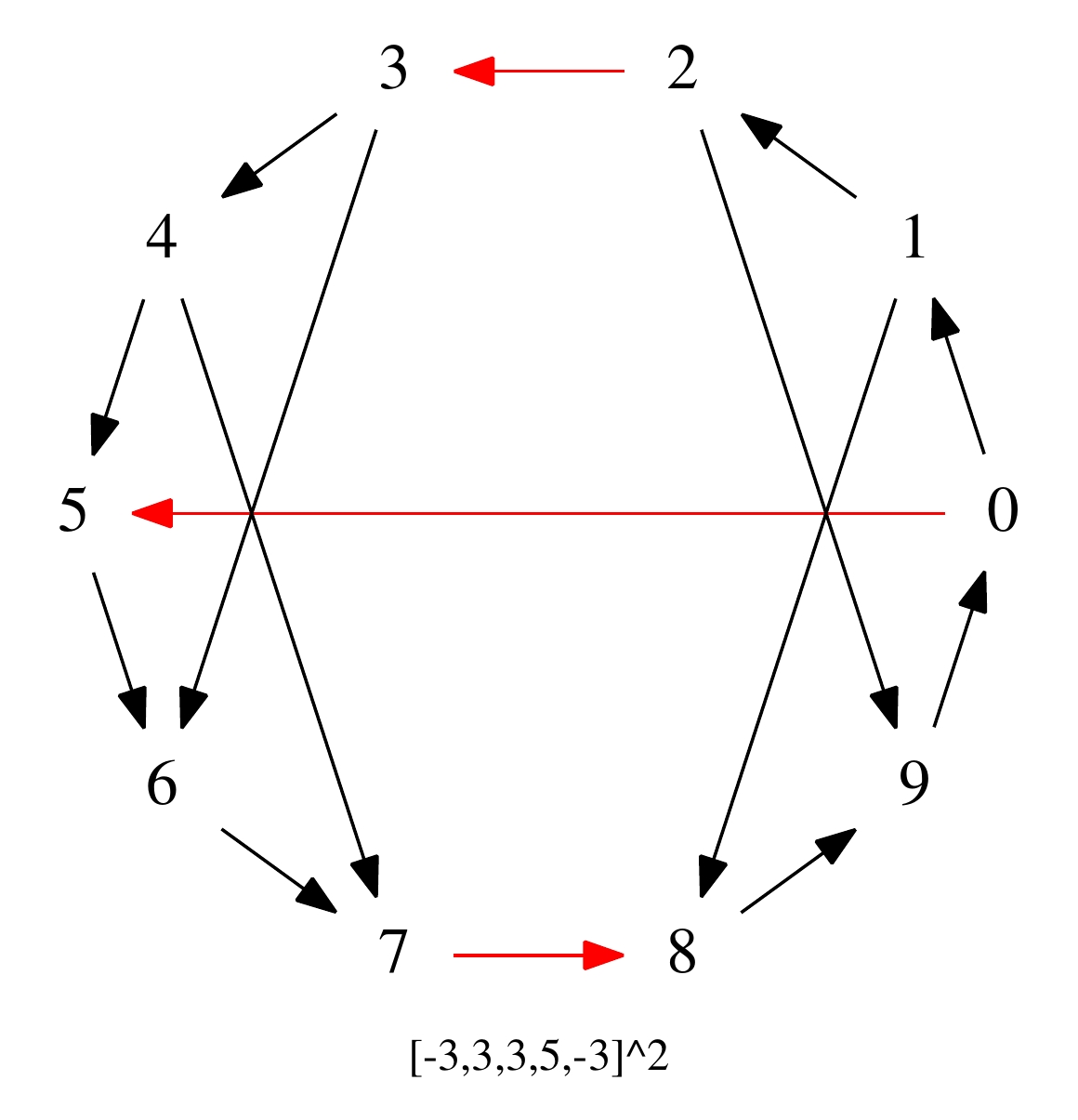}
\caption{The $3$-connected reducible graphs on $n=10$ vertices.
}
\label{fig.10n3}
\end{figure}

\begin{figure}
\includegraphics[scale=0.45]{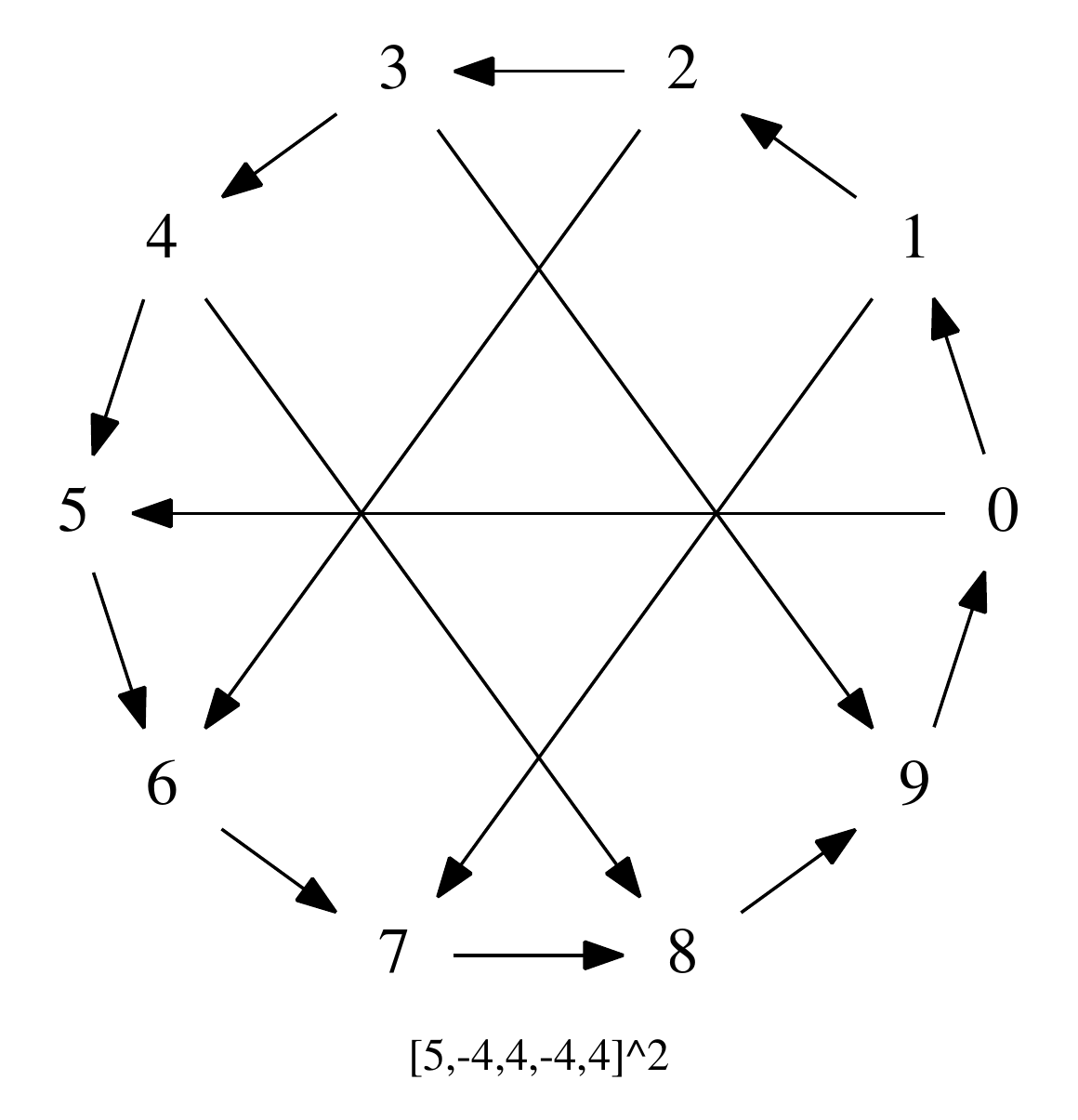}
\includegraphics[scale=0.45]{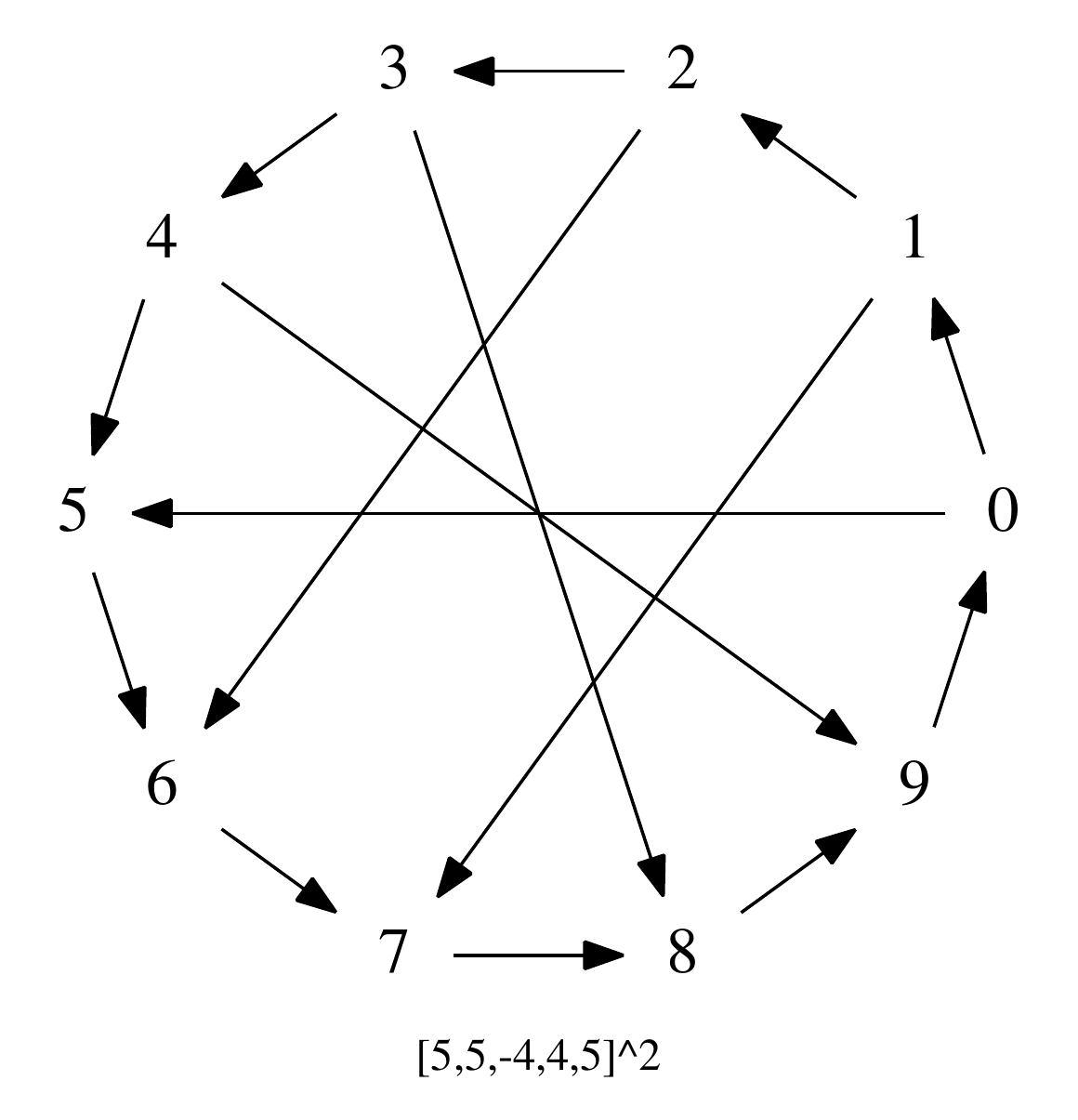}
\includegraphics[scale=0.45]{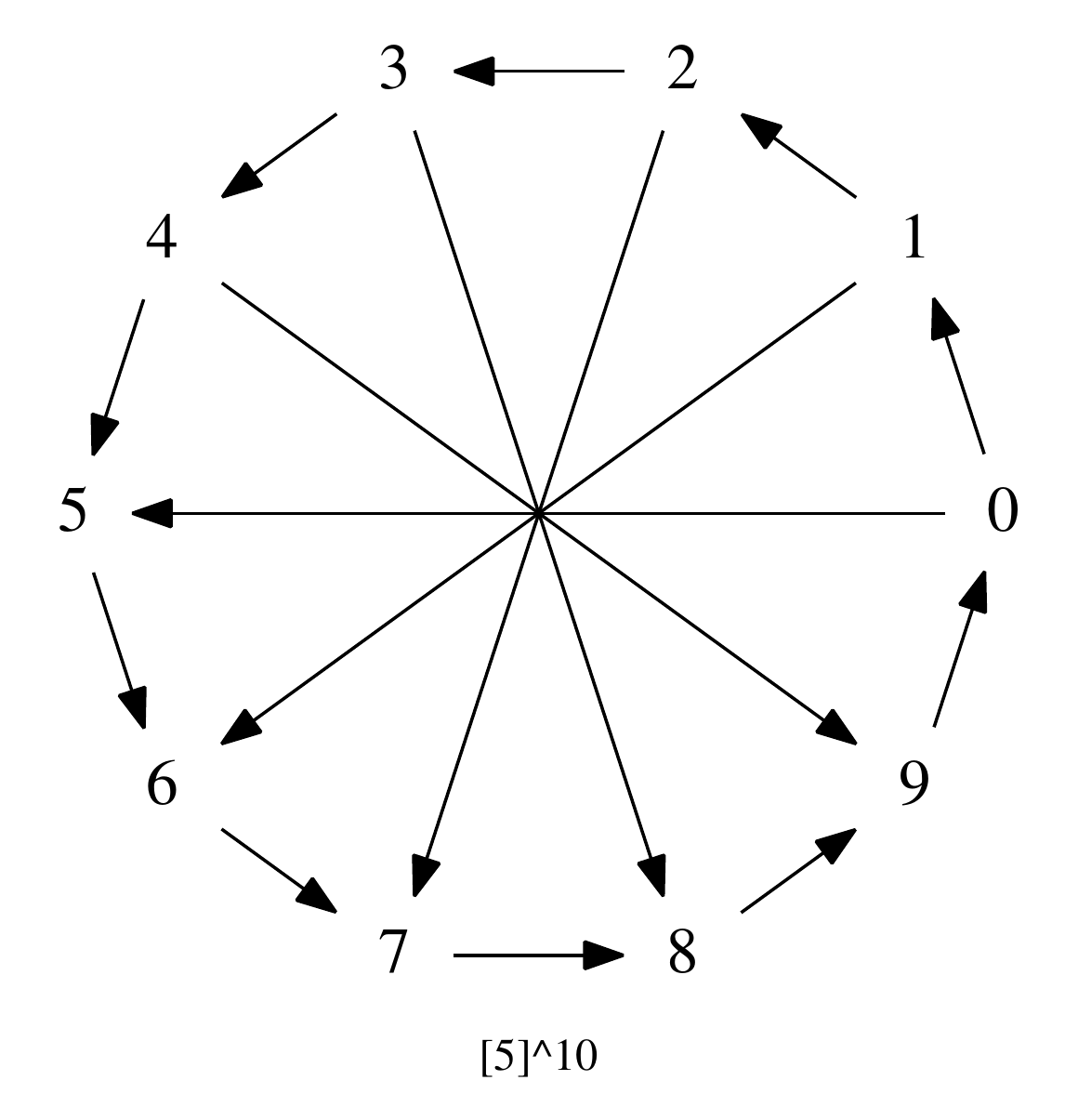}
\includegraphics[scale=0.45]{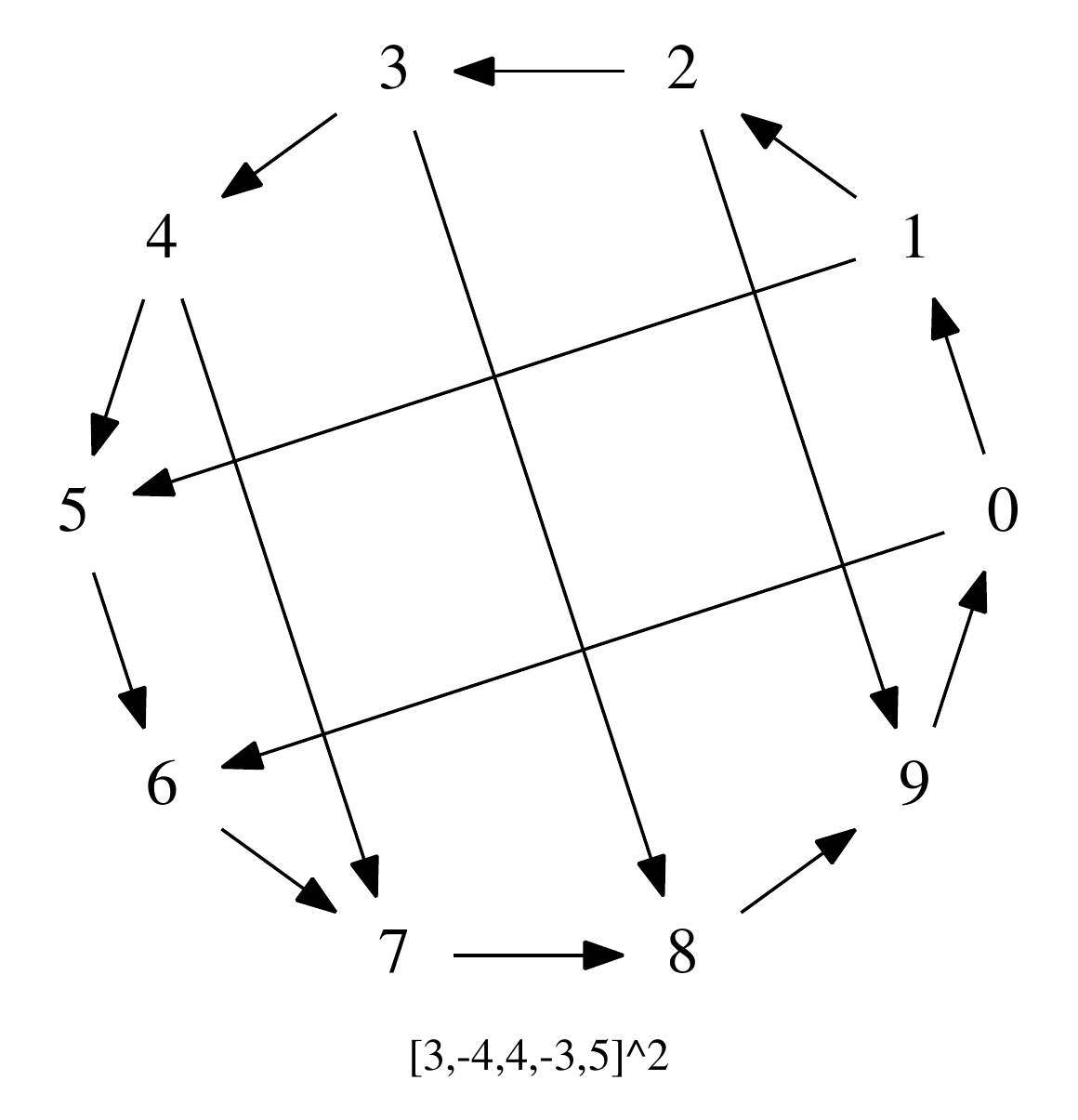}
\includegraphics[scale=0.45]{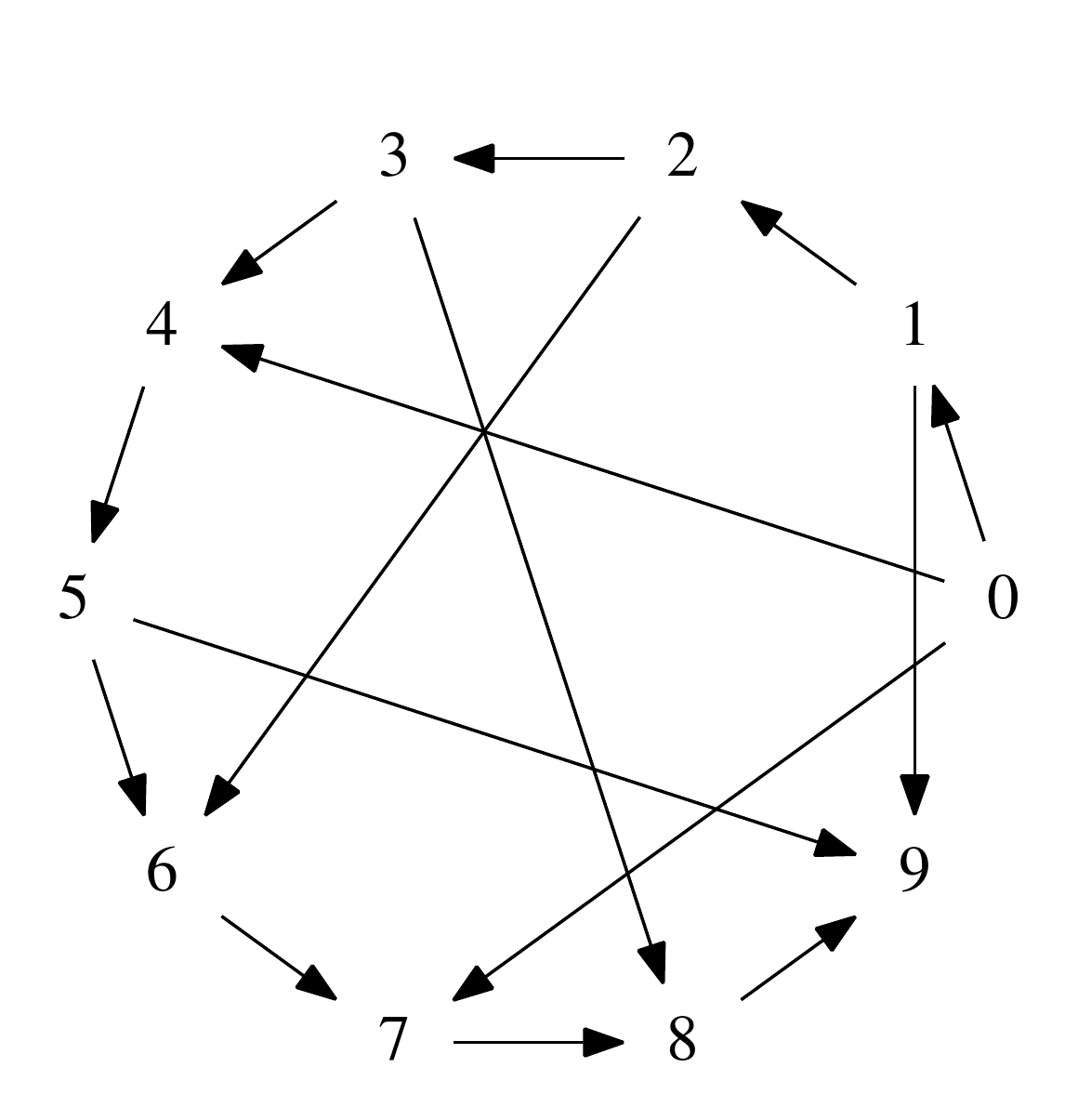}
\caption{Graphs on $n=10$ vertices which define the $15j$-symbols.
Four are cyclically 4-connected, one is cyclically 5-connected.
The one without a LCF name
is the Petersen graph.}
\label{fig.10n4}
\end{figure}
\clearpage

\section{$12$ vertices} 
The 85 graphs with 12 vertices (18 edges)
are 1-connected (Figure \ref{fig.12n1}),
2-connected (Figures \ref{fig.12n2s}--\ref{fig.12n2e})
3-connected reducible (Figures \ref{fig.12n3s}--\ref{fig.12n3e})
and cyclically 4- or 5-connected (Figures \ref{fig.12n4_s}--\ref{fig.12n4_e}).

The four graphs in Figure \ref{fig.12n1}
and one graph in Figure \ref{fig.12n3e} are not Hamiltonian, which
leaves us with 80 lines of LCF notations:

Figure \ref{fig.12n2s}:

\begin{Verbatim}
   LCF [3,-2,-4,-3,4,2]^2  = [4,2,3,-2,-4,-3;-] W150 d5 g3 EE45.12486   
   LCF [3,-2,-4,-3,3,3,3,-3,-3,-3,4,2]  W149 d4 g3 EE44.63116   
   LCF [-3,2,3,-2,2,-3,-2,4,2,3,-2,-4]  W149 d4 g3 EE46.12066   
   LCF [3,3,-3,-3,-3,3]^2  W152 d4 g4 EE44.14446   
   LCF [2,-3,-2,3,3,3;-]  W152 d4 g3 EE45.63732   
   LCF [3,-2,2,-3,-2,2]^2  W152 d4 g3 EE47.13249   
   LCF [-2,3,6,3,-3,2,-3,-2,6,2,2,-2]  = [4,2,-4,-2,-4,6,2,2,-2,-2,4,6] W149 d4 
g3 EE45.89062   
   LCF [3,4,-3,-3,6,-4,2,2,-2,-2,6,3]  W146 d4 g3 EE44.94265   
   LCF [3,-2,-4,-3,5,2,2,-2,-2,-5,4,2]  W154 d5 g3 EE46.30261   
   LCF [-3,-3,-3,5,2,2;-]  W153 d4 g3 EE45.76519   
   LCF [2,-3,-2,5,2,2;-]  W153 d4 g3 EE47.22986   
   LCF [2,4,-2,3,-5,-4,-3,2,2,-2,-2,5]  = [5,2,-4,-2,-5,-5,2,2,-2,-2,4,5] W143 
d4 g3 EE45.58501   
\end{Verbatim}

Figure \ref{fig.12n2e}:

\begin{Verbatim}
   LCF [-2,-2,4,4,4,4;-]  = [3,-4,-4,-3,2,2;-] = 
[5,3,4,4,-3,-5,-4,-4,2,2,-2,-2] W145 d4 g3 EE44.90052   
   LCF [4,-2,4,2,-4,-2,-4,2,2,-2,-2,2]  = [5,-2,2,3,-2,-5,-3,2,2,-2,-2,2] W148 
d4 g3 EE46.95537   
   LCF [2,2,-2,-2,-5,5]^2  W160 d5 g3 EE47.72073   
   LCF [-2,-2,4,5,3,4;-]  W141 d4 g3 EE44.63910   
   LCF [5,2,-3,-2,6,-5,2,2,-2,-2,6,3]  W146 d4 g3 EE45.63214   
   LCF [4,-2,3,3,-4,-3,-3,2,2,-2,-2,2]  W150 d4 g3 EE46.28096   
   LCF [-2,-2,5,3,5,3;-]  = [-2,-2,3,5,3,-3;-] W147 d4 g3 EE45.05416   
   LCF [2,2,-2,-2,6,6]^2  W158 d5 g3 EE47.35563   
   LCF [-3,2,-3,-2,2,2;-]  W152 d4 g3 EE47.39504   
   LCF [-2,-2,5,2,5,-2;-]  W143 d4 g3 EE46.51523   
   LCF [6,-2,2,2,-2,-2,6,2,2,-2,-2,2]  W153 d4 g3 EE48.40271   
   LCF [-2,2,2,-2]^3  W162 d4 g3 EE50.42874   
\end{Verbatim}

Figure \ref{fig.12n3s}:

\begin{Verbatim}
   LCF [2,3,-2,3,-3,3;-]  = [-4,6,4,2,6,-2]^2 W144 d4 g3 EE44.66589   
   LCF [-4,6,3,3,6,-3,-3,6,4,2,6,-2]  = [-2,3,-3,4,-3,3,3,-4,-3,-3,2,3] W140 d4 
g3 EE43.61888   
   LCF [-5,2,-3,-2,6,4,2,5,-2,-4,6,3]  = [-2,3,-3,4,-3,4,2,-4,-2,-4,2,3] = 
[3,-2,3,-3,5,-3,2,3,-2,-5,-3,2] W142 d4 g3 EE44.32053   
   LCF [-5,-5,4,2,6,-2,-4,5,5,2,6,-2]  = [4,-2,3,4,-4,-3,3,-4,2,-3,-2,2] W136 
d3 g3 EE44.01162   
   LCF [-5,-5,3,3,6,-3,-3,5,5,2,6,-2]  = [2,4,-2,3,5,-4,-3,3,3,-5,-3,-3] W136 
d3 g3 EE43.11500   
   LCF [2,4,-2,3,6,-4,-3,2,3,-2,6,-3]  = [2,4,-2,3,5,-4,-3,4,2,-5,-2,-4] = 
[-5,2,-3,-2,5,5,2,5,-2,-5,-5,3] W138 d4 g3 EE43.87324   
   LCF [-5,2,-3,-2,6,3,3,5,-3,-3,6,3]  = [4,-2,-4,4,-4,3,3,-4,-3,-3,4,2] = 
[-3,3,3,4,-3,-3,5,-4,2,3,-2,-5] W139 d4 g3 EE43.30141   
   LCF [2,3,-2,4,-3,6,3,-4,2,-3,-2,6]  = [-4,5,-4,2,3,-2,-5,-3,4,2,4,-2] W139 
d4 g3 EE44.05952   
   LCF [6,3,-4,-4,-3,3,6,2,-3,-2,4,4]  = [-5,-4,4,2,6,-2,-4,5,3,4,6,-3] = 
[3,4,4,-3,4,-4,-4,3,-4,2,-3,-2] = [4,5,-4,-4,-4,3,-5,2,-3,-2,4,4] = 
[4,5,-3,-5,-4,3,-5,2,-3,-2,5,3] W136 d4 g3 EE42.91096   
   LCF [4,6,-4,-4,-4,3,3,6,-3,-3,4,4]  = [-5,-4,3,3,6,-3,-3,5,3,4,6,-3] = 
[4,-3,5,-4,-4,3,3,-5,-3,-3,3,4] W135 d3 g4 EE42.08576   
   LCF [3,3,4,-3,-3,4;-]  = [3,6,-3,-3,6,3]^2 W136 d3 g4 EE42.58760   
   LCF [4,-2,5,2,-4,-2,3,-5,2,-3,-2,2]  = [5,-2,2,4,-2,-5,3,-4,2,-3,-2,2] = 
[2,-5,-2,-4,2,5,-2,2,5,-2,-5,4] W139 d4 g3 EE44.95991   
\end{Verbatim}

Figure \ref{fig.12n32}:

\begin{Verbatim}
   LCF [-2,6,2,-4,-2,3,3,6,-3,-3,2,4]  = [-2,2,5,-2,-5,3,3,-5,-3,-3,2,5] W139 
d4 g3 EE44.12975   
   LCF [2,4,-2,6,2,-4,-2,4,2,6,-2,-4]  = [2,5,-2,2,6,-2,-5,2,3,-2,6,-3] W139 d4 
g3 EE44.87532   
   LCF [6,3,-3,-5,-3,3,6,2,-3,-2,5,3]  = [3,5,3,-3,4,-3,-5,3,-4,2,-3,-2] = 
[-5,-3,4,2,5,-2,-4,5,3,-5,3,-3] W140 d4 g3 EE43.12097   
   LCF [3,-3,5,-3,-5,3,3,-5,-3,-3,3,5]  W142 d4 g4 EE42.31141   
   LCF [4,2,4,-2,-4,4;-]  = [3,5,2,-3,-2,5;-] = [6,2,-3,-2,6,3]^2 W141 d4 g3 
EE44.00528   
   LCF [3,6,4,-3,6,3,-4,6,-3,2,6,-2]  = [4,-4,5,3,-4,6,-3,-5,2,4,-2,6] = 
[-5,5,3,-5,4,-3,-5,5,-4,2,5,-2] W137 d4 g3 EE42.72638   
   LCF [6,-5,2,6,-2,6,6,3,5,6,-3,6]  = [6,2,-5,-2,4,6,6,3,-4,5,-3,6] = 
[5,5,6,4,6,-5,-5,-4,6,2,6,-2] = [-4,4,-3,3,6,-4,-3,2,4,-2,6,3] = 
[6,2,-4,-2,4,4,6,4,-4,-4,4,-4] = [-3,2,5,-2,-5,3,4,-5,-3,3,-4,5] = 
[-5,2,-4,-2,4,4,5,5,-4,-4,4,-5] W133 d3 g3 EE42.37675   
   LCF [2,6,-2,5,6,4,5,6,-5,-4,6,-5]  = [5,6,-4,-4,5,-5,2,6,-2,-5,4,4] = 
[2,4,-2,-5,4,-4,3,4,-4,-3,5,-4] = [2,-5,-2,4,-5,4,4,-4,5,-4,-4,5] W131 d3 g3 
EE42.19745   
   LCF [2,4,-2,-5,5,-4]^2  = [-5,2,4,-2,6,3,-4,5,-3,2,6,-2] W135 d4 g3 
EE43.48153   
   LCF [-4,-4,4,2,6,-2,-4,4,4,4,6,-4]  = [-4,-3,4,2,5,-2,-4,4,4,-5,3,-4] = 
[-3,5,3,4,-5,-3,-5,-4,2,3,-2,5] W137 d4 g3 EE42.85630   
   LCF [2,5,-2,4,4,5;-]  = [2,4,-2,4,4,-4;-] = [-5,5,6,2,6,-2]^2 = 
[5,-2,4,6,3,-5,-4,-3,2,6,-2,2] W134 d3 g3 EE43.48061   
   LCF [3,6,-4,-3,5,6,2,6,-2,-5,4,6]  = [2,-5,-2,4,5,6,4,-4,5,-5,-4,6] = 
[5,-4,4,-4,3,-5,-4,-3,2,4,-2,4] W131 d3 g3 EE42.11275   
\end{Verbatim}

Figure \ref{fig.12n33}:

\begin{Verbatim}
   LCF [6,-5,2,4,-2,5,6,-4,5,2,-5,-2]  = [-2,4,5,6,-5,-4,2,-5,-2,6,2,5] = 
[5,-2,4,-5,4,-5,-4,2,-4,-2,5,2] W133 d4 g3 EE43.16541   
   LCF [2,-5,-2,6,3,6,4,-3,5,6,-4,6]  = [6,3,-3,4,-3,4,6,-4,2,-4,-2,3] = 
[5,-4,6,-4,2,-5,-2,3,6,4,-3,4] = [5,-3,5,6,2,-5,-2,-5,3,6,3,-3] = 
[-5,2,-5,-2,6,3,5,5,-3,5,6,-5] = [-3,4,5,-5,-5,-4,2,-5,-2,3,5,5] = 
[5,5,5,-5,4,-5,-5,-5,-4,2,5,-2] W134 d4 g3 EE42.32276   
   LCF [5,-3,6,3,-5,-5,-3,2,6,-2,3,5]  = [2,6,-2,-5,5,3,5,6,-3,-5,5,-5] = 
[5,5,5,6,-5,-5,-5,-5,2,6,-2,5] = [4,-3,5,2,-4,-2,3,-5,3,-3,3,-3] = 
[5,5,-3,-5,4,-5,-5,2,-4,-2,5,3] W135 d3 g3 EE42.67156   
   LCF [2,4,-2,5,3,-4;-]  = [5,-3,2,5,-2,-5;-] = [3,6,3,-3,6,-3,2,6,-2,2,6,-2] 
W138 d4 g3 EE43.74286   
   LCF [6,2,-4,-2,-5,3,6,2,-3,-2,4,5]  = [2,3,-2,4,-3,4,5,-4,2,-4,-2,-5] = 
[-5,2,-4,-2,-5,4,2,5,-2,-4,4,5] W136 d4 g3 EE43.61258   
   LCF [5,2,5,-2,5,-5;-]  = [6,2,-4,-2,4,6]^2 = [2,-5,-2,6,2,6,-2,3,5,6,-3,6] = 
[-5,-2,6,6,2,5,-2,5,6,6,-5,2] W134 d3 g3 EE43.34214   
   LCF [-3,4,5,-5,2,-4,-2,-5,3,3,5,-3]  W134 d3 g3 EE42.79794   
   LCF [6,-4,3,4,-5,-3,6,-4,2,4,-2,5]  = [-4,6,-4,2,5,-2,5,6,4,-5,4,-5] = 
[5,-5,4,-5,3,-5,-4,-3,5,2,5,-2] W131 d3 g3 EE42.05815   
   LCF [-5,2,4,-2,-5,4;-]  W135 d4 g3 EE43.25057   
   LCF [2,5,-2,5,3,5;-]  = [6,-2,6,6,6,2]^2 = [5,-2,6,6,2,-5,-2,3,6,6,-3,2] 
W136 d3 g3 EE43.60342   
   LCF [6,-2,6,4,6,4,6,-4,6,-4,6,2]  = [5,6,-3,3,5,-5,-3,6,2,-5,-2,3] W133 d3 
g3 EE42.23739   
   LCF [4,-2,4,6,-4,2,-4,-2,2,6,-2,2]  = [5,-2,5,6,2,-5,-2,-5,2,6,-2,2] W135 d3 
g3 EE44.43130   
\end{Verbatim}

Figure \ref{fig.12n3e}:

\begin{Verbatim}
   LCF [6,-2,2]^4  W138 d3 g3 EE45.76235   
   LCF [2,6,-2,6]^3  W135 d3 g3 EE44.26200   
\end{Verbatim}

Figure \ref{fig.12n4_s}:

\begin{Verbatim}
   LCF [-3,3]^6  = [3,-5,5,-3,-5,5]^2 W144 d4 g4 EE42.27027   
   LCF [6,-3,6,6,3,6]^2  = [6,6,-5,5,6,6]^2 = [3,-3,4,-3,3,4;-] = 
[5,-3,6,6,3,-5]^2 = [5,-3,-5,4,4,-5;-] = [6,6,-3,-5,4,4,6,6,-4,-4,5,3] W134 d3 
g4 EE41.69366   
   LCF [-4,4,4,6,6,-4]^2  = [6,-5,5,-5,5,6]^2 = [4,-3,3,5,-4,-3;-] = 
[-4,-4,4,4,-5,5]^2 W132 d3 g4 EE41.28733   
   LCF [-4,6,3,6,6,-3,5,6,4,6,6,-5]  = [-5,4,6,6,6,-4,5,5,6,6,6,-5] = 
[5,-3,4,6,3,-5,-4,-3,3,6,3,-3] = [4,-4,6,4,-4,5,5,-4,6,4,-5,-5] = 
[4,-5,-3,4,-4,5,3,-4,5,-3,-5,3] W132 d3 g4 EE41.34305   
   LCF [3,4,5,-3,5,-4;-]  = [3,6,-4,-3,4,6]^2 = [-4,5,5,-4,5,5;-] = 
[3,6,-4,-3,4,4,5,6,-4,-4,4,-5] = [4,-5,5,6,-4,5,5,-5,5,6,-5,-5] = 
[4,-4,5,-4,-4,3,4,-5,-3,4,-4,4] W130 d3 g4 EE41.02128   
   LCF [4,-4,6]^4  = [3,6,3,-3,6,-3]^2 = [-3,6,4,-4,6,3,-4,6,-3,3,6,4] W134 d3 
g4 EE41.66461   
   LCF [6,-5,5]^4  = [3,4,-4,-3,4,-4]^2 W130 d3 g4 EE41.16056   
   LCF [-3,5,-3,4,4,5;-]  = [4,-5,5,6,-4,6]^2 = [-3,4,-3,4,4,-4;-] = 
[5,6,-3,-5,4,-5,3,6,-4,-3,5,3] = [5,6,4,-5,5,-5,-4,6,3,-5,5,-3] W132 d3 g4 
EE41.28805   
   LCF [4,-3,4,5,-4,4;-]  = [4,5,-5,5,-4,5;-] = [-5,-3,4,5,-5,4;-] W128 d3 g4 
EE40.61559   
   LCF [6,-4,6,-4,3,5,6,-3,6,4,-5,4]  = [6,-4,3,-4,4,-3,6,3,-4,4,-3,4] = 
[5,6,-4,3,5,-5,-3,6,3,-5,4,-3] = [5,-5,4,6,-5,-5,-4,3,5,6,-3,5] = 
[5,5,-4,4,5,-5,-5,-4,3,-5,4,-3] W130 d3 g4 EE40.93704   
   LCF [6,-3,5,6,-5,3,6,-5,-3,6,3,5]  = [3,-4,5,-3,4,6,4,-5,-4,4,-4,6] W130 d3 
g4 EE40.99207   
   LCF [6,-4,5,-5,4,6,6,-5,-4,4,5,6]  W128 d3 g4 EE40.72559   
\end{Verbatim}

Figure \ref{fig.12n4_e}:

\begin{Verbatim}
   LCF [4,-5,4,-5,-4,4;-]  W126 d3 g5 EE40.34891   
   LCF [6,4,6,6,6,-4]^2  = [-3,4,-3,5,3,-4;-] = [-5,3,6,6,-3,5,5,5,6,6,-5,-5] = 
[-3,3,6,4,-3,5,5,-4,6,3,-5,-5] W134 d3 g4 EE41.55455   
   LCF [3,5,5,-3,5,5;-]  = [-3,5,-3,5,3,5;-] = [5,-3,5,5,5,-5;-] W136 d4 g4 
EE41.45861   
   LCF [-5,5]^6  = [5,-5,-3,3]^3 W132 d3 g4 EE41.05212   
   LCF [6]^12  = [6,6,-3,-5,5,3]^2 W138 d3 g4 EE42.25614   
   LCF [6,-5,-4,4,-5,4,6,-4,5,-4,4,5]  W126 d3 g5 EE40.40388   
\end{Verbatim}

\subsection{1-connected} 
\begin{figure}
\includegraphics[scale=0.45]{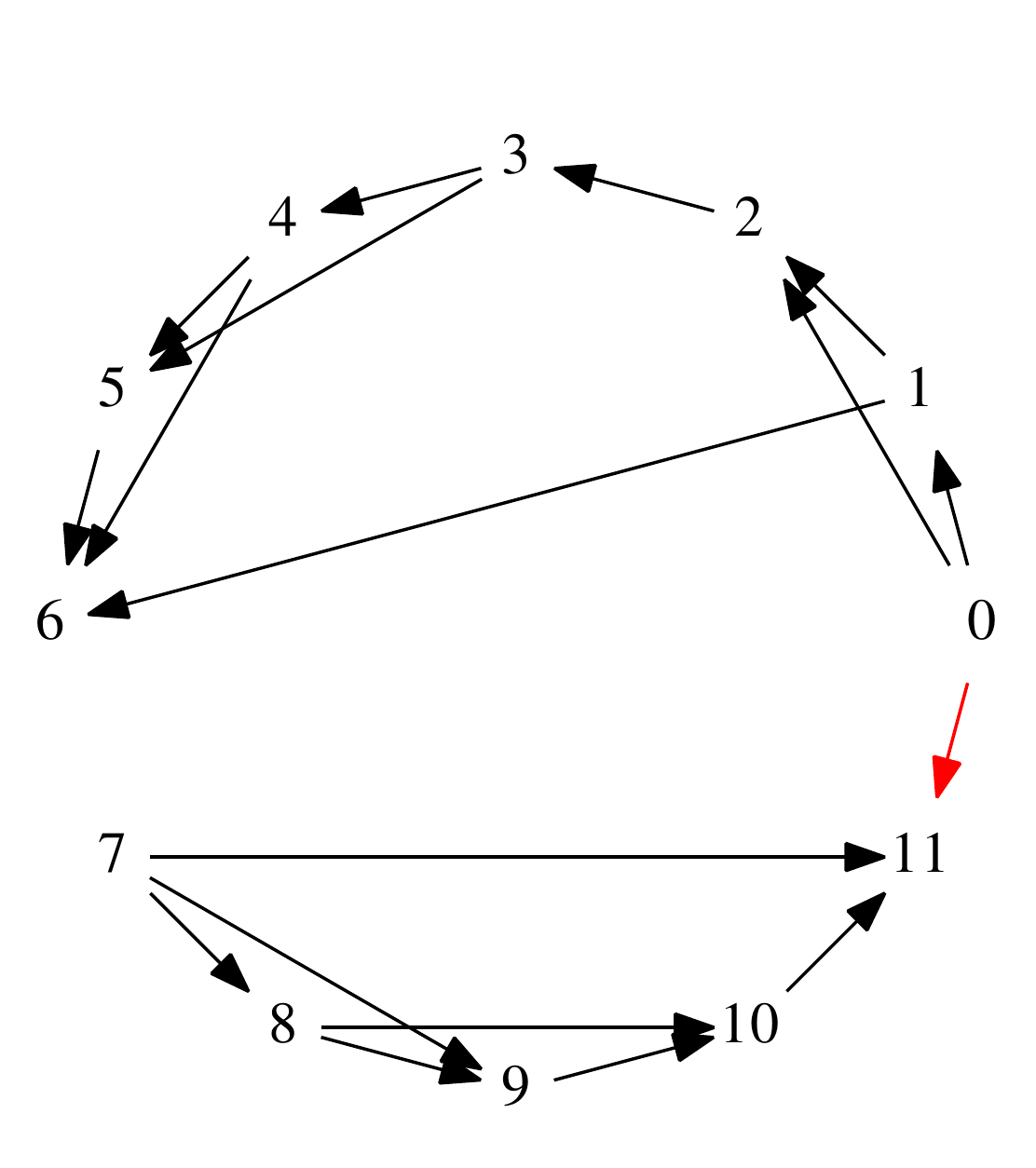}
\includegraphics[scale=0.45]{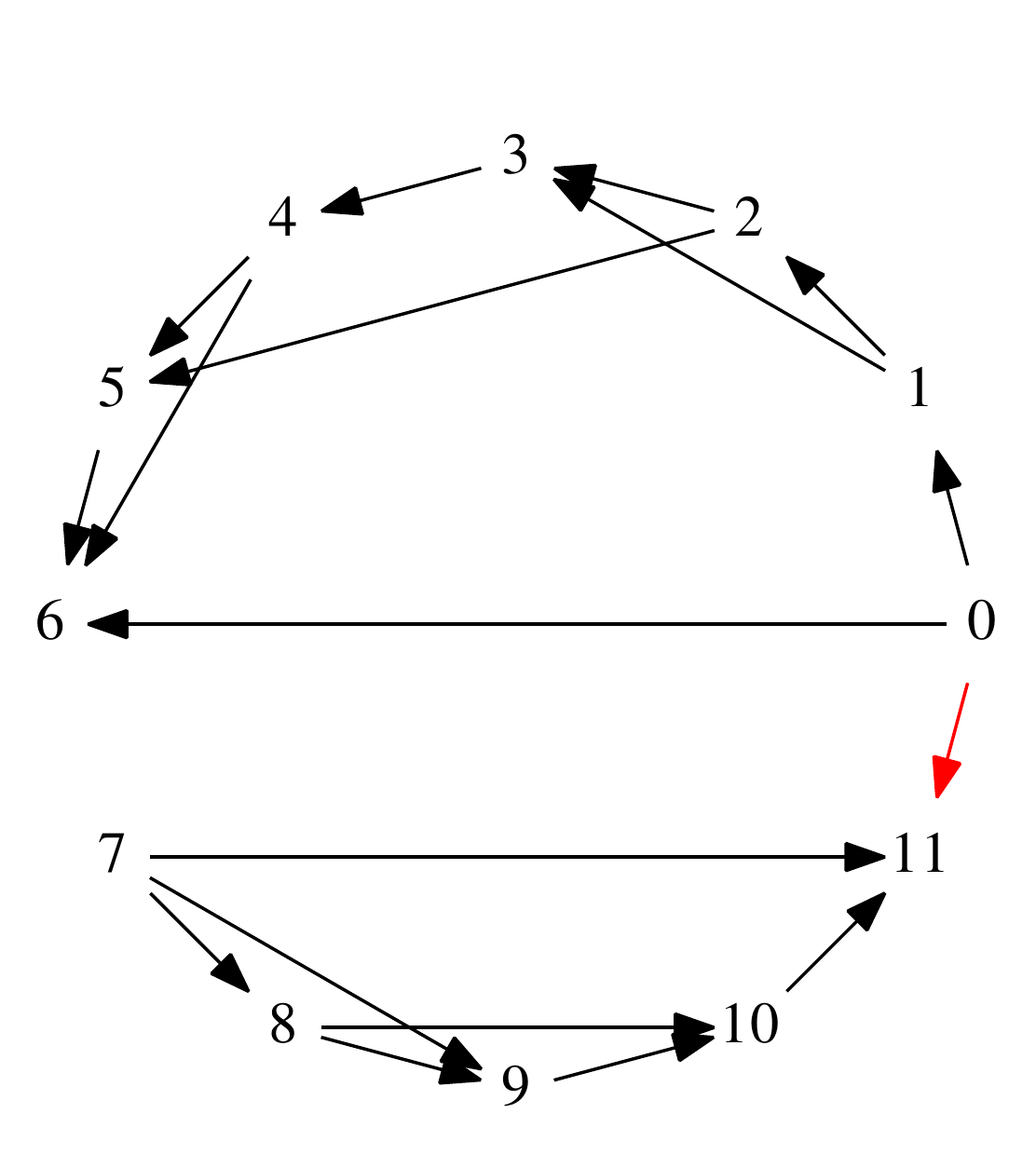}
\includegraphics[scale=0.45]{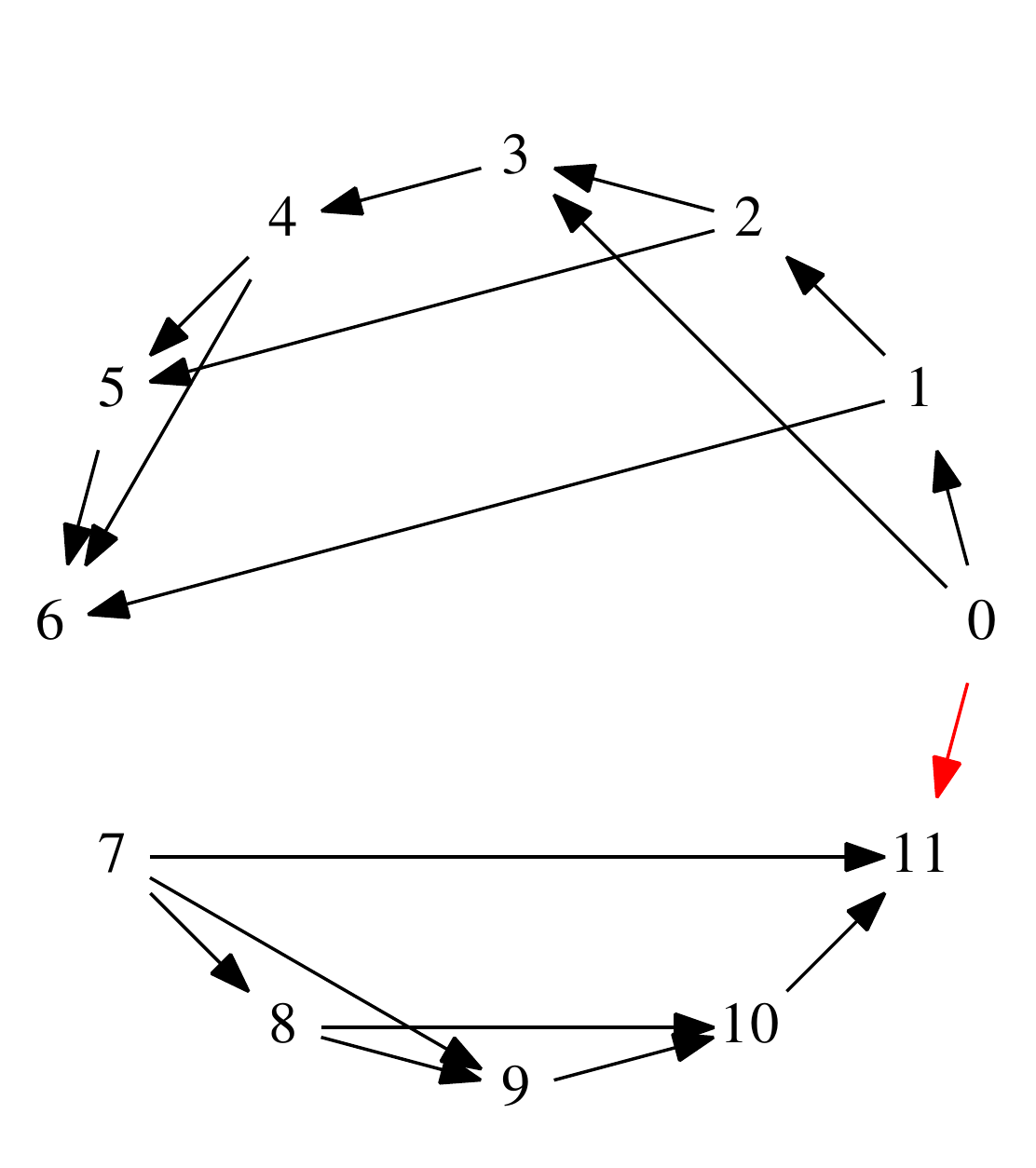}
\includegraphics[scale=0.45]{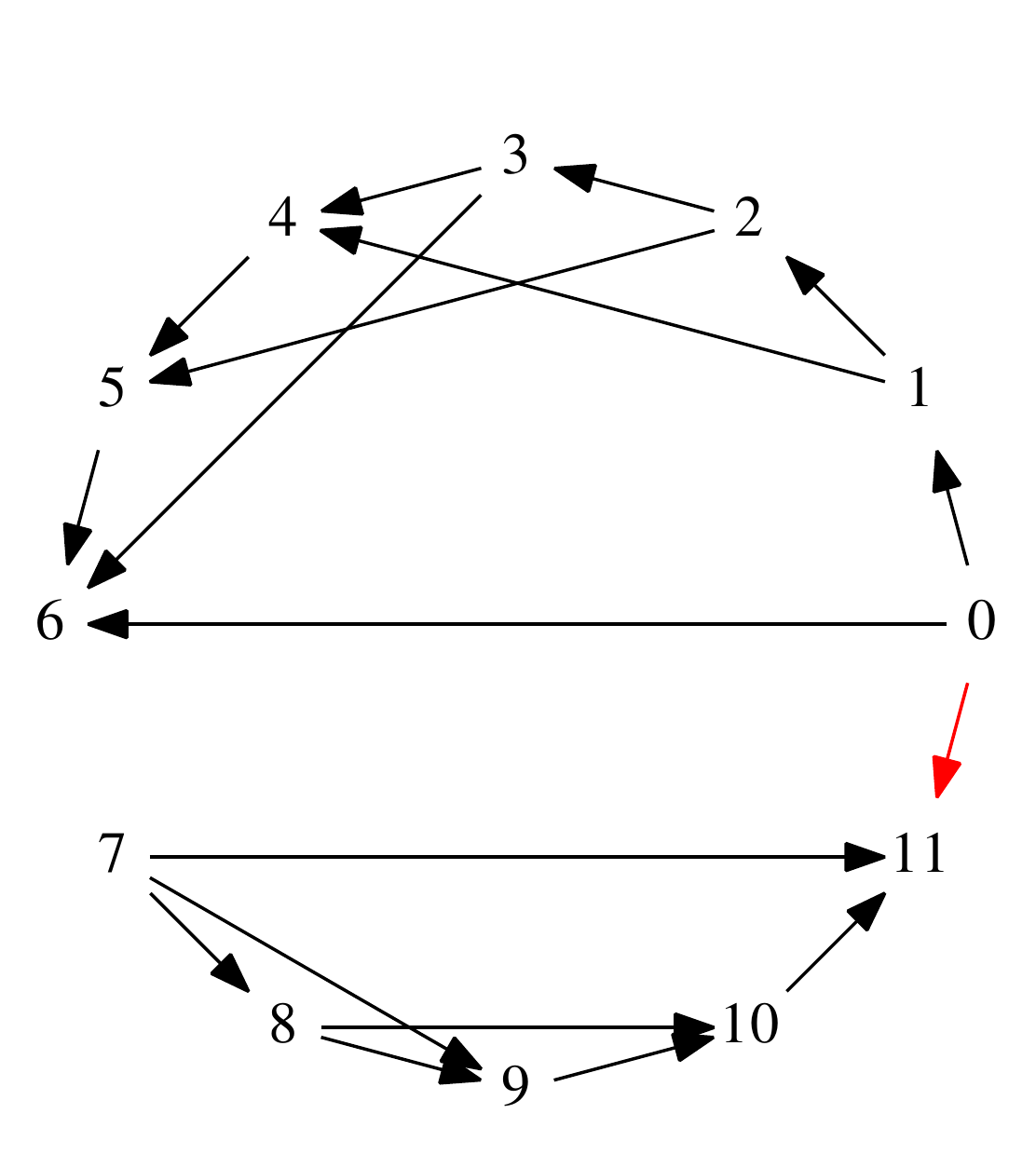}
\caption{$1$-connected graphs on $n=12$ vertices.
\texttt{W184 d6 g3 EE49.84524},
\texttt{W172 d5 g3 EE48.45339},
\texttt{W178 d6 g3 EE47.78916},
and \texttt{W172 d5 g3 EE47.10611}
in that order.
}
\label{fig.12n1}
\end{figure}
\clearpage

\subsection{2-connected} 
\begin{figure}
\includegraphics[scale=0.45]{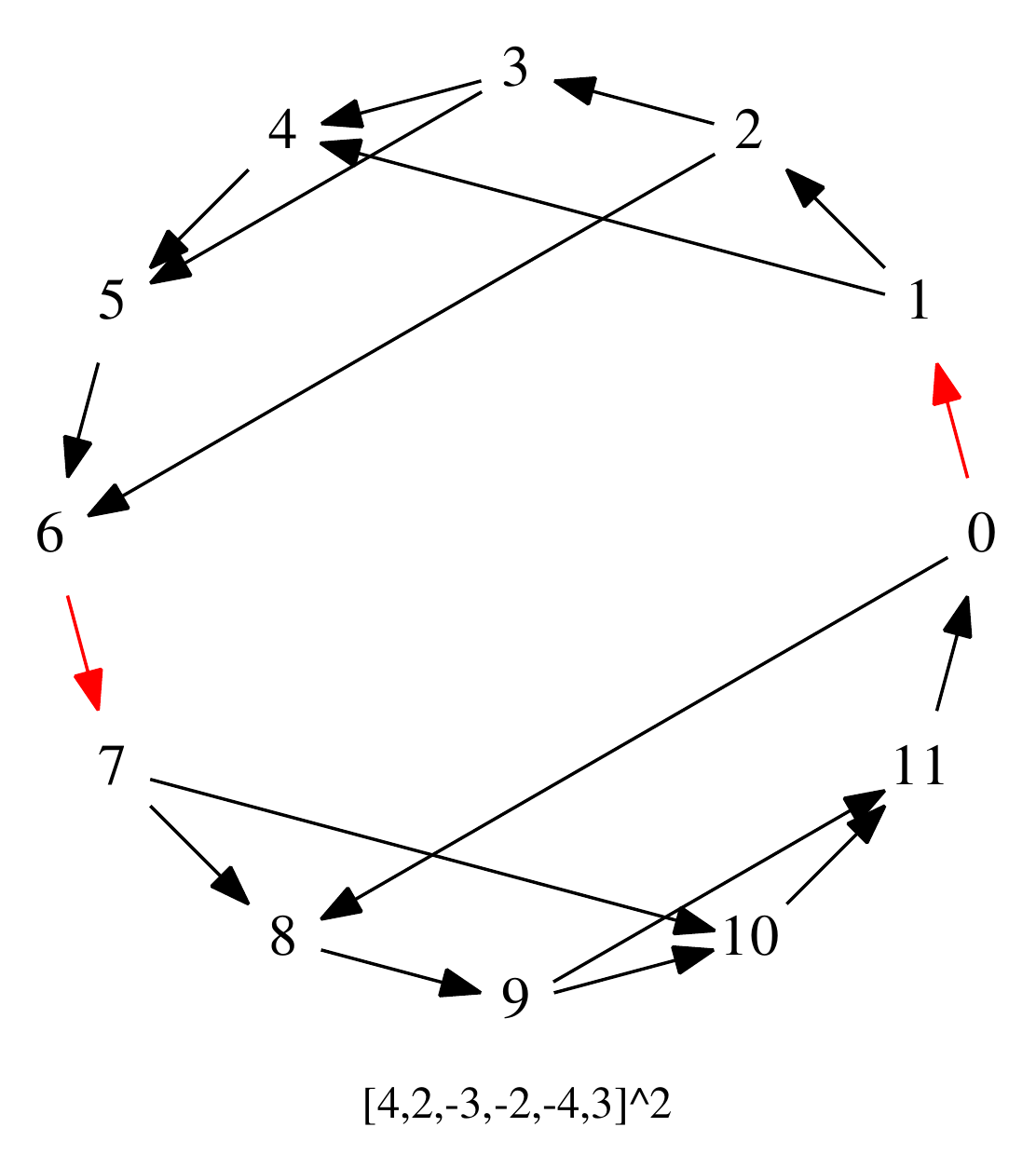}
\includegraphics[scale=0.45]{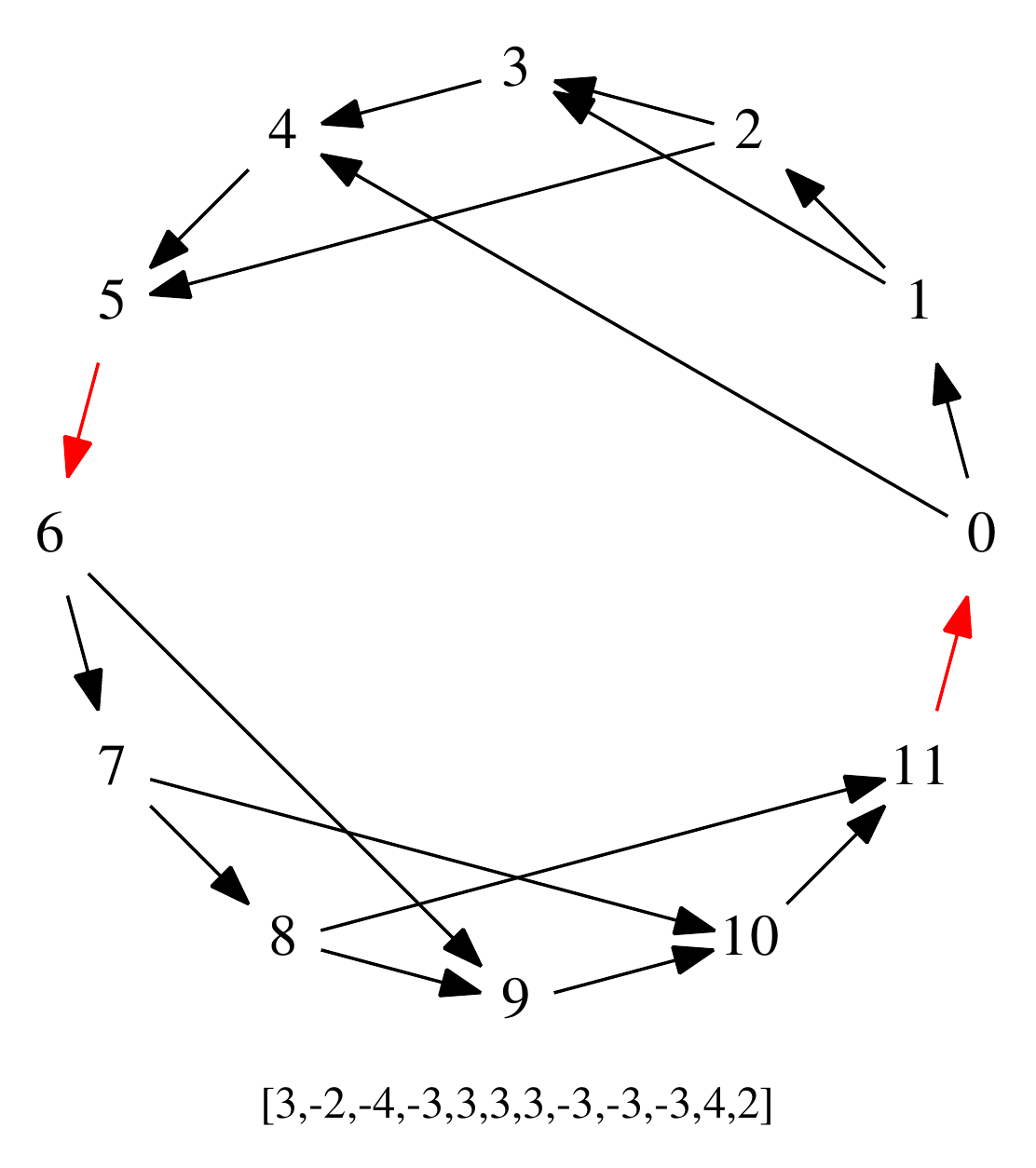}
\includegraphics[scale=0.45]{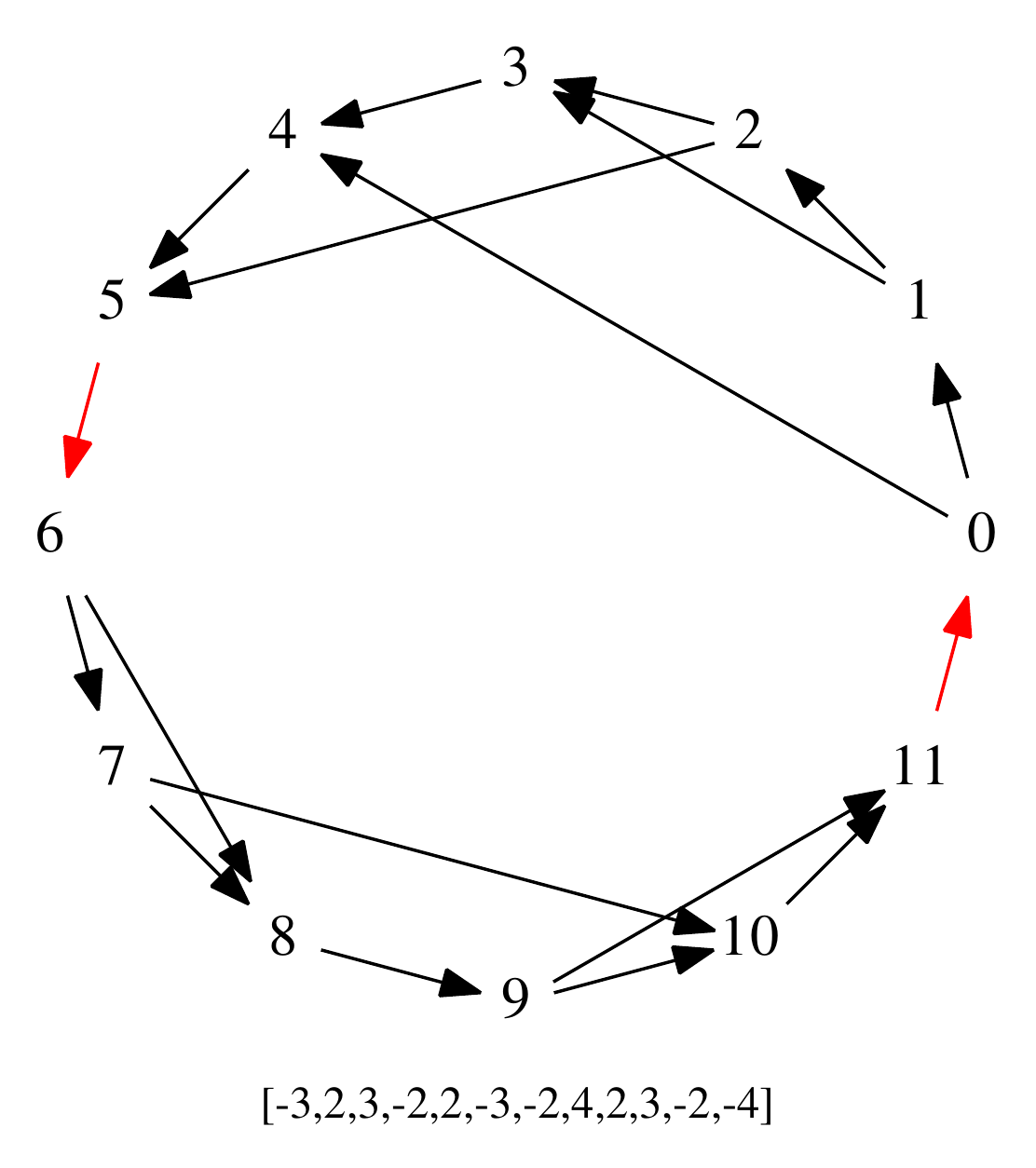}
\includegraphics[scale=0.45]{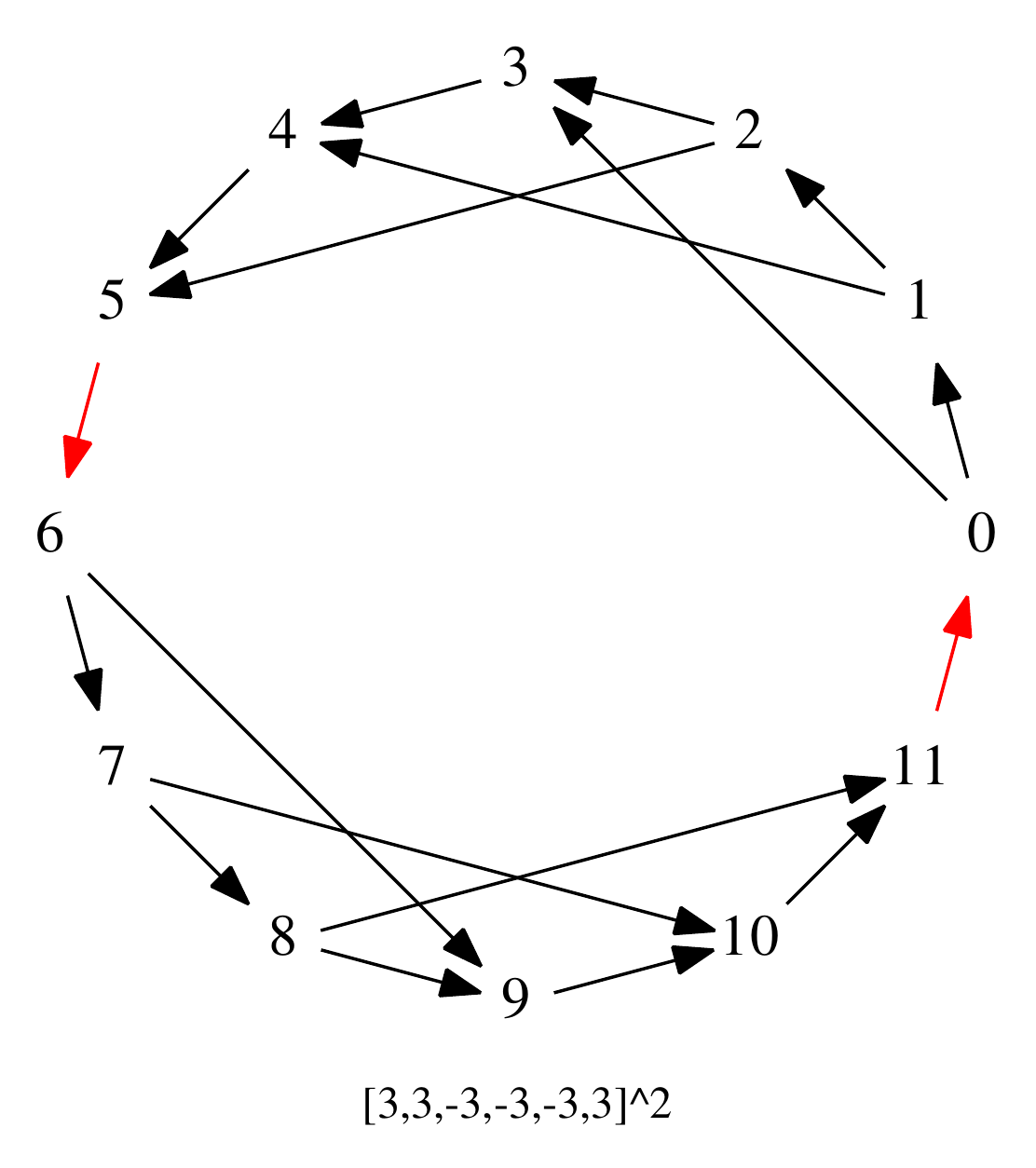}
\includegraphics[scale=0.45]{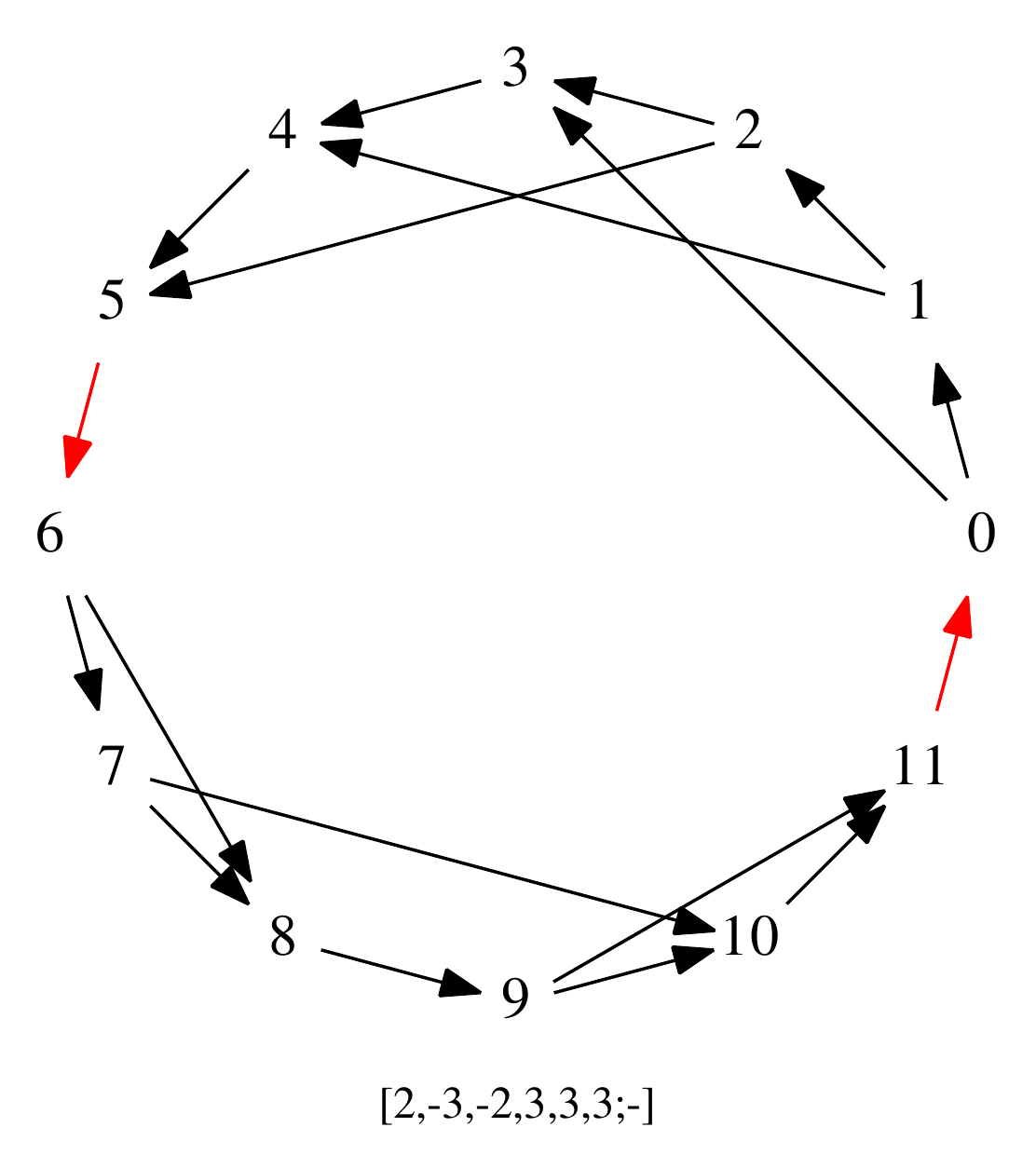}
\includegraphics[scale=0.45]{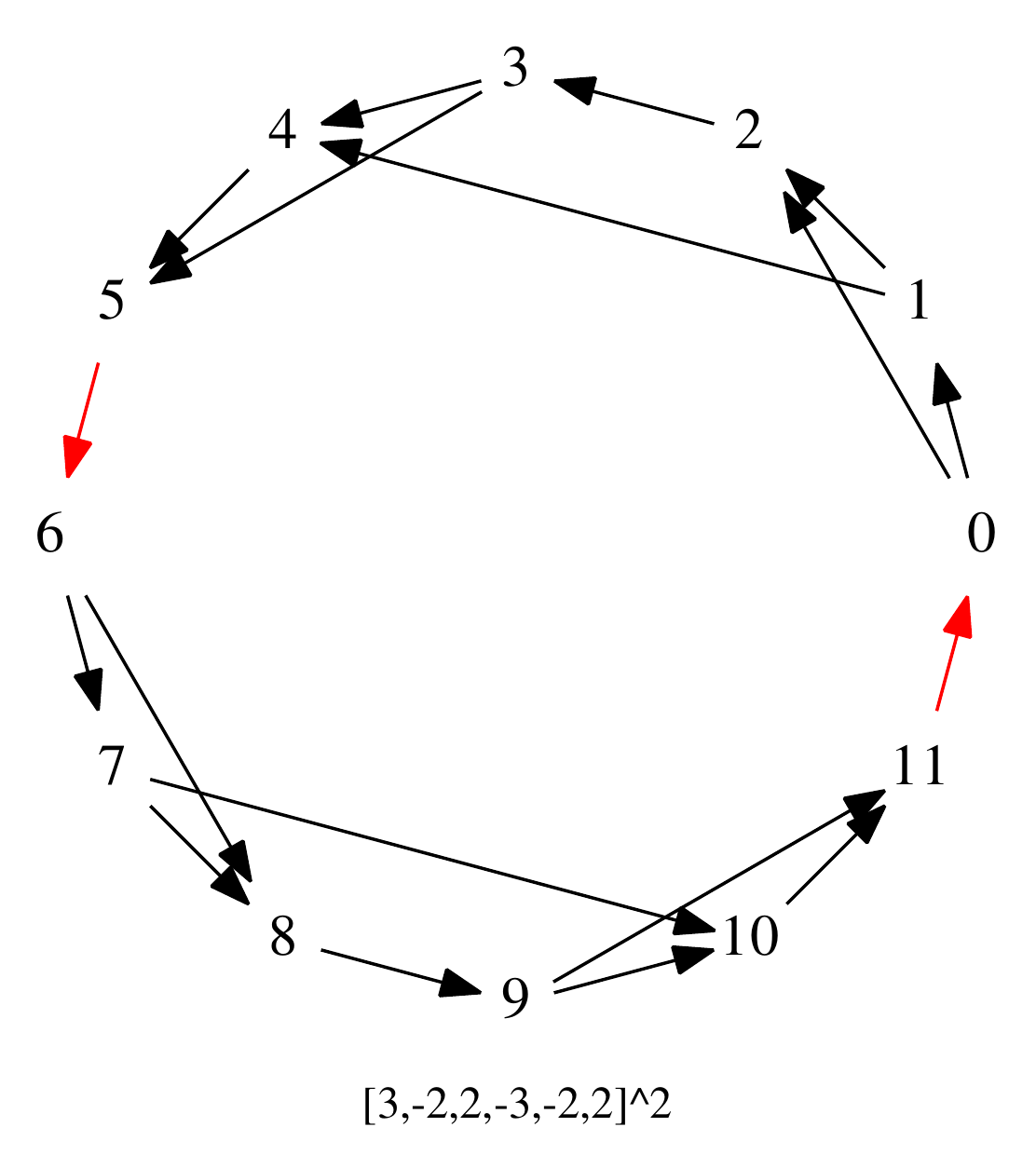}
\includegraphics[scale=0.45]{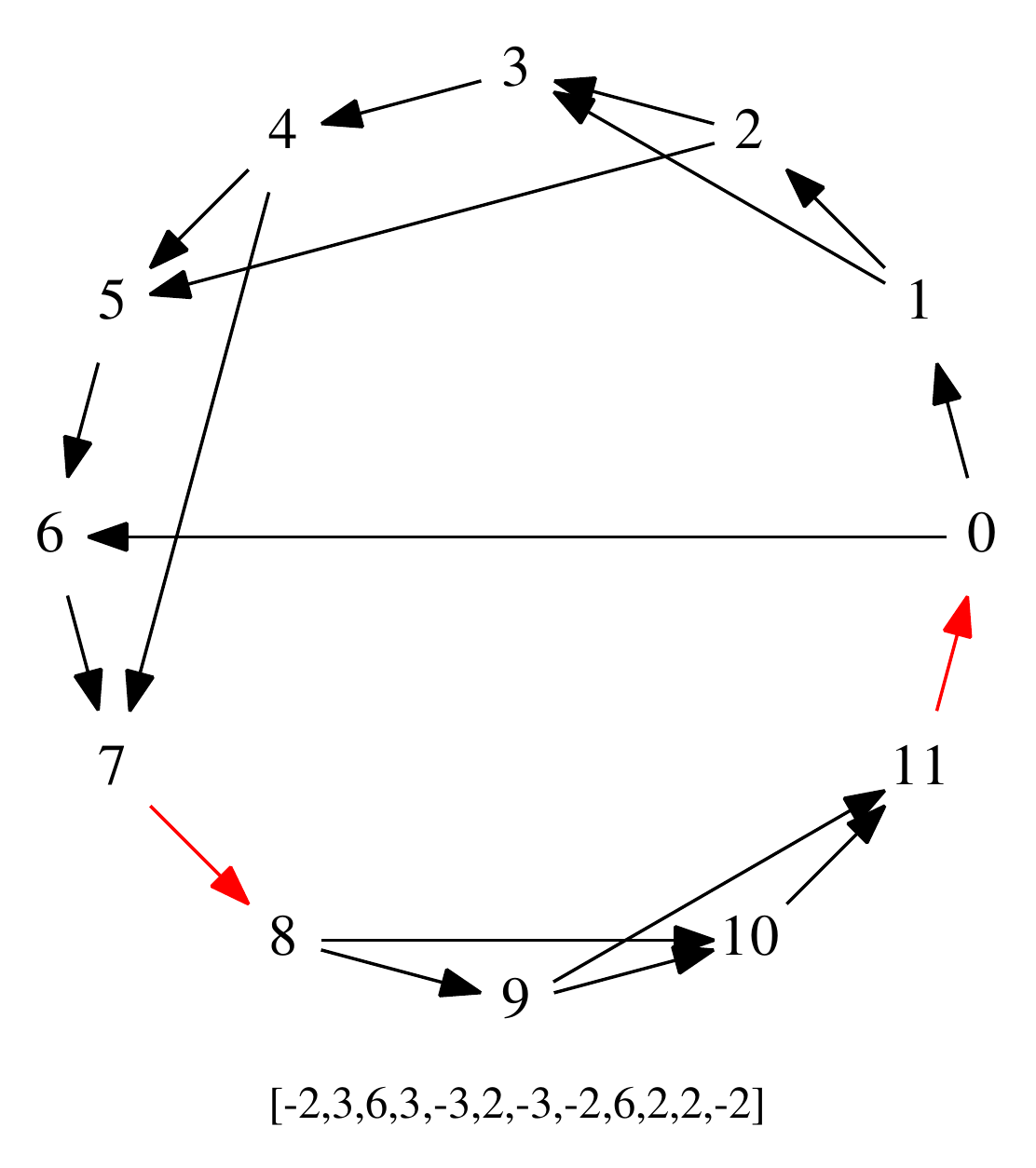}
\includegraphics[scale=0.45]{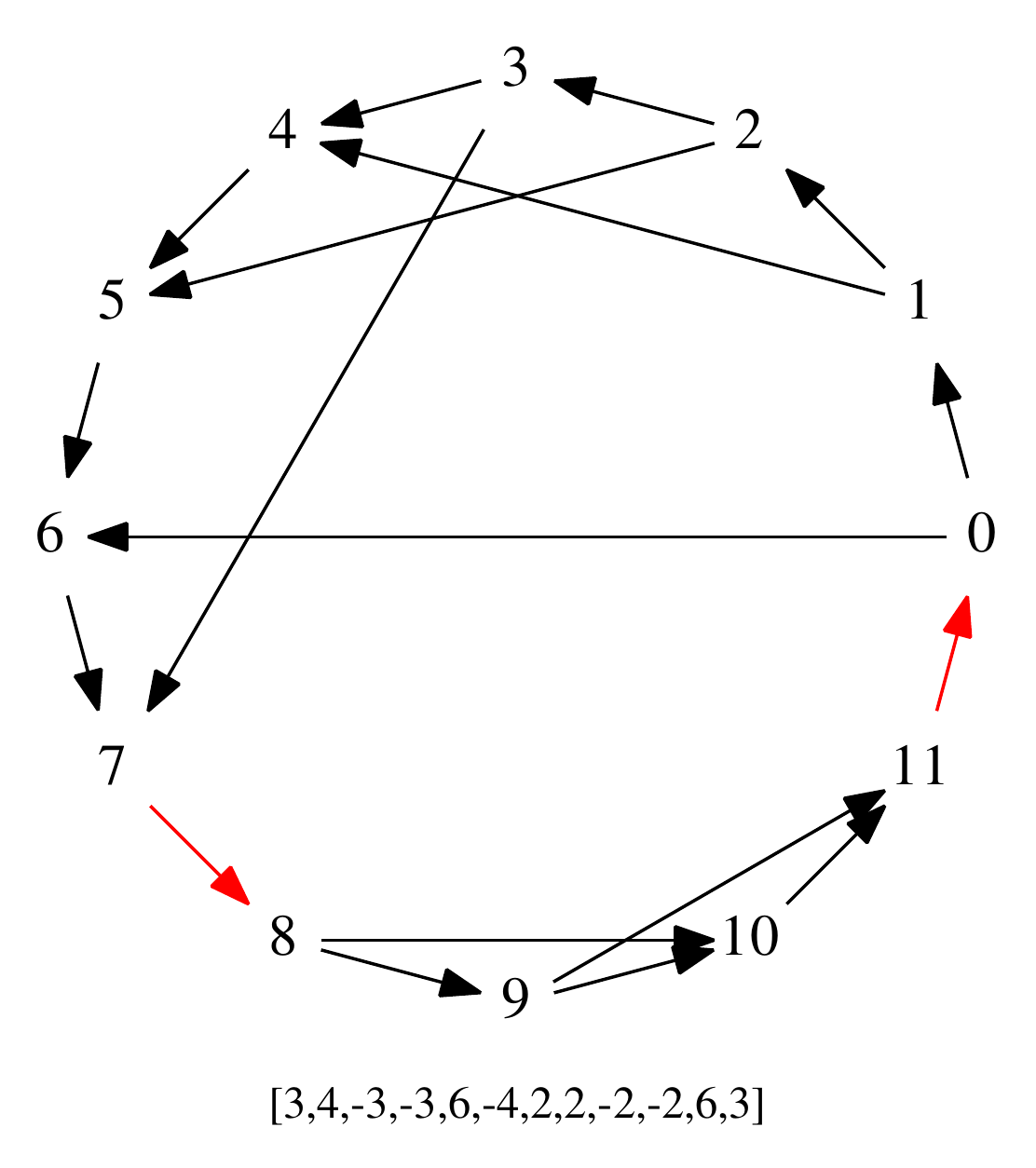}
\includegraphics[scale=0.45]{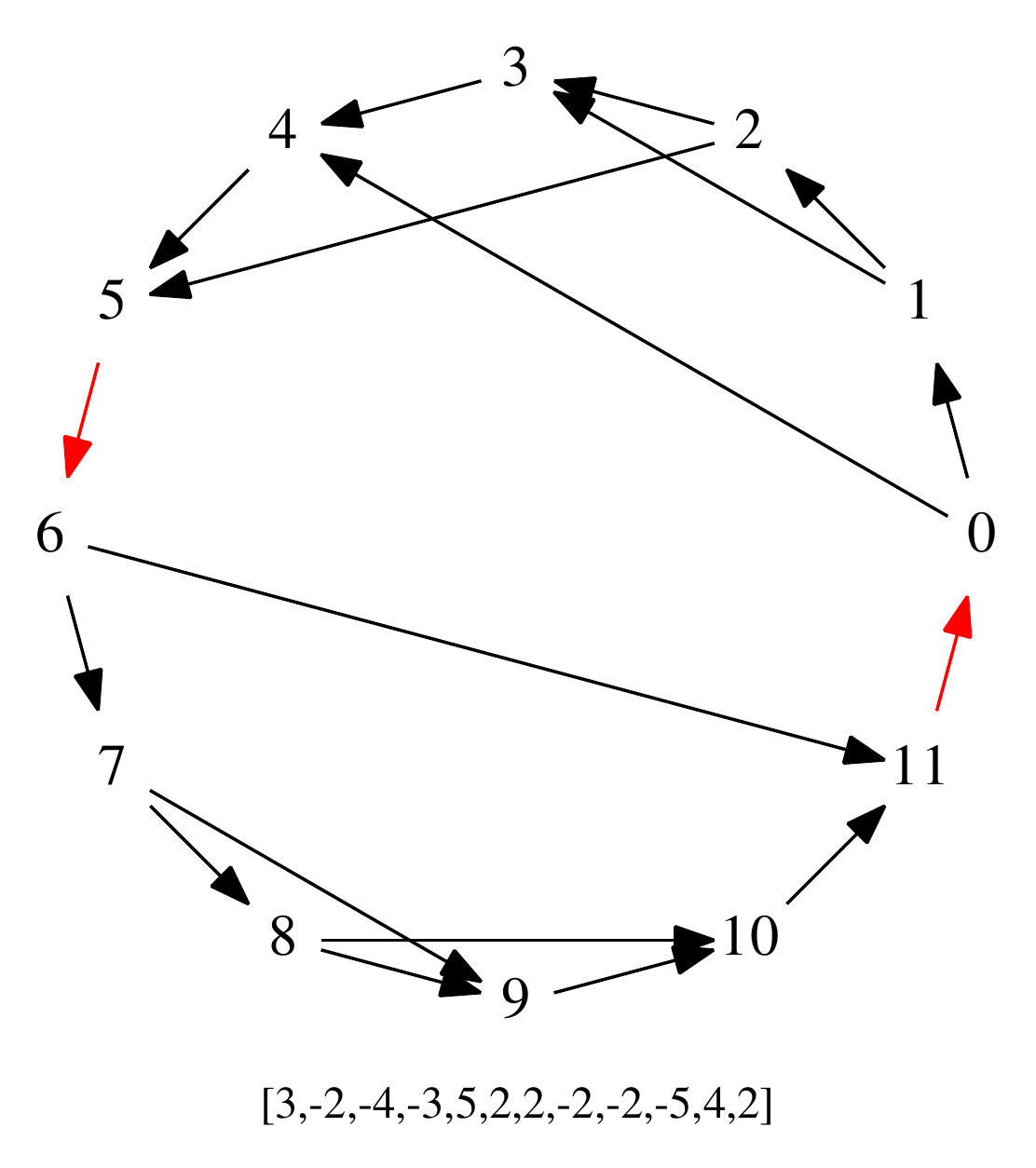}
\includegraphics[scale=0.45]{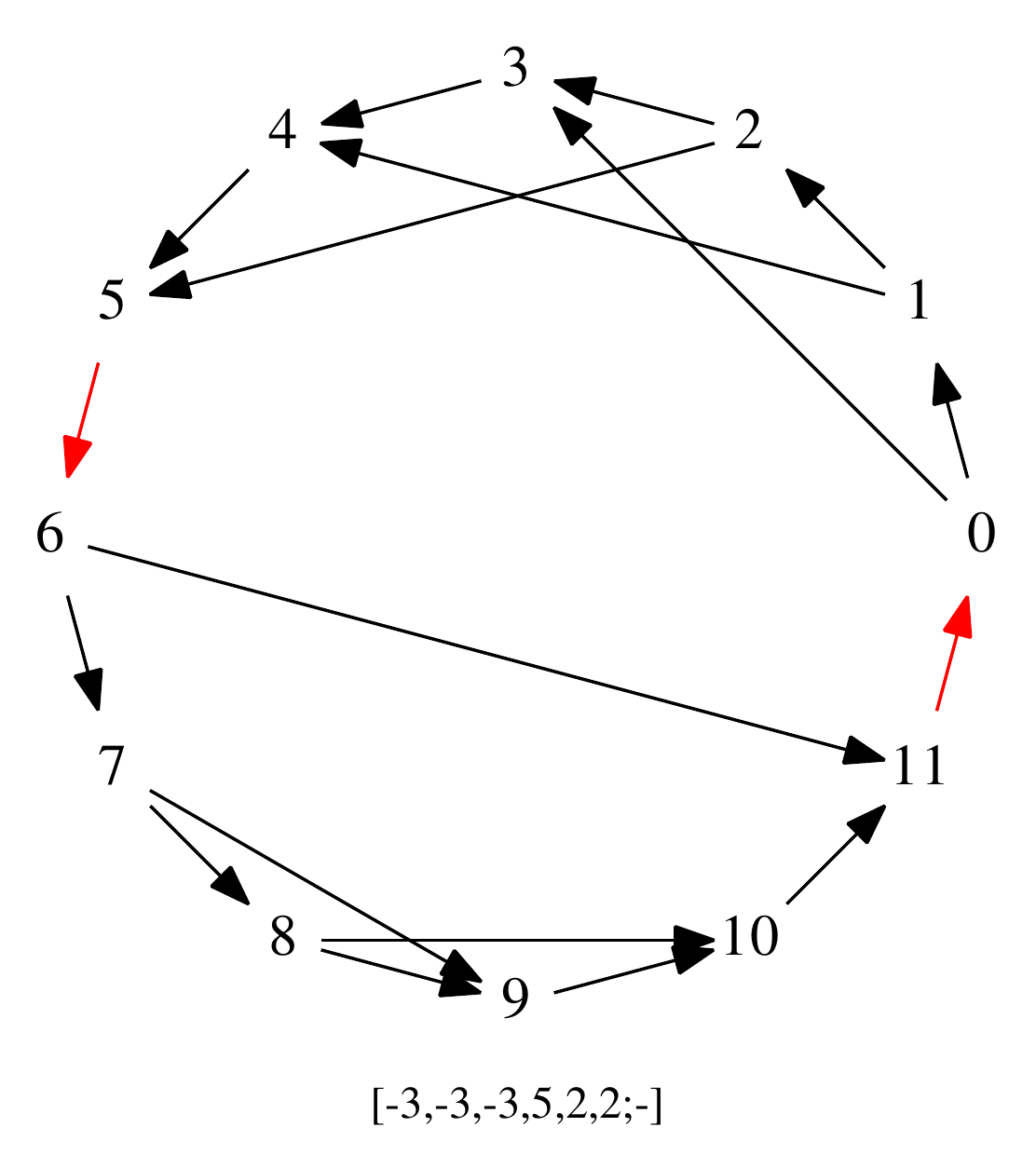}
\includegraphics[scale=0.45]{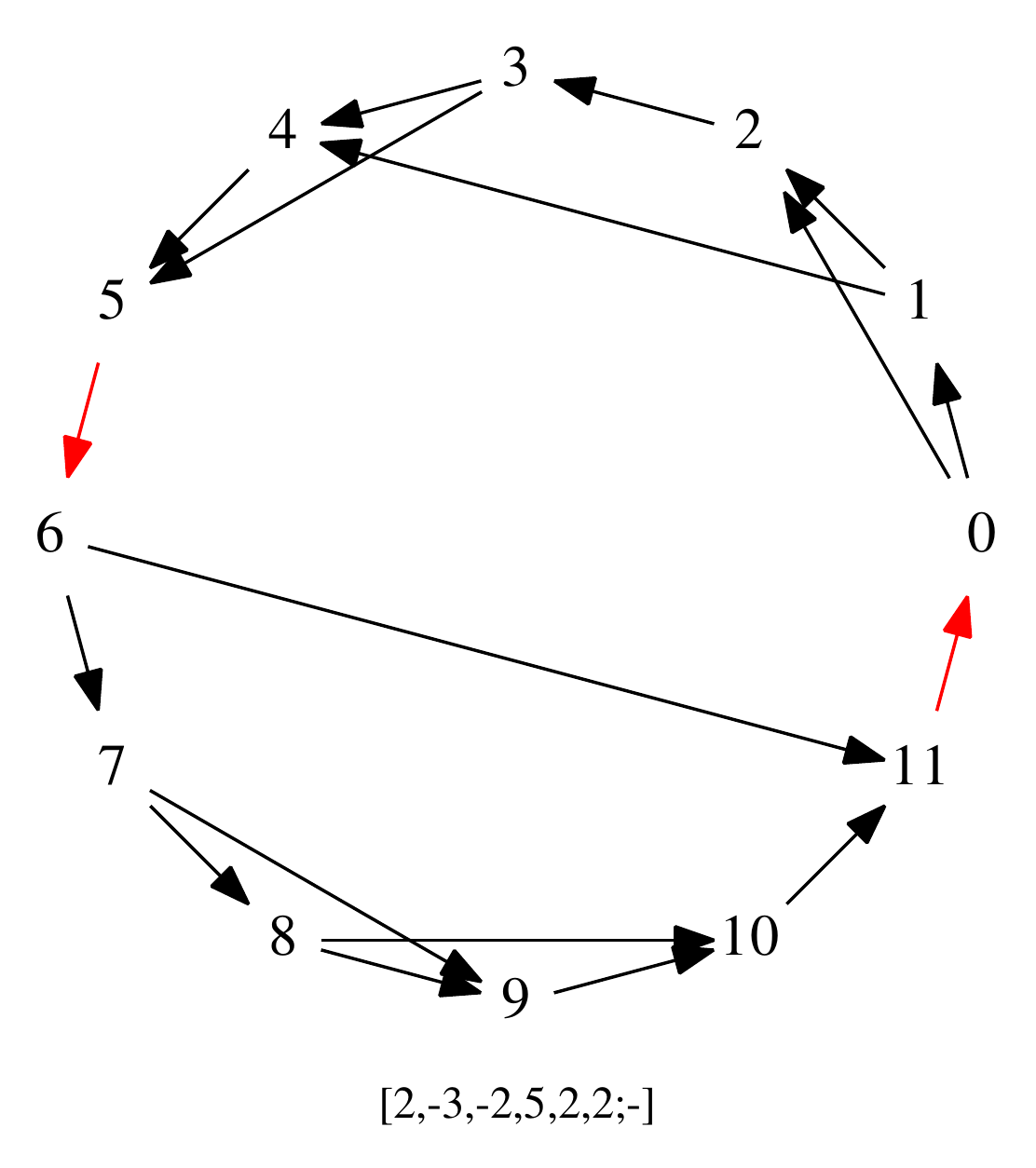}
\includegraphics[scale=0.45]{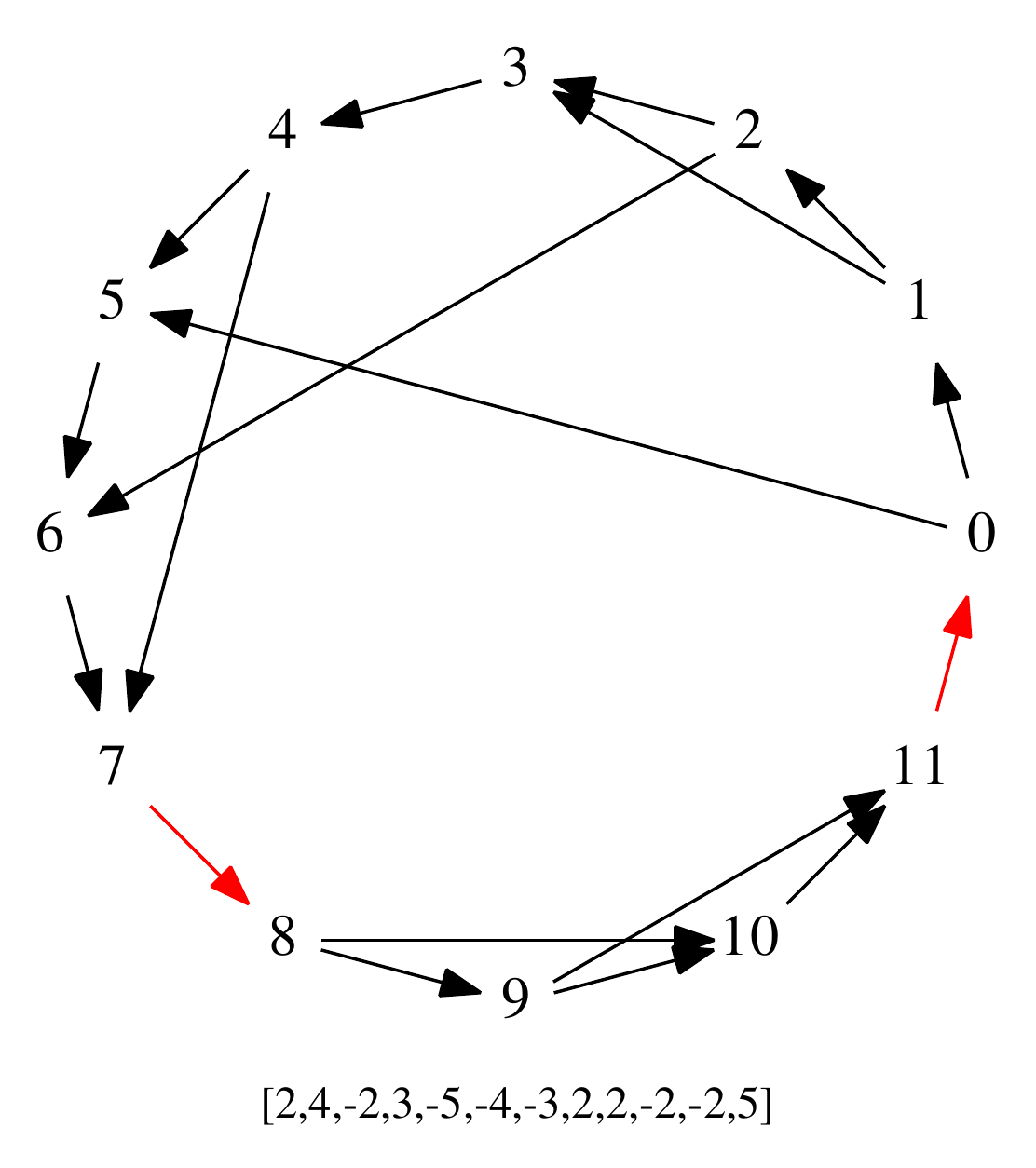}
\caption{$2$-connected graphs on $n=12$ vertices (start).
}
\label{fig.12n2s}
\end{figure}
\begin{figure}
\includegraphics[scale=0.45]{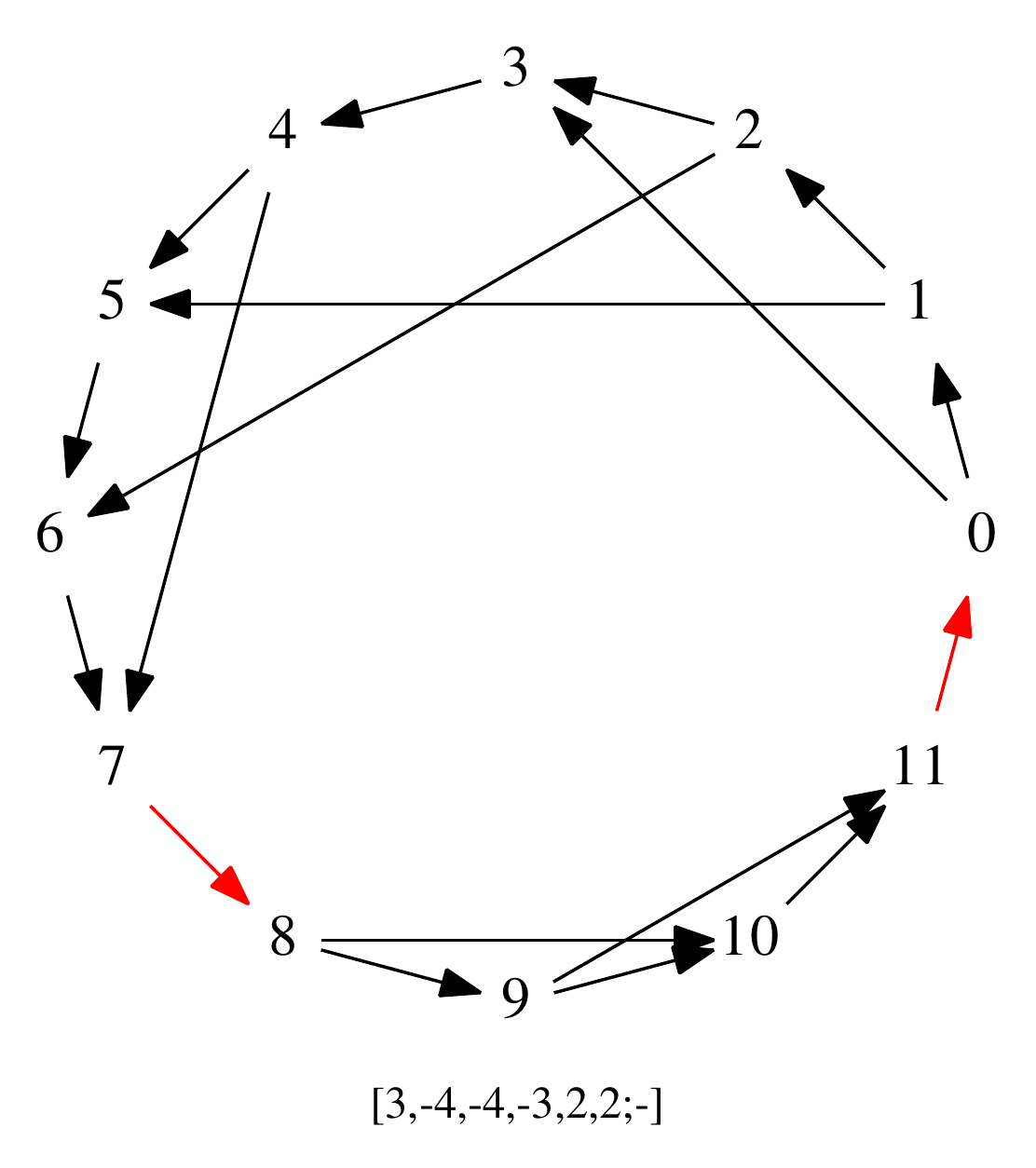}
\includegraphics[scale=0.45]{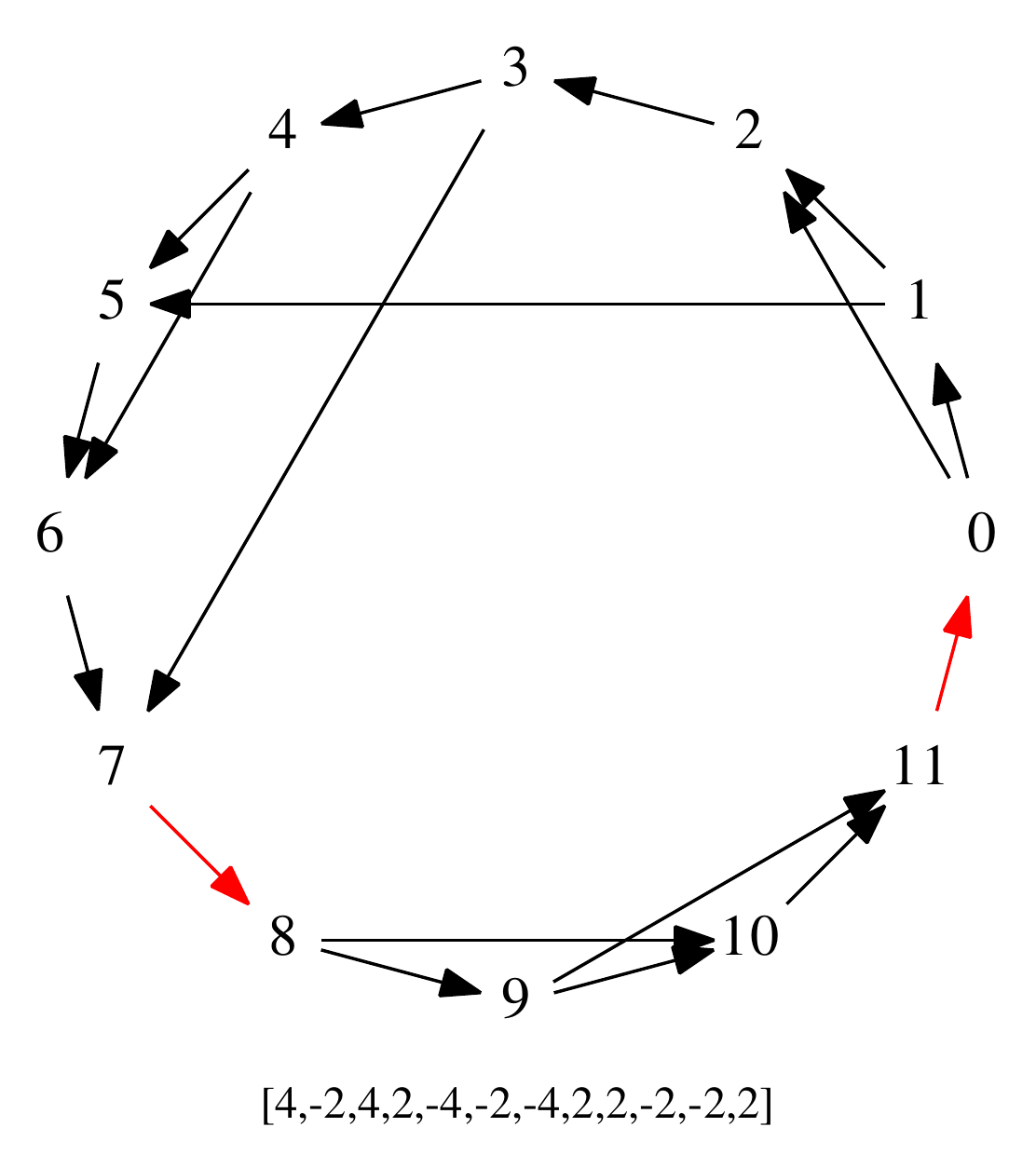}
\includegraphics[scale=0.45]{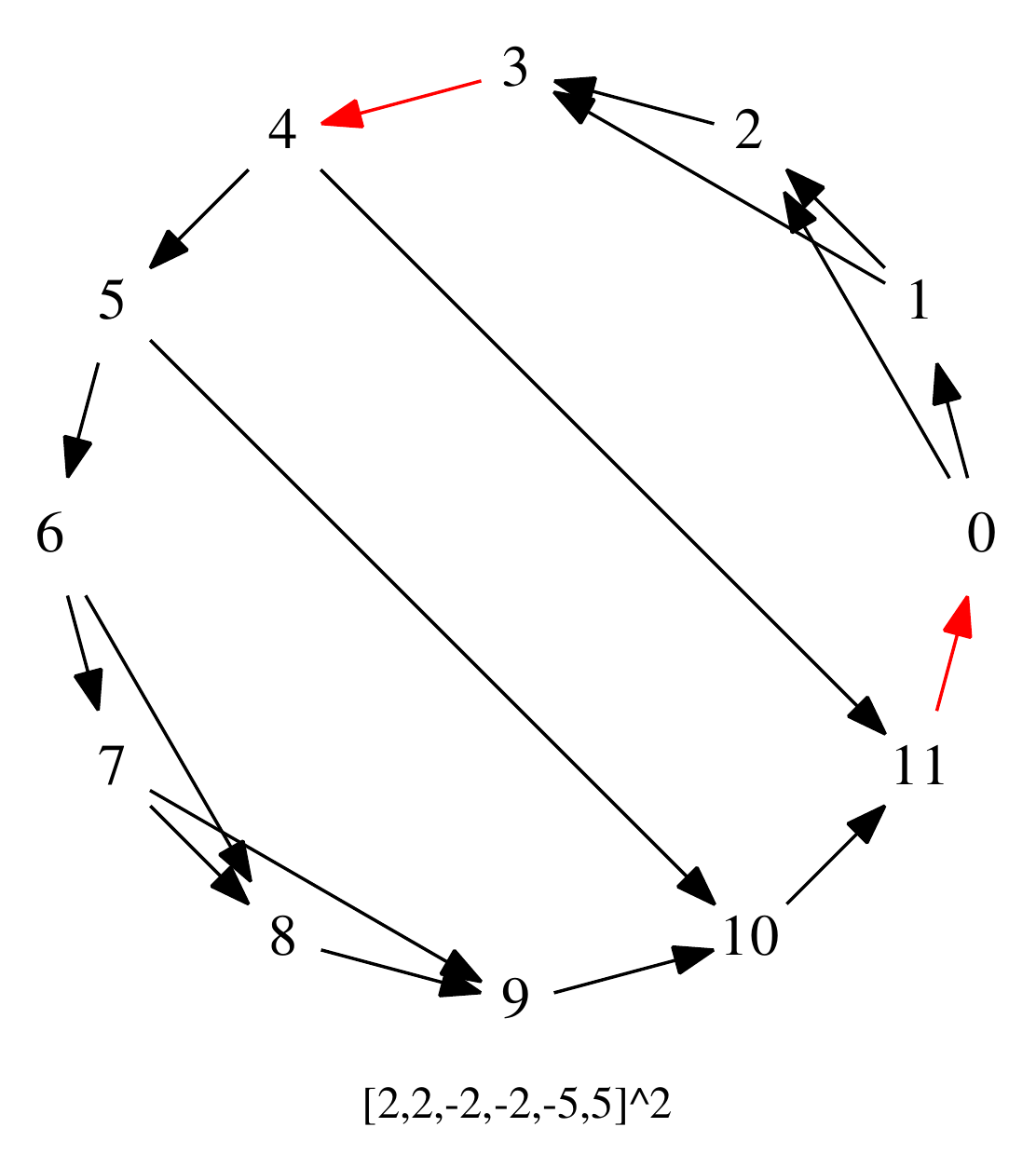}
\includegraphics[scale=0.45]{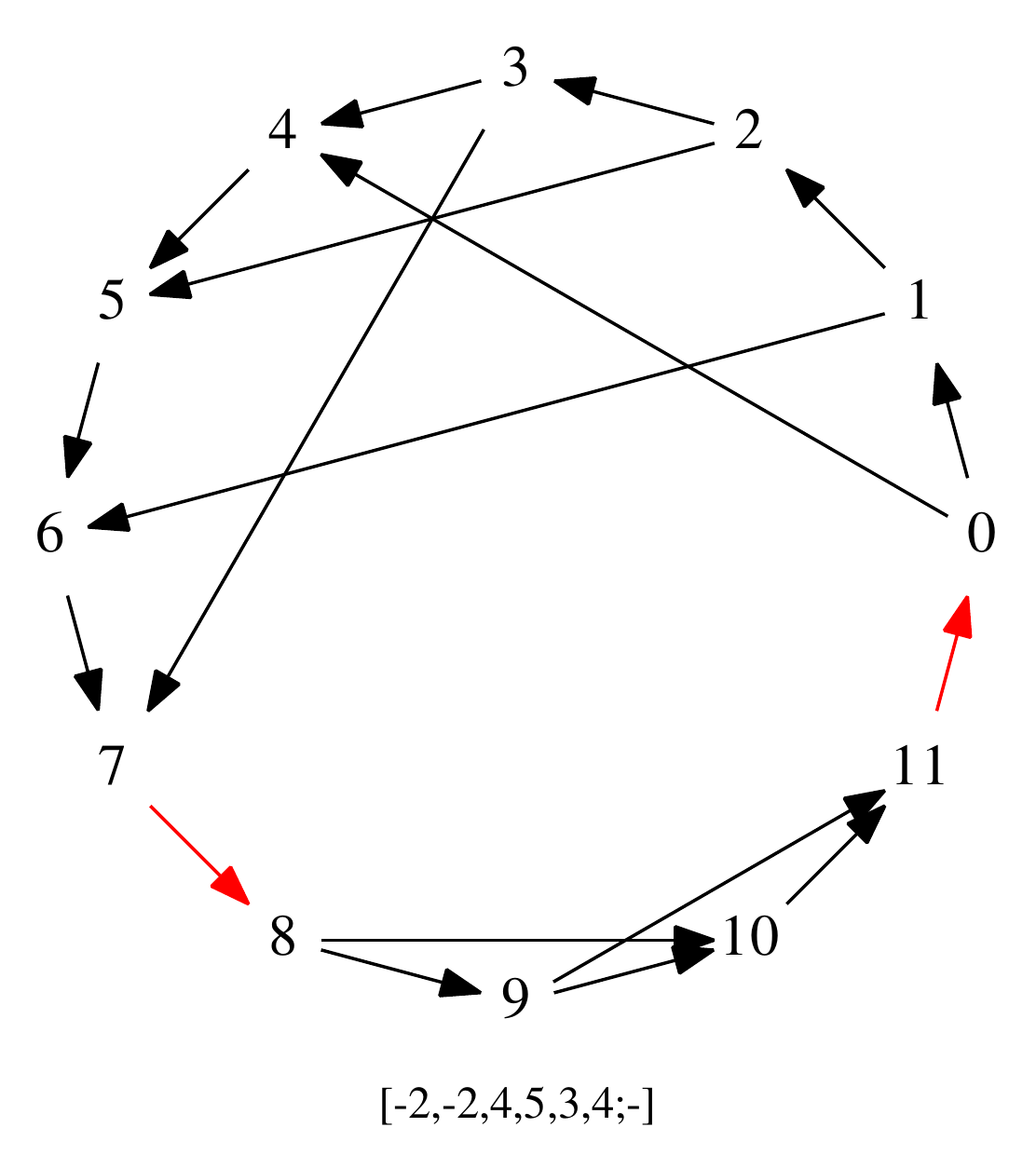}
\includegraphics[scale=0.45]{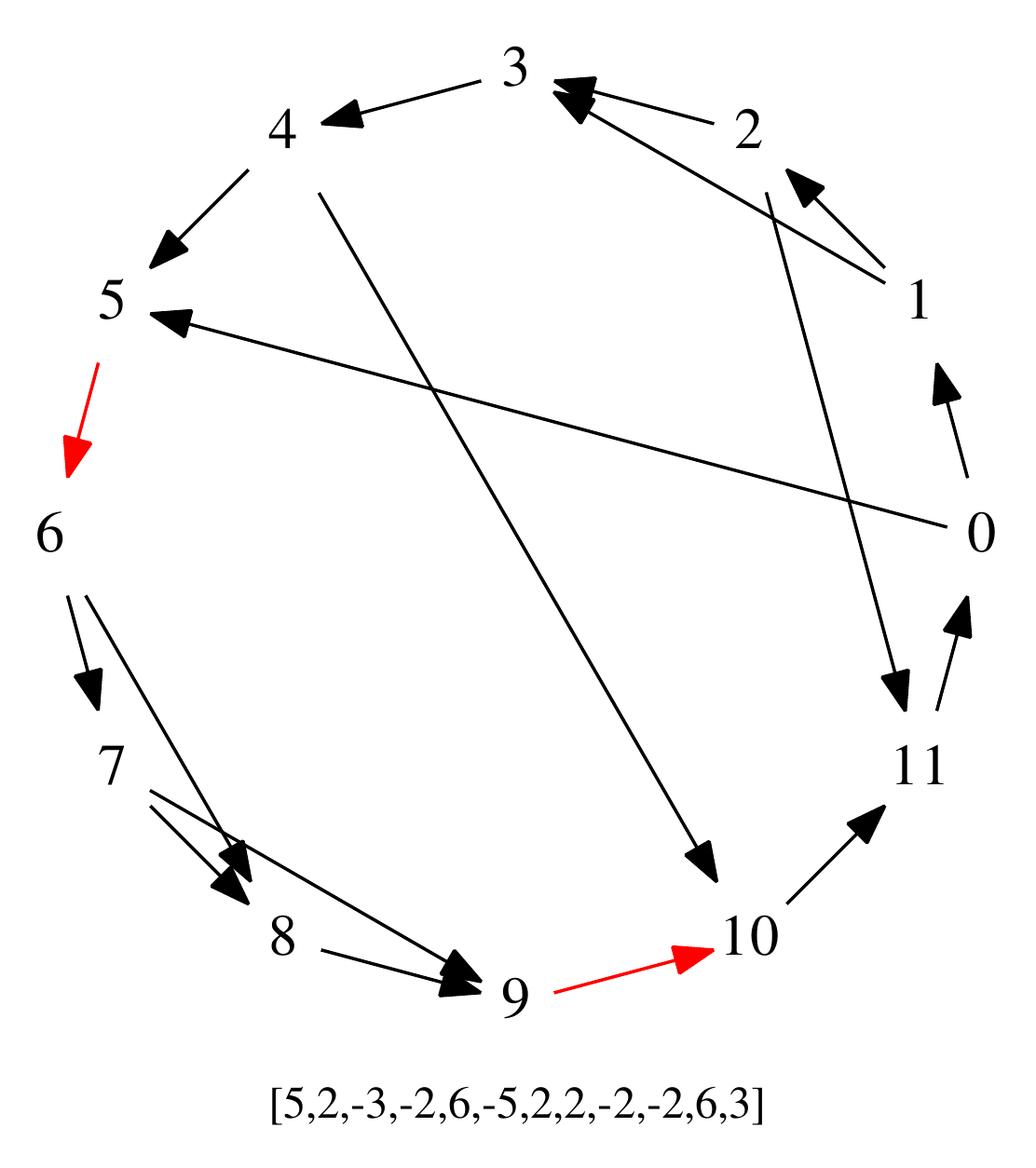}
\includegraphics[scale=0.45]{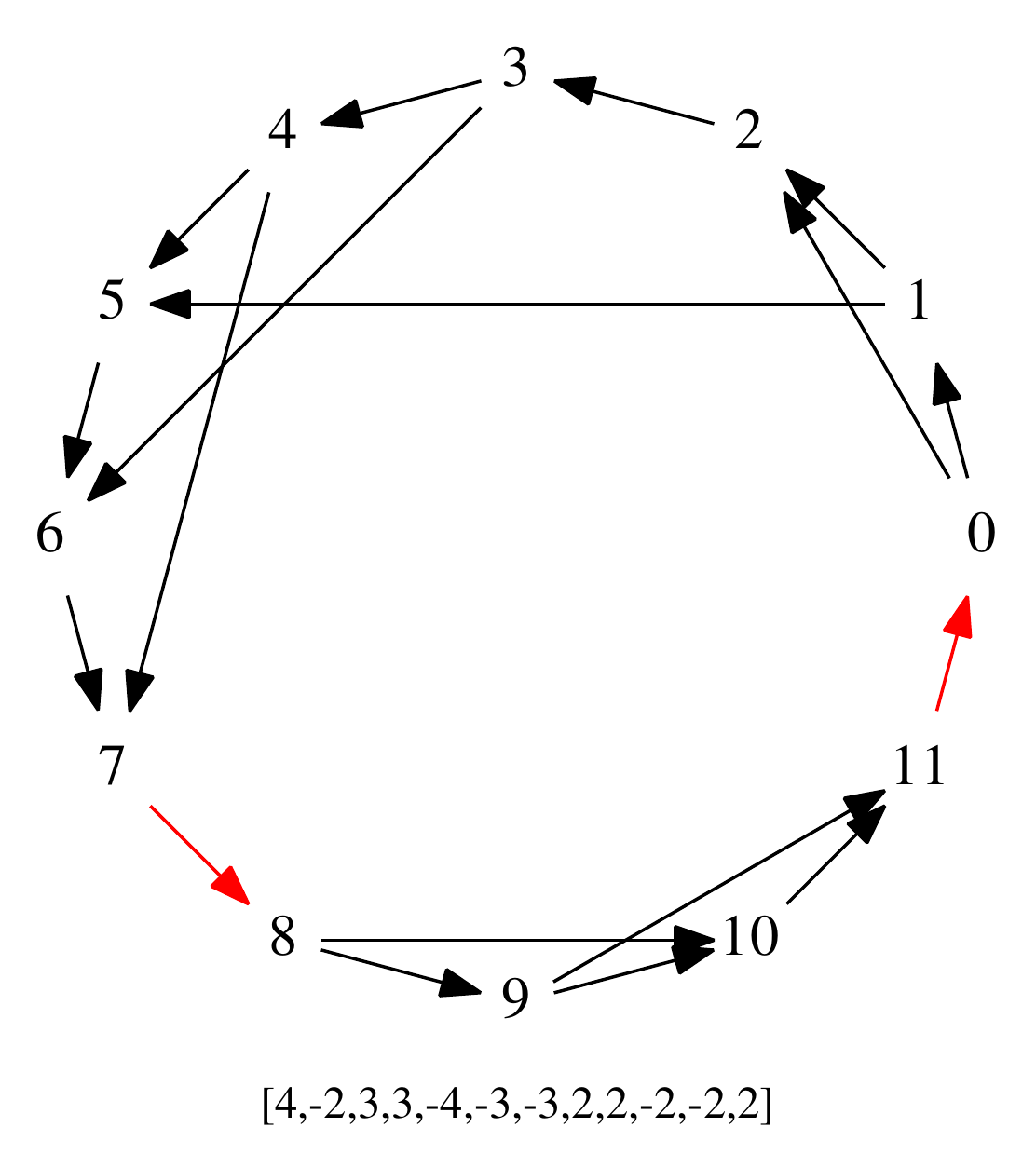}
\includegraphics[scale=0.45]{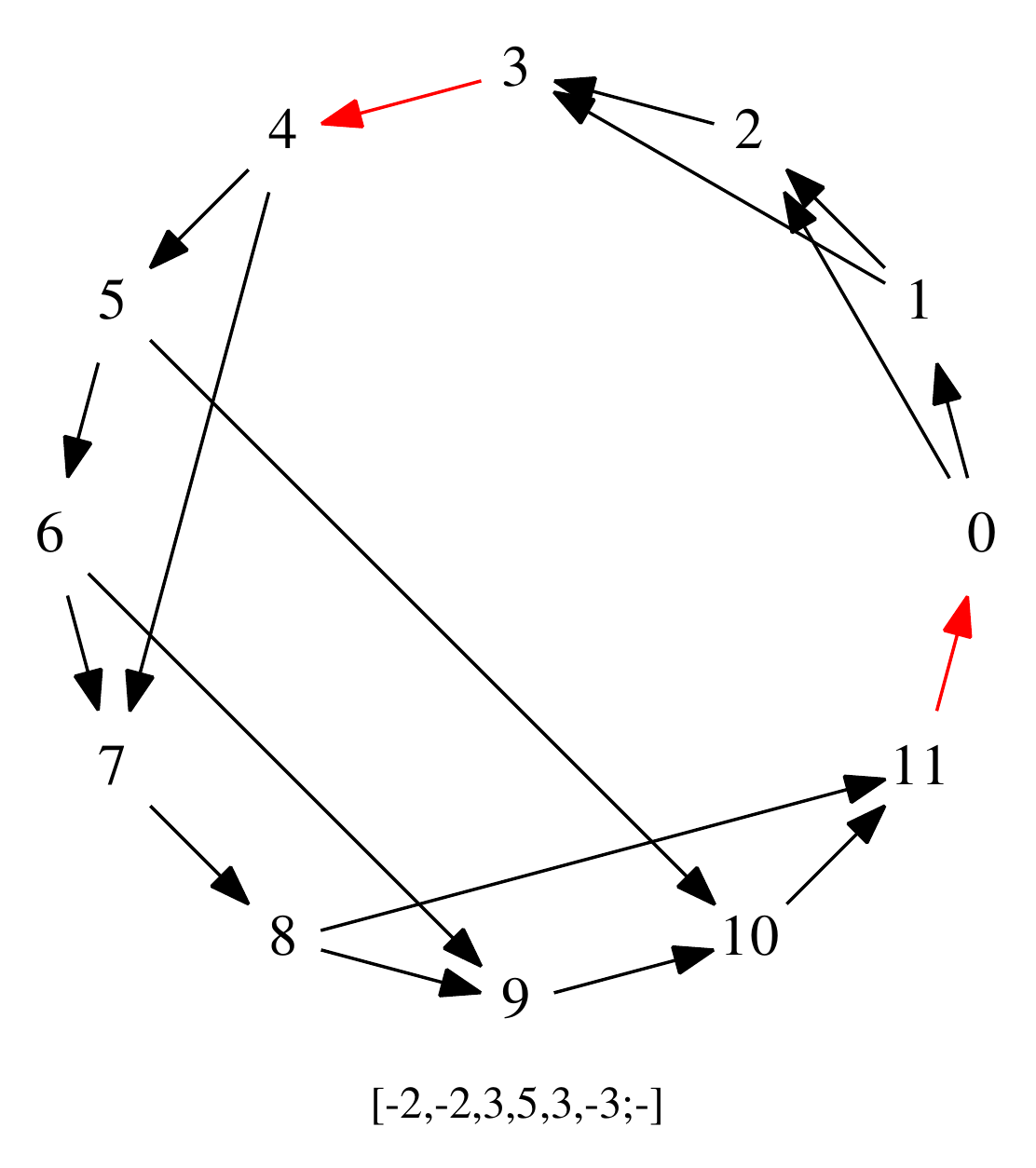}
\includegraphics[scale=0.45]{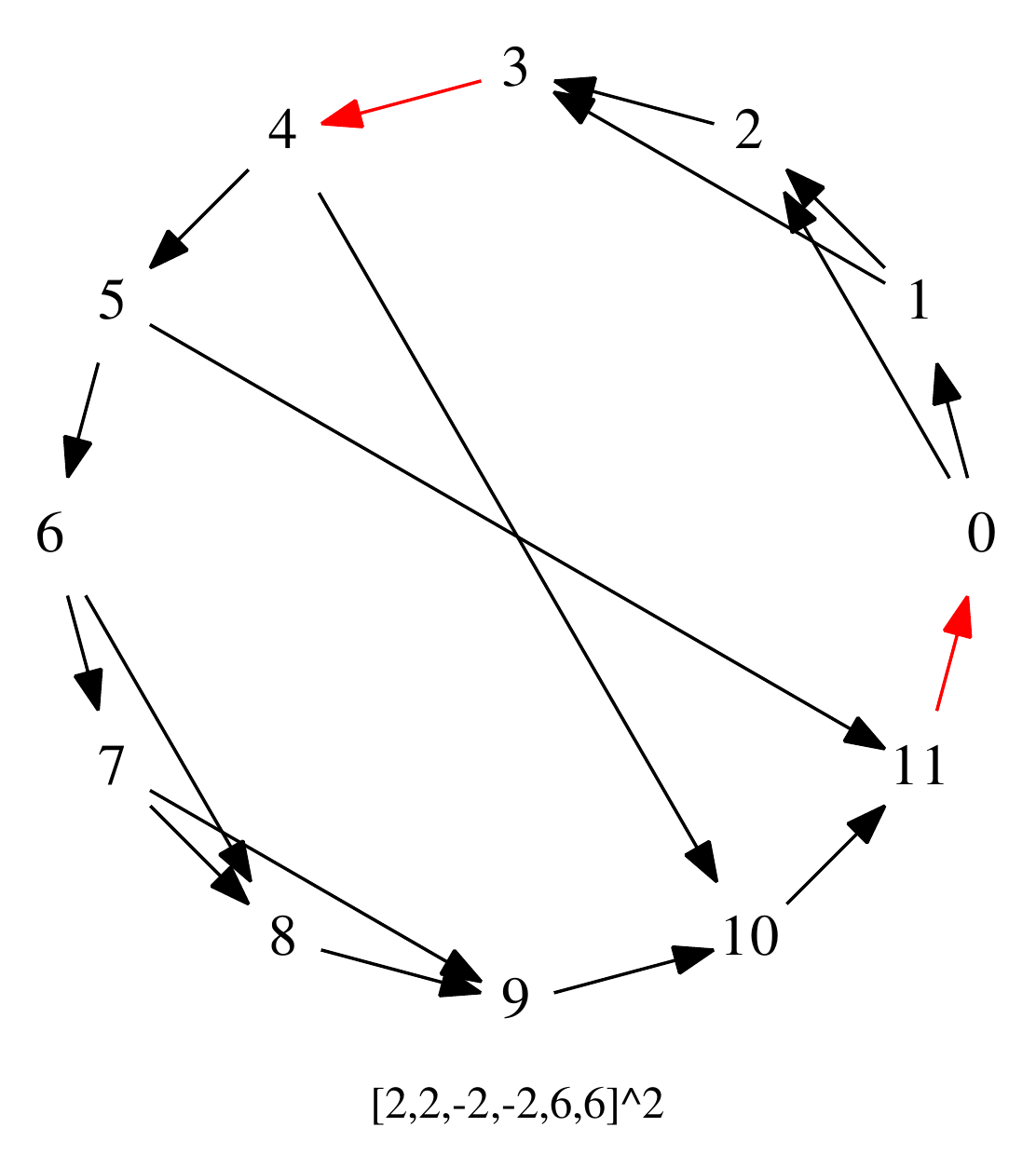}
\includegraphics[scale=0.45]{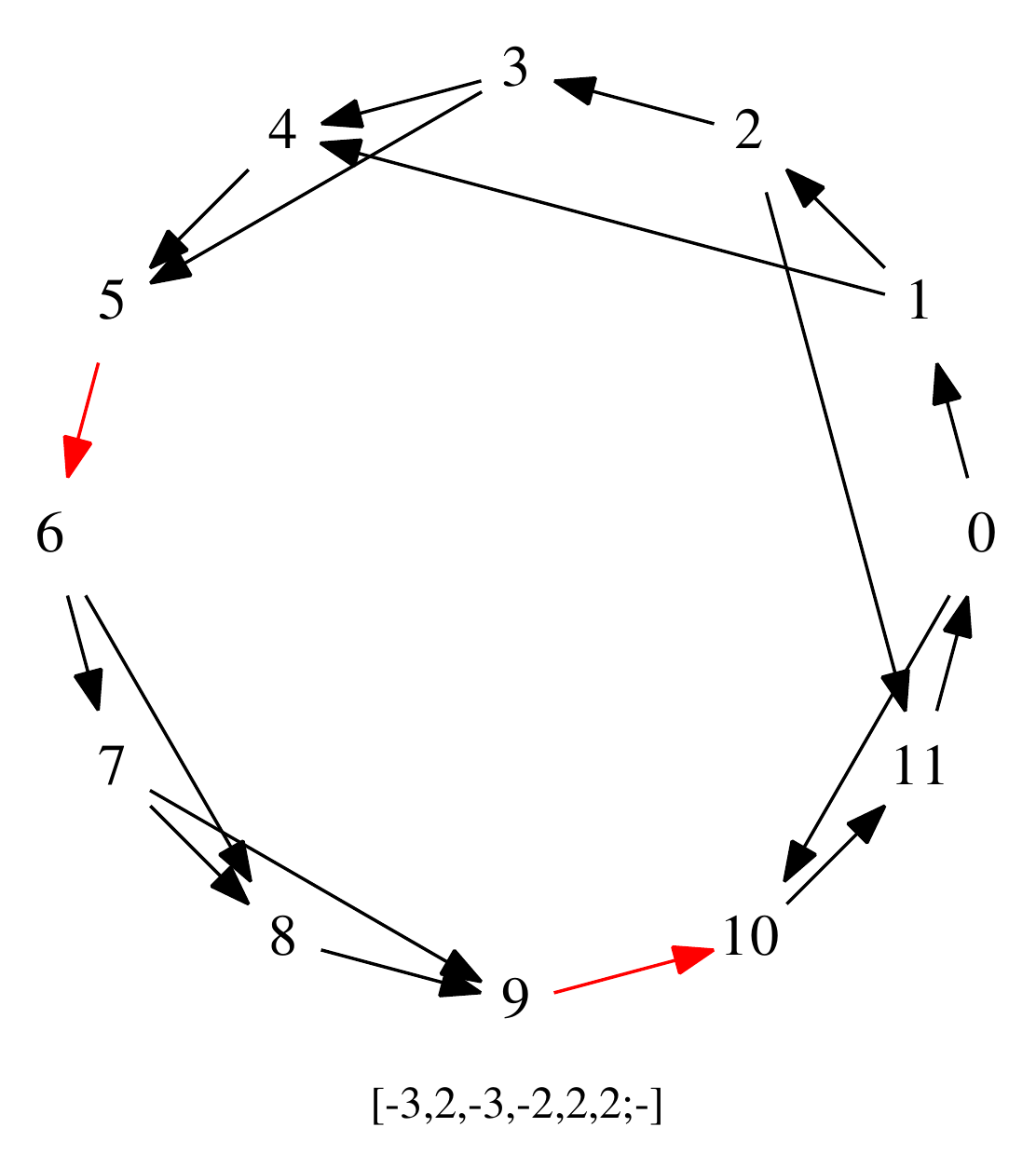}
\includegraphics[scale=0.45]{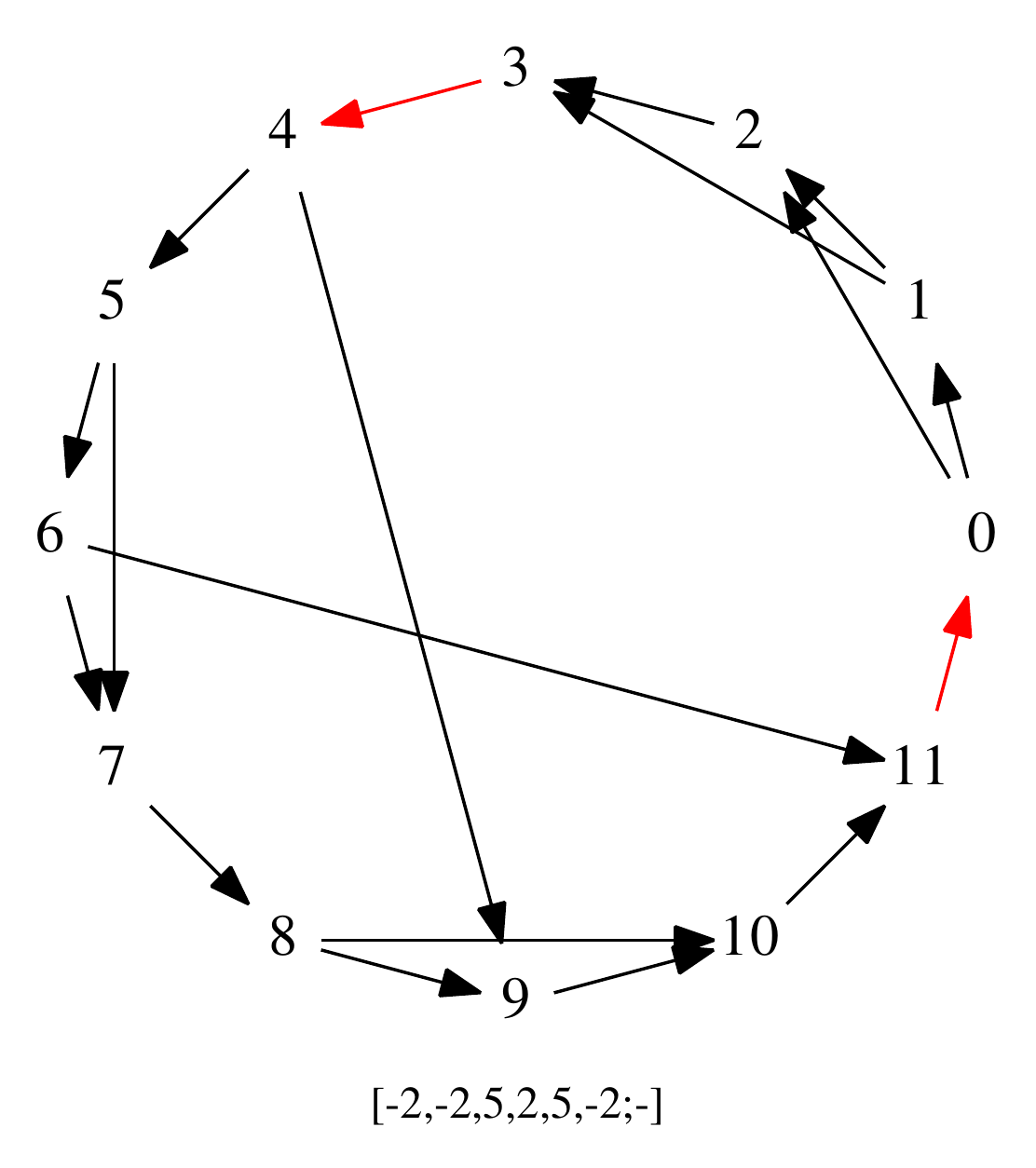}
\includegraphics[scale=0.45]{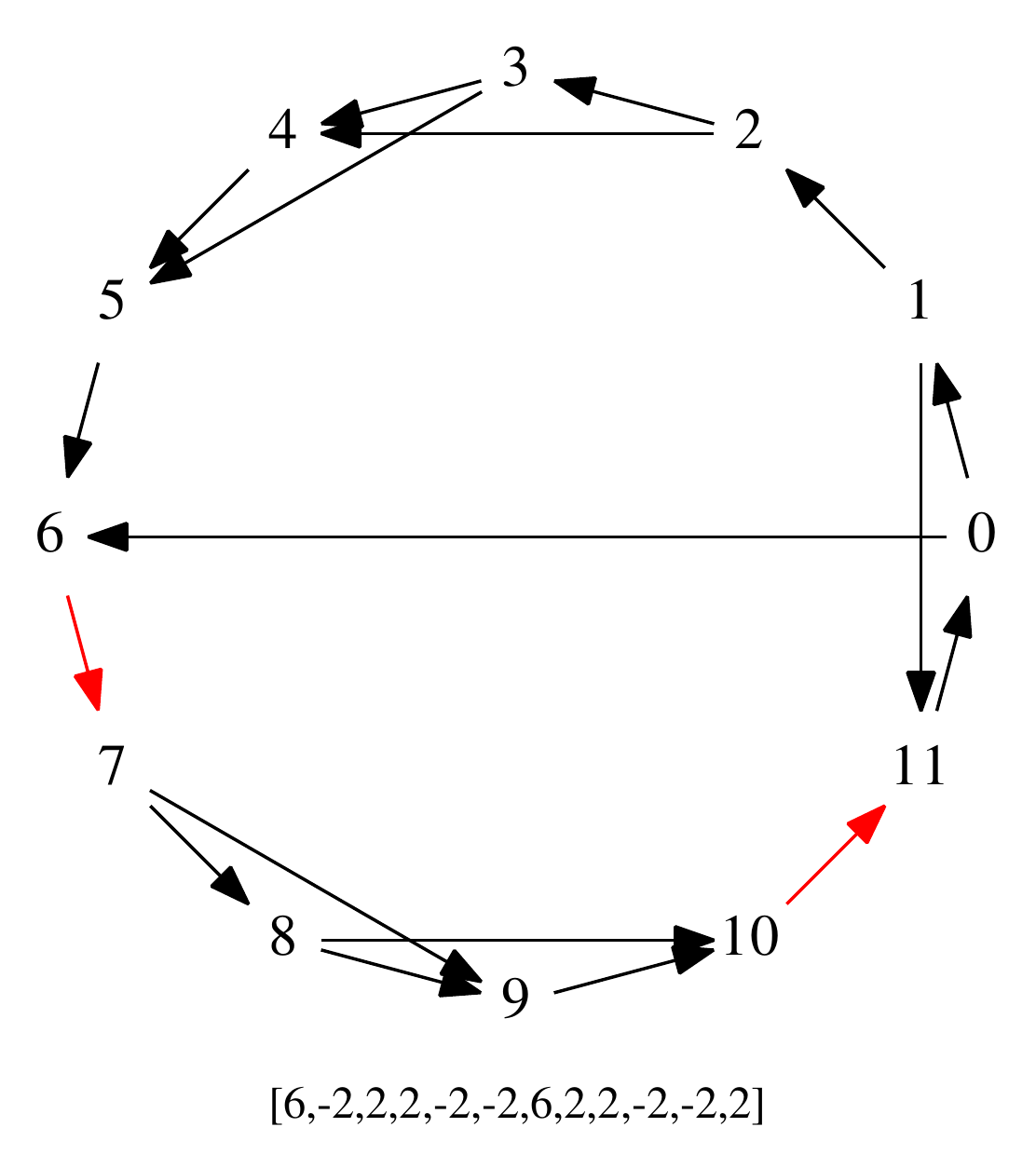}
\includegraphics[scale=0.45]{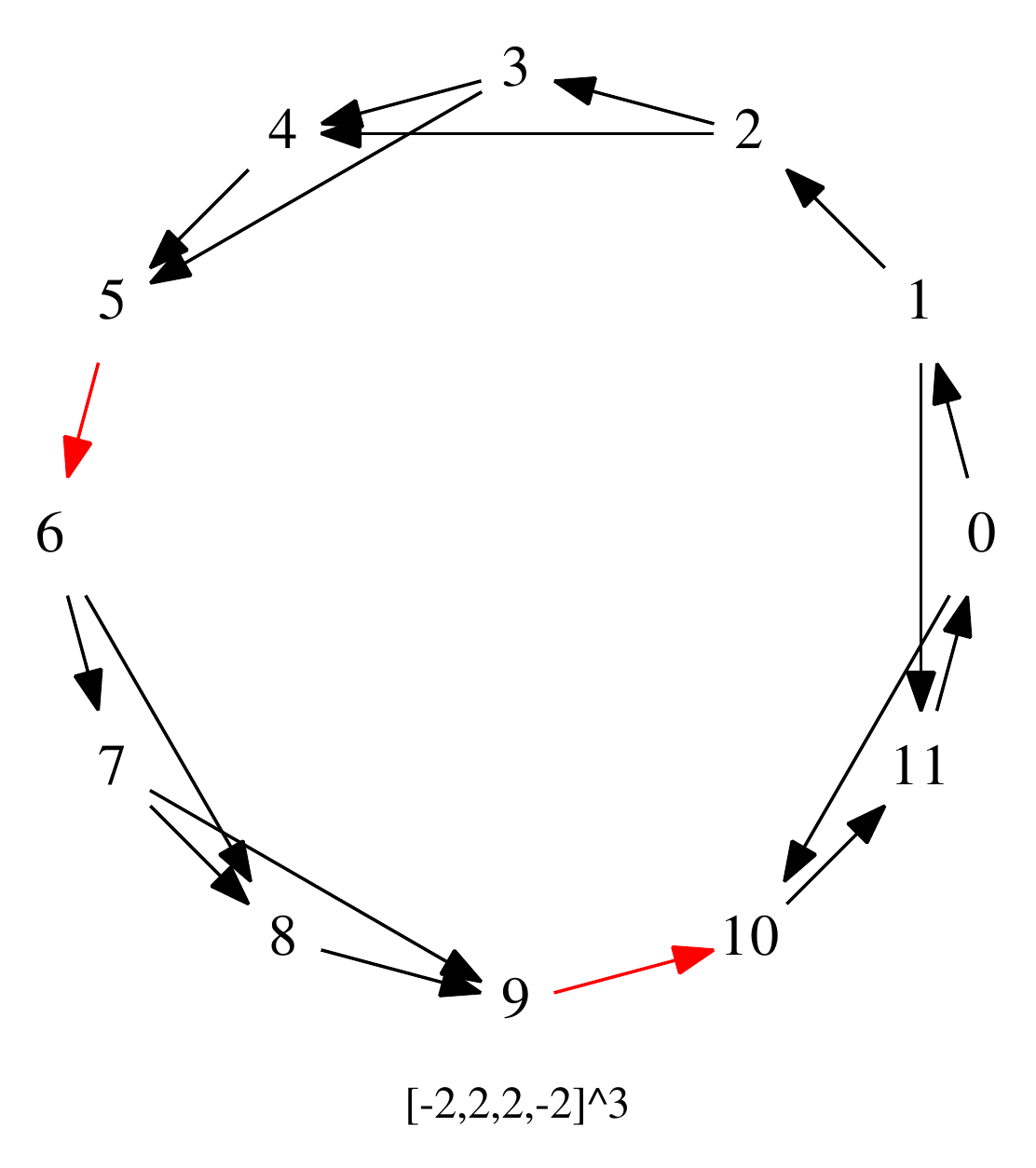}
\caption{$2$-connected graphs on $n=12$ vertices (end).
}
\label{fig.12n2e}
\end{figure}
\clearpage

\subsection{3-connected reducible} 
\begin{figure}
\includegraphics[scale=0.45]{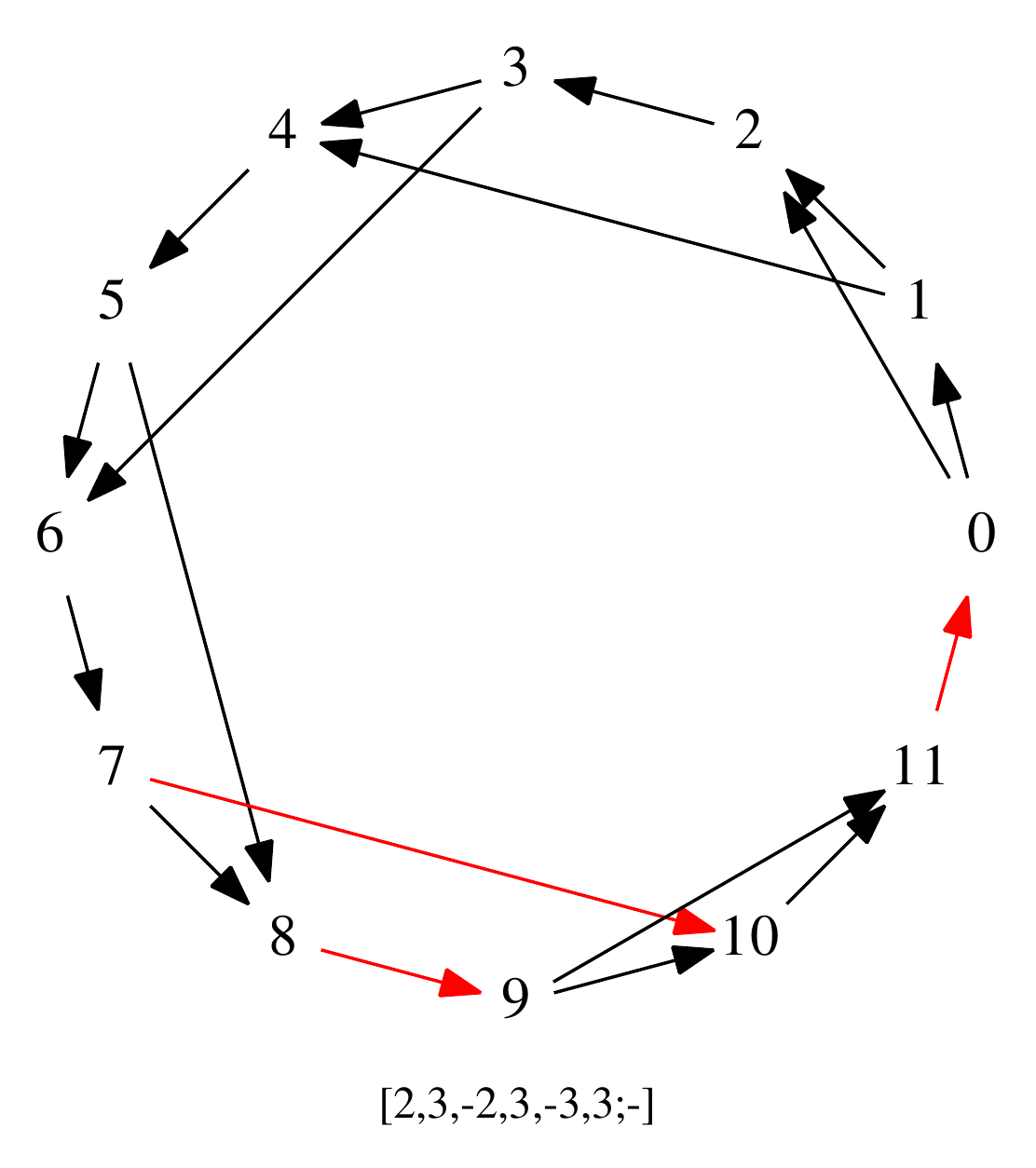}
\includegraphics[scale=0.45]{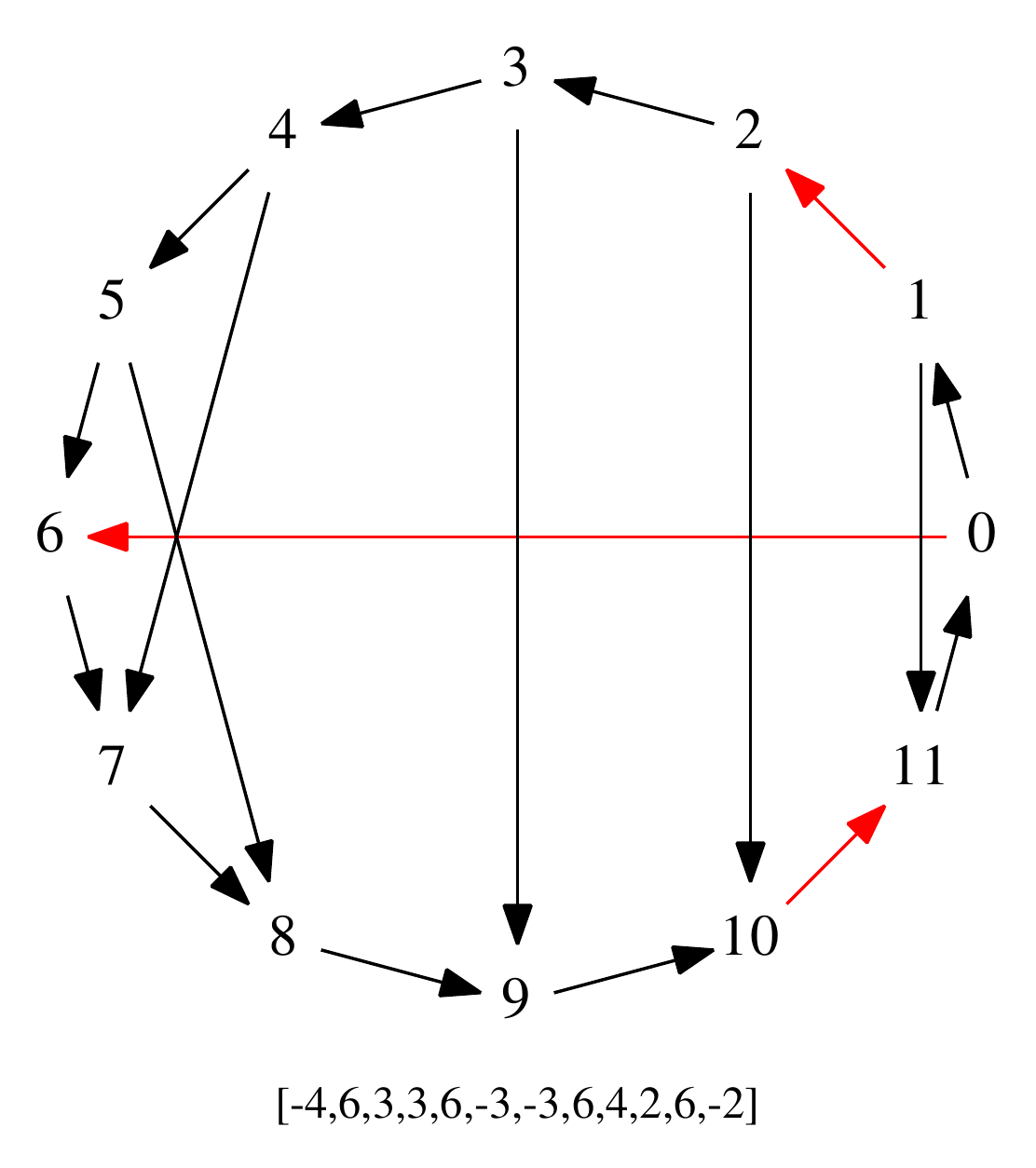}
\includegraphics[scale=0.45]{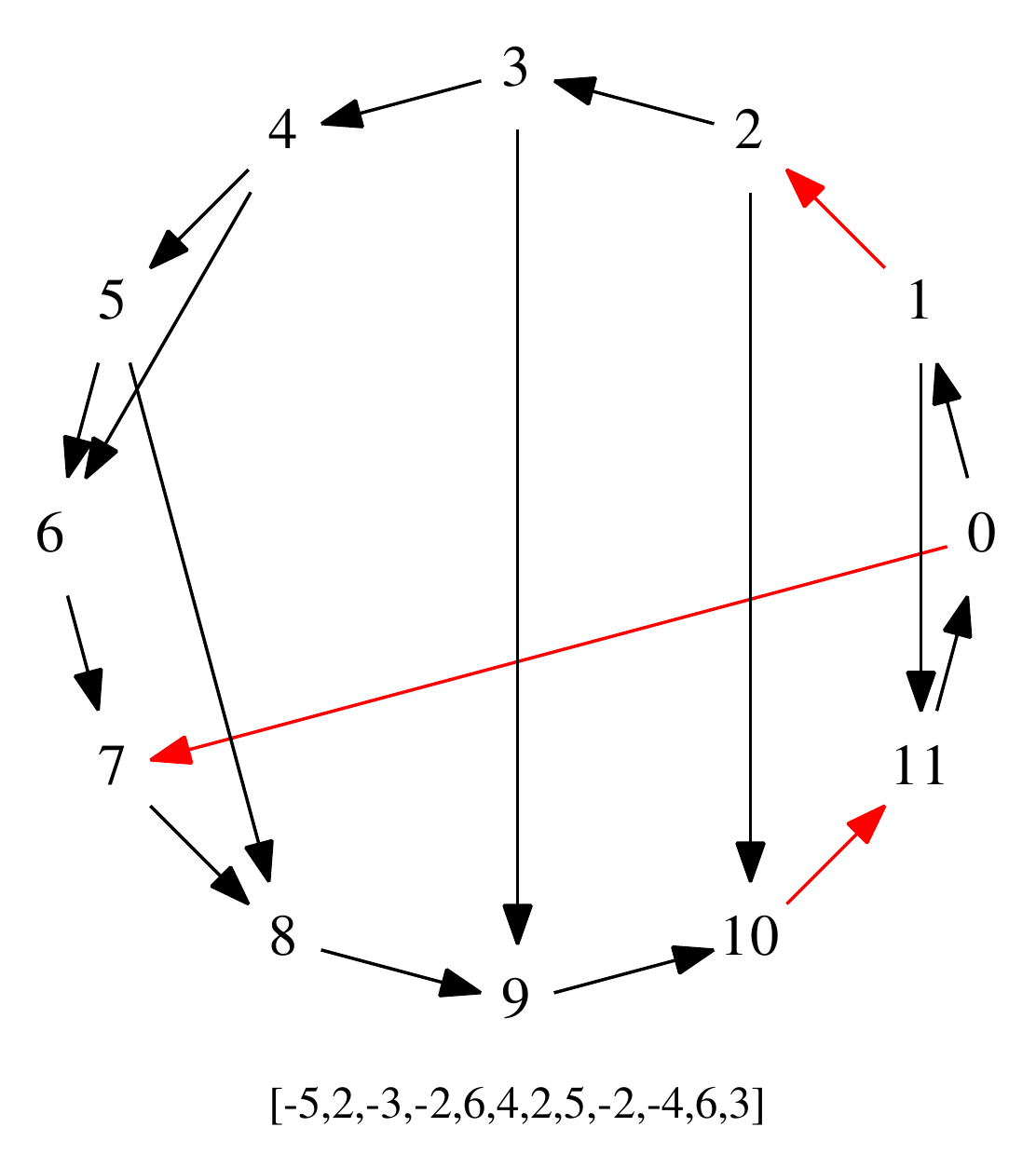}
\includegraphics[scale=0.45]{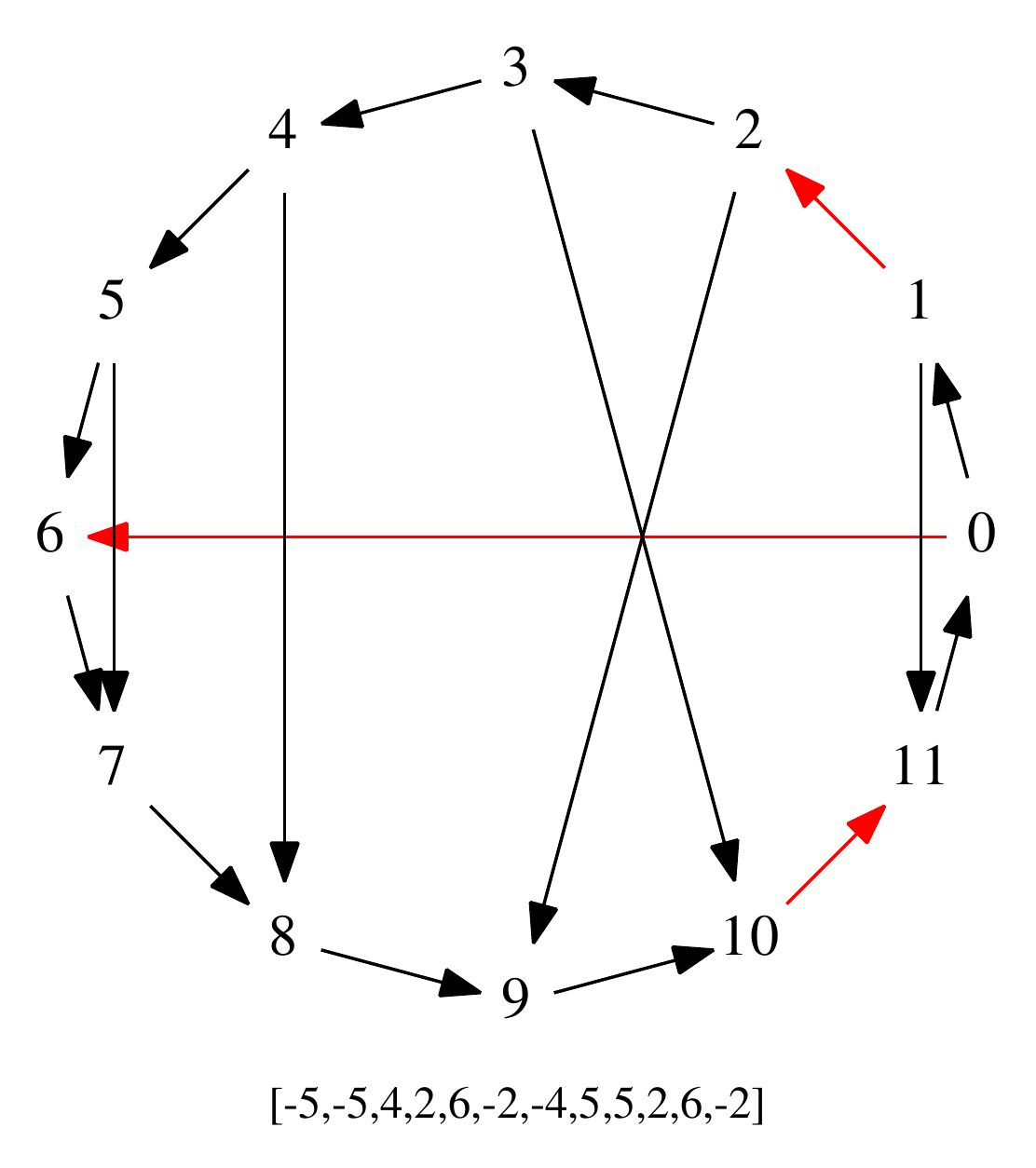}
\includegraphics[scale=0.45]{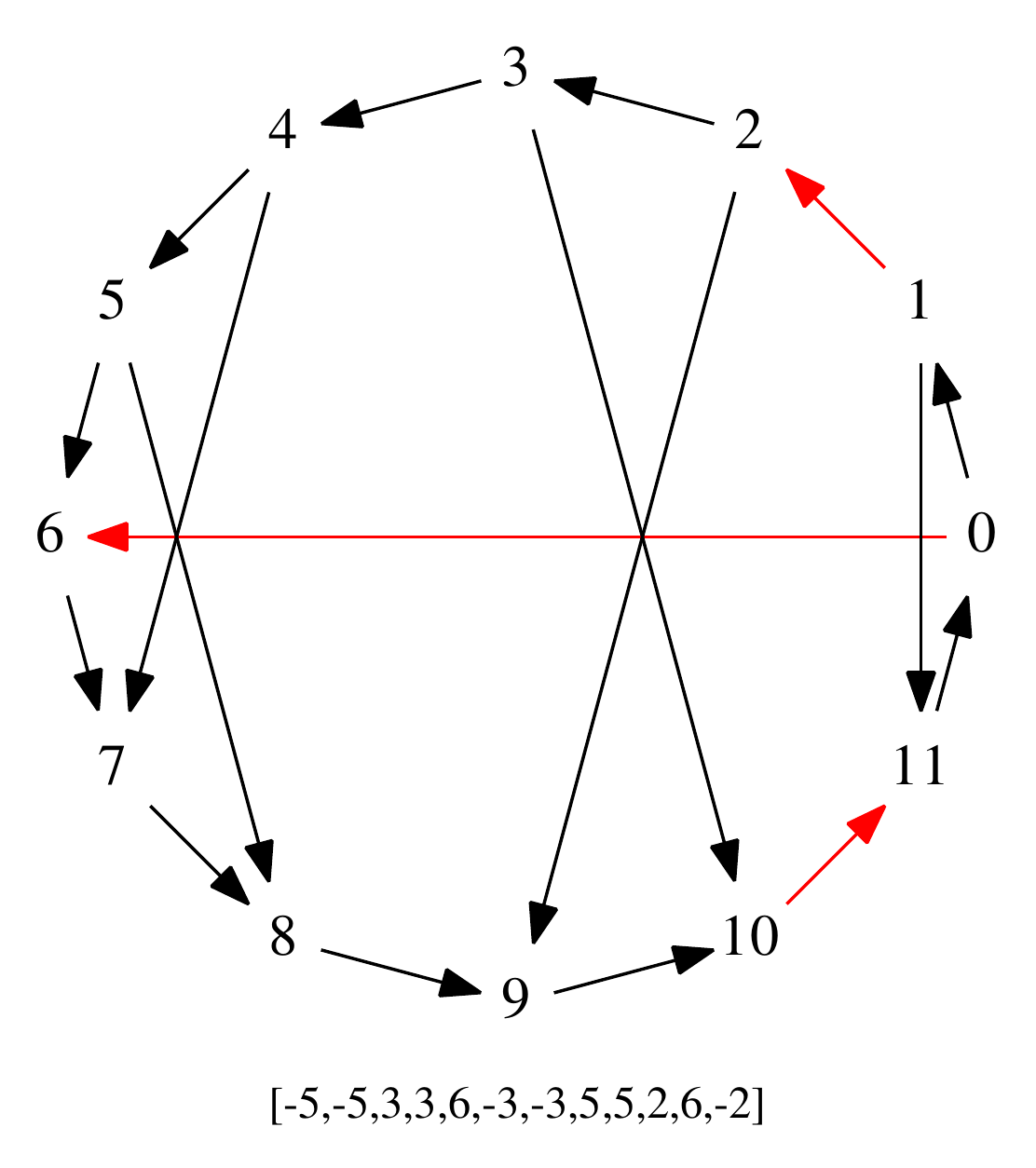}
\includegraphics[scale=0.45]{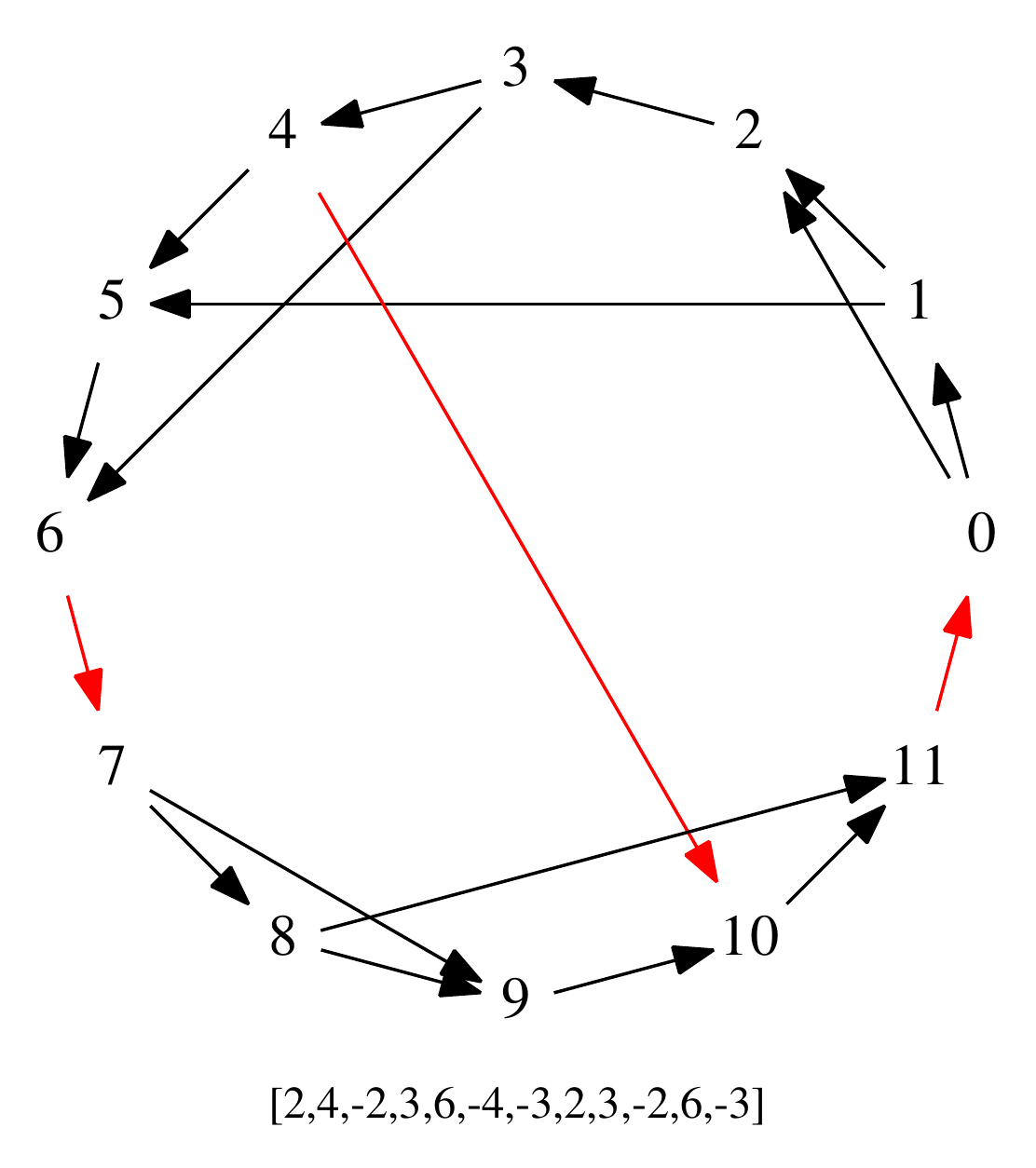}
\includegraphics[scale=0.45]{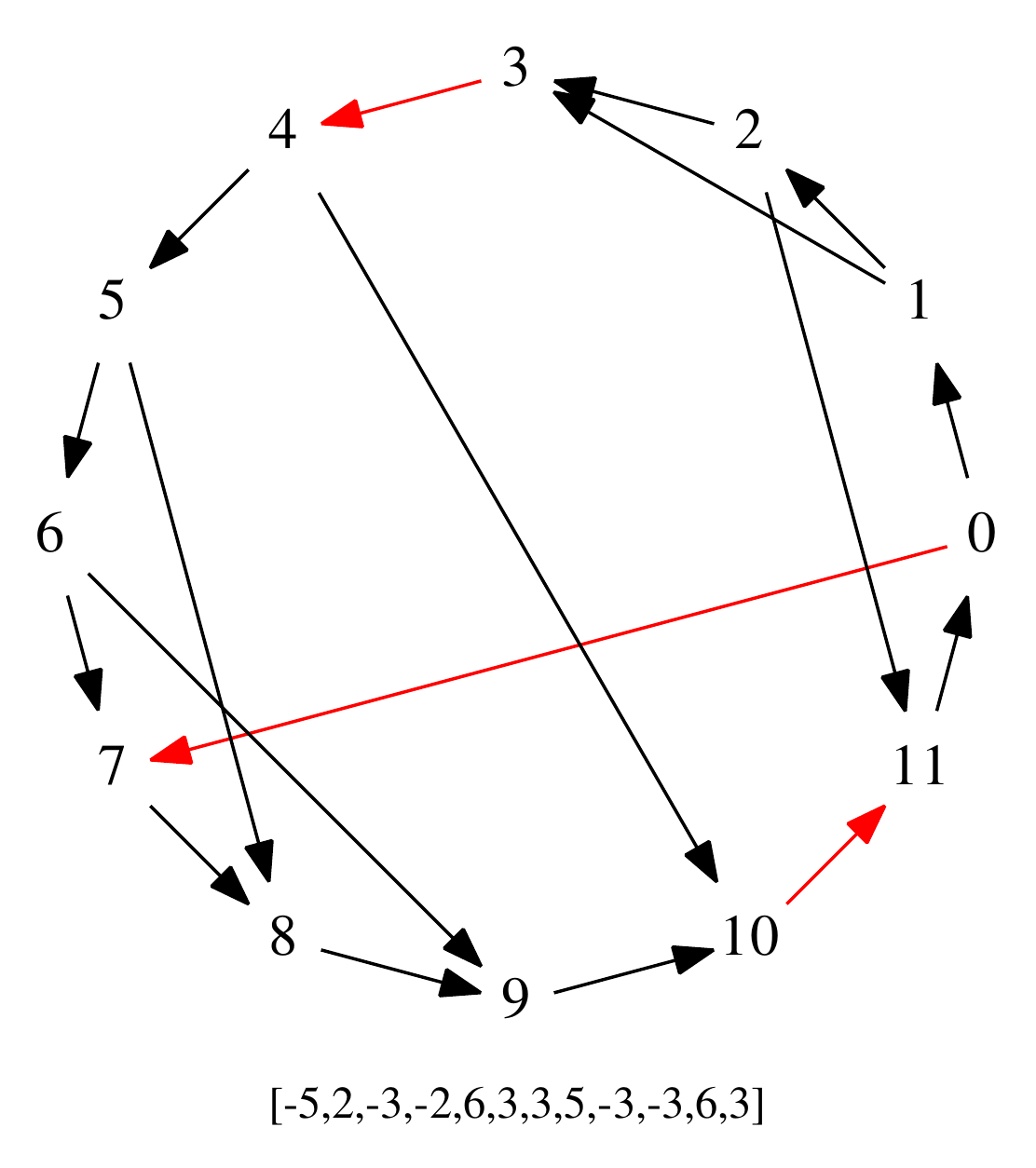}
\includegraphics[scale=0.45]{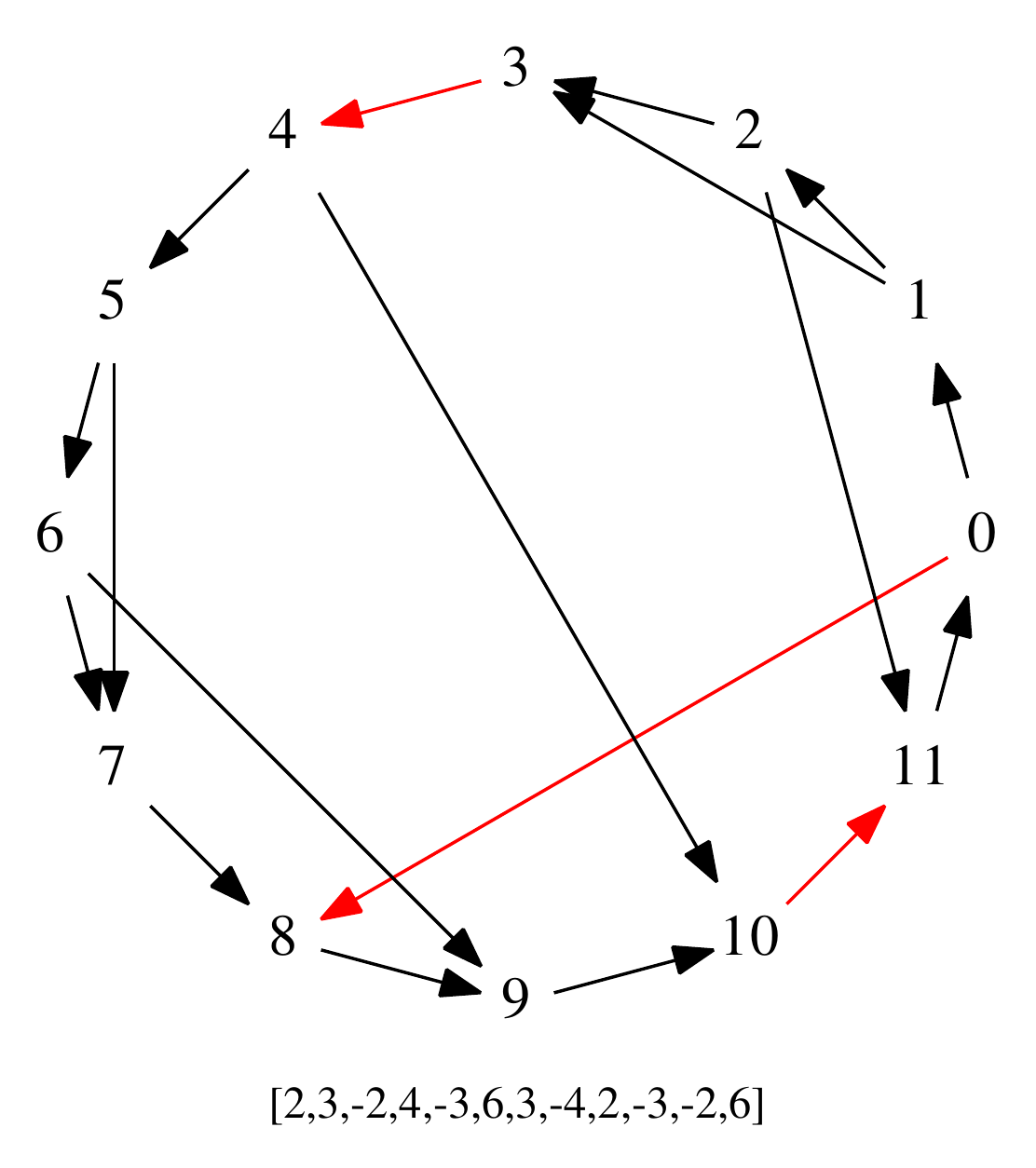}
\includegraphics[scale=0.45]{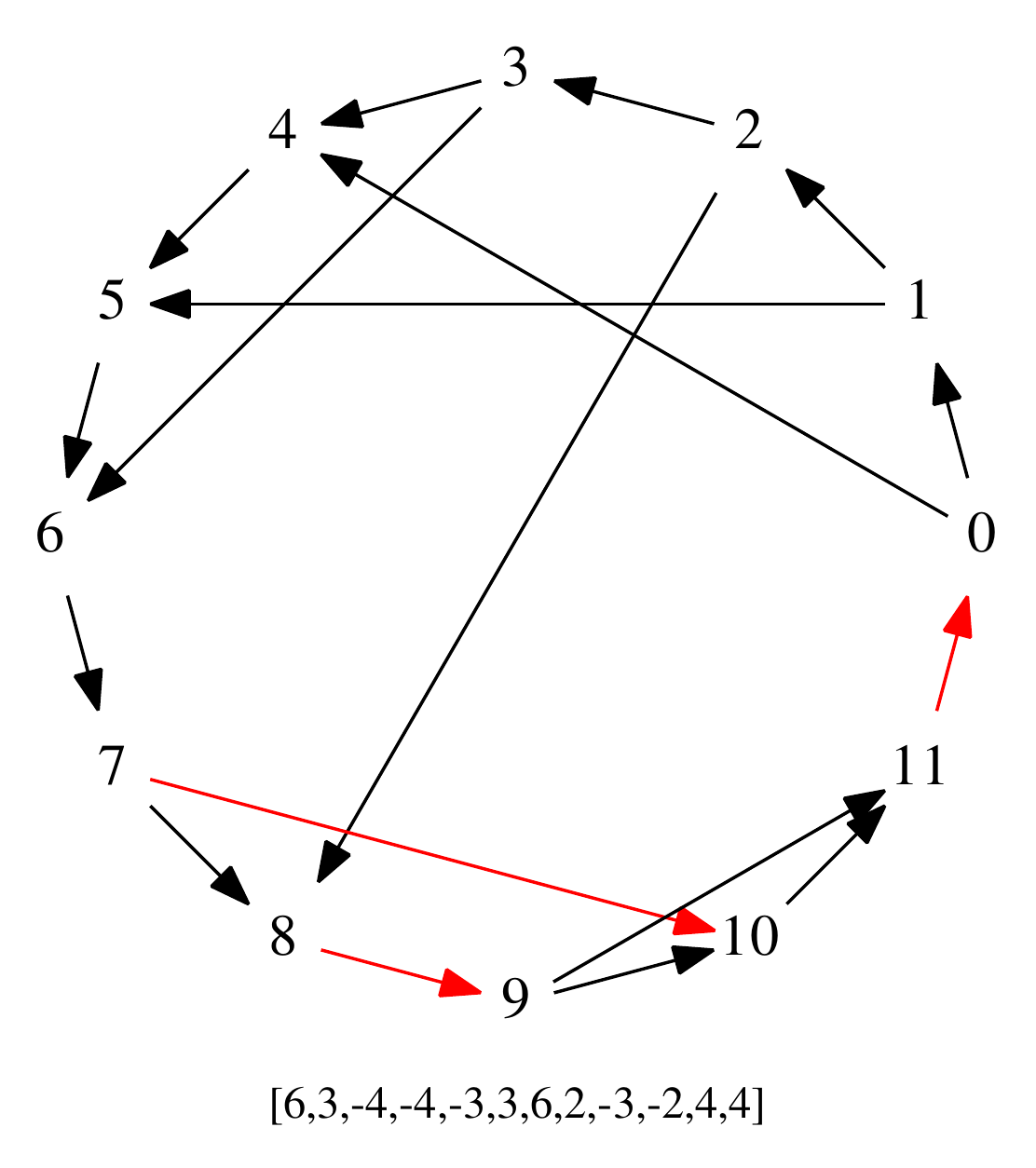}
\includegraphics[scale=0.45]{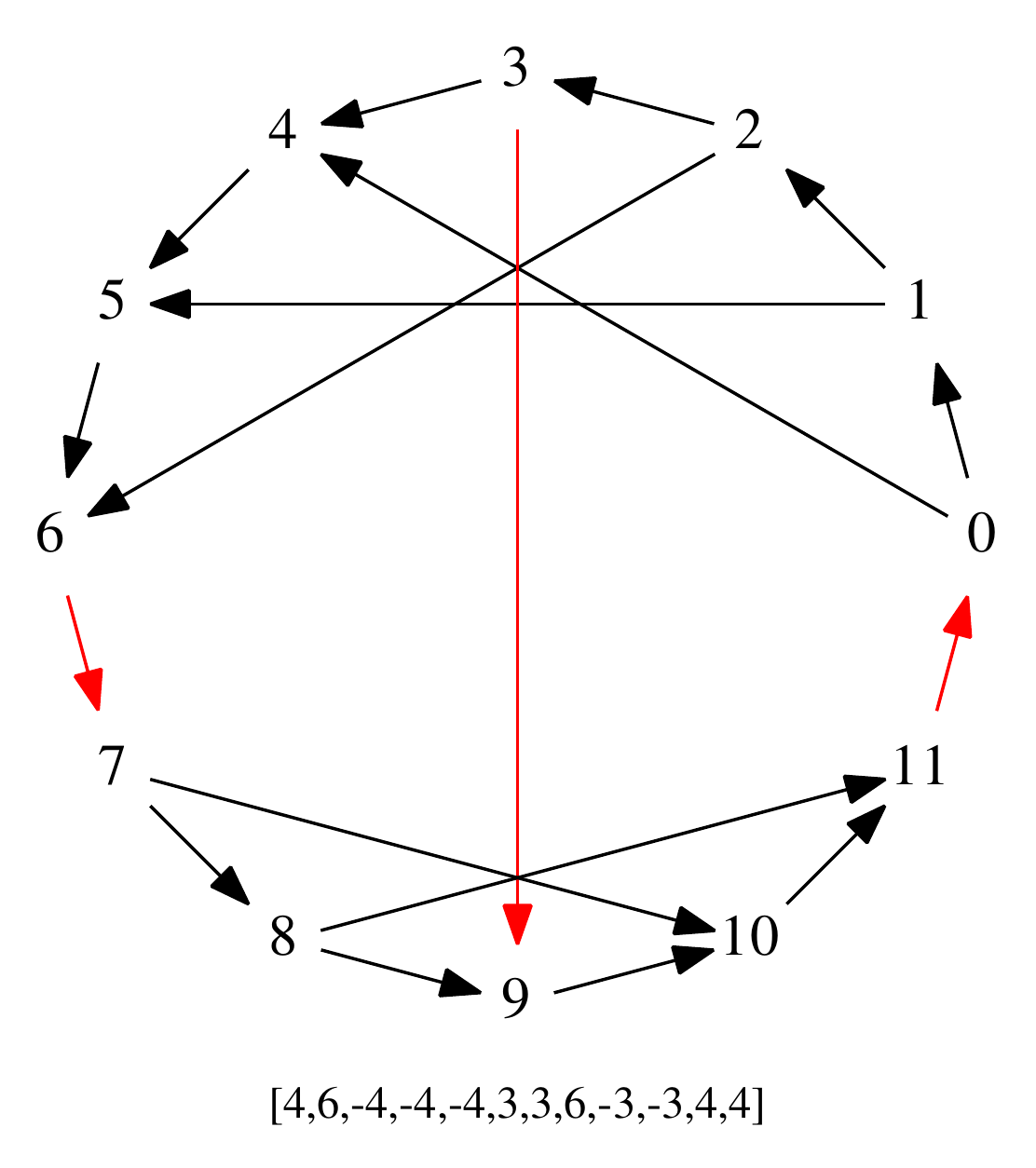}
\includegraphics[scale=0.45]{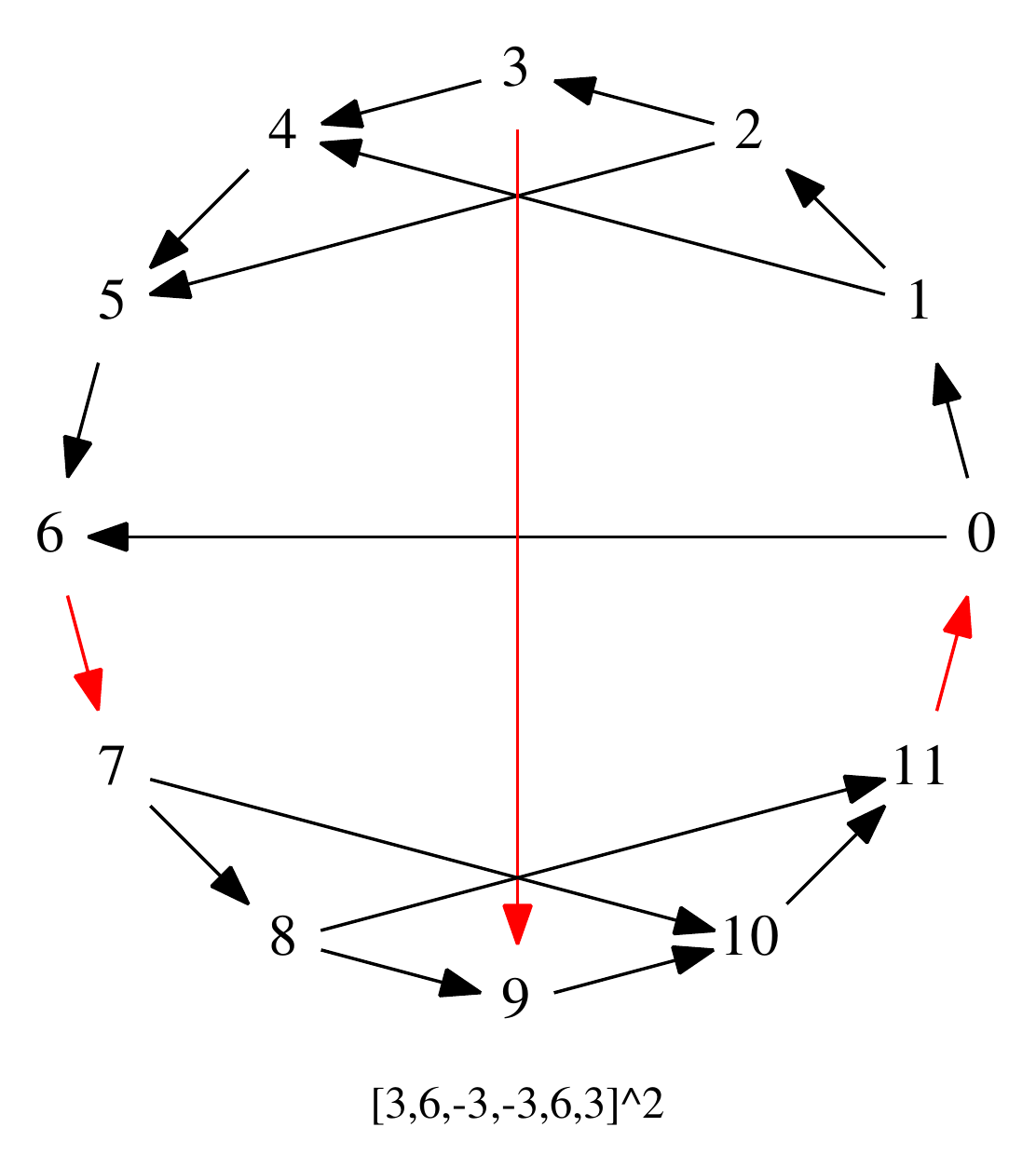}
\includegraphics[scale=0.45]{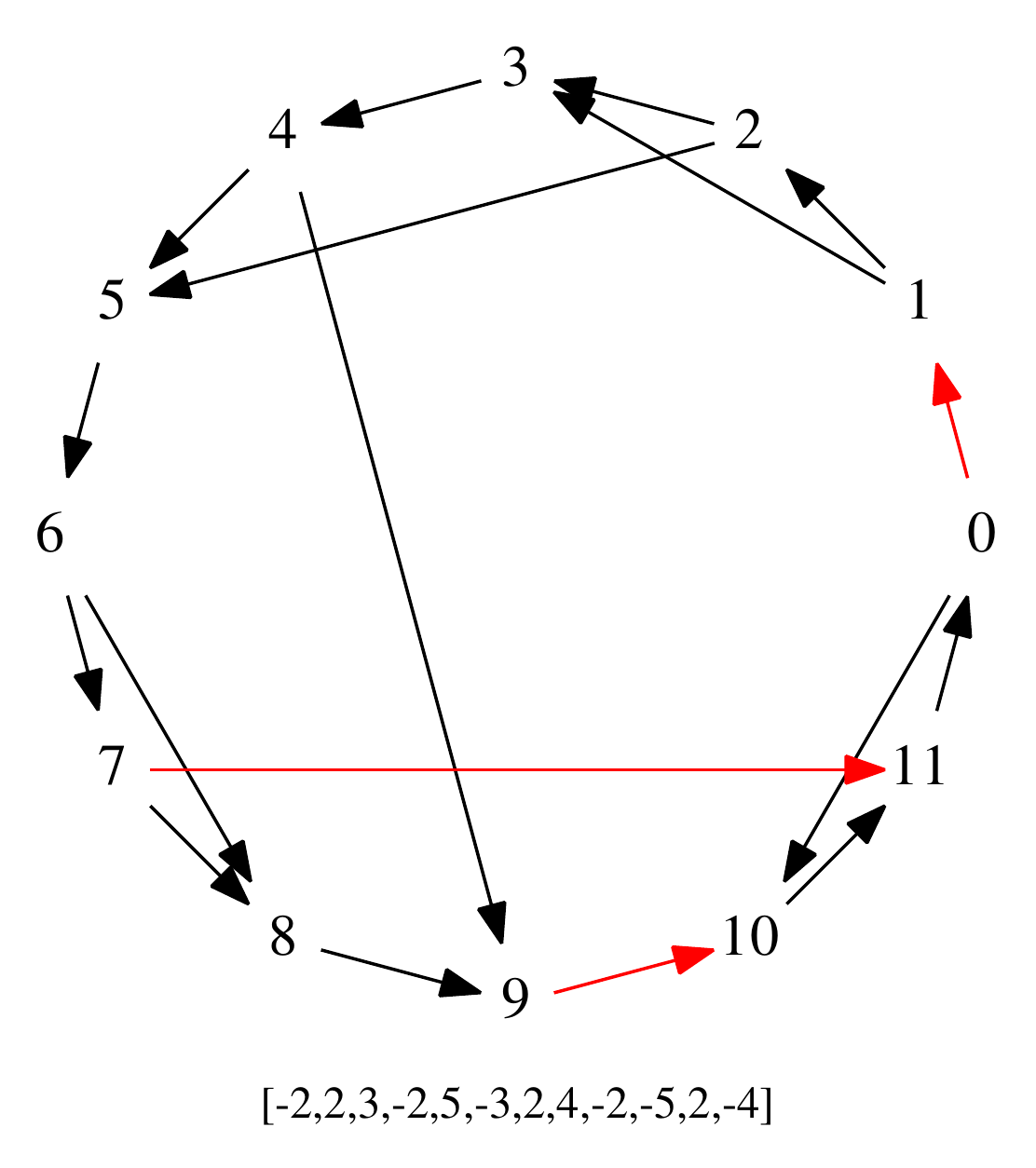}
\caption{$3$-connected graphs on $n=12$ vertices (start).
}
\label{fig.12n3s}
\end{figure}
\begin{figure}
\includegraphics[scale=0.45]{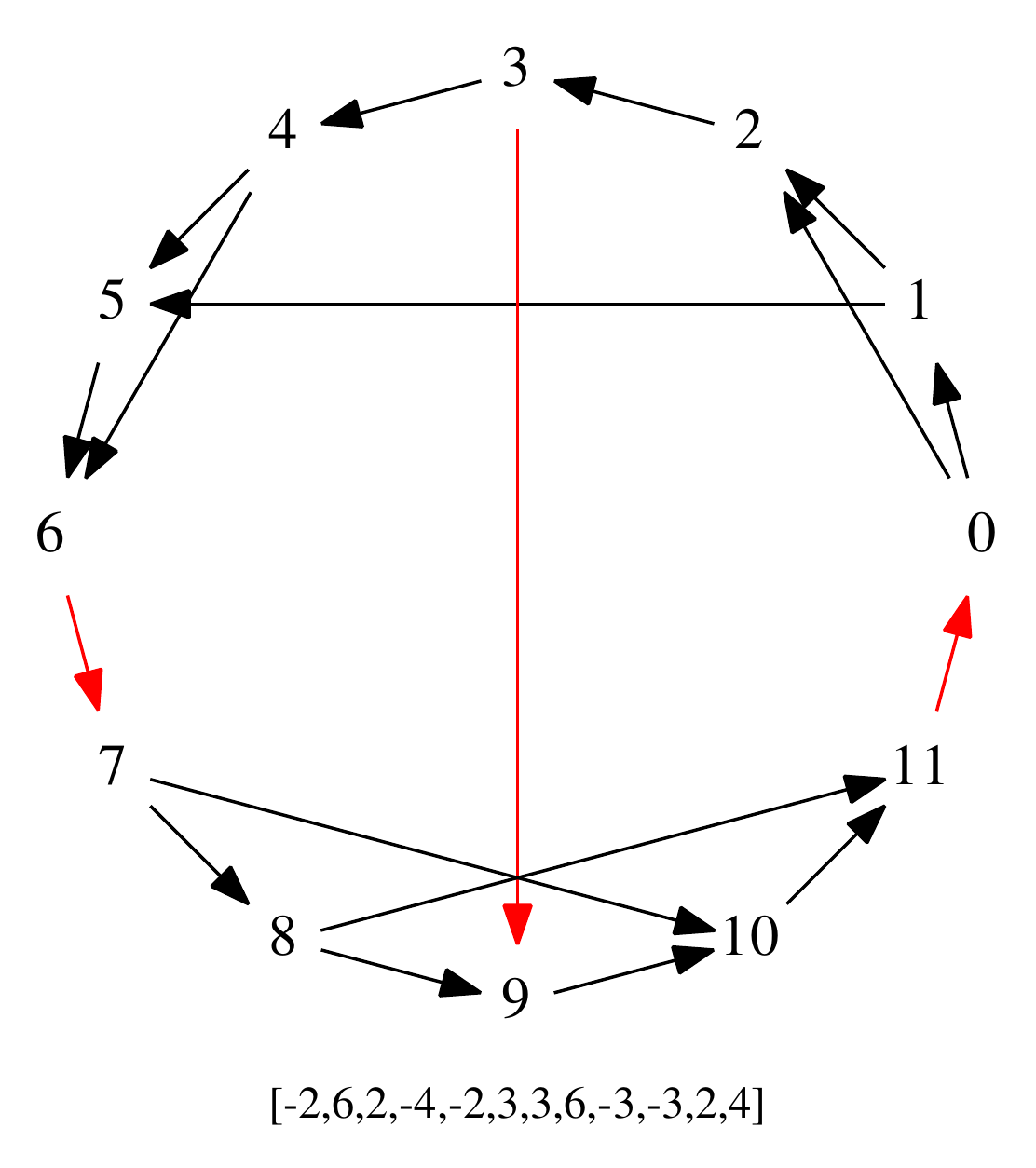}
\includegraphics[scale=0.45]{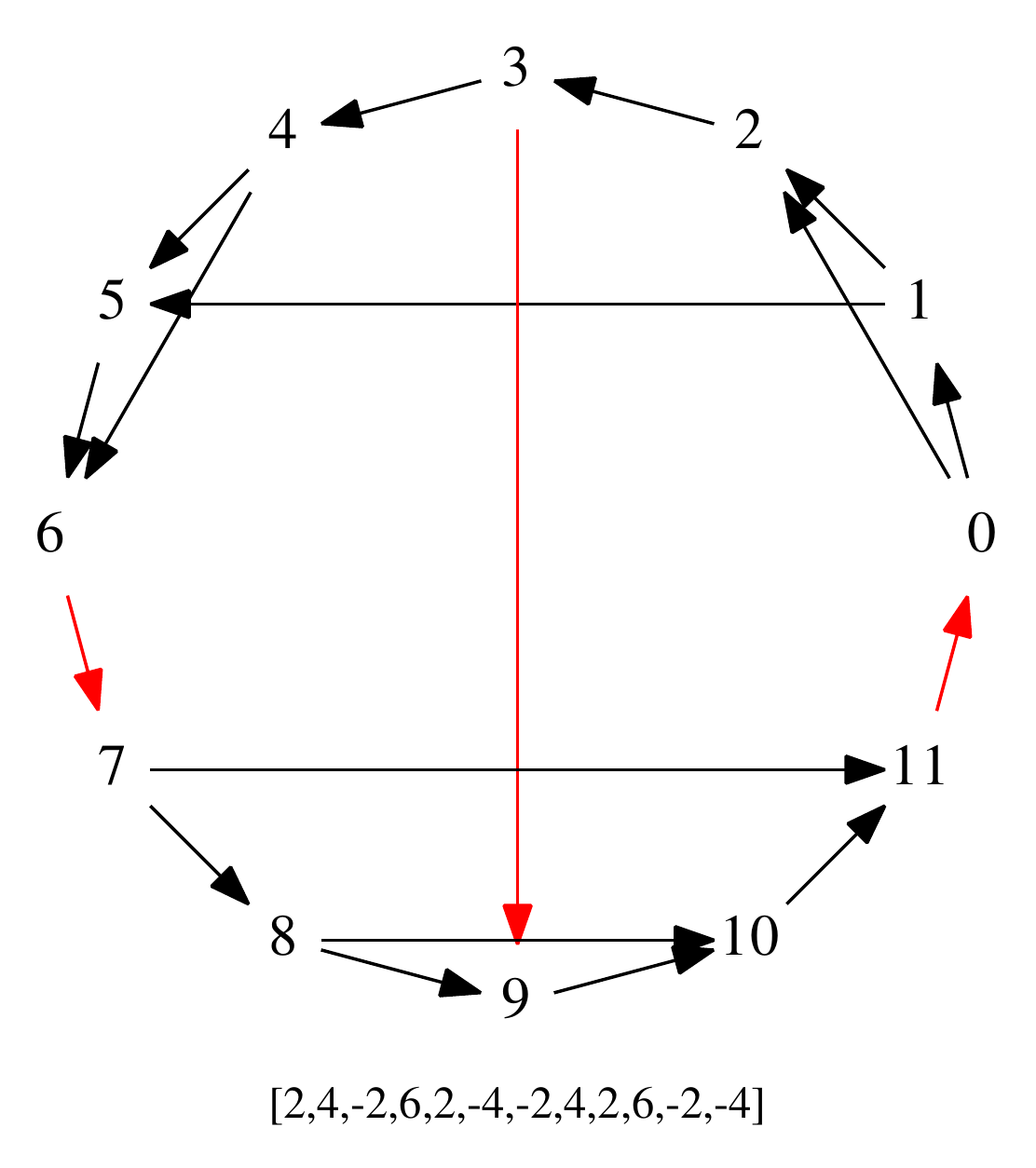}
\includegraphics[scale=0.45]{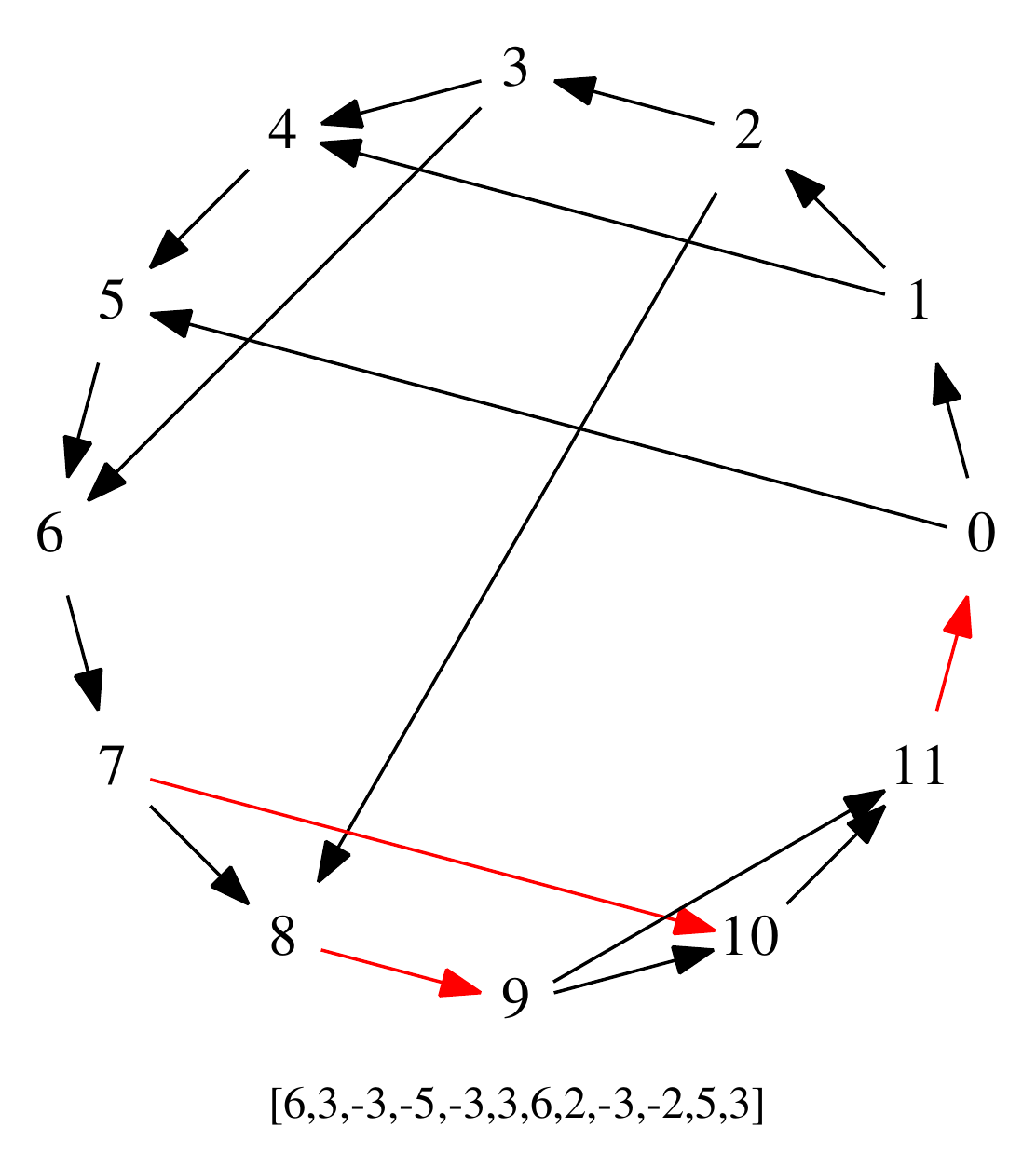}
\includegraphics[scale=0.45]{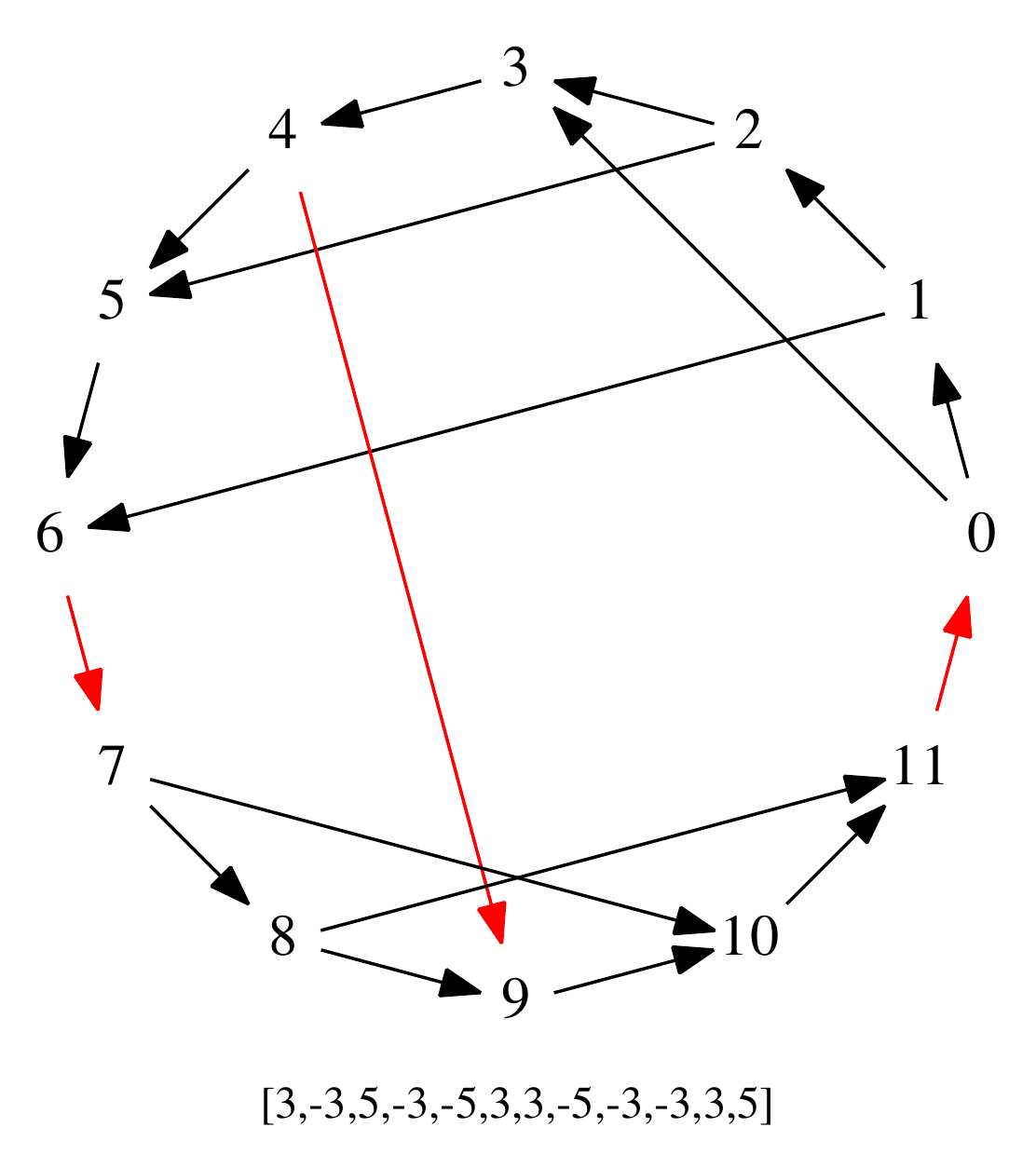}
\includegraphics[scale=0.45]{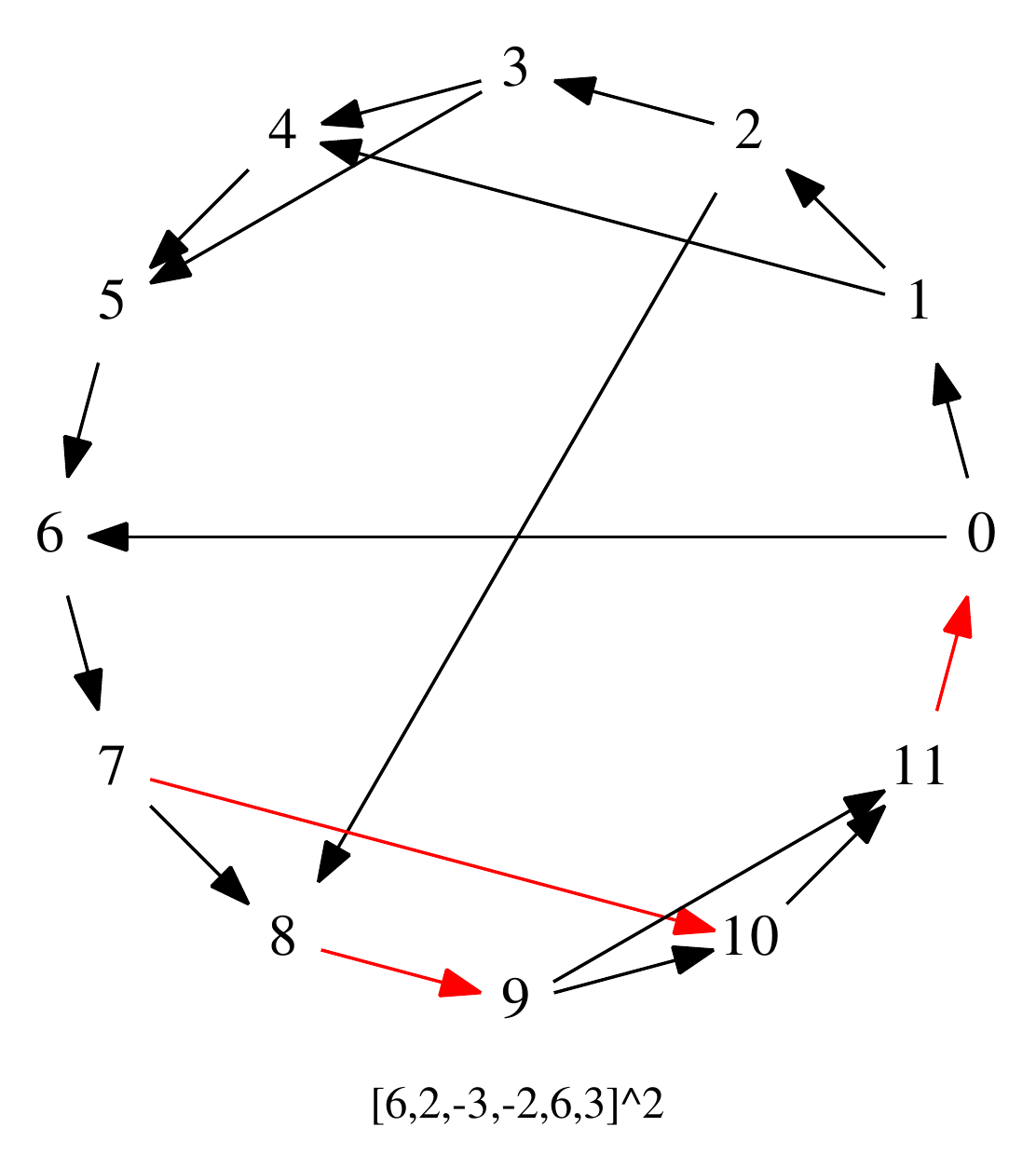}
\includegraphics[scale=0.45]{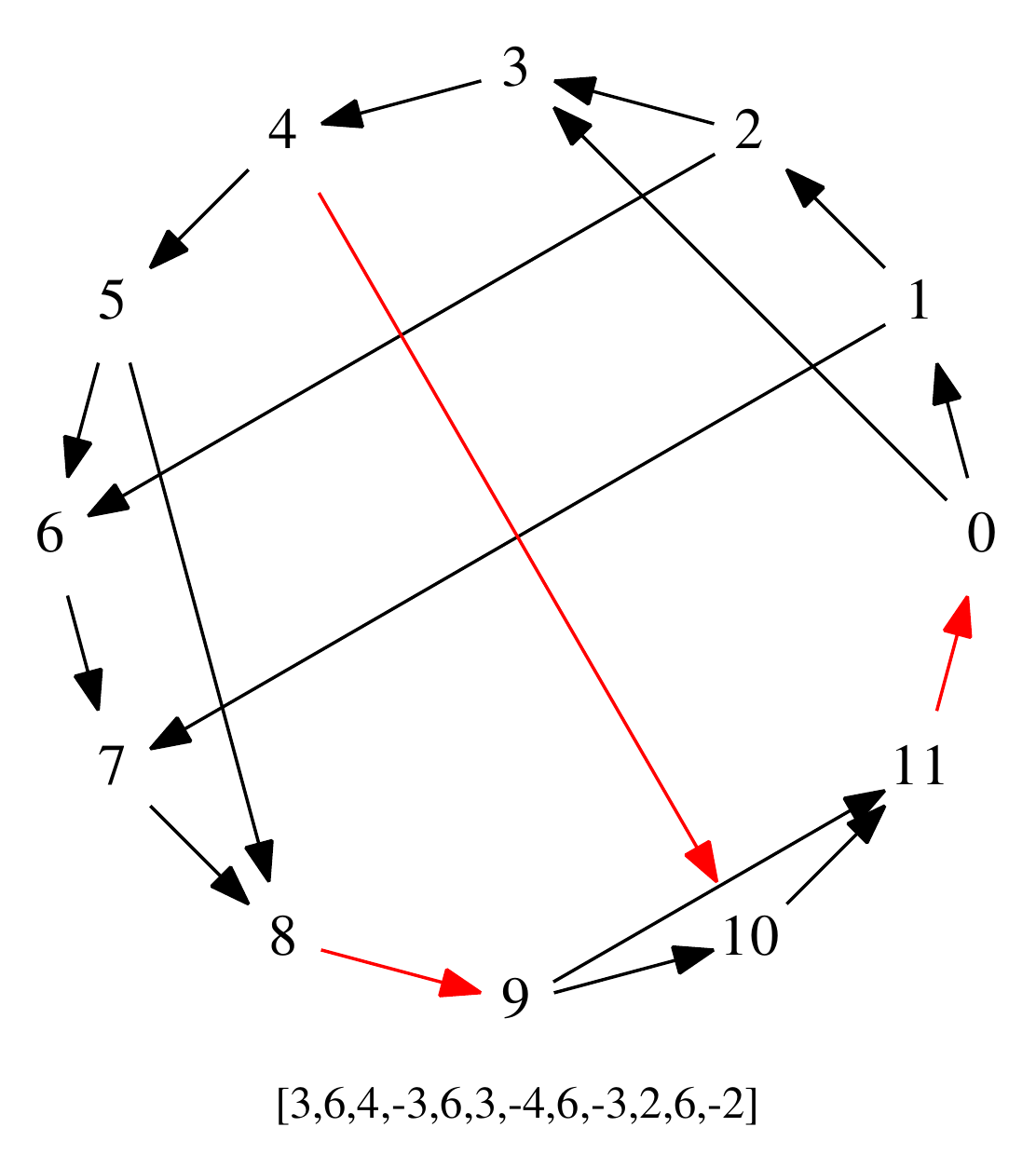}
\includegraphics[scale=0.45]{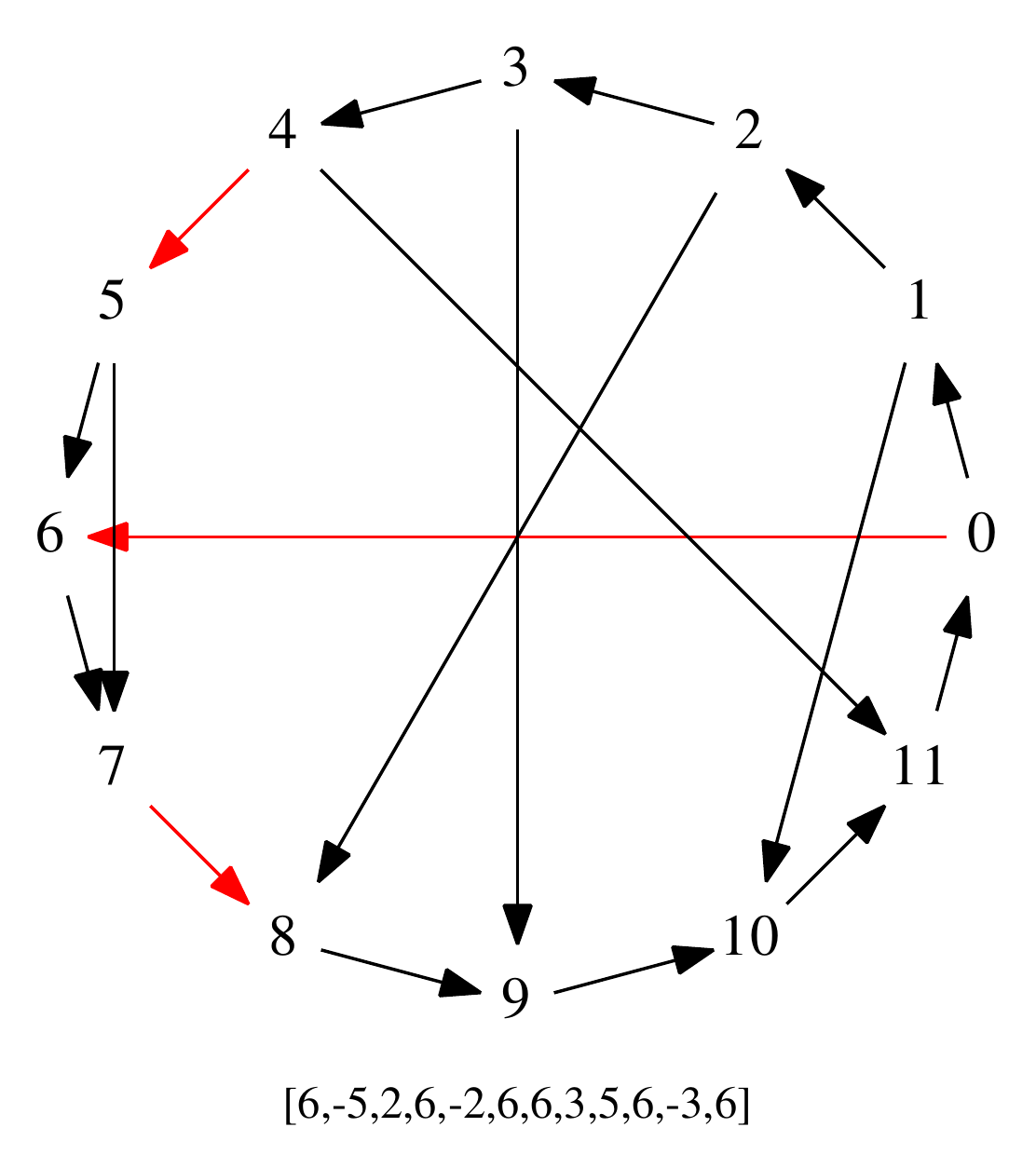}
\includegraphics[scale=0.45]{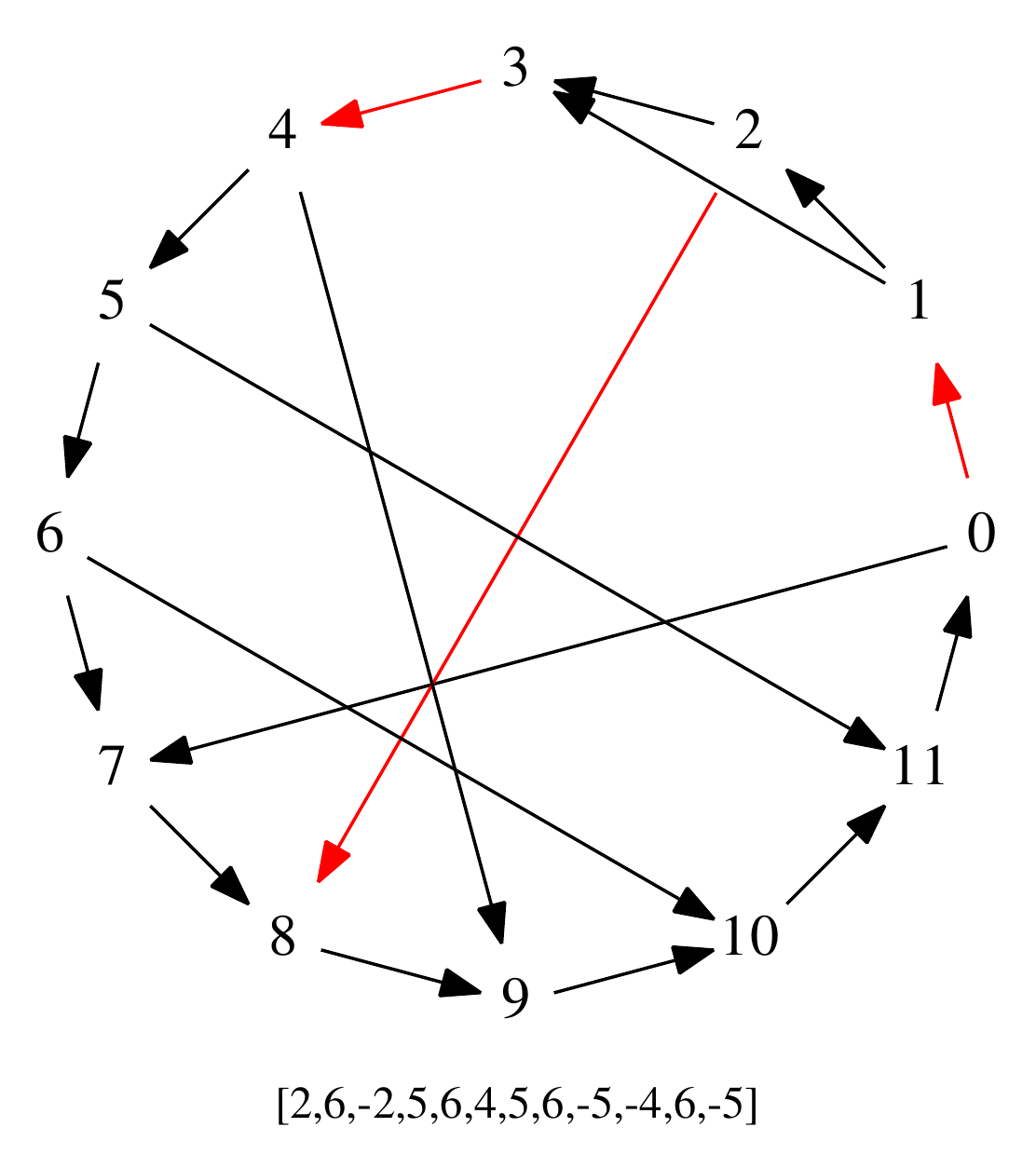}
\includegraphics[scale=0.45]{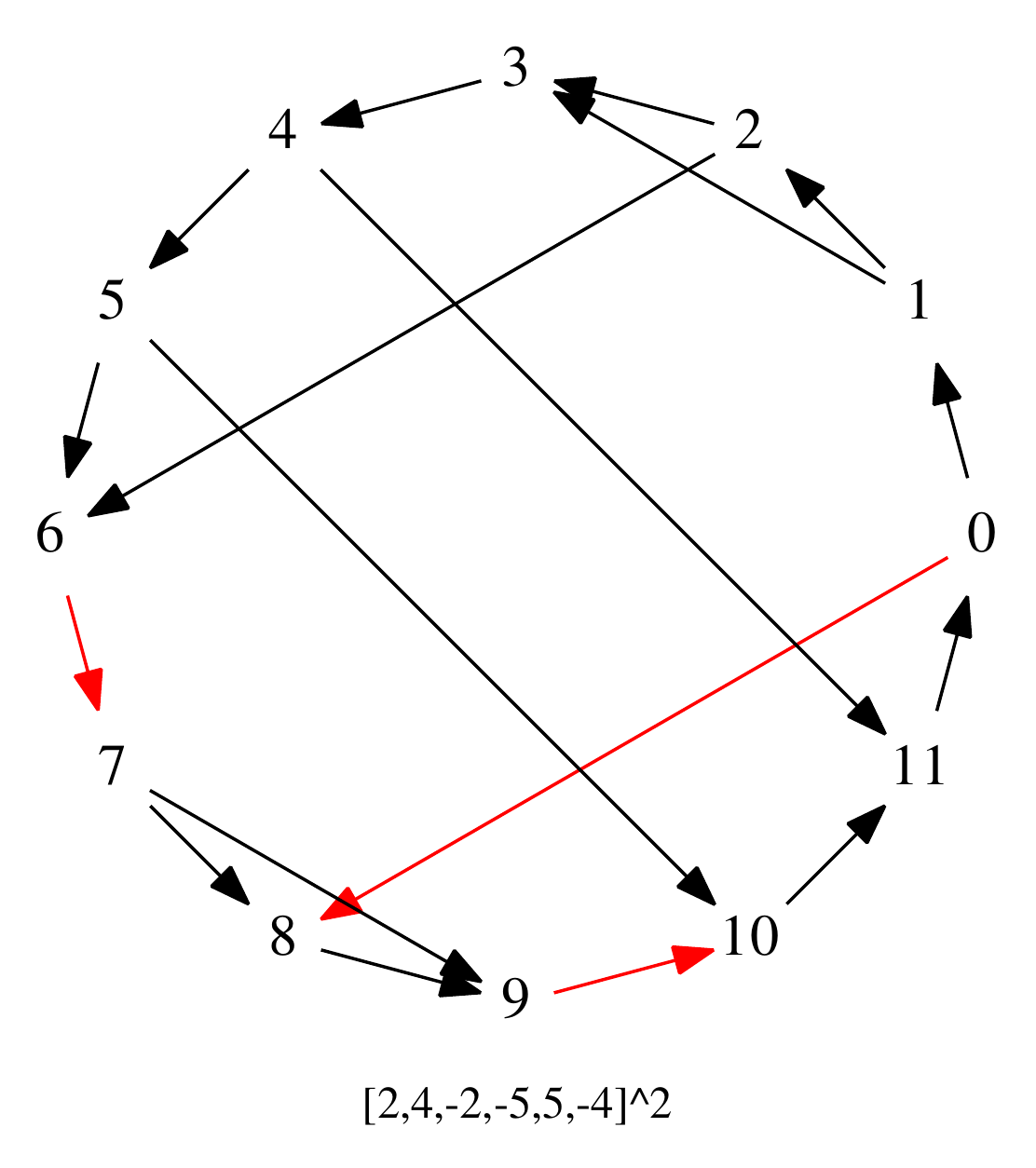}
\includegraphics[scale=0.45]{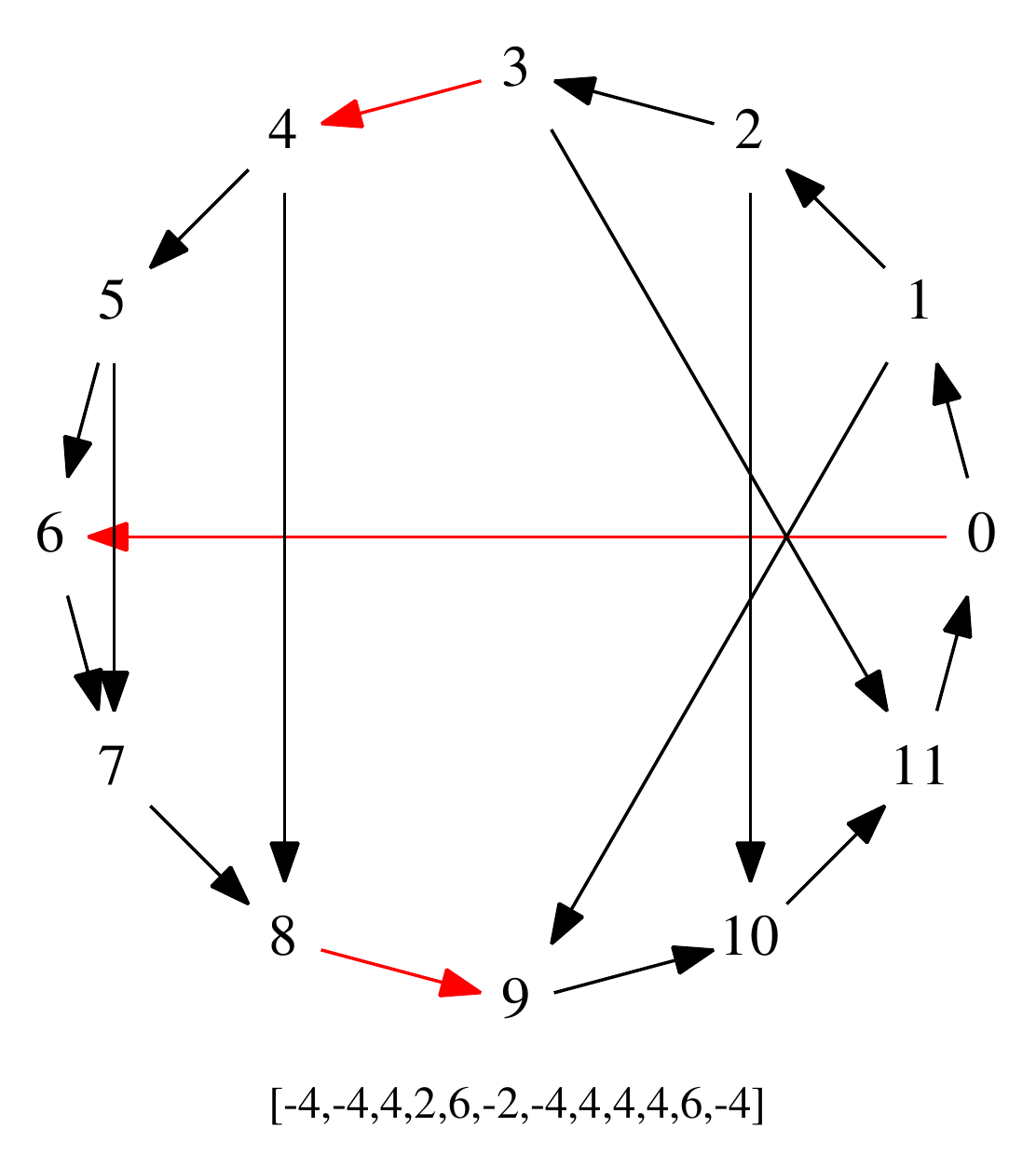}
\includegraphics[scale=0.45]{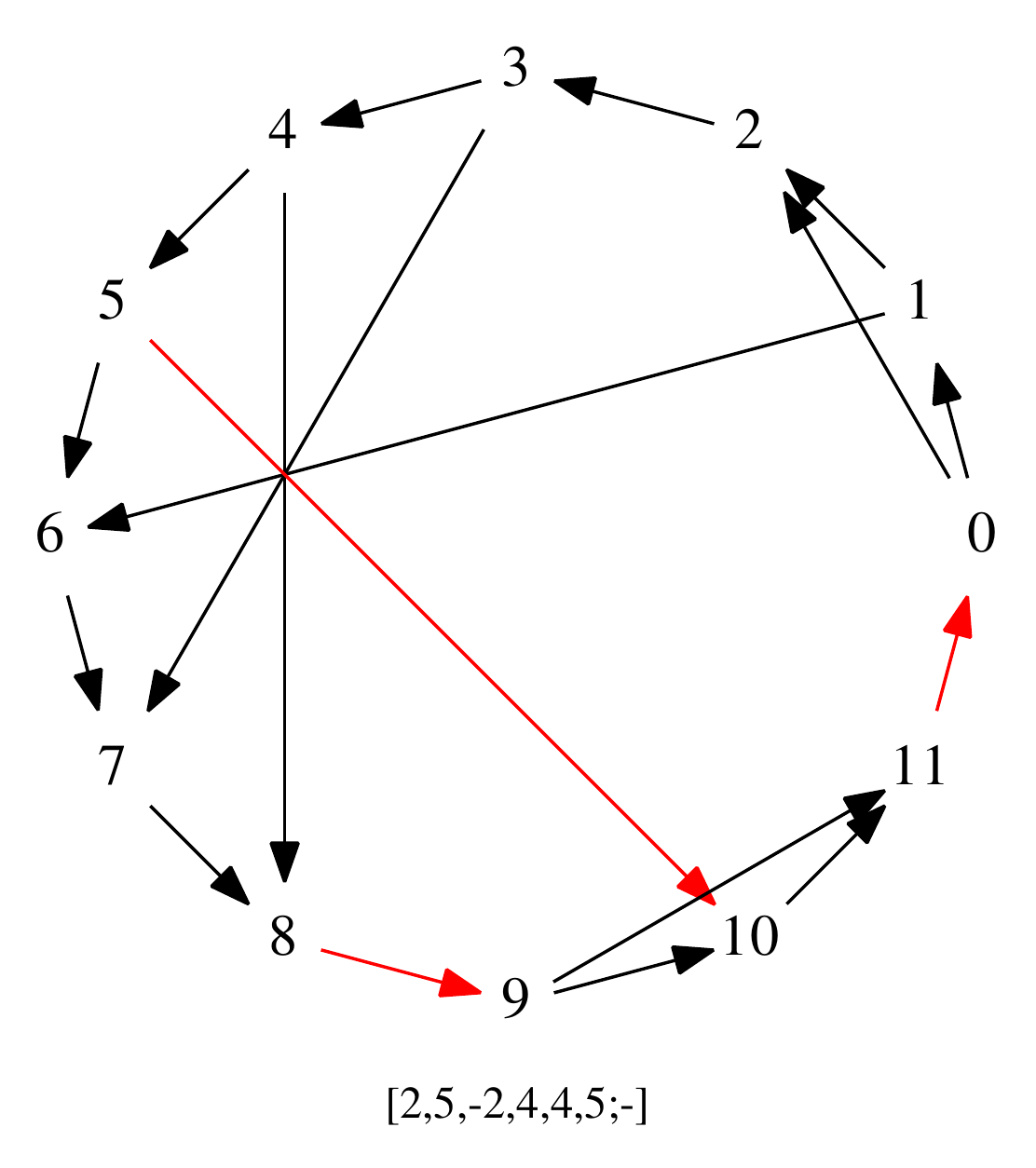}
\includegraphics[scale=0.45]{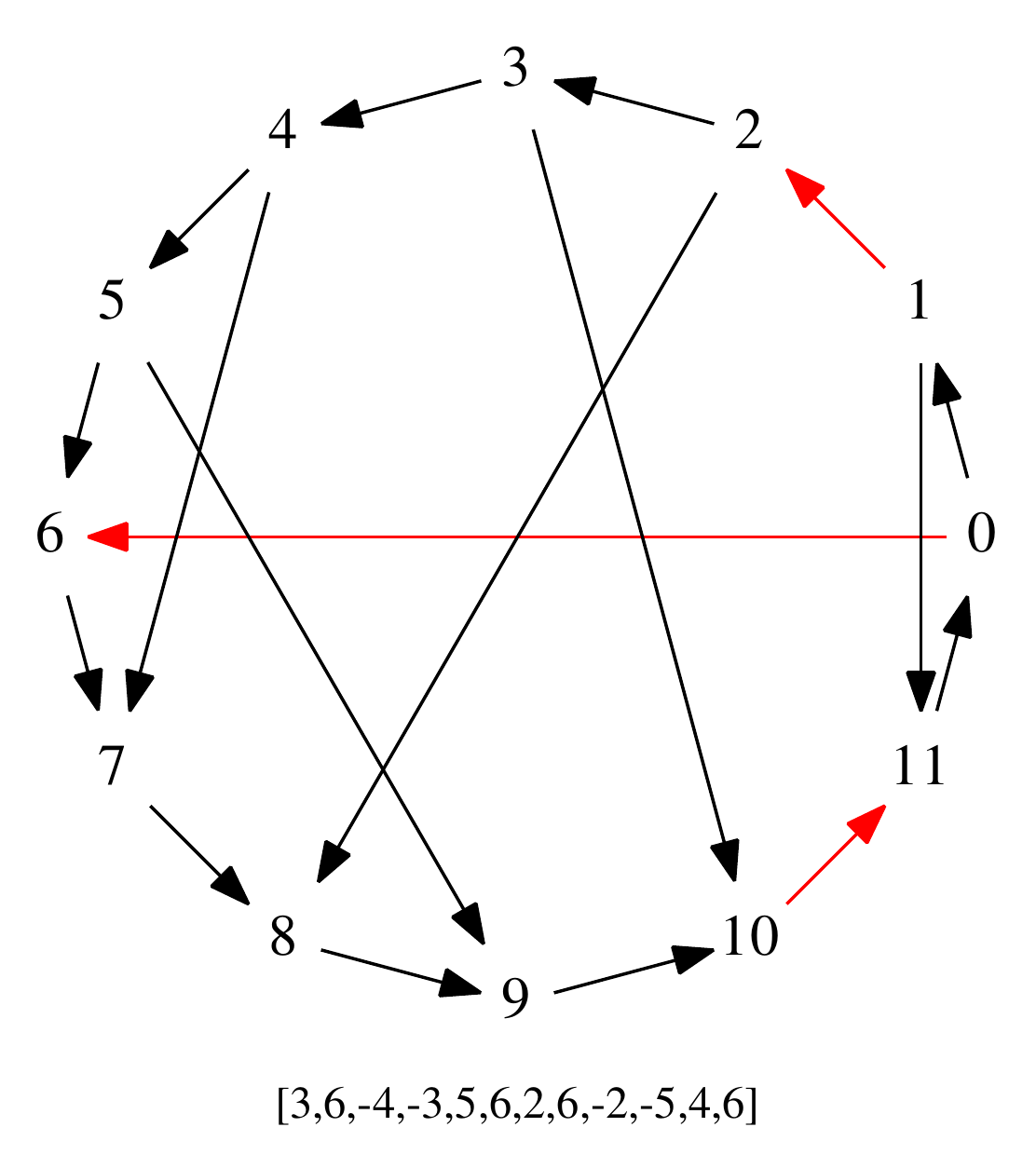}
\caption{$3$-connected graphs on $n=12$ vertices (continued).
}
\label{fig.12n32}
\end{figure}
\begin{figure}
\includegraphics[scale=0.45]{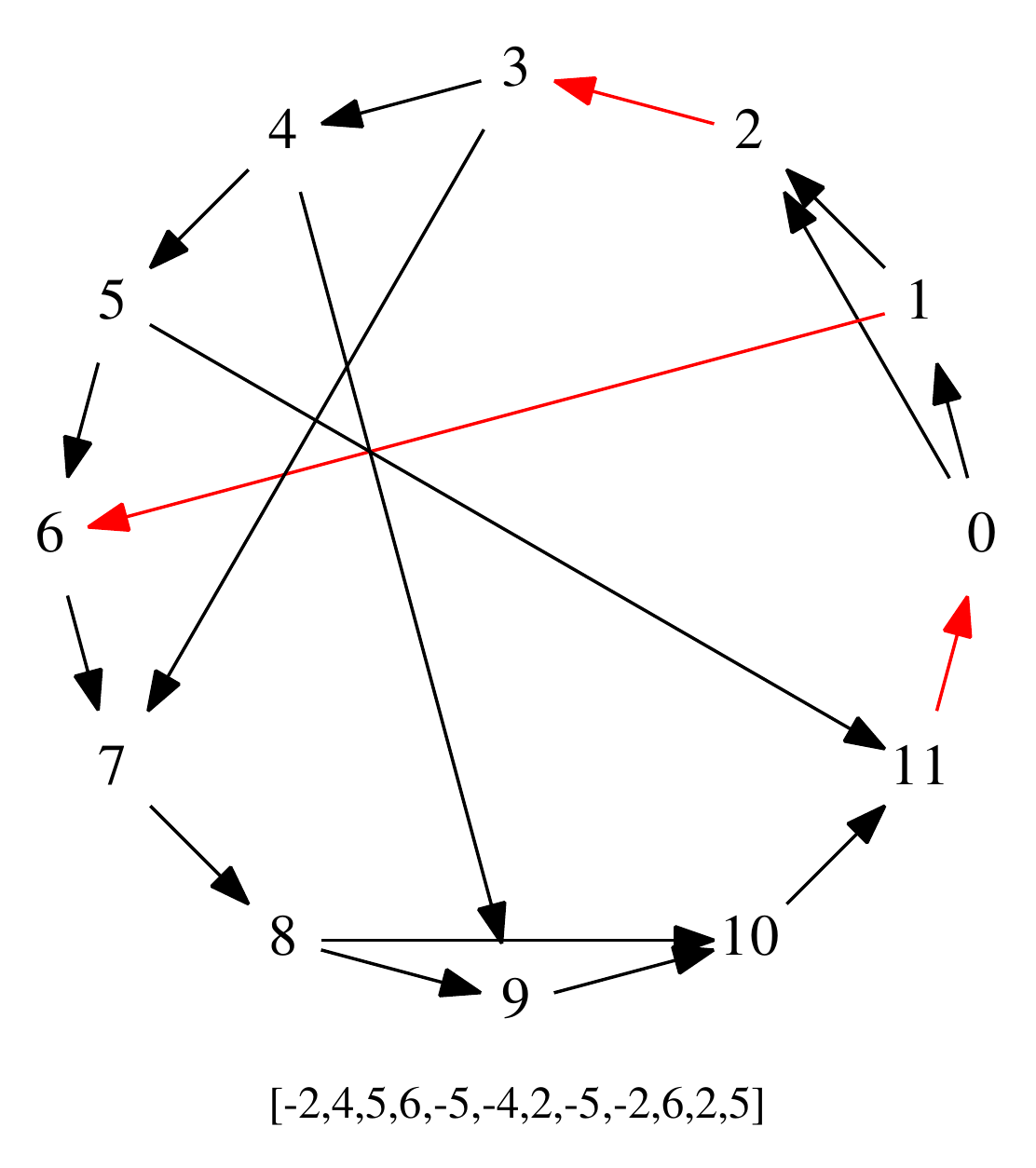}
\includegraphics[scale=0.45]{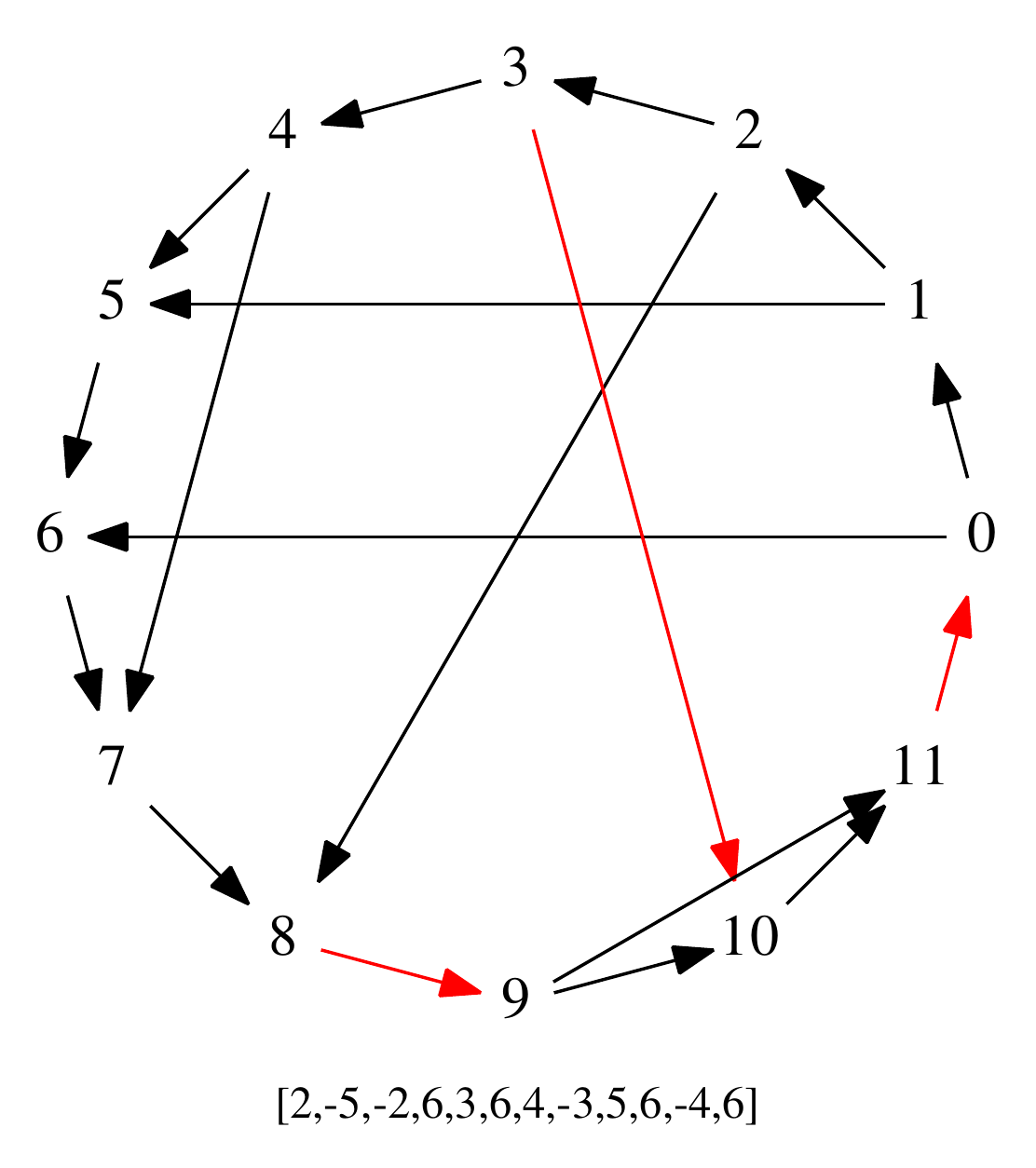}
\includegraphics[scale=0.45]{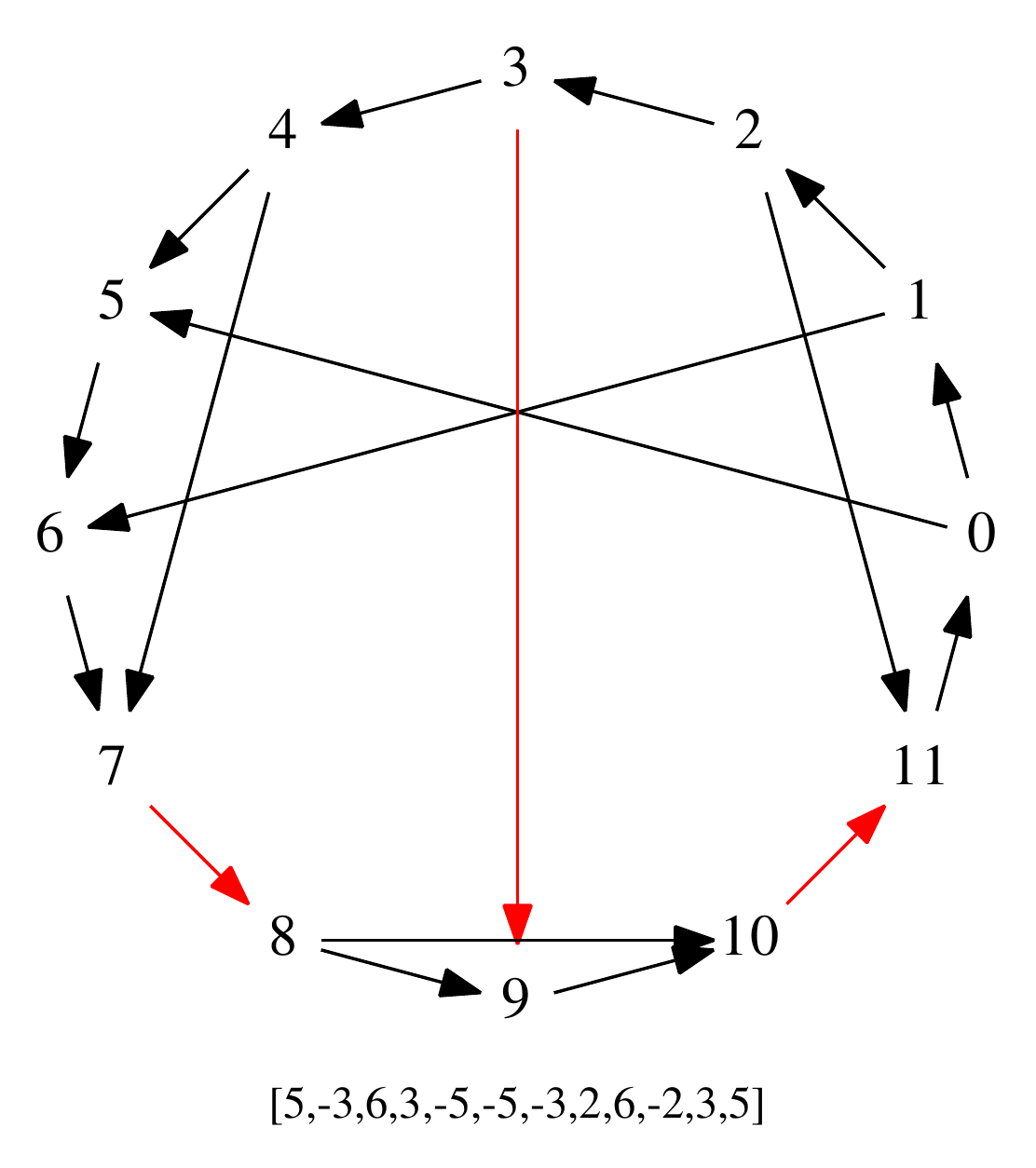}
\includegraphics[scale=0.45]{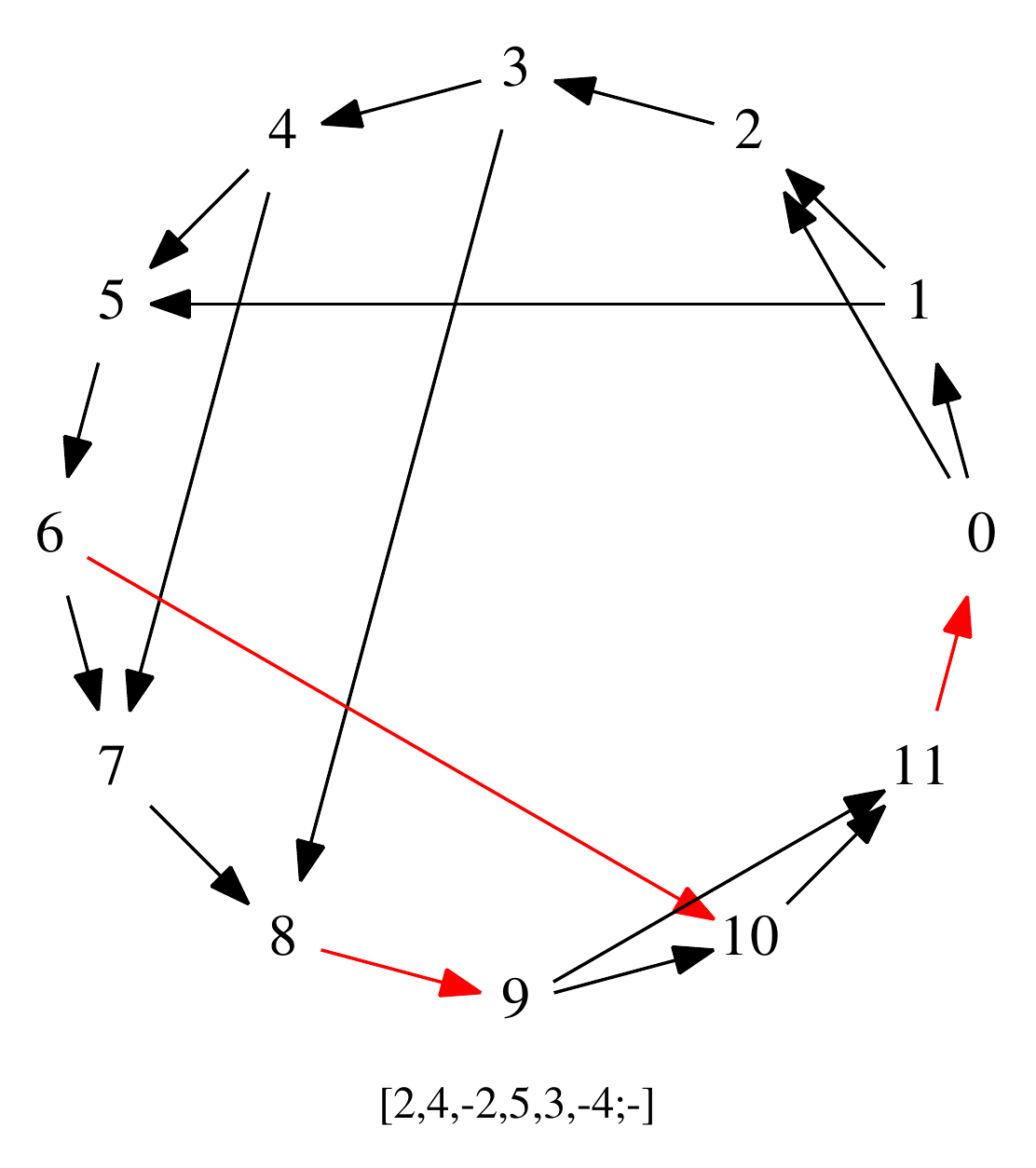}
\includegraphics[scale=0.45]{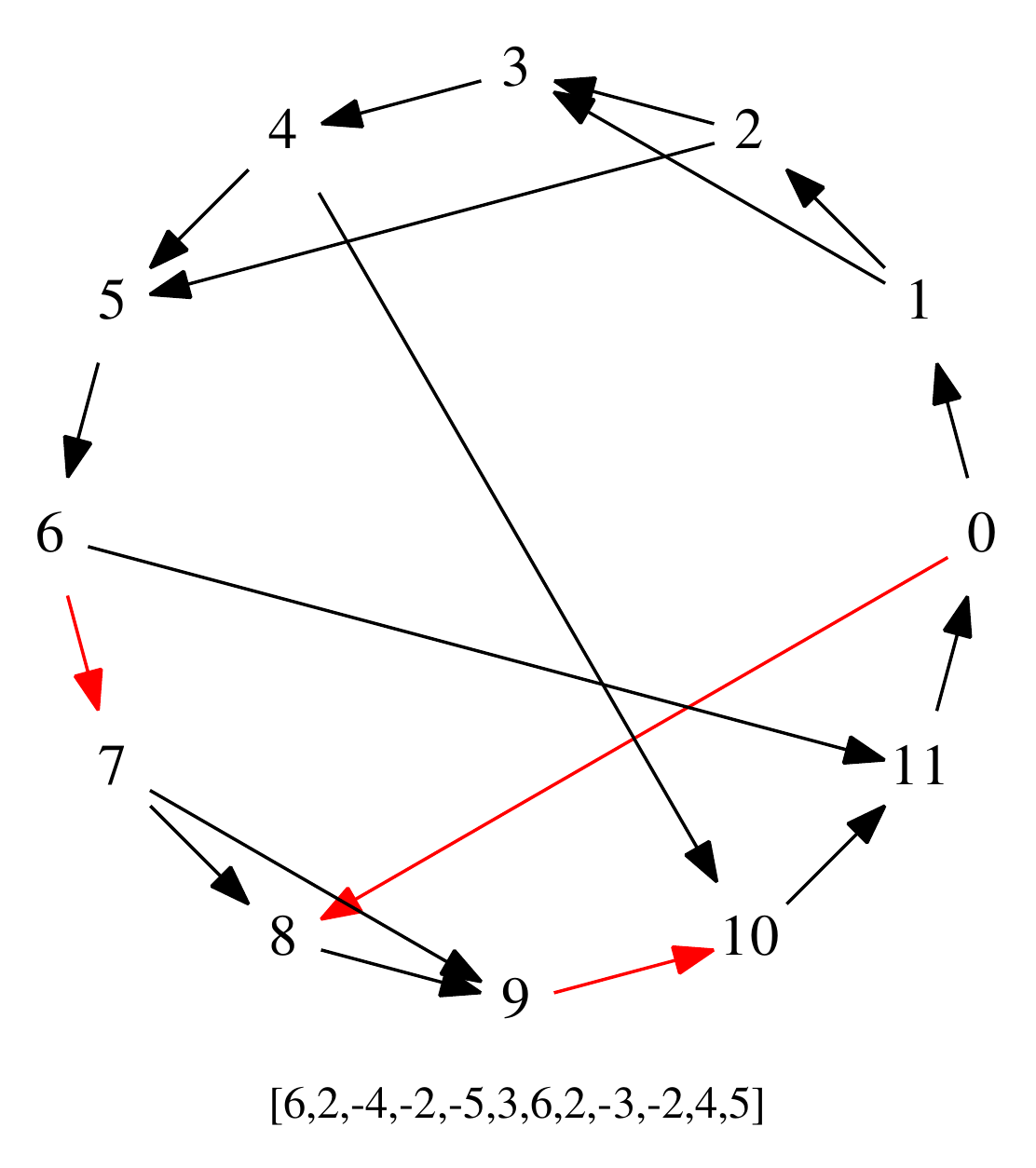}
\includegraphics[scale=0.45]{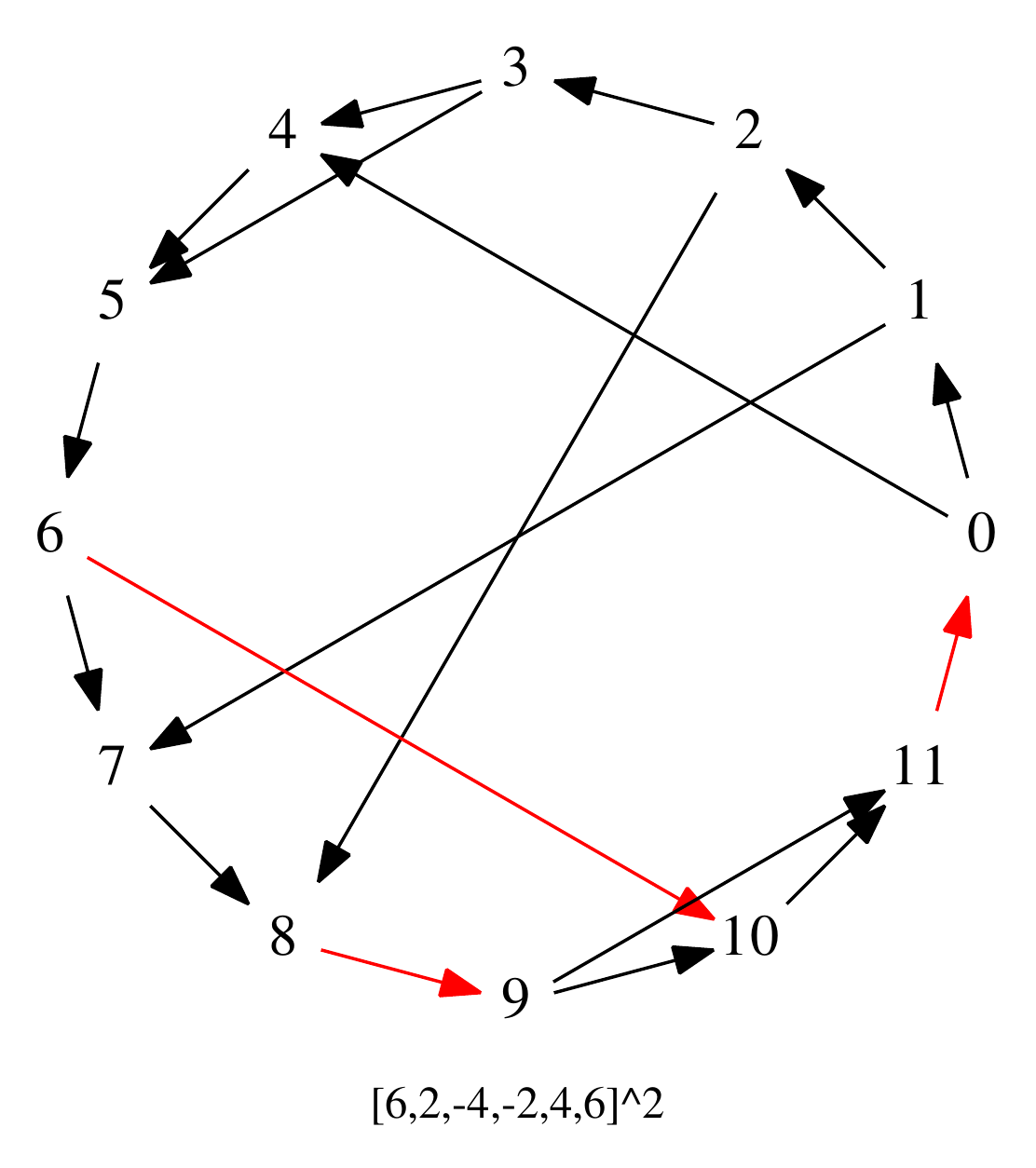}
\includegraphics[scale=0.45]{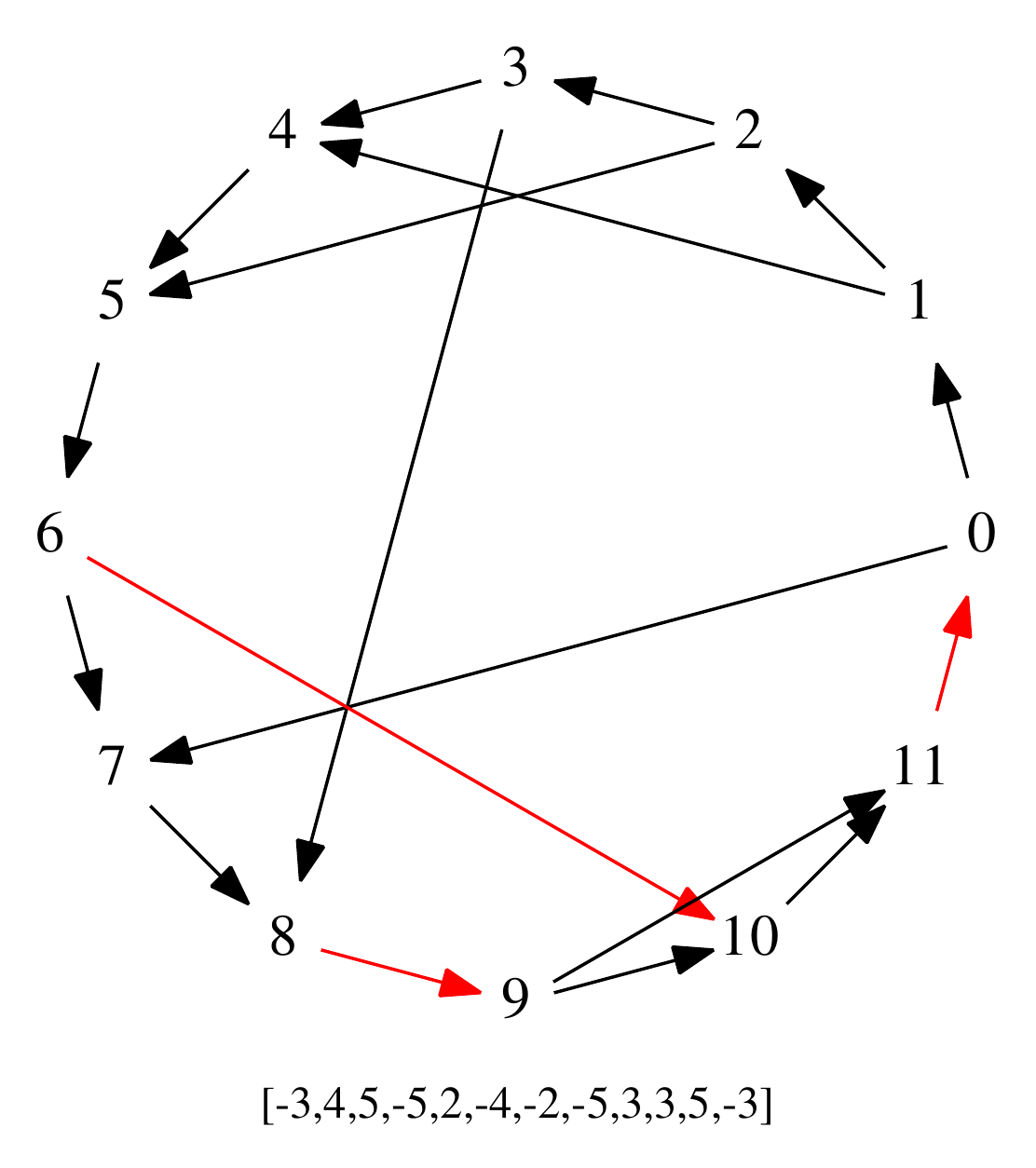}
\includegraphics[scale=0.45]{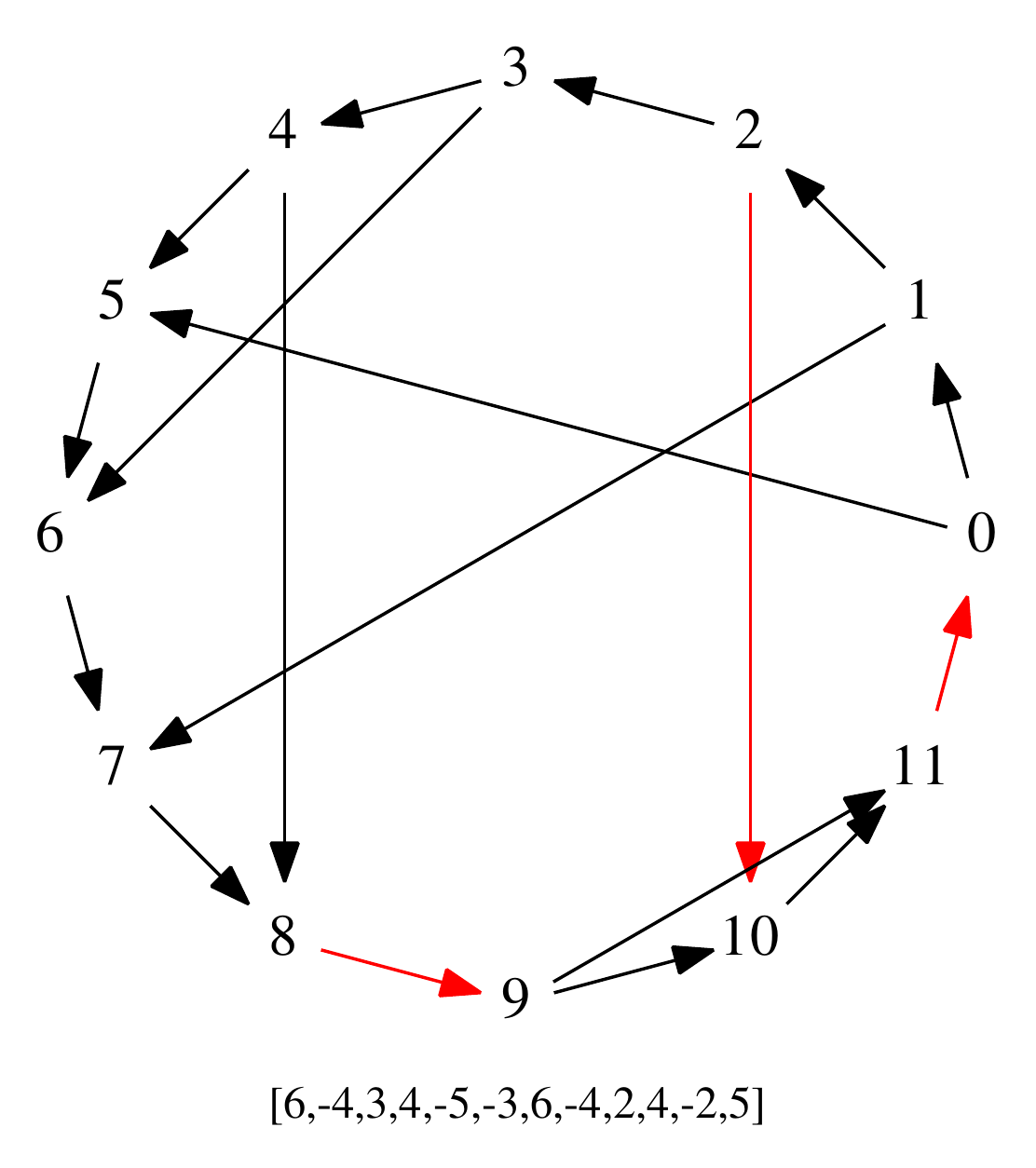}
\includegraphics[scale=0.45]{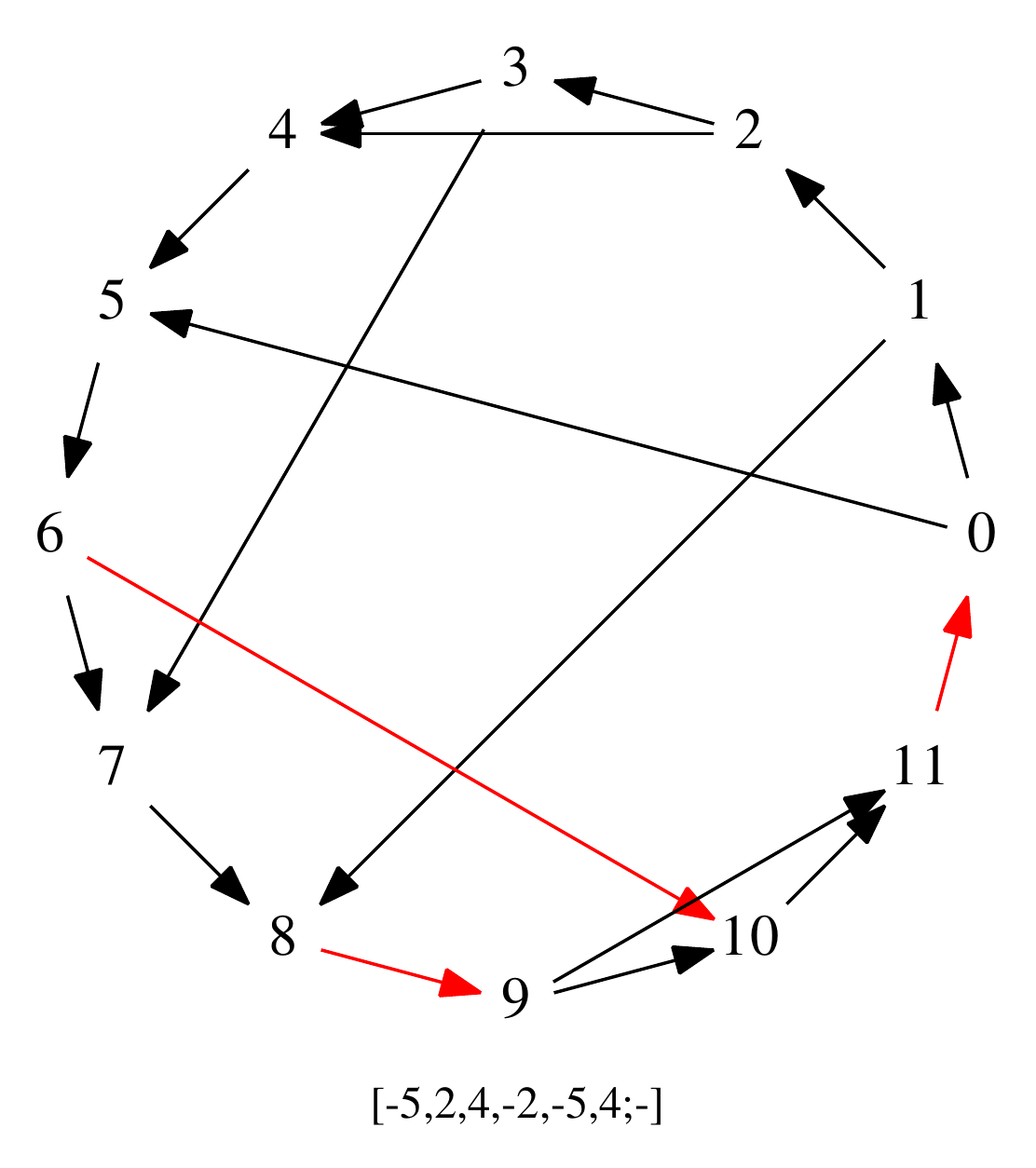}
\includegraphics[scale=0.45]{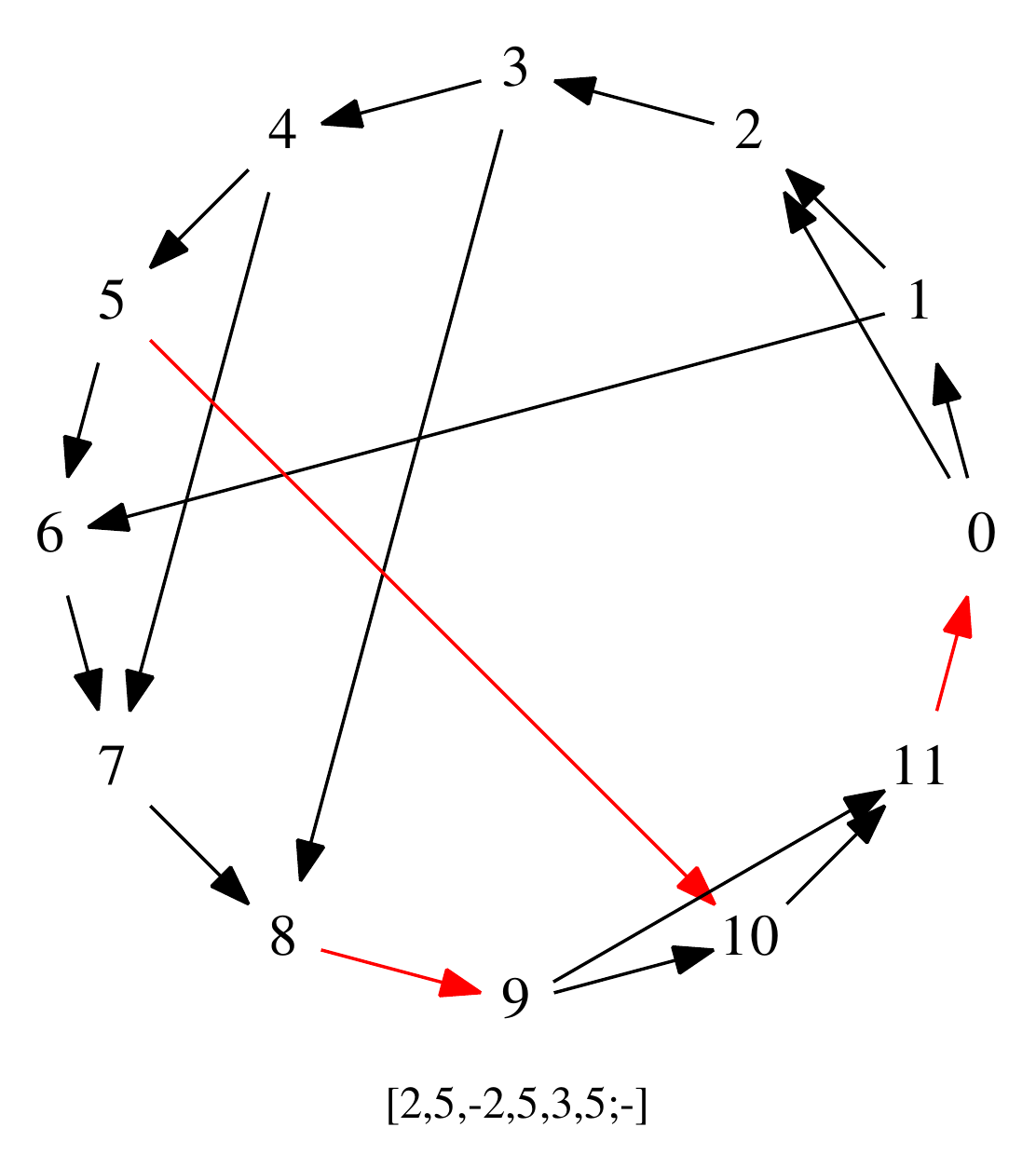}
\includegraphics[scale=0.45]{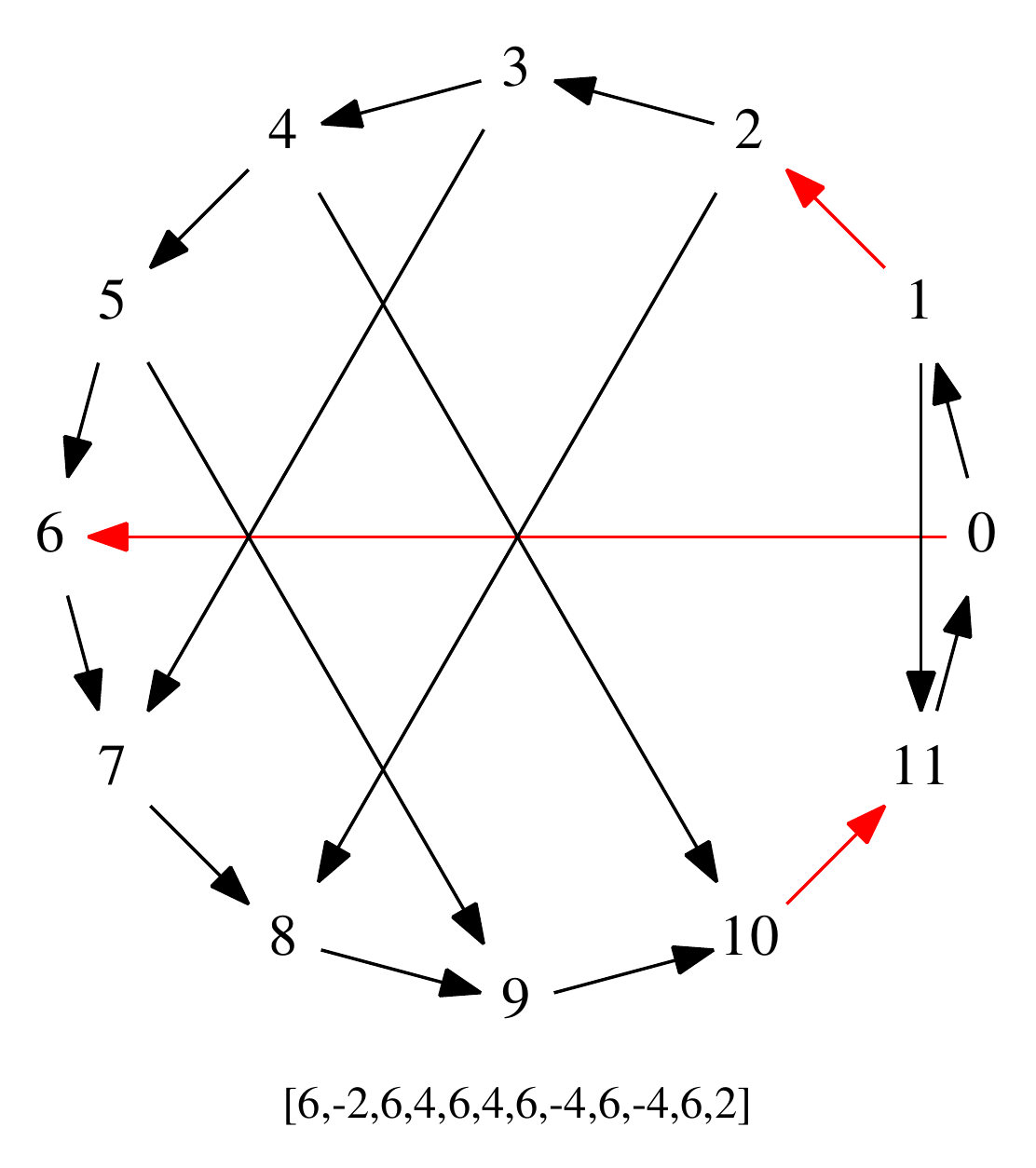}
\includegraphics[scale=0.45]{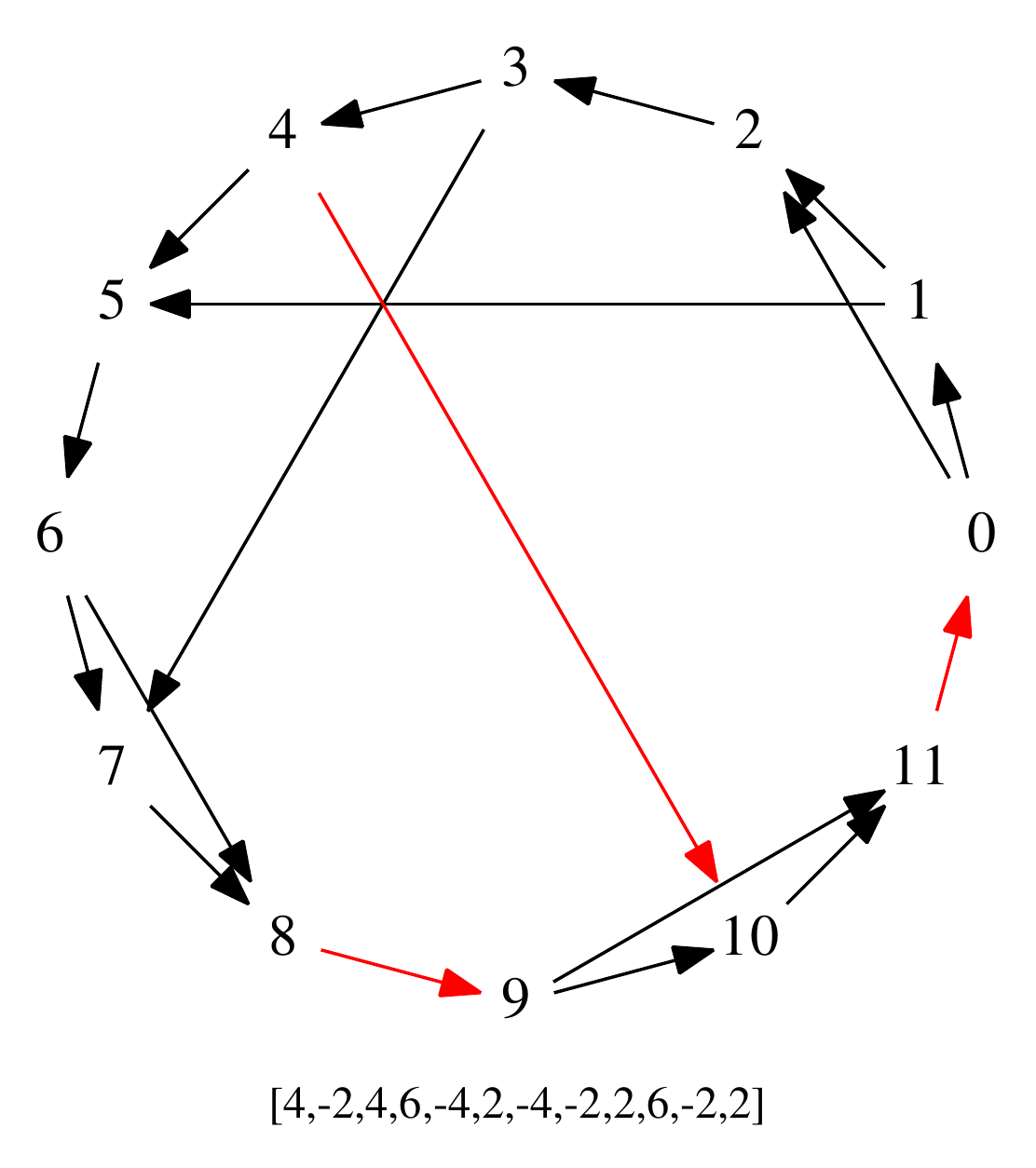}
\caption{$3$-connected graphs on $n=12$ vertices (continued).
}
\label{fig.12n33}
\end{figure}
\begin{figure}
\includegraphics[scale=0.45]{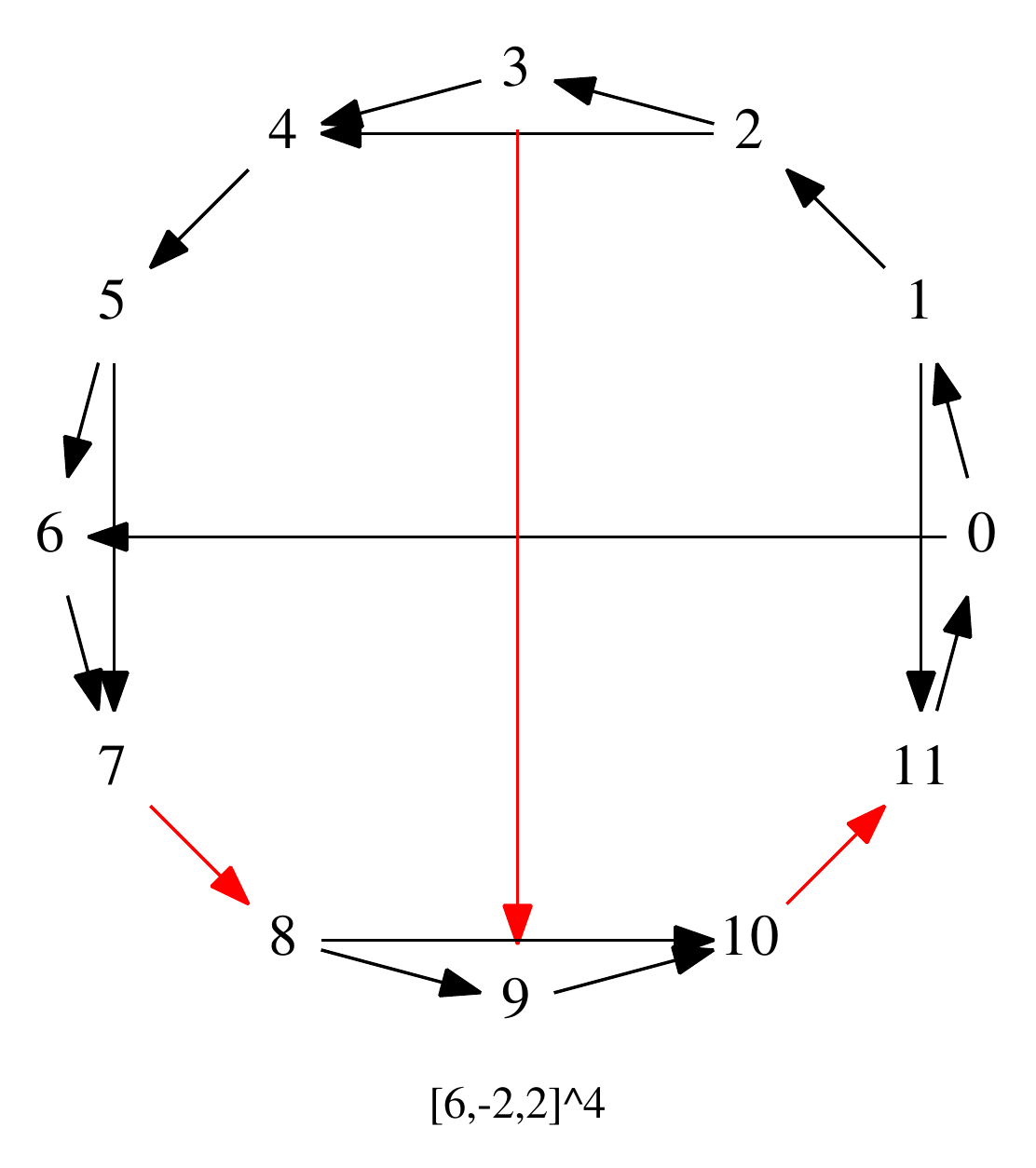}
\includegraphics[scale=0.45]{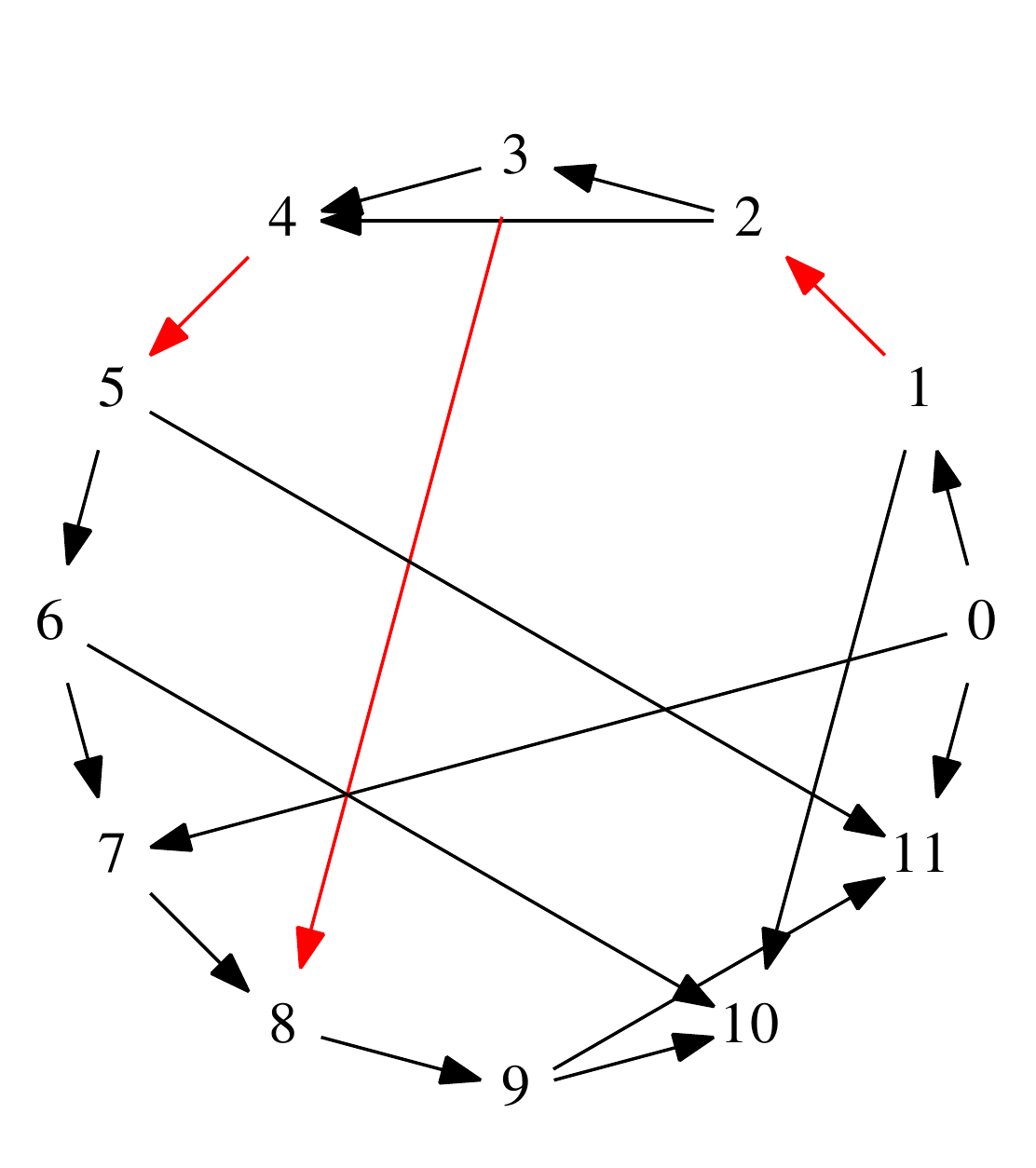}
\includegraphics[scale=0.45]{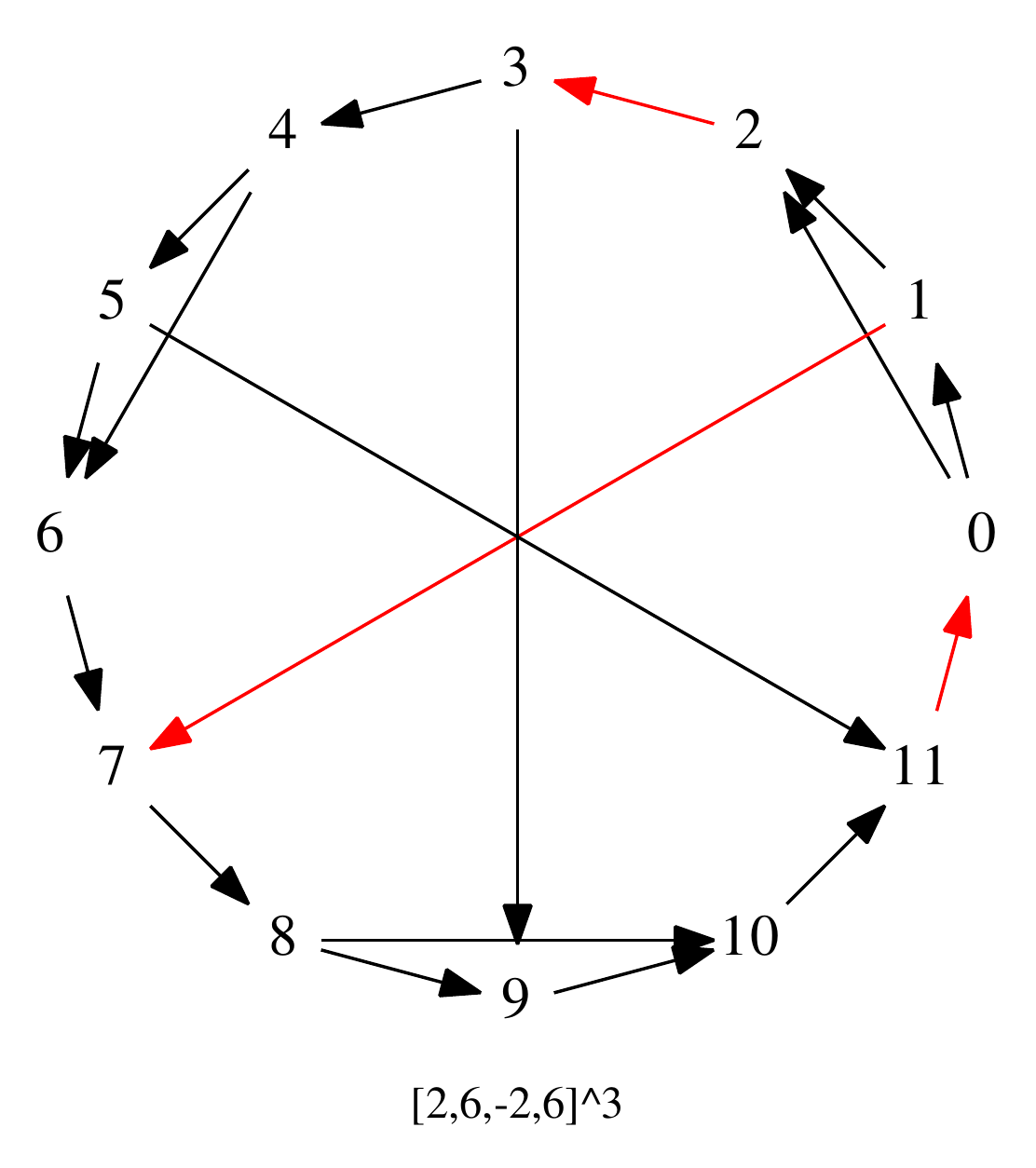}
\caption{$3$-connected graphs on $n=12$ vertices (end).
Tietze's graph (\texttt{W129 d3 g3 EE41.70908}) does not
have a Hamiltonian cycle.
}
\label{fig.12n3e}
\end{figure}
\clearpage

\subsection{Irreducible} 
The 18 graphs on $n=12$ vertices, which are cyclically 4-connected
or 5-connected
and define classes of $18j$-symbols, follow
in Figures \ref{fig.12n4_s}--\ref{fig.12n4_e}.
The translation to the enumeration by 18 capital letters
in the reference work
\cite[App.\ 3]{Yutsis} is:
\begin{itemize}
\item
A \verb+[6]^12+ 
\item
B  \verb+[-3,3]^6+
\item
C  \verb+[-5,5]^6+
\item
D  \verb+[4,-4,6]^4+.
This representation is found by walking $j_1$, $s_1$, $j_2$, $j_2'$ $s_1'$, $j_1'$, $j_4'$, $s_2$,
$j_3'$, $j_3$, $s_2$, $j_4$ in \cite[Fig. A 3.4]{Yutsis}.
\item
E  \verb+[3,5,5,-3,5,5;-]+
This connection is found by walking
$j_3$, $l_2$, $j_3'$, $k_1'$ $s_2$, $k_1$, $s_1$,
$k_2'$,
$j_4'$, $l_1$, $j_4$, $k_2$ in \cite[Fig. A 3.5]{Yutsis}.
\item
F  \verb+[4,-5,4,-5,-4,4;-]+
\cite[Fig. A 3.6]{Yutsis}.
\item
G  \verb+[6,-5,5]^4+
\cite[Fig. A 3.7]{Yutsis}.
\item
H  \verb+[6,-5,-4,4,-5,4,6,-4,5,-4,4,5]+
\cite[Fig. A 3.8]{Yutsis}.
\item
I \verb+[6,-4,5,-5,4,6,6,-5,-4,4,5,6]+
\cite[Fig. A 3.9]{Yutsis}.
\item
K \verb+[-4,4,4,6,6,-4]^2+
\cite[Fig. A 3.10]{Yutsis}.
\item
L \verb+[6,-3,6,6,3,6]^2+
\cite[Fig. A 3.12]{Yutsis}.
\item
M \verb+[6,4,6,6,6,-4]^2+
\cite[Fig. A 3.13]{Yutsis}.
\item
N \verb+[4,-3,4,5,-4,4;-]+
\cite[Fig. A 3.15]{Yutsis}.
\item
P \verb+[6,-3,5,6,-5,3,6,-5,-3,6,3,5]+
\cite[Fig. A 3.16]{Yutsis}.
\item
R \verb+[3,4,5,-3,5,-4;-]+
\cite[Fig. A 3.17]{Yutsis}.
\item
S \verb+[-3,5,-3,4,4,5;-]+
\cite[Fig. A 3.18]{Yutsis}.
\item
T \verb+[-4,6,3,6,6,-3,5,6,4,6,6,-5]+
walking for example $r$, $l$, $s$, $u$, $n$, $p$, $j$, $r'$,
$l'$, $m'$, $p'$, $j'$ in
\cite[Fig. A 3.11]{Yutsis}.
\item
V \verb+[6,-4,6,-4,3,5,6,-3,6,4,-5,4]+
\cite[Fig. A 3.14]{Yutsis}.
\end{itemize}

\begin{figure}[h]
\includegraphics[scale=0.45]{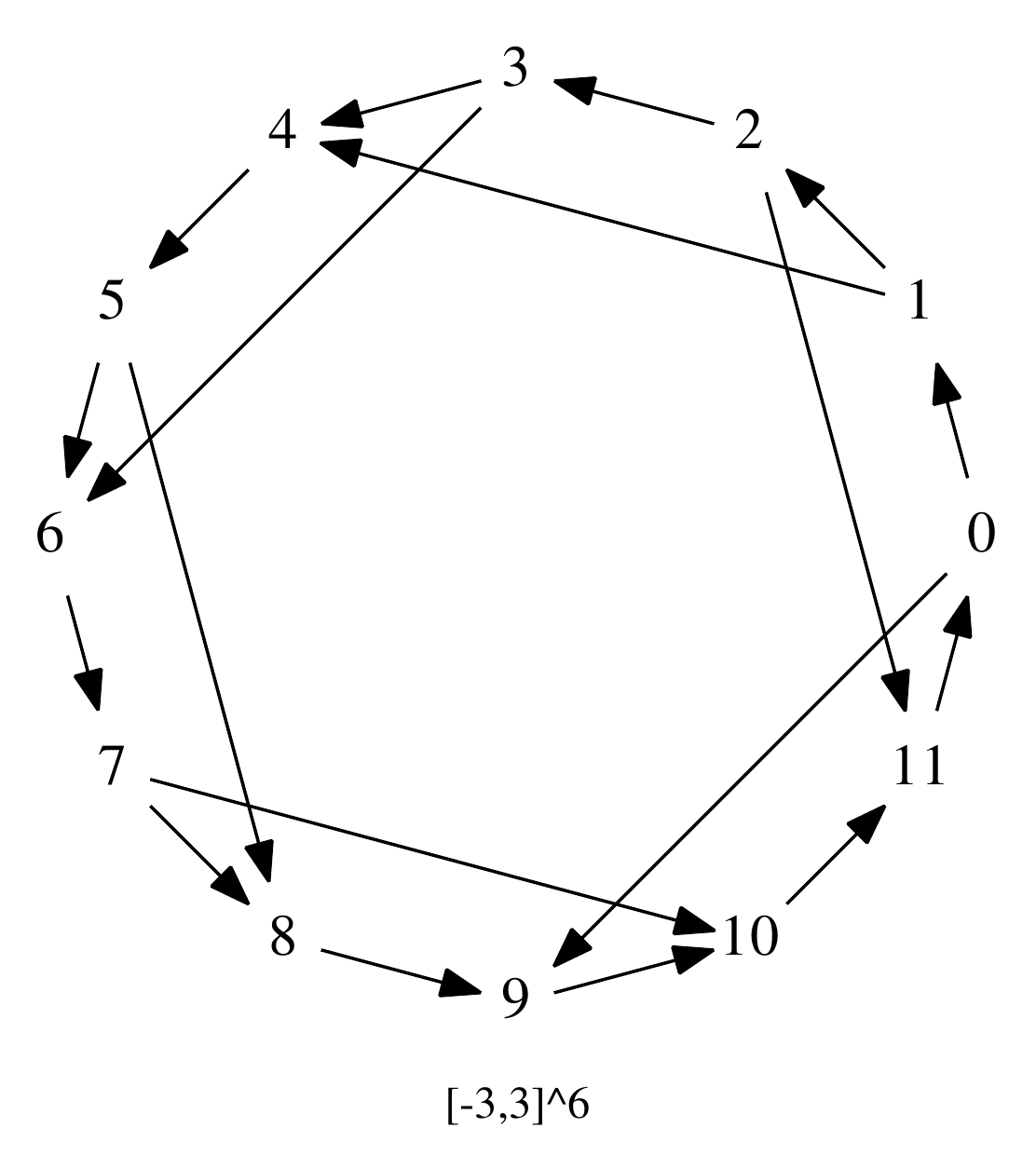}
\includegraphics[scale=0.45]{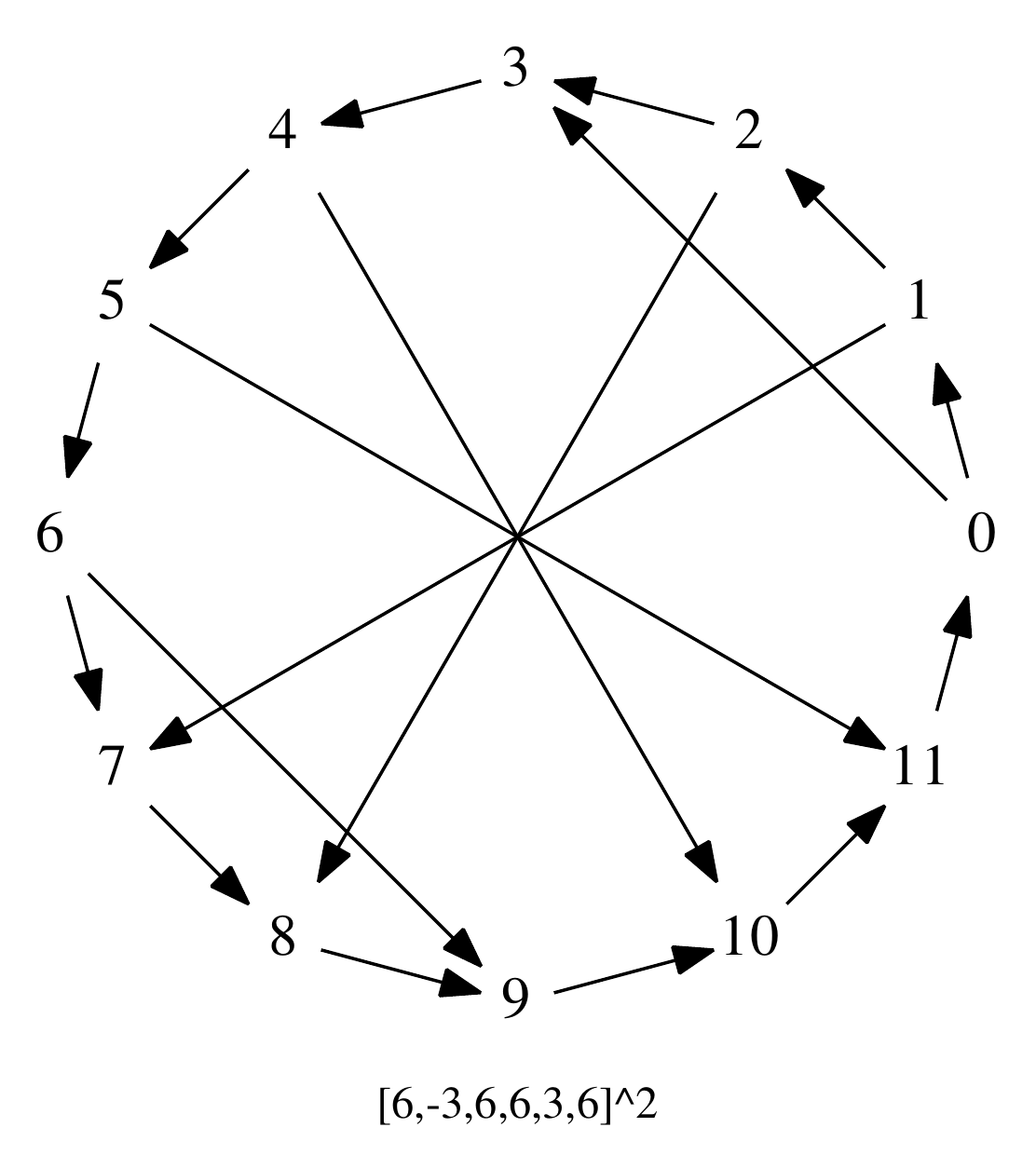}
\includegraphics[scale=0.45]{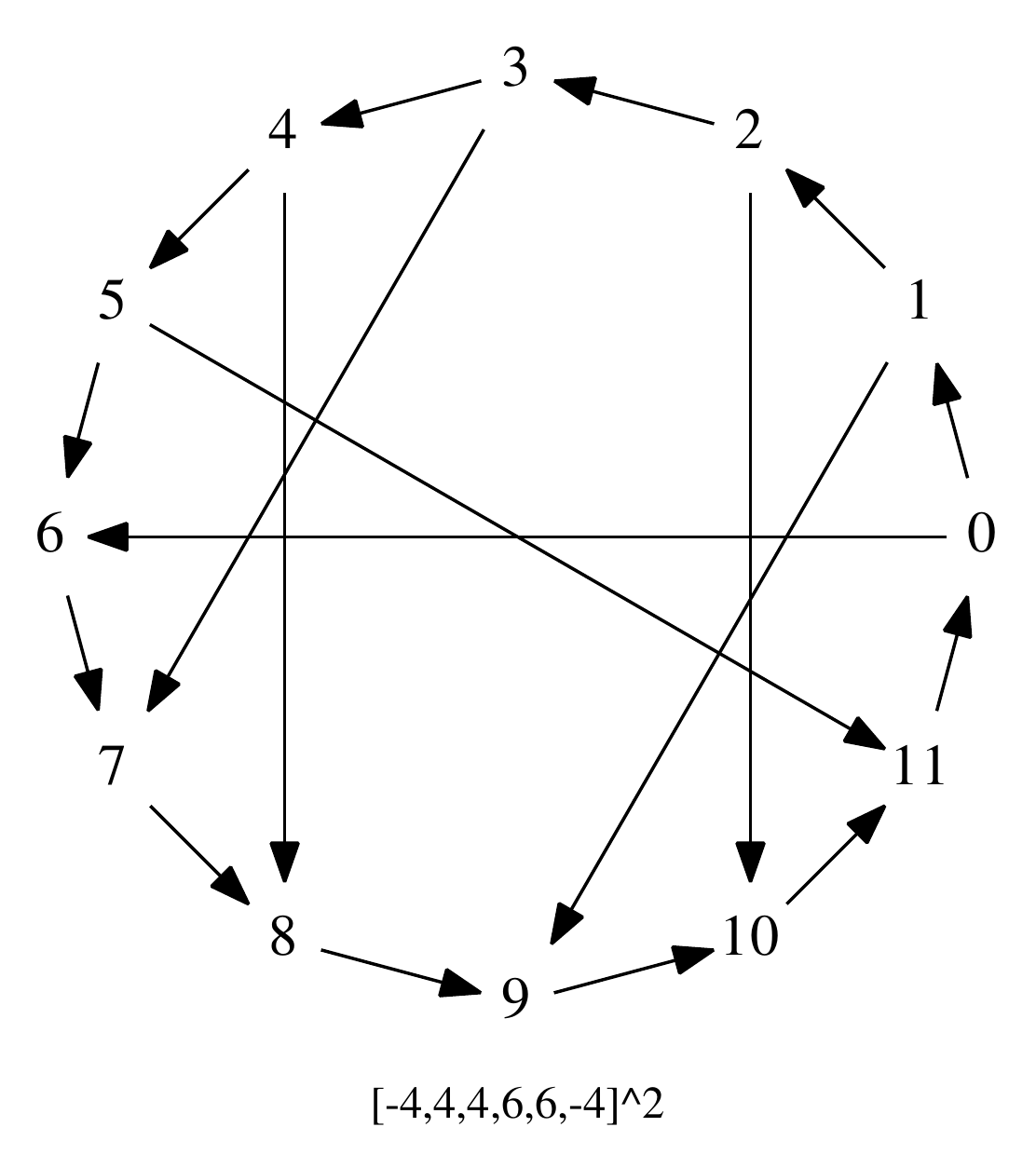}
\includegraphics[scale=0.45]{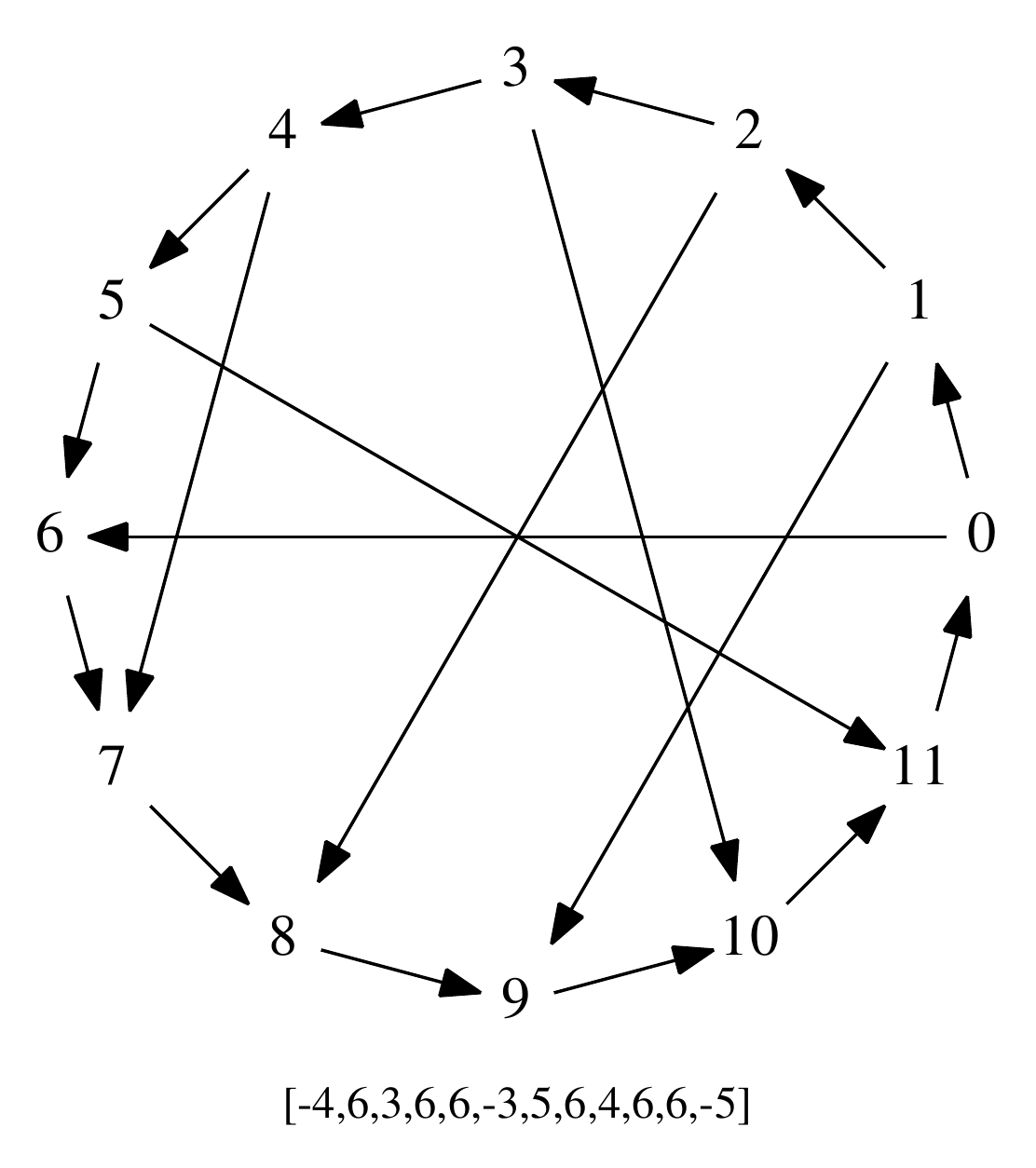}
\includegraphics[scale=0.45]{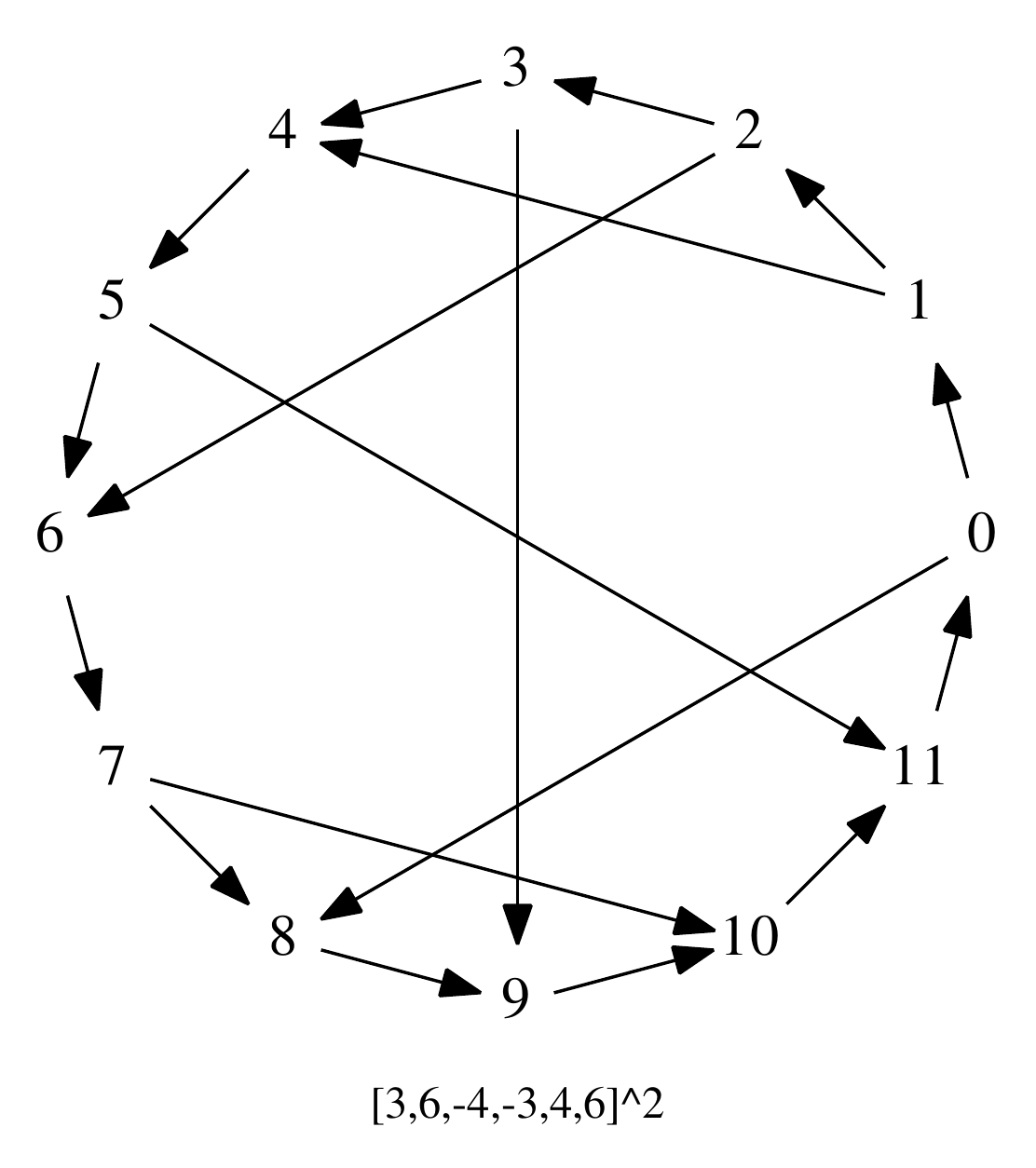}
\includegraphics[scale=0.45]{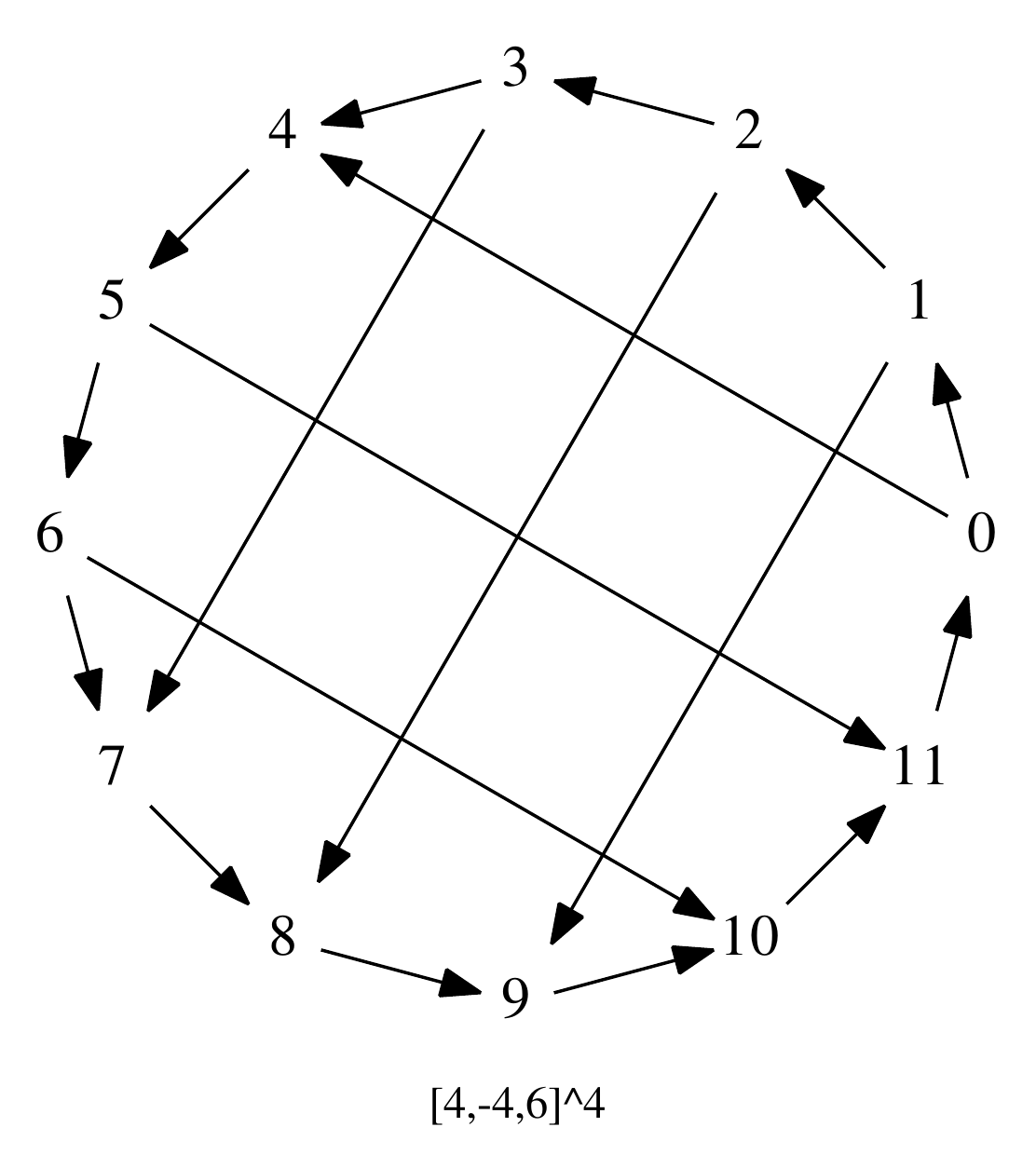}
\includegraphics[scale=0.45]{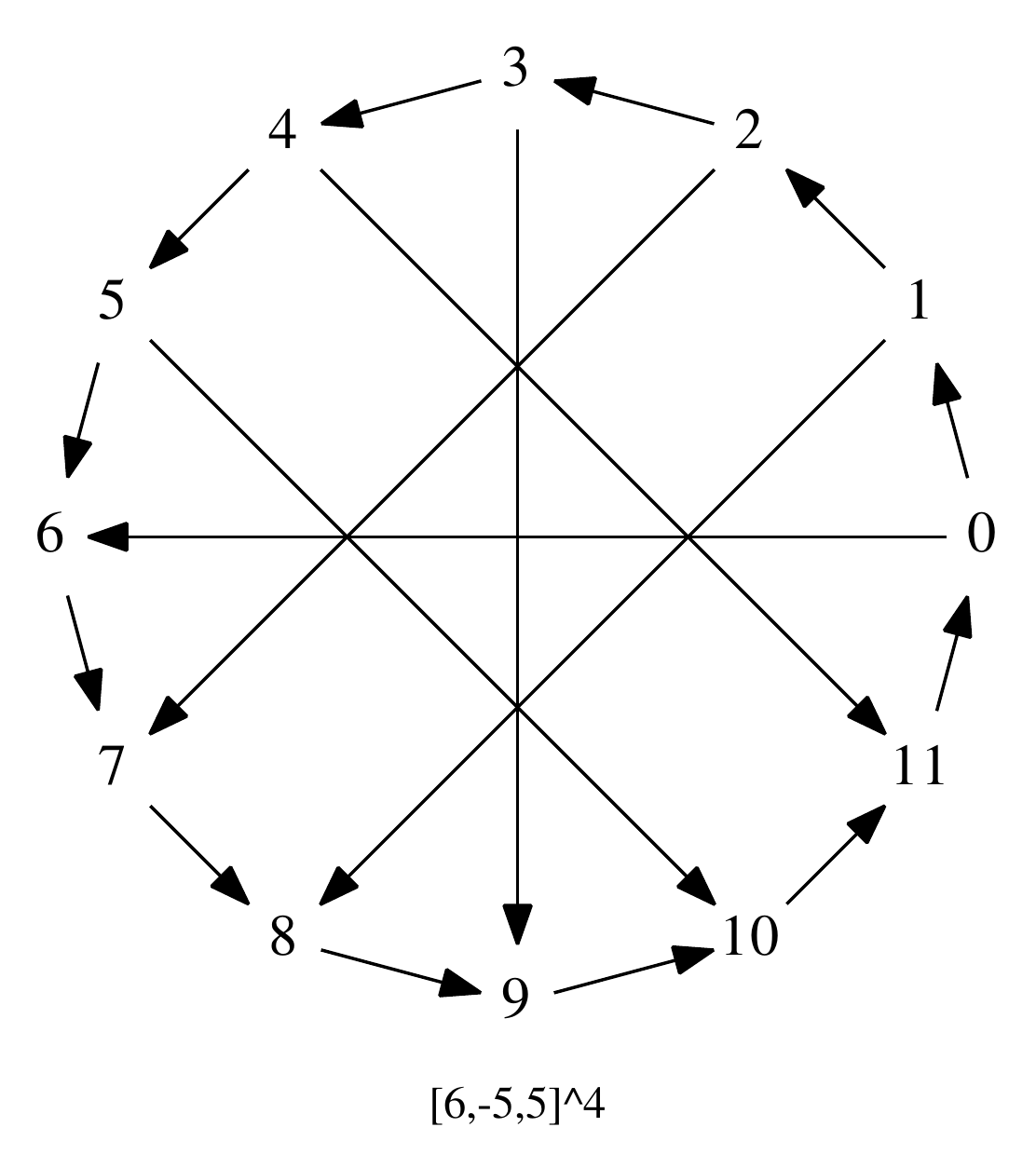}
\includegraphics[scale=0.45]{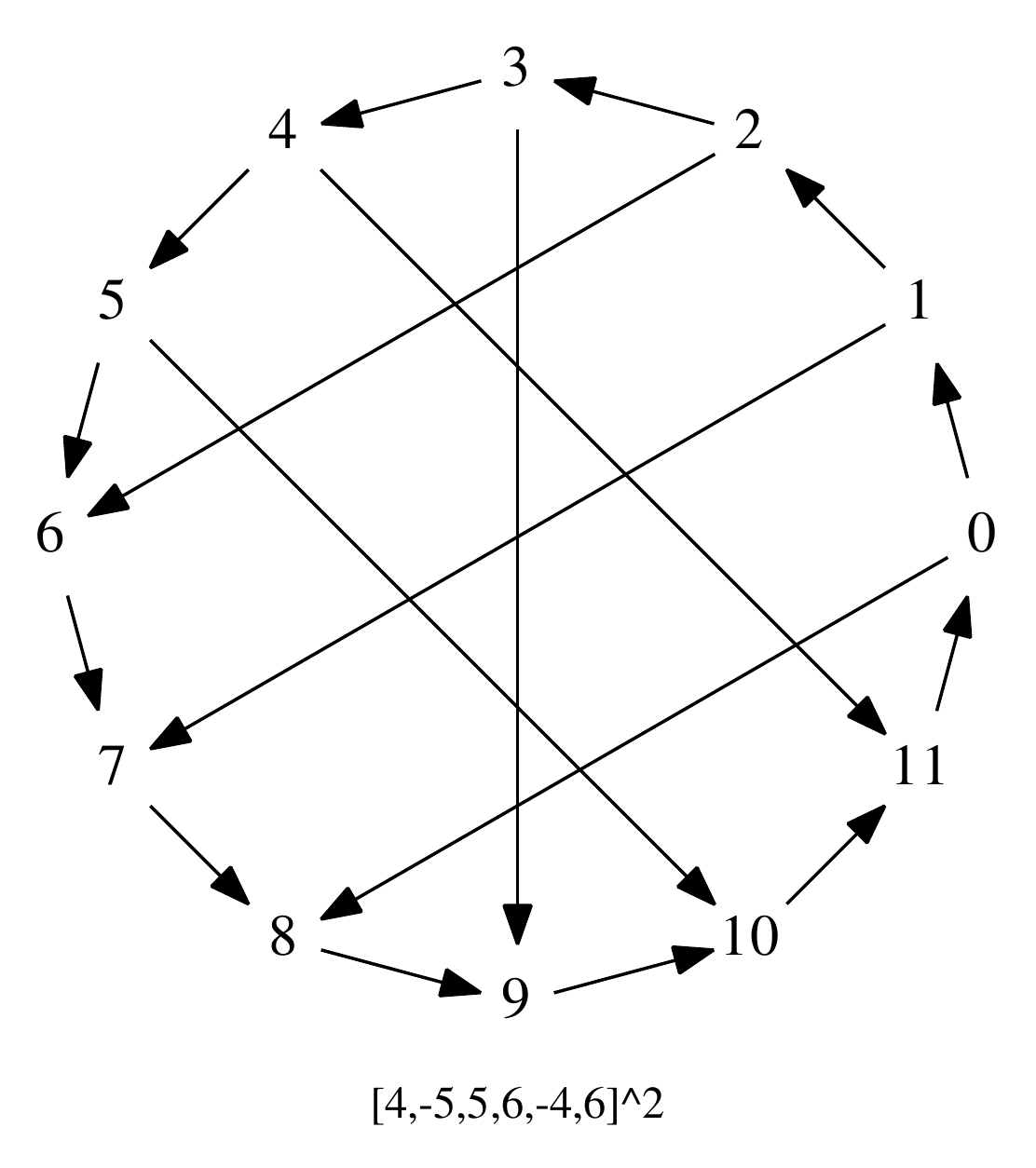}
\includegraphics[scale=0.45]{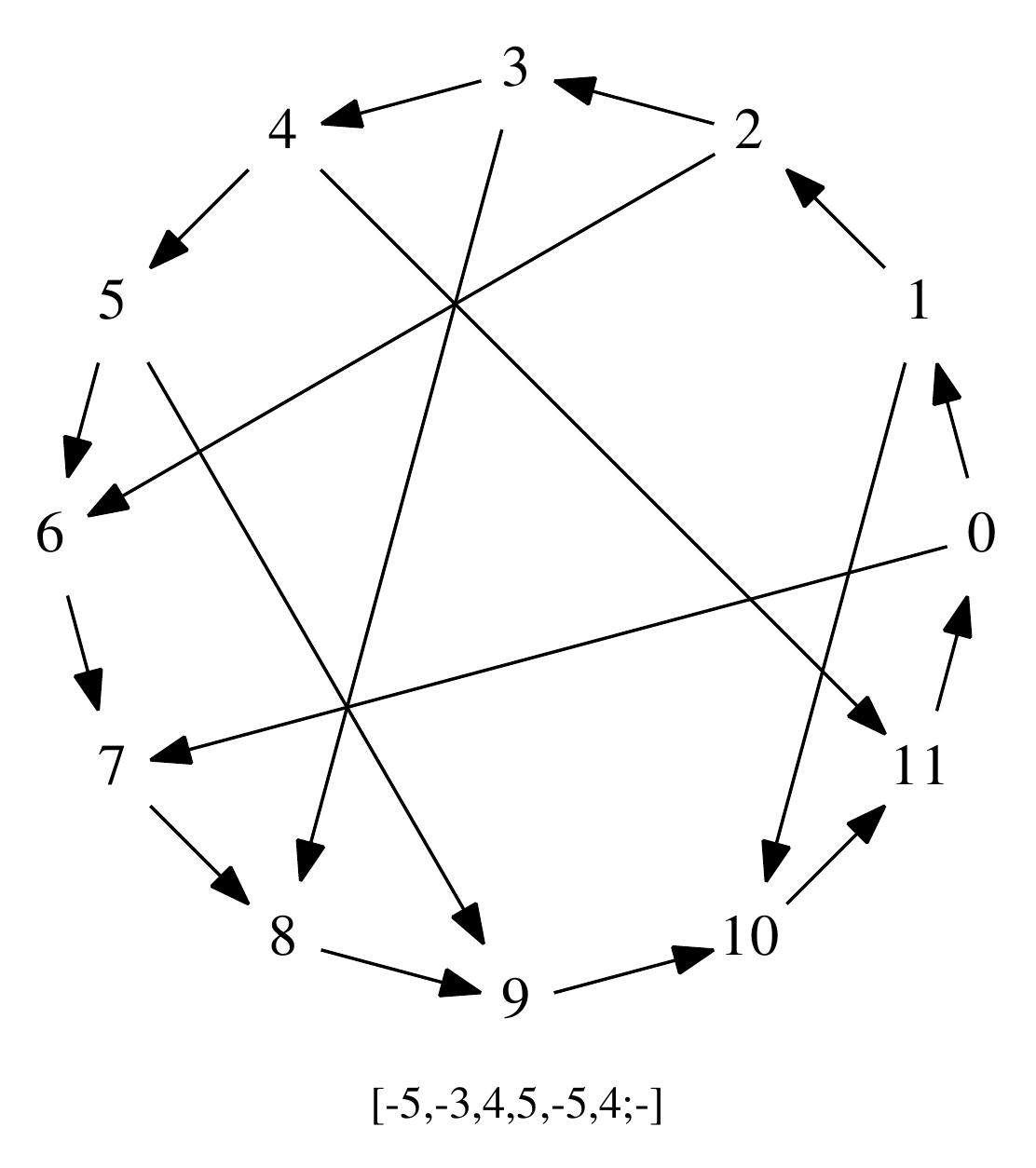}
\includegraphics[scale=0.45]{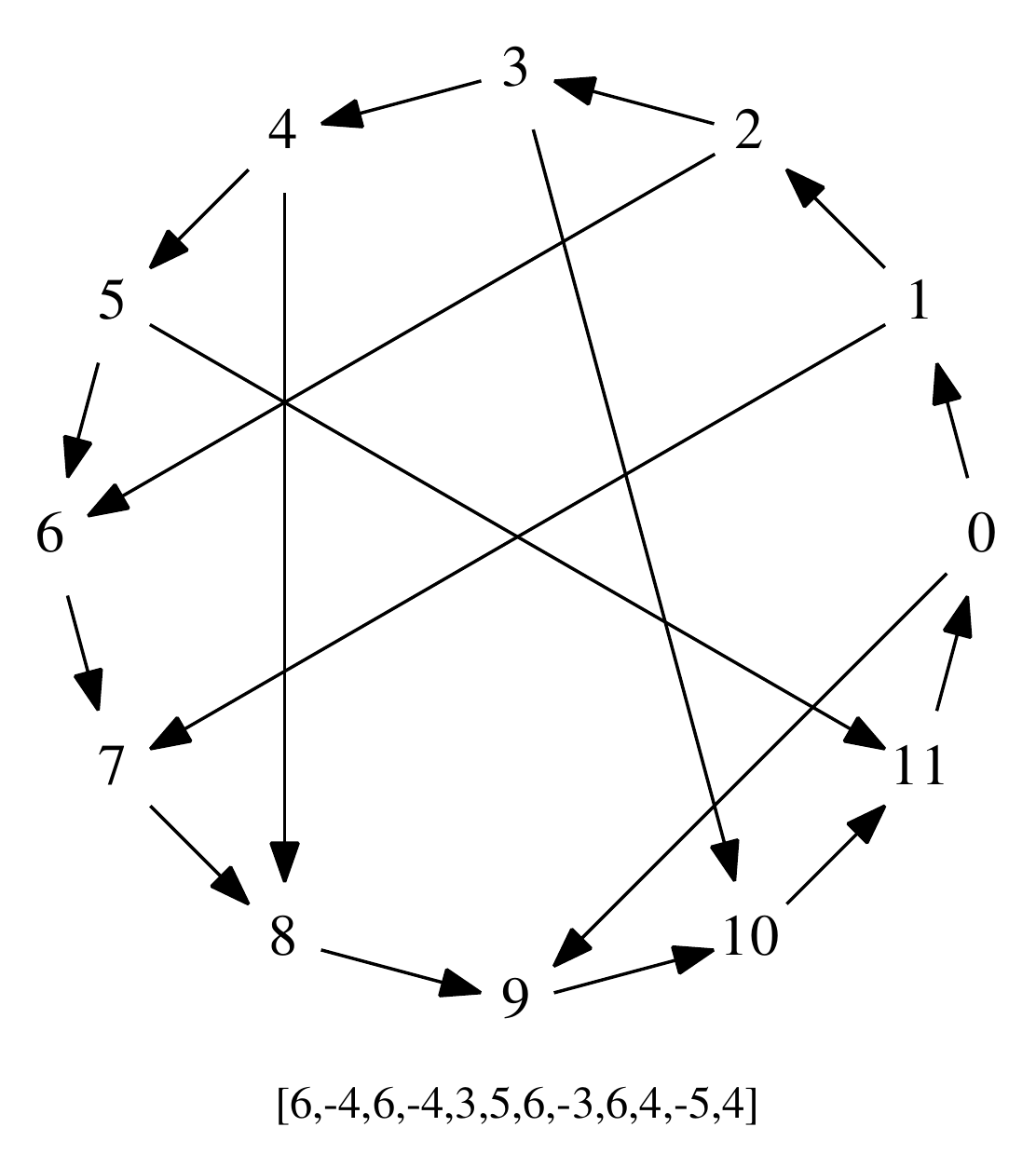}
\includegraphics[scale=0.45]{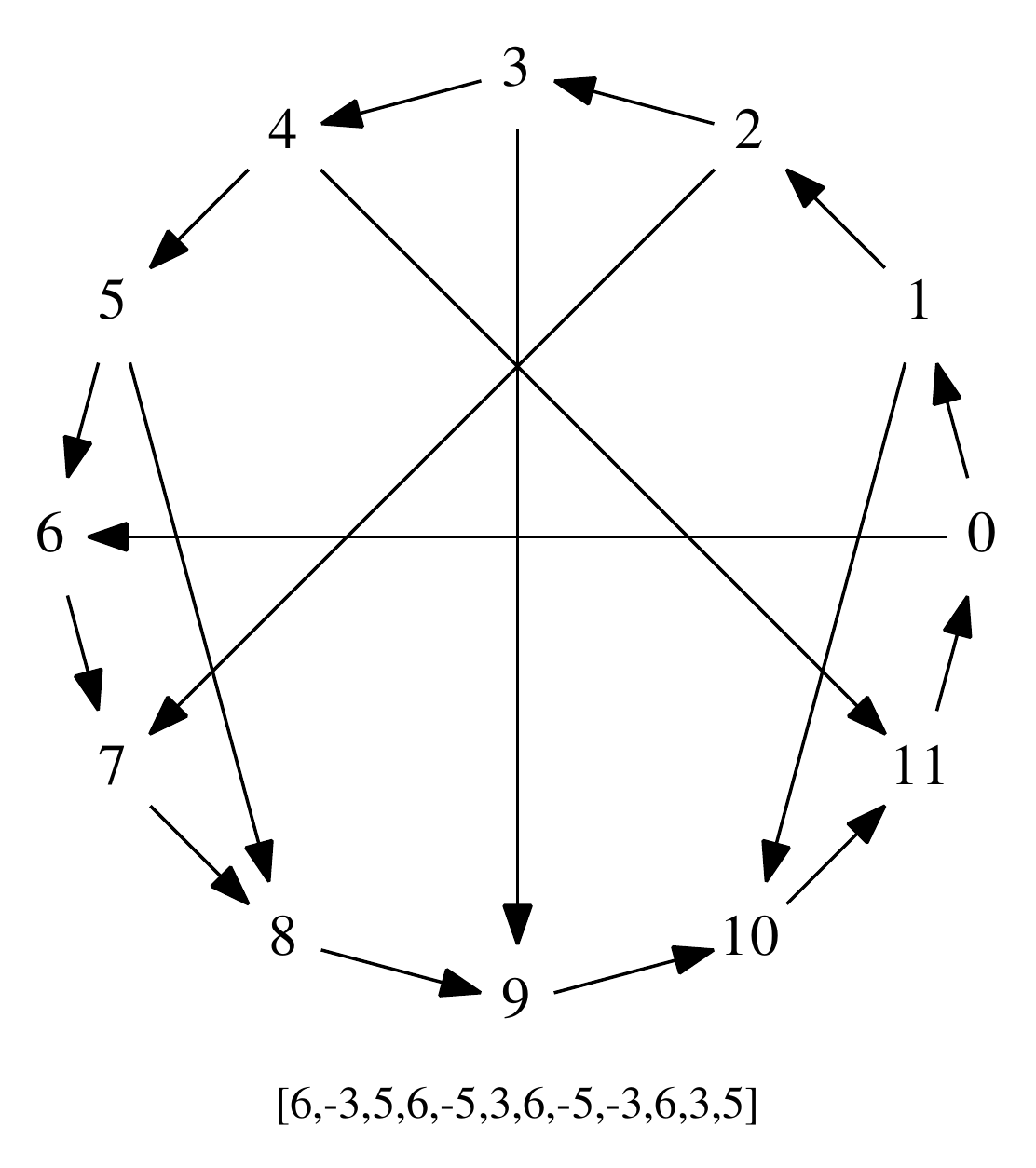}
\includegraphics[scale=0.45]{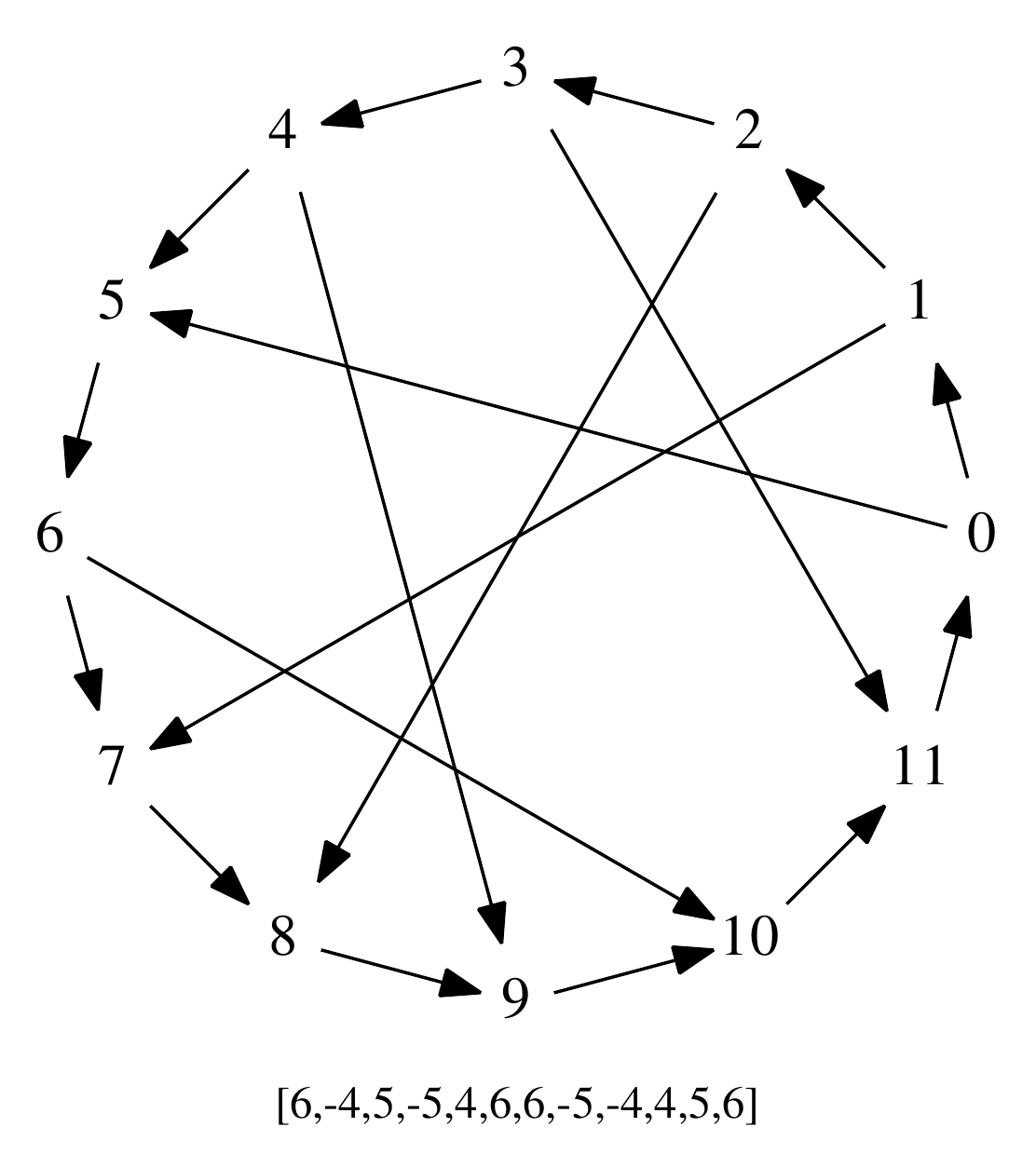}
\caption{12 of the 18 graphs on $n=12$ vertices which are irreducible.
}
\label{fig.12n4_s}
\end{figure}
\begin{figure}
\includegraphics[scale=0.45]{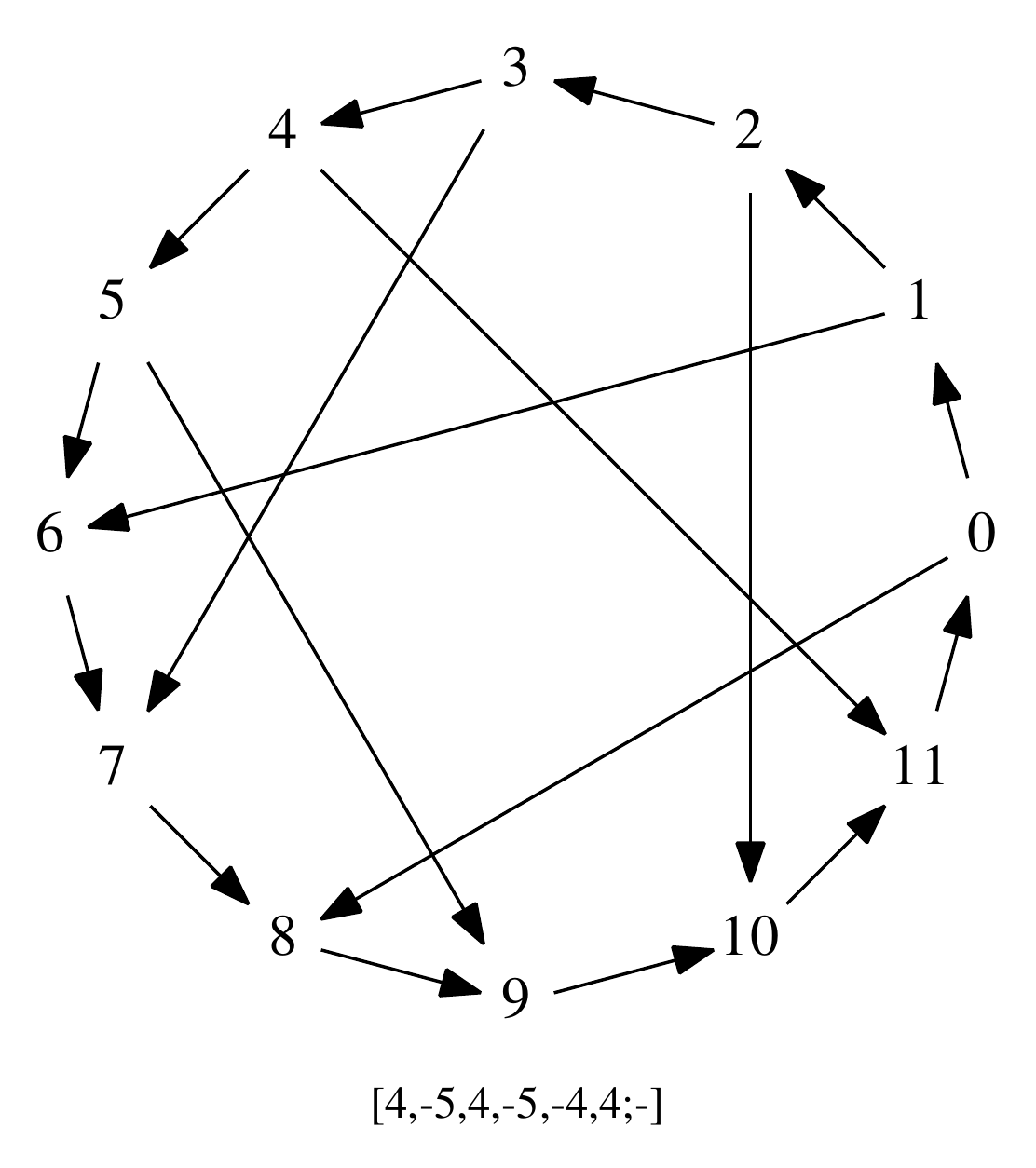}
\includegraphics[scale=0.45]{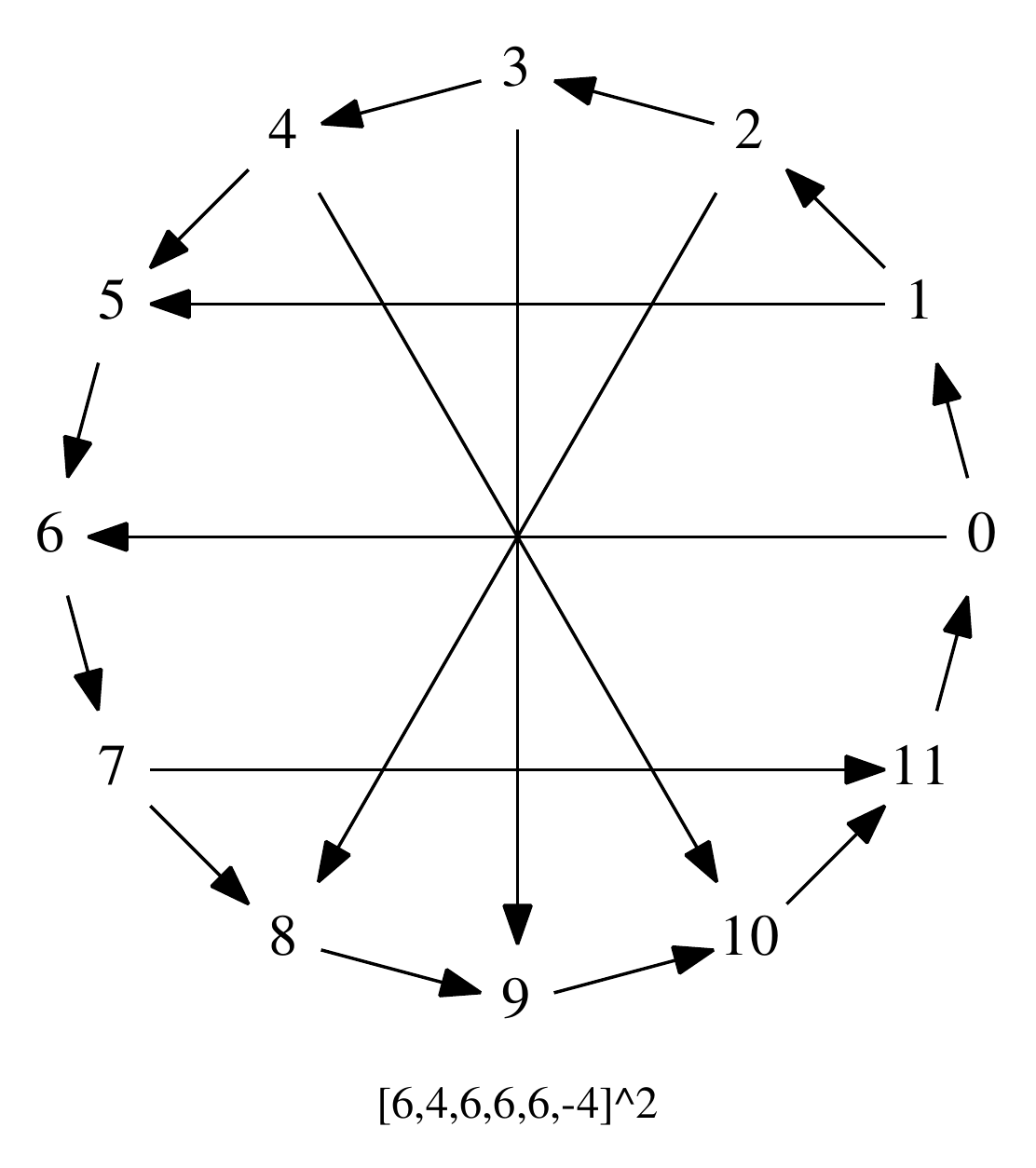}
\includegraphics[scale=0.45]{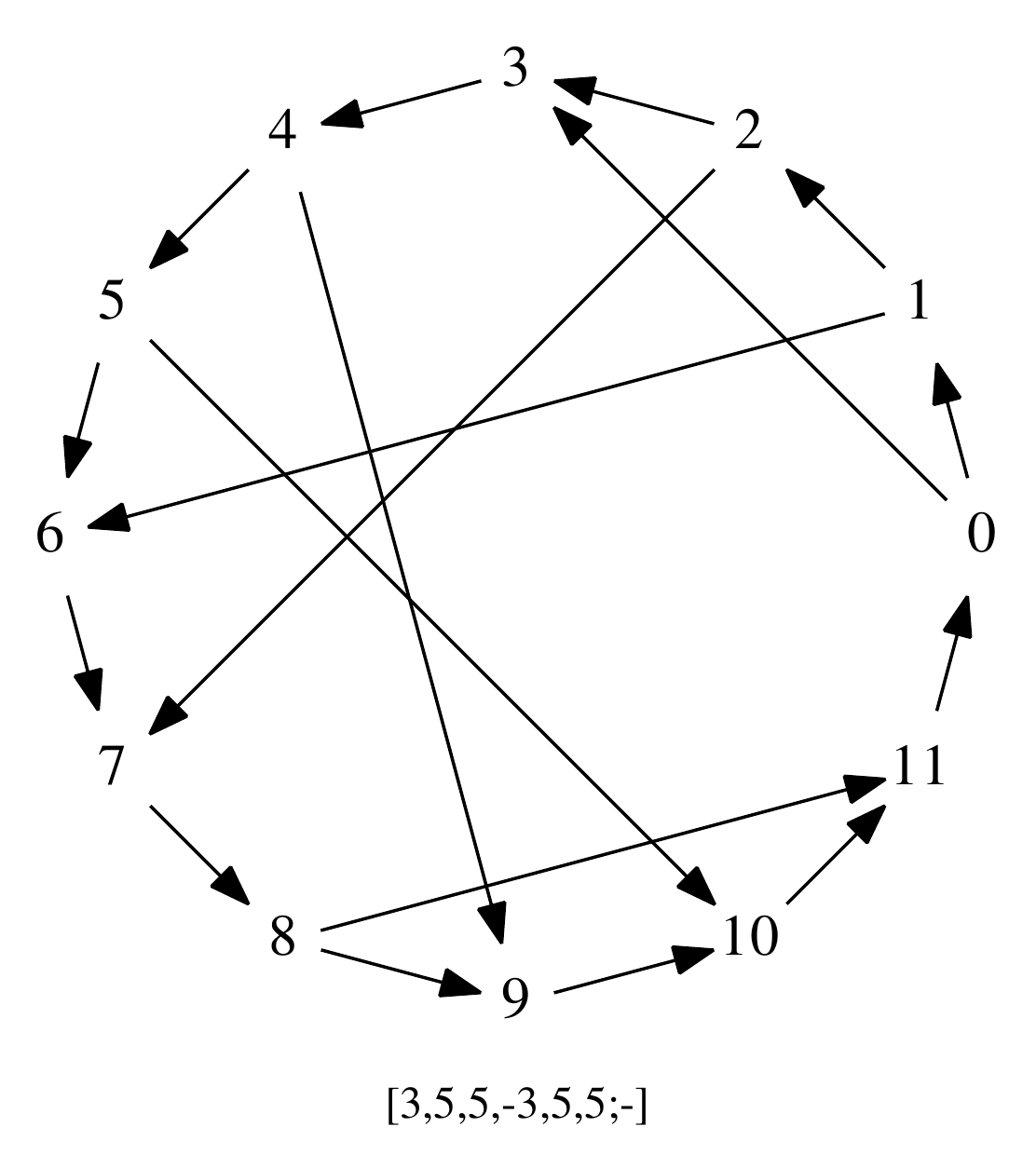}
\includegraphics[scale=0.45]{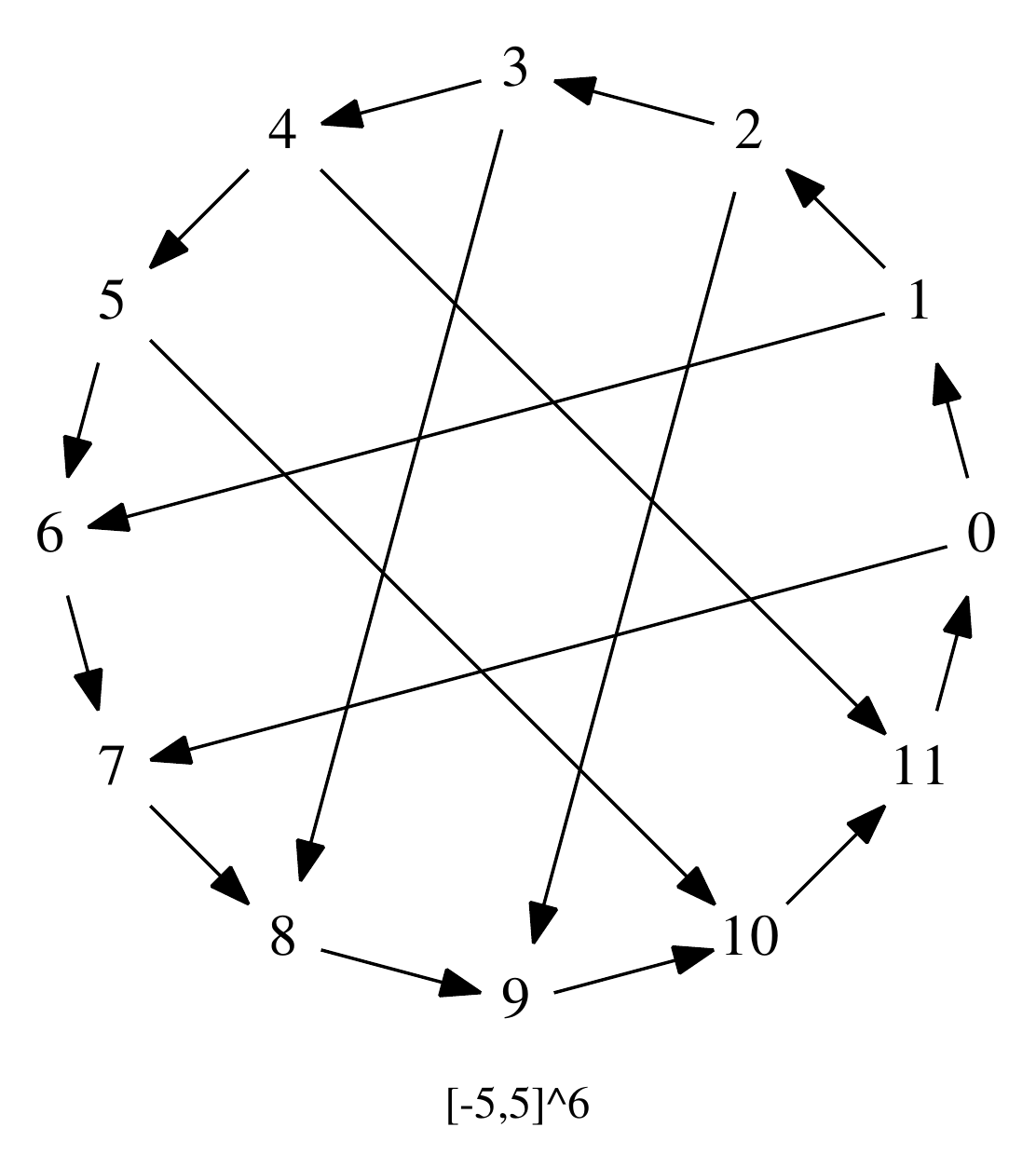}
\includegraphics[scale=0.45]{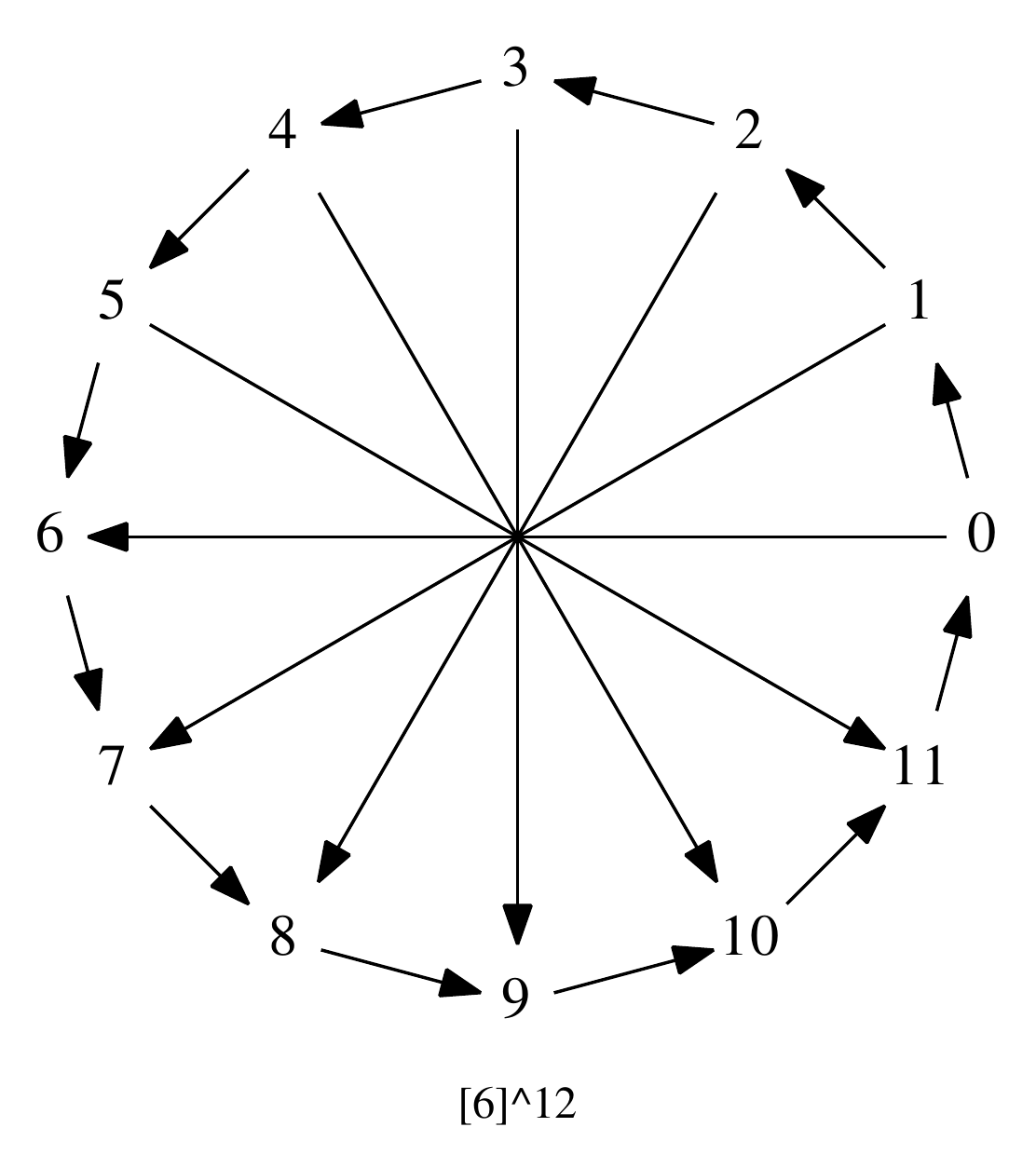}
\includegraphics[scale=0.45]{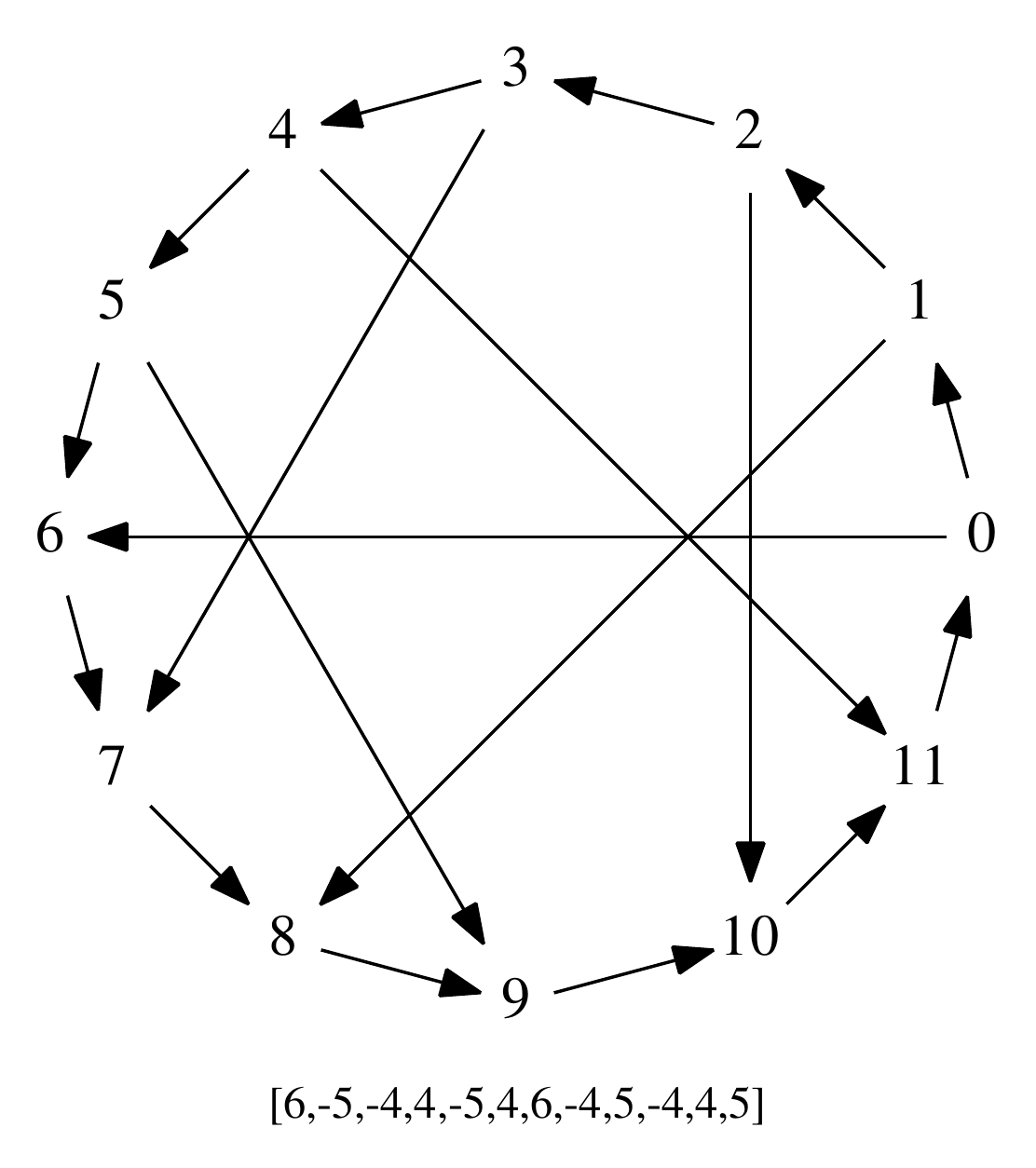}
\caption{The remaining 6 of the 18 graphs on $n=12$ vertices which are irreducible.
\texttt{[-5,5]\char94 6} is the Franklin graph;
\texttt{[6]\char94 12} is the 6-prism graph.
}
\label{fig.12n4_e}
\end{figure}
\clearpage
\section{$14$ vertices} 

The total number of graphs on $14$ vertices is 509 \cite{BrinkmannJGT23,MeringerJGT30,vanDyckCPC173,BrinkmannDMTCS13}
\cite[A002851]{EIS}.

Only the 84 of the diagrams which are
irreducible
are finally shown in Figures \ref{fig.14ns}--\ref{fig.14ne}, each representing a $21j$-symbol.
The 84 graphs can be characterized by the following
Hamiltonian cycles:

Figure \ref{fig.14ns}:

\begin{Verbatim}
   LCF [3,-3,4,-3,5,3,-4;-]  = [-6,6,4,-5,7,5,-4]^2 = 
[-5,6,4,-5,7,3,-4,-6,-3,5,3,7,5,-3] = [-5,7,-3,3,-6,6,-3,3,7,5,-3,-6,6,3] W209 
d4 g4 EE48.48328   
   LCF [3,7,7,-3,7,7,7]^2  = [7,7,7,-5,7,5,7]^2 = [3,-3,5,-3,5,3,5;-] = 
[7,3,-5,7,-3,3,7,7,-3,3,7,5,-3,7] = [7,7,-3,-5,5,5,5,7,7,-5,-5,-5,5,3] = 
[5,3,-5,5,-3,-5,7,3,-5,3,-3,5,-3,7] = [-5,5,5,-5,7,3,-5,-5,-3,5,3,7,5,-3] W209 
d4 g4 EE48.36236   
   LCF [7,3,-4,7,-3,-6,3,7,3,-3,7,-3,4,6]  = [-3,-6,3,7,4,-3,7,5,-4,6,7,3,-5,7] 
= [-3,5,5,7,4,-6,-5,-5,-4,3,7,3,-3,6] = [-5,7,-4,3,-5,6,-3,3,7,5,-3,-6,4,5] = 
[7,-3,6,-6,-5,5,3,7,-6,-3,-5,6,3,5] = [4,6,-4,3,-4,4,-3,-6,3,-4,3,-3,4,-3] = 
[5,-4,4,5,-5,-5,-4,3,-5,3,-3,4,-3,5] = [4,-5,4,6,-4,-6,-4,5,3,-6,5,-3,-5,6] = 
[-5,5,-4,-6,6,3,-5,6,-3,5,-6,6,4,-6] W200 d4 g4 EE47.76064   
   LCF [-3,7,4,6,-5,7,-4,3,7,-6,-3,3,7,5]  = 
[-3,6,4,7,4,-6,-4,-6,-4,3,7,3,-3,6] = [-5,5,5,-6,4,7,-5,-5,-4,5,3,6,7,-3] = 
[5,-3,6,-6,5,-5,7,3,-6,-5,-3,6,3,7] = [-6,-4,3,7,-6,-3,3,6,6,-3,7,4,6,-6] = 
[6,-4,3,5,-5,-3,-6,3,-5,3,-3,4,-3,5] = [-3,-5,4,6,-5,3,-4,5,-3,-6,5,3,-5,5] = 
[5,-5,4,6,-5,-5,-4,5,3,-6,5,-3,-5,5] = [6,3,-6,4,-3,6,-6,-4,5,3,6,-6,-3,-5] = 
[-3,-6,3,5,6,-3,6,6,-5,6,-6,3,-6,-6] = [-5,3,-6,5,-3,6,6,6,-5,5,6,-6,-6,-6] 
W197 d4 g4 EE47.67915   
   LCF [7,3,-3,7,-3,7,3]^2  = [5,-5,7,5,7,-5,7]^2 = [-3,5,-3,5,5,5,-5;-] = 
[5,-3,7,5,7,-5,7,3,-5,7,-3,7,3,7] = [3,7,5,-3,5,7,5,-5,7,-5,3,-5,7,-3] = 
[-5,-3,3,7,3,-3,5,-3,5,5,7,-5,3,-5] = [3,7,-5,-3,-5,5,3,5,7,-3,-5,5,-5,5] W205 
d4 g4 EE47.93708   
   LCF [6,6,-3,7,5,7,-6,-6,3,-5,7,-3,7,3]  = 
[4,-3,5,7,-4,-6,3,-5,3,-3,7,-3,3,6] = [6,-3,5,7,5,-6,-6,-5,3,-5,7,-3,3,6] = 
[-6,-6,4,-5,7,5,-4,6,6,6,-5,7,5,-6] = [-3,-5,5,3,-6,4,-3,-5,5,-4,5,3,6,-5] = 
[6,6,6,-6,-6,4,-6,-6,-6,-4,3,6,6,-3] W201 d4 g4 EE47.97647   
   LCF [4,-3,6,4,-4,6,4;-]  = [7,-4,3,7,4,-3,7]^2 = [4,-3,6,3,-4,6,-3;-] = 
[3,-6,5,-3,-6,3,5;-] = [-5,-3,6,4,-5,6,4;-] = [7,-3,6,7,7,-6,3,7,-6,-3,7,7,3,6] 
= [7,-3,6,-6,6,-6,3,7,-6,-3,-6,6,3,6] = [3,-6,-5,-3,4,-6,6,3,-4,6,-3,5,-6,6] 
W197 d4 g4 EE47.72988   
   LCF [-3,3,6,7,-3,-6,3,5,-6,-3,7,3,-5,6]  = 
[6,-5,6,4,7,-6,-6,-4,-6,3,5,7,-3,6] = [-4,-6,3,-4,4,-3,5,5,-4,6,4,-5,-5,4] = 
[-6,3,-4,6,-3,5,5,6,6,-6,-5,-5,4,-6] = [-6,5,-3,6,-6,6,-5,3,6,-6,-3,-6,6,3] 
W195 d3 g4 EE47.64838   
   LCF [-3,6,-3,6,6,3,6;-]  = [7,7,-6,6,-5,7,5]^2 = [-3,5,-3,6,6,3,-5;-] = 
[6,-4,-4,5,5,5,-6;-] = [-5,-5,4,4,7,7,-4,-4,5,5,5,7,7,-5] = 
[-5,-3,4,4,7,5,-4,-4,5,5,-5,7,3,-5] = [-6,3,-6,6,-3,7,5,6,6,-6,6,-5,7,-6] = 
[5,3,6,-5,-3,-5,3,4,-6,-3,3,-4,5,-3] = [-6,3,-5,-5,-3,4,4,6,6,-4,-4,5,5,-6] 
W199 d4 g4 EE48.03267   
   LCF [-3,-6,3,7,3,-3,6,-3,5,6,7,3,-6,-5]  = 
[-4,-6,5,3,7,-6,-3,-5,4,6,4,7,-4,6] = [6,-4,6,7,3,-6,-6,-3,-6,3,7,4,-3,6] = 
[5,-6,6,4,7,-5,6,-4,-6,6,3,7,-6,-3] = [6,6,7,-5,6,-6,-6,-6,3,7,-6,-3,5,6] = 
[-6,3,-4,3,-3,5,-3,4,6,4,-5,-4,4,-4] = [5,3,-6,5,-3,-5,5,6,-5,3,6,-5,-3,-6] = 
[6,6,-3,-6,4,5,-6,-6,-4,3,-5,6,-3,3] W197 d4 g4 EE47.75062   
   LCF [-4,-6,4,4,7,7,-4,-4,5,6,4,7,7,-5]  = [5,-6,6,7,7,-5,6,6,-6,6,7,7,-6,-6] 
= [4,-3,3,7,-4,-3,3,4,5,-3,7,-4,3,-5] = [-6,-6,3,4,7,-3,6,-4,6,6,3,7,-6,-3] = 
[-6,5,5,-6,6,7,-5,-5,6,4,-6,6,7,-4] = [4,4,-4,5,-4,-4,3,4,-5,-3,3,-4,4,-3] = 
[4,4,6,-5,-4,-4,3,4,-6,-3,3,-4,5,-3] = [3,4,6,-3,-6,-4,3,4,-6,-3,3,-4,6,-3] = 
[-6,3,4,-6,-3,5,-4,6,6,3,-5,6,-3,-6] = [-5,4,4,-6,6,-4,-4,6,4,5,-6,6,-4,-6] = 
[4,-6,-6,5,-4,-6,5,5,-5,6,6,-5,-5,6] W197 d4 g4 EE47.83708   
   LCF [3,-6,6,-3,7,-6,6]^2  = [-6,6,4,-6,6,7,-4]^2 = 
[4,7,-3,6,-4,7,5,3,7,-6,-3,-5,7,3] = [-3,3,4,5,-3,-6,-4,3,-5,3,-3,3,-3,6] = 
[6,4,5,-6,6,-4,-6,-5,5,3,-6,6,-3,-5] W202 d4 g4 EE48.20639   
\end{Verbatim}

Figure \ref{fig.14n2}:

\begin{Verbatim}
   LCF [3,-3,5,-3,4,4,5;-]  = [-6,6,7,3,7,7,-3]^2 = [3,-3,4,-3,4,4,-4;-] = 
[6,4,-5,6,6,-4,-6;-] = [6,6,7,7,7,7,-6,-6,5,7,7,7,7,-5] = 
[-5,5,6,4,7,7,-5,-4,-6,5,3,7,7,-3] = [4,-5,3,6,-4,-3,7,5,3,-6,5,-3,-5,7] W205 
d4 g4 EE48.21641   
   LCF [-3,5,3,7,4,-3,-5,6,-4,3,7,3,-3,-6]  = 
[5,-3,-6,6,4,-5,7,4,-4,-6,6,-4,3,7] W204 d4 g4 EE48.18647   
   LCF [6,7,7,-4,7,7,-6,3,7,7,-3,7,7,4]  = [6,7,5,7,7,7,-6,-5,7,4,7,7,7,-4] = 
[7,4,-5,7,4,-4,7,7,-4,3,7,5,-3,7] = [7,7,-6,-5,-5,6,3,7,7,-3,6,-6,5,5] = 
[4,5,-3,7,-4,4,-5,5,3,-4,7,-3,-5,3] = [4,4,-4,7,-4,-4,3,6,3,-3,7,-3,4,-6] = 
[-3,3,4,7,-3,4,-4,6,4,-4,7,3,-4,-6] = [5,-3,-6,5,5,-5,7,4,-5,-5,6,-4,3,7] = 
[4,-3,-6,5,-4,-6,3,4,-5,-3,6,-4,3,6] = [-5,6,-4,3,4,6,-3,-6,-4,5,3,-6,4,-3] = 
[3,-3,6,-3,5,-6,4,4,-6,-5,-4,-4,3,6] W200 d4 g4 EE47.84785   
   LCF [3,-4,7,-3,4,-6,6]^2  W199 d4 g4 EE47.86854   
   LCF [-4,7,4,4,7,7,-4]^2  = [7,-6,6,7,7,-6,6]^2 = [4,-3,3,4,-4,-3,4;-] = 
[-4,7,4,4,-6,6,-4]^2 W197 d4 g4 EE48.08295   
   LCF [7,7,7,-6,6,7,7]^2  = [-3,3,-3,4,-3,3,4;-] = [6,-3,-5,6,6,3,-6;-] = 
[-4,7,5,3,7,7,-3,-5,7,4,4,7,7,-4] = [-5,6,6,3,7,7,-3,-6,-6,5,3,7,7,-3] = 
[4,4,-6,6,-4,-4,7,5,3,-6,6,-3,-5,7] W207 d4 g4 EE48.66179   
   LCF [6,6,7,7,7,-6,-6,-6,4,7,7,7,-4,6]  = [7,7,-6,-4,-6,6,3,7,7,-3,6,-6,6,4] 
= [-6,3,-4,3,-3,6,-3,3,6,4,-3,-6,4,-4] = [5,6,-3,-5,6,-5,3,-6,3,-3,-6,-3,5,3] = 
[-6,6,-4,3,6,6,-3,-6,6,4,-6,-6,4,-4] W201 d4 g4 EE48.11933   
   LCF [7,7,-6,6,-6,6,7]^2  = [3,4,-5,-3,3,-4,3;-] = [6,-4,-4,6,6,3,-6;-] = 
[-6,-6,4,4,7,7,-4,-4,6,6,3,7,7,-3] W201 d4 g4 EE48.21995   
   LCF [7]^14  = [-3,3]^7 = [-5,7,5,3,7,7,-3]^2 W217 d4 g4 EE49.25357   
   LCF [3,6,-5,-3,5,3,6;-]  = [5,-6,6,-6,6,-5,7]^2 = 
[3,-6,6,-3,7,5,7,5,-6,6,-5,7,-5,7] W197 d3 g4 EE47.87399   
   LCF [4,-3,5,5,-4,5,5;-]  = [6,3,-5,5,-3,5,-6;-] = 
[-3,7,3,7,-6,-3,3,6,7,-3,7,3,6,-6] = [-6,5,-5,7,3,7,-5,-3,6,4,7,5,7,-4] = 
[3,-6,4,-3,7,5,-4,5,5,6,-5,7,-5,-5] = [-6,3,-6,4,-3,7,5,-4,6,4,6,-5,7,-4] W197 
d4 g4 EE47.58692   
   LCF [-5,6,4,7,7,7,-4,-6,5,5,7,7,7,-5]  = [6,6,7,5,7,7,-6,-6,-5,7,3,7,7,-3] = 
[-3,7,-3,3,-6,4,-3,3,7,-4,-3,3,6,3] = [-3,7,5,-4,-6,4,4,-5,7,-4,-4,3,6,4] = 
[-5,6,4,-5,7,5,-4,-6,5,5,-5,7,5,-5] = [-3,5,3,-5,4,-3,-5,3,-4,4,-3,3,5,-4] = 
[-4,-4,3,6,4,-3,6,6,-4,-6,4,4,-6,-6] W203 d4 g4 EE48.11421   
\end{Verbatim}

Figure \ref{fig.14n3}:

\begin{Verbatim}
   LCF [7,-3,-6,6,-5,3,5]^2  = [7,3,-4,6,-3,5,7,7,3,-6,-5,-3,4,7] = 
[7,-5,3,6,-6,-3,7,7,3,-6,5,-3,6,7] = [-4,5,-3,6,3,7,-5,-3,3,-6,4,-3,7,3] = 
[-5,-5,3,5,7,-3,6,6,-5,5,5,7,-6,-6] = [5,-4,-6,5,-5,-5,3,5,-5,-3,6,4,-5,5] = 
[-6,3,-4,6,-3,6,4,6,6,-6,-4,-6,4,-6] W197 d4 g4 EE47.60125   
   LCF [7,-5,7,-4,7,3,6,7,-3,7,5,7,-6,4]  = [7,7,-6,-4,5,7,5,7,7,-5,6,-5,7,4] = 
[7,-4,3,7,-5,-3,3,7,4,-3,7,4,-4,5] = [-5,5,-6,4,7,7,-5,-4,5,5,6,7,7,-5] = 
[-3,6,4,7,-5,4,-4,-6,4,-4,7,3,-4,5] = [-5,4,5,-5,7,-4,4,-5,5,5,-4,7,5,-5] = 
[7,3,6,-6,-3,-6,4,7,-6,3,-4,6,-3,6] = [7,-6,-6,3,-5,6,-3,7,4,6,6,-6,-4,5] = 
[-5,5,-3,4,-6,5,-5,-4,3,5,-5,-3,6,3] = [-5,-4,3,5,6,-3,6,6,-5,5,-6,4,-6,-6] 
W195 d4 g4 EE47.32016   
   LCF [7,-4,6,7,-6,-6,3,7,-6,-3,7,4,6,6]  = 
[4,-4,-6,5,-4,4,7,5,-5,-4,6,4,-5,7] = [-5,-4,3,7,4,-3,6,6,-4,5,7,4,-6,-6] W194 
d4 g4 EE47.44879   
   LCF [-5,5,5,7,7,7,-5]^2  = [3,5,-5,-3,5,3,-5;-] = [-5,5,5,-5,7,5,-5]^2 = 
[7,7,-3,7,5,7,5,7,7,-5,7,-5,7,3] = [7,3,-3,7,-3,3,5,7,-3,3,7,-5,-3,3] W205 d4 
g4 EE47.91606   
   LCF [7,-5,3,5,7,-3,7]^2  = [-5,-3,5,5,-5,5,5;-] = 
[-3,3,5,7,-3,7,3,-5,5,-3,7,3,7,-5] = [7,-5,3,-5,5,-3,7,7,3,-5,5,-3,5,7] = 
[-5,5,-5,7,3,7,-5,-3,5,5,7,5,7,-5] W201 d4 g4 EE47.49035   
   LCF [7,4,7,-5,6,-4,7,7,3,7,-6,-3,5,7]  = [-5,-4,4,7,3,7,-4,-3,5,5,7,4,7,-5] 
= [4,7,-3,7,-4,6,3,5,7,-3,7,-6,-5,3] = [7,4,-5,5,6,-4,7,7,-5,3,-6,5,-3,7] = 
[-5,7,3,7,-6,-3,5,6,7,5,7,-5,6,-6] = [-3,-6,4,7,3,-6,-4,-3,4,6,7,3,-4,6] = 
[-4,-6,4,4,7,-6,-4,-4,4,6,4,7,-4,6] = [-6,-4,4,-4,3,5,-4,-3,6,3,-5,4,-3,4] = 
[-5,-4,4,-5,3,5,-4,-3,5,5,-5,4,5,-5] = [-6,-4,4,-5,3,5,-4,-3,6,4,-5,4,5,-4] = 
[4,6,4,-6,-4,5,-4,-6,4,4,-5,6,-4,-4] W197 d4 g4 EE47.42171   
   LCF [3,-3,6,-3,3,6,3;-]  = [-6,6,4,7,7,7,-4]^2 = [6,-3,3,6,6,-3,-6;-] = 
[-6,-6,3,7,7,-3,6,6,6,6,7,7,-6,-6] = [-3,-5,3,6,4,-3,6,6,-4,-6,5,3,-6,-6] W203 
d4 g4 EE48.46324   
   LCF [7,-4,3,-4,4,-3,4]^2  = [-5,5,-3,4,7,5,-5,-4,4,5,-5,7,-4,3] W197 d4 g4 
EE47.32024   
   LCF [5,-3,5,6,6,-5,5;-]  = [4,6,-6,-6,-4,3,6;-] = 
[7,-4,3,5,-5,-3,4,7,-5,3,-4,4,-3,5] = [6,-6,-3,4,7,5,-6,-4,4,6,-5,7,-4,3] = 
[7,5,-6,5,-5,6,-5,7,-5,3,6,-6,-3,5] W193 d3 g4 EE47.18167   
   LCF [-5,7,3,-5,7,-3,4,6,7,5,-4,7,5,-6]  = [3,7,5,-3,6,7,5,-5,7,4,-6,-5,7,-4] 
= [5,-6,-5,7,4,-5,7,5,-4,6,7,5,-5,7] = [-3,6,4,7,-6,3,-4,-6,-3,4,7,3,6,-4] = 
[-4,4,-4,5,3,-4,5,-3,-5,4,4,-5,4,-4] = [-3,5,-3,5,-6,4,-5,3,-5,-4,-3,3,6,3] = 
[-4,-4,5,3,-6,4,-3,-5,5,-4,4,4,6,-5] = [-4,-3,3,6,4,-3,5,6,-4,-6,4,-5,3,-6] = 
[-3,6,-6,3,-6,3,-3,-6,-3,4,6,3,6,-4] W199 d4 g4 EE47.76644   
   LCF [4,6,-3,7,-4,7,3,-6,3,-3,7,-3,7,3]  = [6,-5,7,5,7,-6,-6,5,-5,7,5,7,-5,6] 
= [-4,-4,5,3,5,7,-3,-5,5,-5,4,4,7,-5] = [5,-3,6,7,-5,-5,3,4,-6,-3,7,-4,3,5] = 
[-5,7,-4,3,6,6,-3,6,7,5,-6,-6,4,-6] = [6,6,-3,7,5,6,-6,-6,4,-5,7,-6,-4,3] = 
[4,-4,6,-4,-4,5,3,5,-6,-3,-5,4,-5,4] = [-5,3,-5,-5,-3,3,4,6,-3,5,-4,5,5,-6] = 
[-6,3,-5,3,-3,5,-3,5,6,4,-5,5,-5,-4] = [5,5,5,-5,6,-5,-5,-5,3,4,-6,-3,5,-4] = 
[6,-5,6,6,-6,-6,-6,4,-6,-6,5,-4,6,6] W198 d4 g4 EE47.70929   
   LCF [3,-6,5,-3,7,5,7,-5,4,6,-5,7,-4,7]  = 
[-5,4,-4,7,3,-4,5,-3,5,5,7,-5,4,-5] = [6,-6,-3,7,3,6,-6,-3,4,6,7,-6,-4,3] = 
[5,-6,5,-6,6,-5,7,-5,4,6,-6,6,-4,7] = [-5,3,-4,5,-3,5,5,6,-5,5,-5,-5,4,-6] = 
[6,-6,-3,-6,3,5,-6,-3,4,6,-5,6,-4,3] W193 d3 g4 EE47.23870   
\end{Verbatim}

Figure \ref{fig.14n4}:

\begin{Verbatim}
   LCF [-3,6,4,7,4,7,-4,-6,-4,4,7,3,7,-4]  = 
[-5,-4,4,4,6,7,-4,-4,5,5,-6,4,7,-5] = [3,-6,5,-3,7,5,6,-5,5,6,-5,7,-6,-5] = 
[3,4,-6,-3,4,-4,5,6,-4,3,6,-5,-3,-6] = [4,4,-6,5,-4,-4,5,5,-5,4,6,-5,-5,-4] 
W194 d4 g4 EE47.39737   
   LCF [6,-4,7,7,3,-6,-6,-3,4,7,7,4,-4,6]  = [7,3,-6,6,-3,6,7,7,4,-6,6,-6,-4,7] 
= [-3,7,3,-5,3,-3,4,-3,7,4,-4,3,5,-4] = [5,5,-4,7,3,-5,-5,-3,3,4,7,-3,4,-4] = 
[5,-6,5,3,7,-5,-3,-5,4,6,3,7,-4,-3] = [-4,5,5,5,7,-6,-5,-5,-5,3,4,7,-3,6] = 
[6,6,7,-6,6,-6,-6,-6,4,7,-6,6,-4,6] = [5,-3,-6,5,-5,-5,3,4,-5,-3,6,-4,3,5] = 
[-4,-3,5,3,5,6,-3,-5,5,-5,4,-6,3,-5] = [-6,3,-6,3,-3,6,-3,5,6,4,6,-6,-5,-4] = 
[-5,6,-4,3,6,6,-3,-6,5,5,-6,-6,4,-5] = [5,-6,6,-6,-5,-5,6,3,-6,6,-3,6,-6,5] 
W197 d4 g4 EE47.78069   
   LCF [-3,6,-3,5,5,5,6;-]  = [3,-4,3,-3,4,-3,7]^2 = 
[3,5,5,-3,-6,4,-5,-5,3,-4,3,-3,6,-3] = [3,-6,6,-3,6,-6,6,4,-6,6,-6,-4,-6,6] 
W201 d4 g4 EE48.06221   
   LCF [6,7,5,-6,5,7,-6,-5,7,-5,3,6,7,-3]  = 
[3,4,6,-3,7,-4,4,6,-6,3,-4,7,-3,-6] = [7,-3,6,-6,-5,4,4,7,-6,-4,-4,6,3,5] = 
[7,-5,-5,6,-5,3,6,7,-3,-6,5,5,-6,5] = [-4,4,-3,5,-6,-4,3,4,-5,-3,4,-4,6,3] = 
[-5,4,-4,3,6,-4,-3,4,5,5,-6,-4,4,-5] = [-6,4,-4,3,6,-4,-3,4,6,4,-6,-4,4,-4] = 
[4,4,-6,5,-4,-4,5,6,-5,3,6,-5,-3,-6] = [-5,-4,3,4,6,-3,6,-4,5,5,-6,4,-6,-5] = 
[6,-3,6,-6,-5,4,-6,3,-6,-4,-3,6,3,5] = [6,-3,5,-6,5,5,-6,-5,5,-5,-5,6,3,-5] 
W194 d4 g4 EE47.41178   
   LCF [4,7,-6,6,-4,7,7]^2  = [4,-4,-4,4,-4,3,4;-] = [-4,-3,5,-4,4,4,5;-] = 
[-3,7,4,4,6,7,-4,-4,7,4,-6,3,7,-4] = [6,-4,7,7,4,-6,-6,5,-4,7,7,4,-5,6] = 
[-6,5,7,-6,6,7,-5,6,6,7,-6,6,7,-6] = [-3,7,4,4,-6,5,-4,-4,7,4,-5,3,6,-4] = 
[5,-3,-6,-4,4,-5,3,4,-4,-3,6,-4,3,4] W195 d4 g4 EE47.65770   
   LCF [3,6,-5,-3,4,4,6;-]  = [3,5,-5,-3,4,4,-5;-] = [5,5,-5,6,6,-5,-5;-] = 
[-5,5,5,-6,6,7,-5]^2 = [-6,6,-5,6,6,-6,6;-] = 
[6,-3,7,-4,3,5,-6,-3,3,7,-5,-3,3,4] W195 d3 g4 EE47.62118   
   LCF [7,7,7,-4,7,7,4]^2  = [5,-4,7,7,4,-5,7]^2 = [4,-3,-5,4,-4,3,4;-] = 
[-4,-3,6,-4,3,6,3;-] = [7,-6,3,7,7,-3,6,7,5,6,7,7,-6,-5] W199 d4 g4 EE47.92512  
   LCF [7,-5,7,5,7,-6,6]^2  = [-4,-3,5,-4,5,3,5;-] = 
[-5,5,-4,-4,7,3,-5,3,-3,5,-3,7,4,4] = [4,4,6,-6,-4,-4,7,3,-6,3,-3,6,-3,7] = 
[7,5,6,-6,6,-6,-5,7,-6,3,-6,6,-3,6] W199 d4 g4 EE47.81083   
   LCF [-5,-4,3,7,4,-3,7,5,-4,5,7,4,-5,7]  = [7,-6,3,5,7,-3,6,7,-5,6,3,7,-6,-3] 
= [7,7,-6,-4,-6,4,5,7,7,-4,6,-5,6,4] = [6,-6,5,7,7,-6,-6,-5,4,6,7,7,-4,6] = 
[-4,7,-3,3,-6,4,-3,4,7,-4,4,-4,6,3] = [-6,5,-4,7,3,6,-5,-3,6,4,7,-6,4,-4] = 
[-3,-6,3,-4,4,-3,4,5,-4,6,-4,3,-5,4] = [4,-4,6,6,-4,-6,4,5,-6,-6,-4,4,-5,6] 
W195 d4 g4 EE47.39167   
   LCF [-4,5,7,4,7,7,-5]^2  = [4,-3,4,5,-4,5,-4;-] = 
[7,-3,7,-6,4,-6,4,7,-4,7,-4,6,3,6] = [6,7,-4,7,3,6,-6,-3,7,4,7,-6,4,-4] = 
[5,-6,5,7,7,-5,6,-5,5,6,7,7,-6,-5] = [5,-4,7,-4,3,-5,4,-3,4,7,-4,4,-4,4] = 
[5,-5,6,4,7,-5,6,-4,-6,4,5,7,-6,-4] W193 d4 g4 EE47.05310   
   LCF [4,-5,5,-5,-4,3,5;-]  = [6,-5,6,-5,5,6,-6;-] = 
[4,-4,5,7,-4,7,3,-5,5,-3,7,4,7,-5] = [-3,4,7,-5,4,-4,4,6,-4,7,-4,3,5,-6] = 
[-4,7,3,-5,5,-3,5,6,7,-5,4,-5,5,-6] = [7,-5,-4,5,5,-6,5,7,-5,-5,5,-5,4,6] = 
[-3,4,6,-5,5,-4,4,6,-6,-5,-4,3,5,-6] = [6,-4,5,-5,5,5,-6,-5,5,-5,-5,4,5,-5] 
W191 d3 g4 EE46.91447   
   LCF [6,7,-6,6,-5,7,-6,4,7,-6,6,-4,7,5]  = 
[5,-4,4,7,3,-5,-4,-3,4,4,7,4,-4,-4] = [6,-4,-3,7,3,4,-6,-3,4,-4,7,4,-4,3] = 
[-4,5,6,4,7,-6,-5,-4,-6,3,4,7,-3,6] = [5,-3,6,-4,5,-5,4,4,-6,-5,-4,-4,3,4] W193 
d3 g4 EE47.55605   
\end{Verbatim}

Figure \ref{fig.14n5}:

\begin{Verbatim}
   LCF [-4,3,5,-4,-3,3,5;-]  = [7,-6,6,-6,6,-6,6]^2 W195 d3 g4 EE47.79409   
   LCF [6,-3,6,6,6,6,-6;-]  = [7,3,-6,6,-3,7,7]^2 = [6,-3,-5,5,5,5,-6;-] = 
[6,6,-3,7,7,3,-6,-6,-3,3,7,7,-3,3] = [3,-6,3,-3,6,-3,7,5,3,6,-6,-3,-5,7] = 
[6,-5,-5,5,3,-6,-6,-3,-5,3,5,5,-3,6] W203 d4 g4 EE48.34463   
   LCF [6,-5,5,-4,7,3,-6,-5,-3,3,5,7,-3,4]  = 
[5,-6,-3,7,3,-5,5,-3,4,6,7,-5,-4,3] = [4,-3,5,7,-4,6,3,-5,5,-3,7,-6,3,-5] = 
[5,-6,-3,7,4,-5,5,5,-4,6,7,-5,-5,3] = [5,-5,5,-6,4,-5,7,-5,-4,3,5,6,-3,7] = 
[-3,-6,3,-5,4,-3,4,6,-4,6,-4,3,5,-6] = [-5,4,-3,5,6,-4,6,4,-5,5,-6,-4,-6,3] = 
[-5,5,-4,4,6,6,-5,-4,5,5,-6,-6,4,-5] = [-5,5,-6,-5,3,6,-5,-3,5,5,6,-6,5,-5] = 
[-5,5,-6,4,-6,6,-5,-4,5,5,6,-6,6,-5] = [-5,4,6,-6,6,-4,6,6,-6,5,-6,6,-6,-6] 
W194 d4 g4 EE47.25303   
   LCF [5,-5,7,4,7,-5,7,-4,4,7,5,7,-4,7]  = [7,-4,4,7,-5,3,-4,7,-3,3,7,4,-3,5] 
= [-5,4,-5,7,4,-4,7,5,-4,5,7,5,-5,7] = [-5,3,-4,7,-3,3,5,6,-3,5,7,-5,4,-6] = 
[7,-5,-3,6,-6,3,5,7,-3,-6,5,-5,6,3] = [6,6,-5,7,-5,4,-6,-6,4,-4,7,5,-4,5] = 
[7,-4,6,6,-6,-6,4,7,-6,-6,-4,4,6,6] = [3,-4,6,-3,6,-6,3,5,-6,-3,-6,4,-5,6] W195 
d4 g4 EE47.32025   
   LCF [3,5,6,-3,5,6,-5;-]  = [-4,6,6,-4,5,6,6;-] = [-4,6,3,-4,5,-3,6;-] = 
[-6,4,7,-5,7,-4,4,6,6,7,-4,7,5,-6] W194 d4 g4 EE47.31023   
   LCF [7,-5,3,5,7,-3,6,7,-5,4,5,7,-6,-4]  = [5,-5,7,4,7,-5,6,-4,5,7,5,7,-6,-5] 
= [7,-5,-4,6,-5,3,5,7,-3,-6,5,-5,4,5] = [6,7,-5,4,-6,4,-6,-4,7,-4,3,5,6,-3] = 
[4,7,-4,6,-4,6,4,6,7,-6,-4,-6,4,-6] W193 d4 g4 EE46.99597   
   LCF [4,-4,4,7,-4,4,-4,5,5,-4,7,4,-5,-5]  = 
[7,-6,3,-4,6,-3,5,7,4,6,-6,-5,-4,4] = [4,7,-4,6,-4,5,5,6,7,-6,-5,-5,4,-6] = 
[-6,-4,4,7,-6,4,-4,6,6,-4,7,4,6,-6] = [-6,4,7,-5,6,-4,5,6,6,7,-6,-5,5,-6] = 
[5,-5,6,-6,3,-5,6,-3,-6,4,5,6,-6,-4] W191 d3 g4 EE47.04310   
   LCF [-3,7,4,-5,3,5,-4]^2  = [4,7,-6,6,-4,6,7,5,7,-6,6,-6,-5,7] = 
[-4,4,6,3,7,-4,-3,5,-6,4,4,7,-5,-4] = [6,7,-5,-4,-6,4,-6,3,7,-4,-3,5,6,4] = 
[3,-6,-4,-3,4,-6,4,4,-4,6,-4,-4,4,6] W196 d4 g4 EE47.51335   
   LCF [-3,5,7,-5,3,5,-5]^2  = [5,-3,5,7,5,-5,5,-5,5,-5,7,-5,3,-5] W197 d4 g4 
EE47.08518   
   LCF [5,-3,6,7,5,-5,5,6,-6,-5,7,-5,3,-6]  = 
[-4,-4,4,6,-5,3,-4,5,-3,-6,4,4,-5,5] = [4,-5,-3,5,-4,6,3,5,-5,-3,5,-6,-5,3] = 
[6,3,-6,4,-3,6,-6,-4,4,4,6,-6,-4,-4] = [3,-6,-4,-3,4,6,4,6,-4,6,-4,-6,4,-6] 
W196 d4 g4 EE47.35467   
   LCF [-5,-4,4,7,-5,3,-4,5,-3,5,7,4,-5,5]  = 
[7,-5,3,6,-5,-3,5,7,4,-6,5,-5,-4,5] = [-5,5,6,-4,7,5,-5,5,-6,5,-5,7,-5,4] = 
[6,-4,5,-4,5,5,-6,-5,4,-5,-5,4,-4,4] = [-5,5,-5,-4,6,3,-5,5,-3,5,-6,5,-5,4] = 
[-4,5,-4,6,3,6,-5,-3,5,-6,4,-6,4,-5] = [5,5,6,-5,6,-5,-5,4,-6,4,-6,-4,5,-4] = 
[-5,4,6,-4,6,-4,6,4,-6,5,-6,-4,-6,4] = [-5,4,-5,5,6,-4,6,6,-5,5,-6,5,-6,-6] = 
[6,-6,-6,4,-5,6,-6,-4,4,6,6,-6,-4,5] W192 d4 g4 EE46.91455   
   LCF [6,-6,5,-4,7,5,-6,-5,4,6,-5,7,-4,4]  = 
[-5,-4,4,-4,6,3,-4,5,-3,5,-6,4,-5,4] = [6,-4,5,-4,6,4,-6,-5,4,-4,-6,4,-4,4] = 
[-4,6,3,6,-6,-3,5,-6,5,-6,4,-5,6,-5] W191 d3 g4 EE46.97169   
\end{Verbatim}

Figure \ref{fig.14n6}:

\begin{Verbatim}
   LCF [3,4,5,-3,5,-4,5;-]  = [4,7,-3,7,-4,7,3]^2 = [4,4,-5,4,-4,-4,4;-] = 
[-4,-4,4,4,5,7,-4,-4,5,-5,4,4,7,-5] = [-4,4,6,3,7,-4,-3,6,-6,3,4,7,-3,-6] = 
[-6,5,-4,7,5,6,-5,6,6,-5,7,-6,4,-6] = [7,-6,-6,5,-5,6,6,7,-5,6,6,-6,-6,5] = 
[4,-4,6,-4,-4,4,4,5,-6,-4,-4,4,-5,4] = [4,-5,3,6,-4,-3,6,4,5,-6,5,-4,-6,-5] 
W195 d4 g4 EE47.49901   
   LCF [-4,3,-5,4,-3,7,3,-4,5,-3,4,5,7,-5]  = 
[7,3,-6,5,-3,-6,5,7,-5,3,6,-5,-3,6] = [6,-3,-6,4,-5,4,-6,-4,3,-4,6,-3,3,5] = 
[-3,6,4,6,-6,5,-4,-6,5,-6,-5,3,6,-5] = [-4,6,4,6,-6,6,-4,-6,5,-6,4,-6,6,-5] 
W193 d3 g4 EE47.32454   
   LCF [4,6,-3,6,-4,7,4,-6,3,-6,-4,-3,7,3]  = 
[6,6,-4,7,5,6,-6,-6,5,-5,7,-6,4,-5] = [-6,5,7,-5,6,6,-5,6,6,7,-6,-6,5,-6] = 
[5,-3,6,-4,6,-5,3,4,-6,-3,-6,-4,3,4] = [6,4,-6,5,-5,-4,-6,4,-5,3,6,-4,-3,5] 
W194 d4 g4 EE47.48901   
   LCF [7,-6,4,-6,6,-6,-4,7,4,6,-6,6,-4,6]  = 
[-4,6,-4,3,5,6,-3,-6,5,-5,4,-6,4,-5] W191 d3 g4 EE47.05742   
   LCF [3,6,6,-3,5,6,6;-]  = [3,6,3,-3,5,-3,6;-] = [5,3,-5,4,-3,-5,4;-] = 
[6,6,-4,7,5,-6,-6,-6,3,-5,7,-3,4,6] = [-3,6,6,7,-6,-6,3,-6,-6,-3,7,3,6,6] W200 
d4 g4 EE48.02357   
   LCF [5,-6,6,-4,7,-5,6,3,-6,6,-3,7,-6,4]  W193 d3 g4 EE47.48376   
   LCF [7,4,-4,6,-5,-4,4,7,3,-6,-4,-3,4,5]  = 
[7,-6,-5,5,6,-6,6,7,-5,6,-6,5,-6,6] = [-5,-4,4,-4,3,5,-4,-3,4,5,-5,4,-4,4] W191 
d3 g4 EE47.14460   
   LCF [5,-3,4,6,6,-5,-4;-]  = [-6,-3,5,6,6,-6,5;-] W193 d3 g4 EE47.30020   
   LCF [5,-6,6,-4,7,-5,4]^2  = [-6,5,-3,7,-6,4,-5,4,6,-4,7,-4,6,3] = 
[-4,7,3,6,-6,-3,5,6,7,-6,4,-5,6,-6] = [6,-5,-4,4,-5,4,-6,-4,3,-4,5,-3,4,5] = 
[-4,6,4,-5,5,6,-4,-6,5,-5,4,-6,5,-5] W191 d3 g4 EE47.04309   
   LCF [7,-4,3,7,4,-3,6,7,-4,4,7,4,-6,-4]  = [5,-6,-4,7,4,-5,7,4,-4,6,7,-4,4,7] 
= [6,4,-4,7,4,-4,-6,4,-4,4,7,-4,4,-4] = [7,-3,6,-6,5,-6,4,7,-6,-5,-4,6,3,6] = 
[-4,6,-5,3,-6,4,-3,-6,5,-4,4,5,6,-5] W191 d3 g4 EE47.11027   
   LCF [5,6,-5,6,6,-5,6;-]  = [5,-4,-4,4,5,-5,4;-] = [-4,6,-5,-4,4,4,6;-] = 
[5,6,-6,-6,5,-5,6;-] W191 d3 g4 EE47.06319   
   LCF [5,-3,-5,4,5,-5,4;-]  = [-4,6,-5,-4,5,3,6;-] = 
[-5,5,6,-5,7,5,-5,6,-6,5,-5,7,5,-6] W195 d4 g4 EE47.34090   
\end{Verbatim}

Figure \ref{fig.14ne}:

\begin{Verbatim}
   LCF [-3,4,-3,4,5,-4,4;-]  = [-6,5,7,4,7,7,-5,-4,6,7,3,7,7,-3] = 
[7,7,-6,-5,5,6,7,7,7,-5,6,-6,5,7] = [7,3,-3,7,-3,4,5,7,4,-4,7,-5,-4,3] W199 d4 
g4 EE47.50380   
   LCF [5,3,-6,6,-3,-5,7]^2  W217 d4 g4 EE49.25588   
   LCF [7,-6,-4,7,4,6,7,7,-4,6,7,-6,4,7]  = [5,-4,7,-4,4,-5,6,3,-4,7,-3,4,-6,4] 
W193 d4 g4 EE47.19176   
   LCF [4,7,-4,7,-4,4,7]^2  = [7,-4,4,7,-5,4,-4,7,4,-4,7,4,-4,5] = 
[7,-6,-5,-4,4,5,6,7,-4,6,-5,5,-6,4] W189 d3 g5 EE46.75706   
   LCF [-5,5,-4,7,4,-6,-5,4,-4,5,7,-4,4,6]  = 
[5,-5,5,-6,4,-5,6,-5,-4,4,5,6,-6,-4] = [6,-6,5,-6,-5,5,-6,-5,4,6,-5,6,-4,5] 
W189 d3 g5 EE46.62280   
   LCF [-4,5,6,-4,5,6,-5;-]  = [-6,-4,5,7,-5,4,6,-5,6,-4,7,4,-6,5] W190 d4 g5 
EE46.60848   
   LCF [5,-6,5,-4,7,-5,4,-5,4,6,-4,7,-4,4]  = 
[-5,5,-4,7,4,6,-5,6,-4,5,7,-6,4,-6] = [7,-5,6,-6,-5,4,6,7,-6,-4,5,6,-6,5] = 
[5,-6,-5,-4,4,-5,4,5,-4,6,-4,5,-5,4] W189 d3 g5 EE46.70425   
   LCF [-5,4,-4,7,4,-4,5,6,-4,5,7,-5,4,-6]  W189 d3 g5 EE46.67561   
   LCF [5,-4,7,-4,4,-5,4]^2  = [4,7,-4,7,-4,4,5,6,7,-4,7,-5,4,-6] = 
[7,-4,6,7,-5,4,6,7,-6,-4,7,4,-6,5] W190 d4 g5 EE46.77138   
   LCF [4,6,-5,5,-4,5,6;-]  = [4,5,-5,5,-4,5,-5;-] = [-5,6,-5,5,-5,5,6;-] W189 
d3 g5 EE46.50858   
   LCF [-4,4,5,-4,5,-4,5;-]  = [7,-5,-4,5,-5,4,5]^2 = 
[7,-6,-5,5,-5,5,6,7,-5,6,-5,5,-6,5] W189 d3 g5 EE46.64714   
   LCF [-5,5]^7  W189 d3 g6 EE46.27353   
\end{Verbatim}

Some of these have appeared in the nuclear physics literature \cite{PonzanoNC36,LevaNCIM51}.
Ponzano's figures (1)--(8) are number 58, 32, 68, 33, 71, 57, 59 and 82 in this list of 84.

\begin{figure}
\includegraphics[scale=0.45]{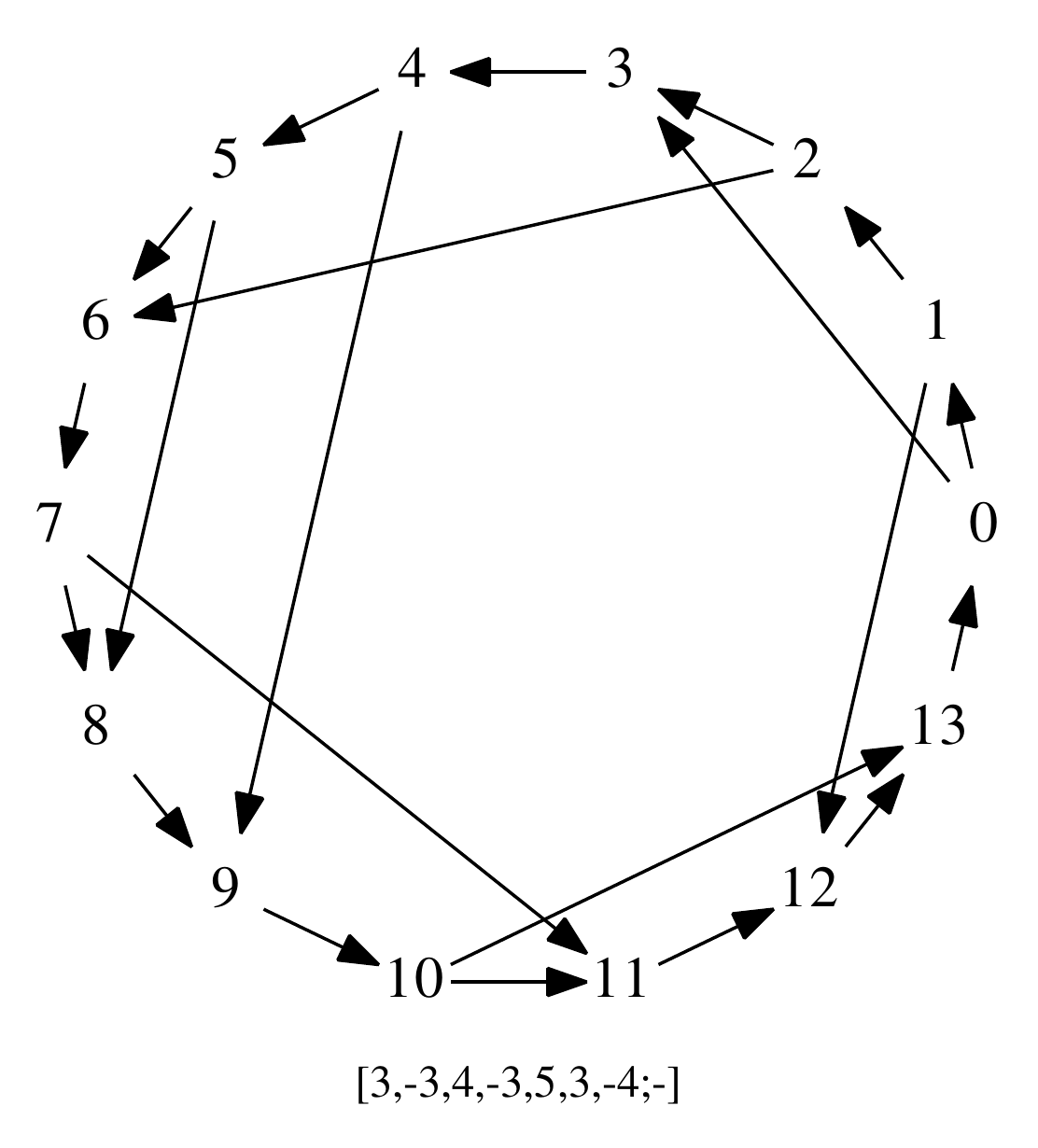}
\includegraphics[scale=0.45]{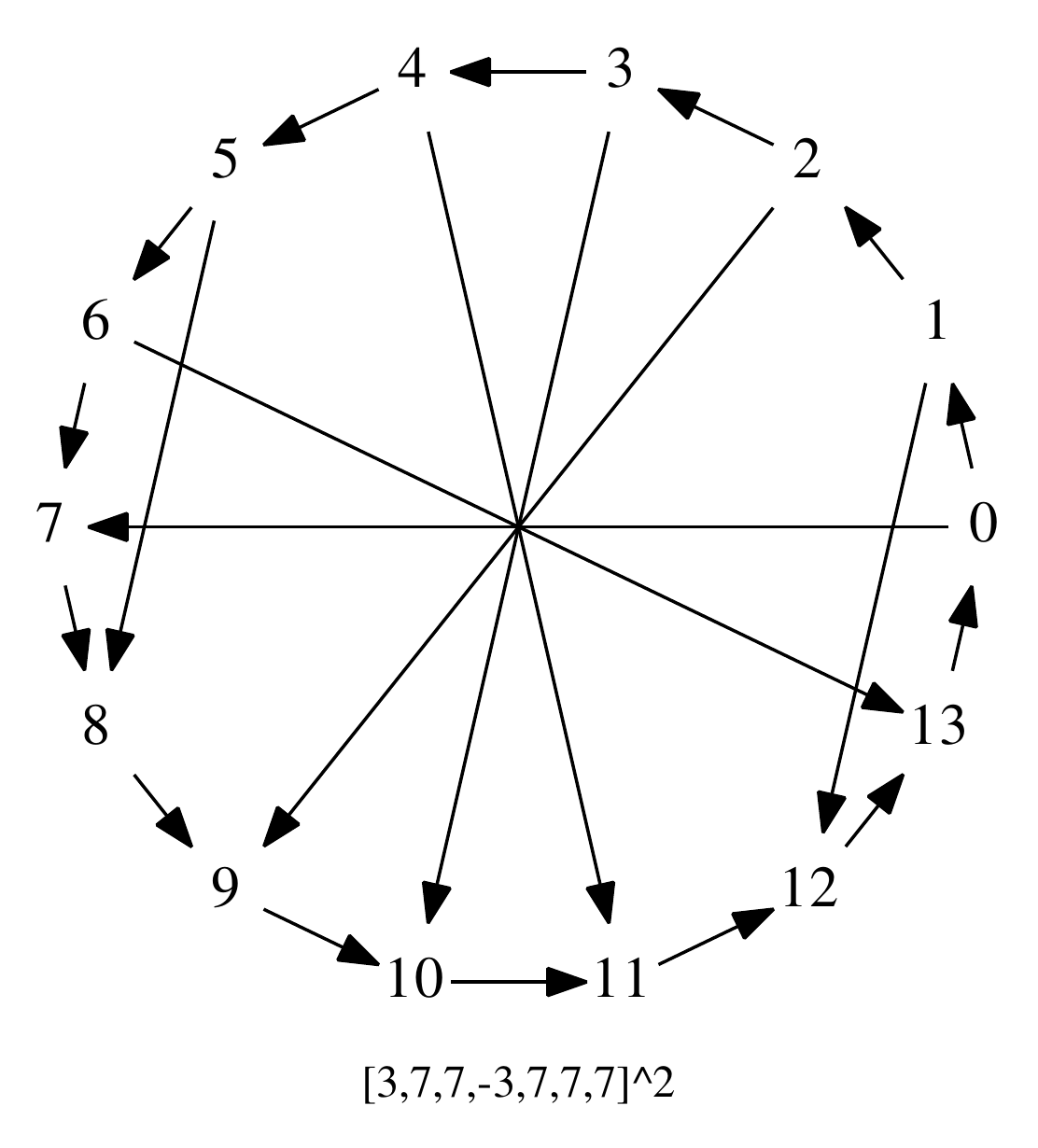}
\includegraphics[scale=0.45]{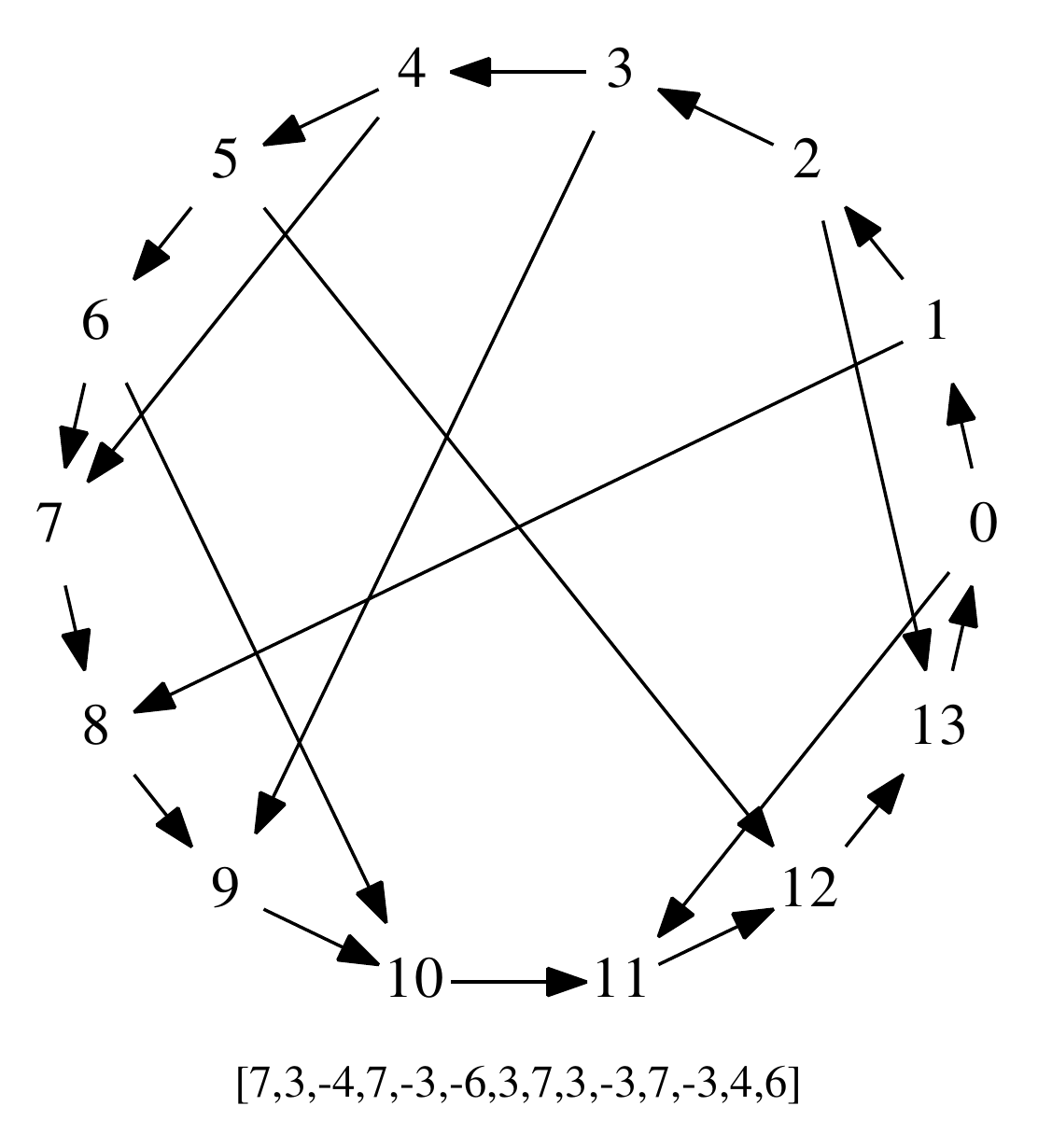}
\includegraphics[scale=0.45]{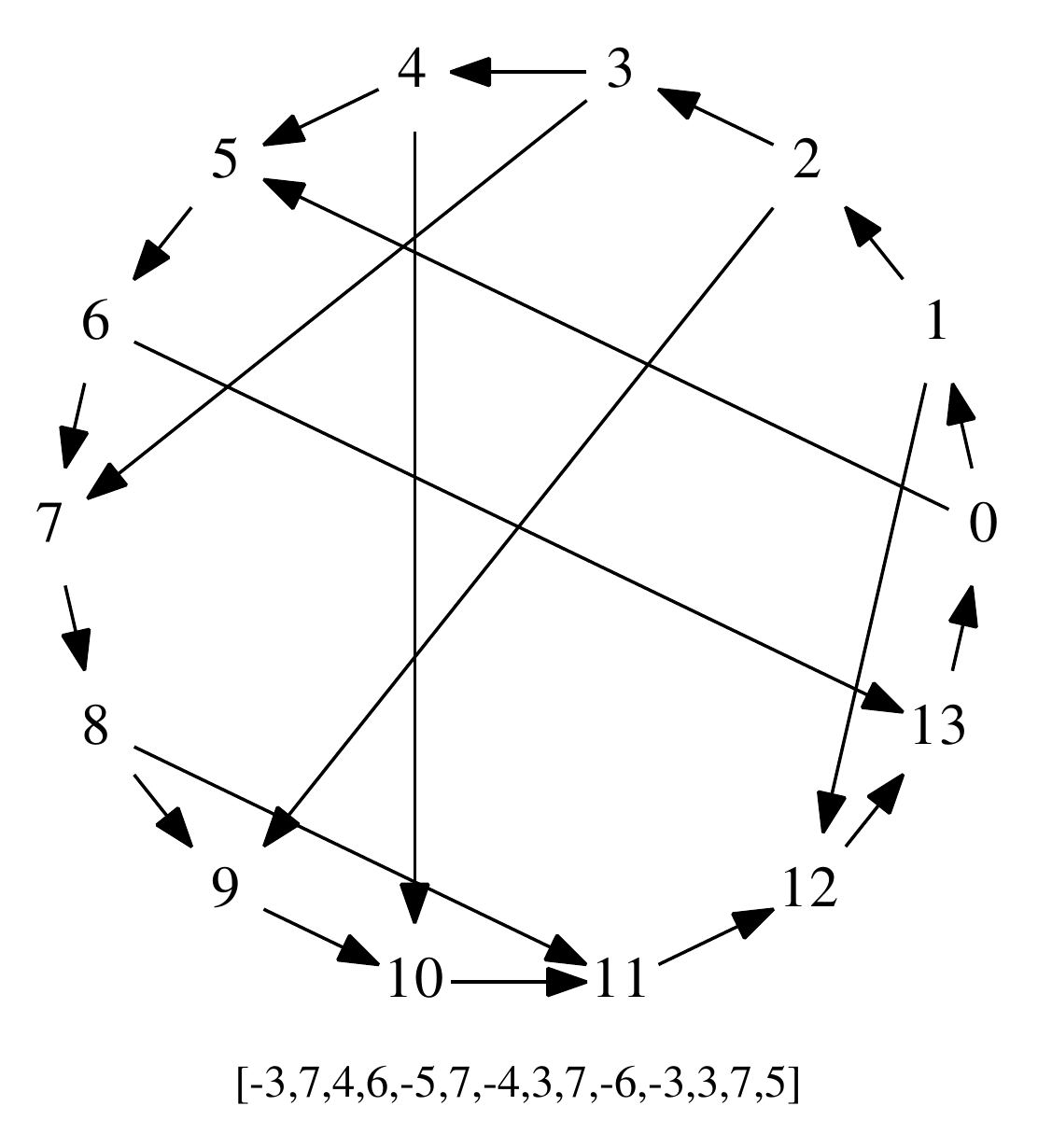}
\includegraphics[scale=0.45]{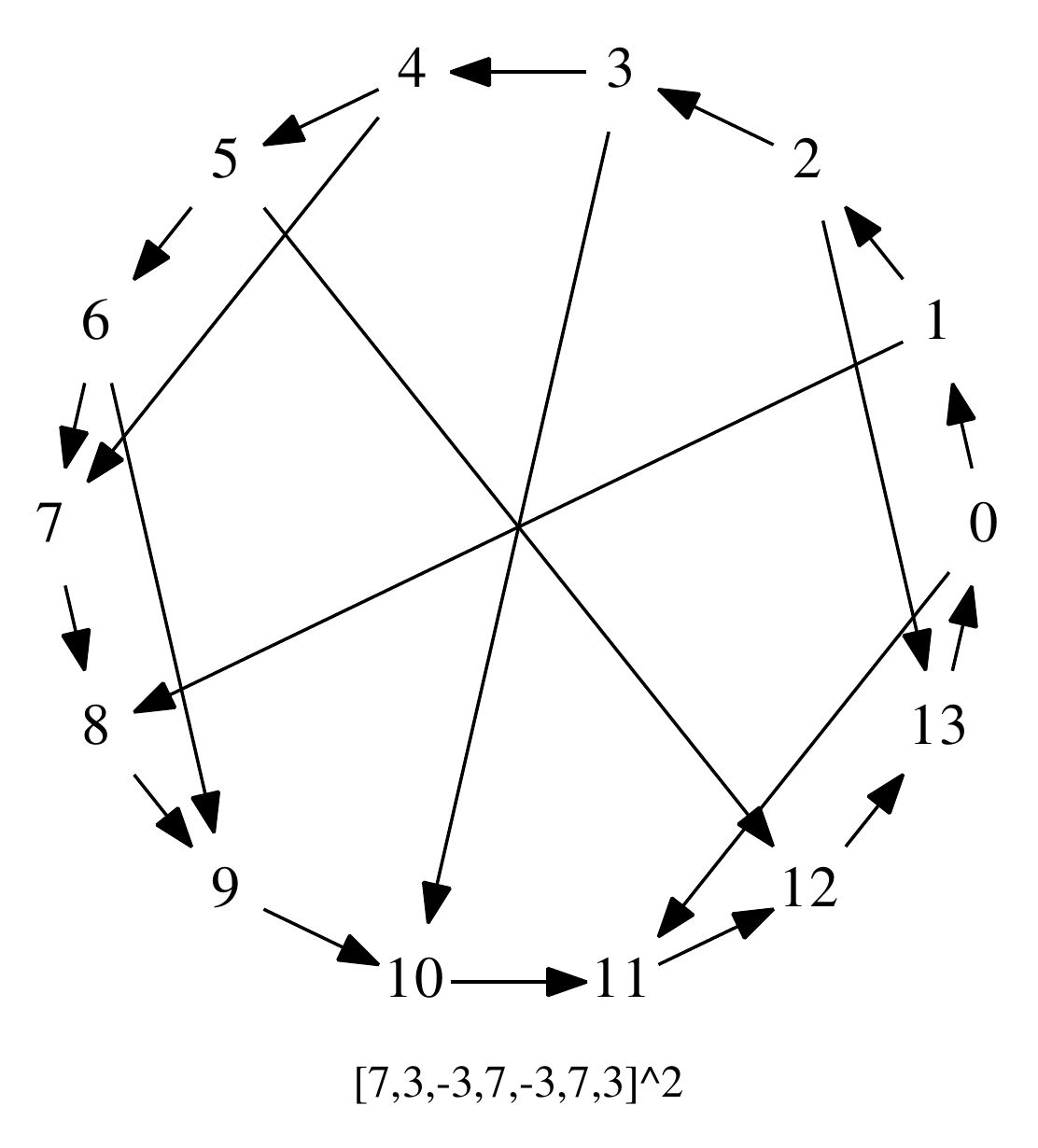}
\includegraphics[scale=0.45]{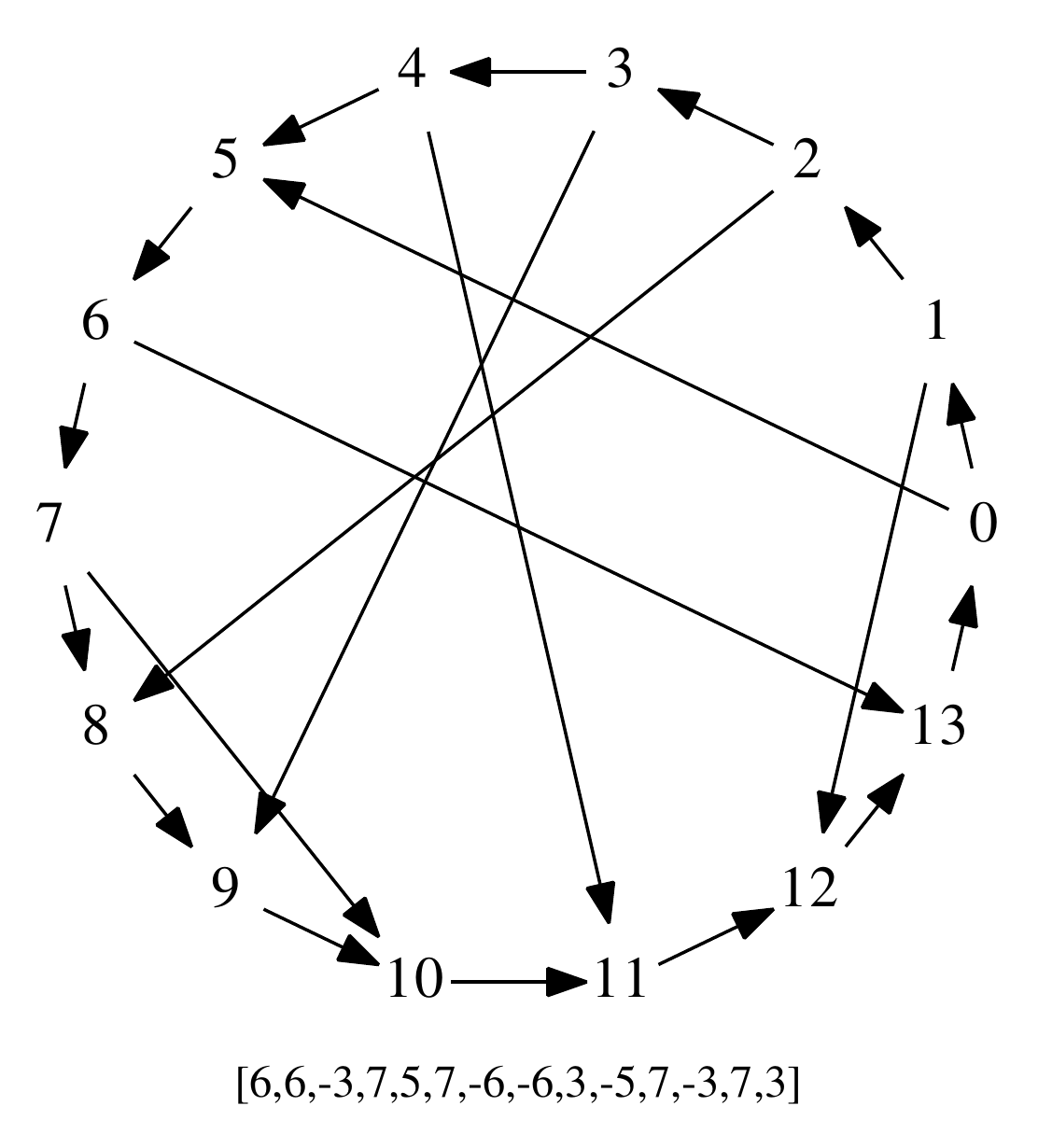}
\includegraphics[scale=0.45]{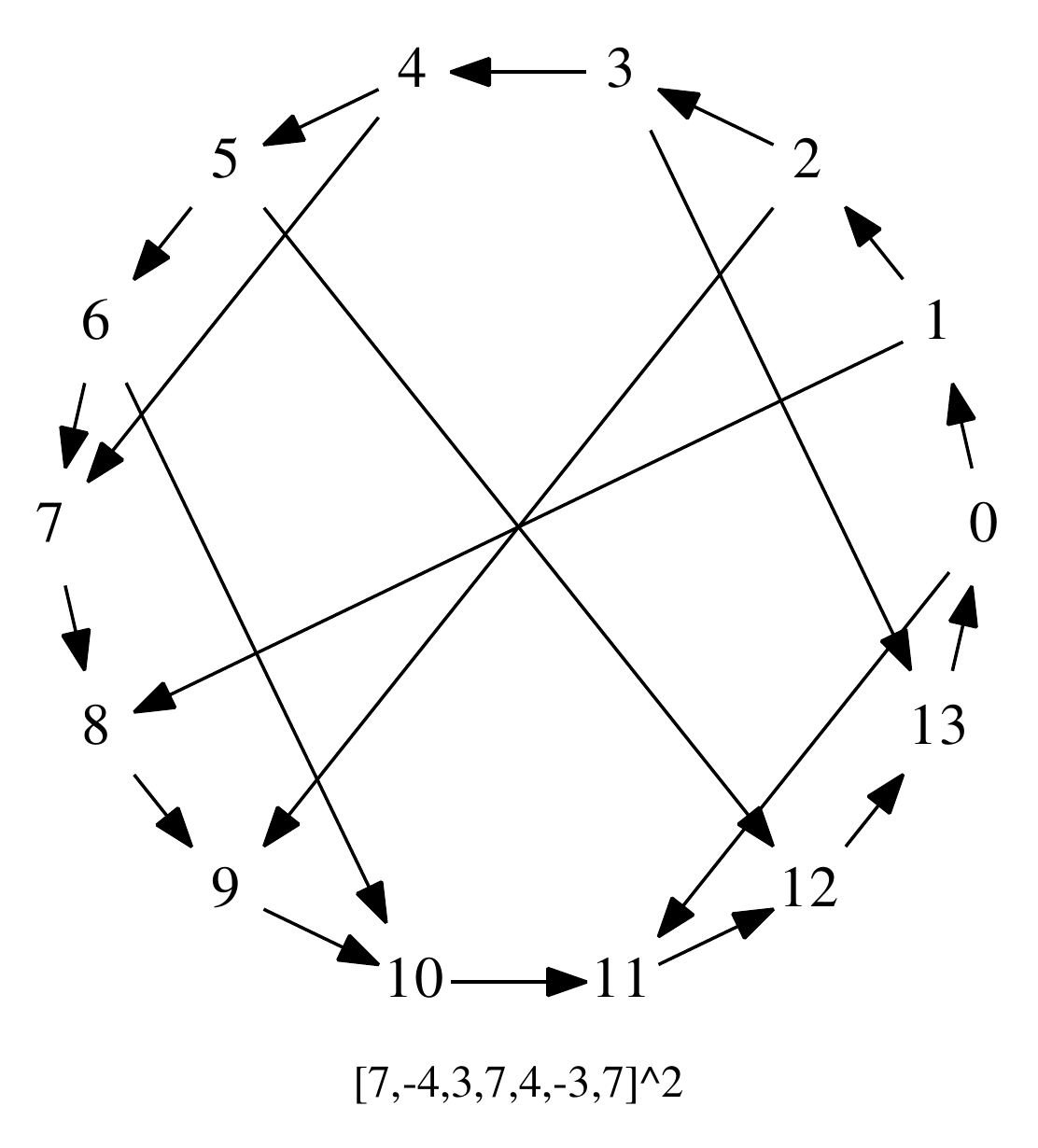}
\includegraphics[scale=0.45]{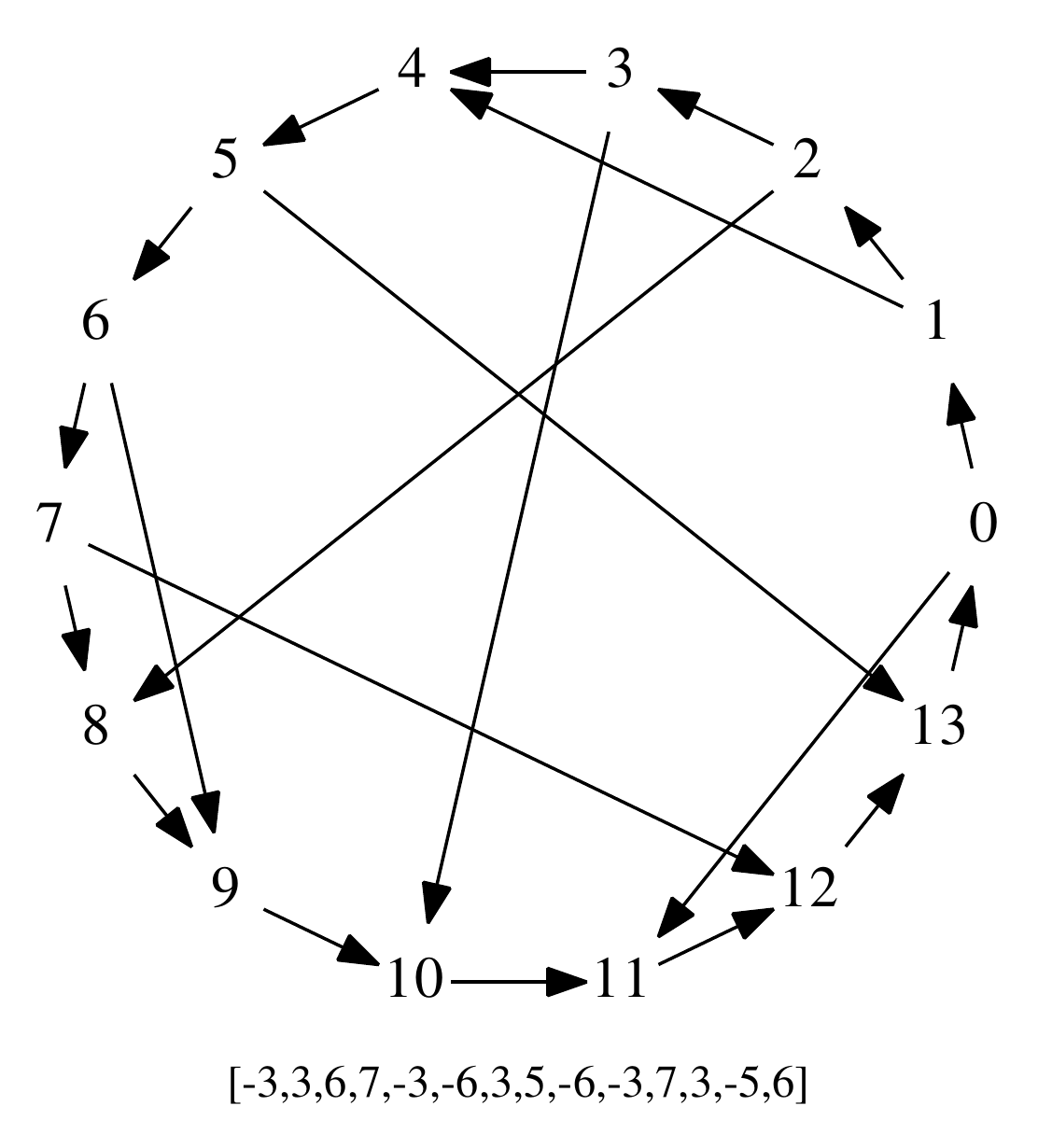}
\includegraphics[scale=0.45]{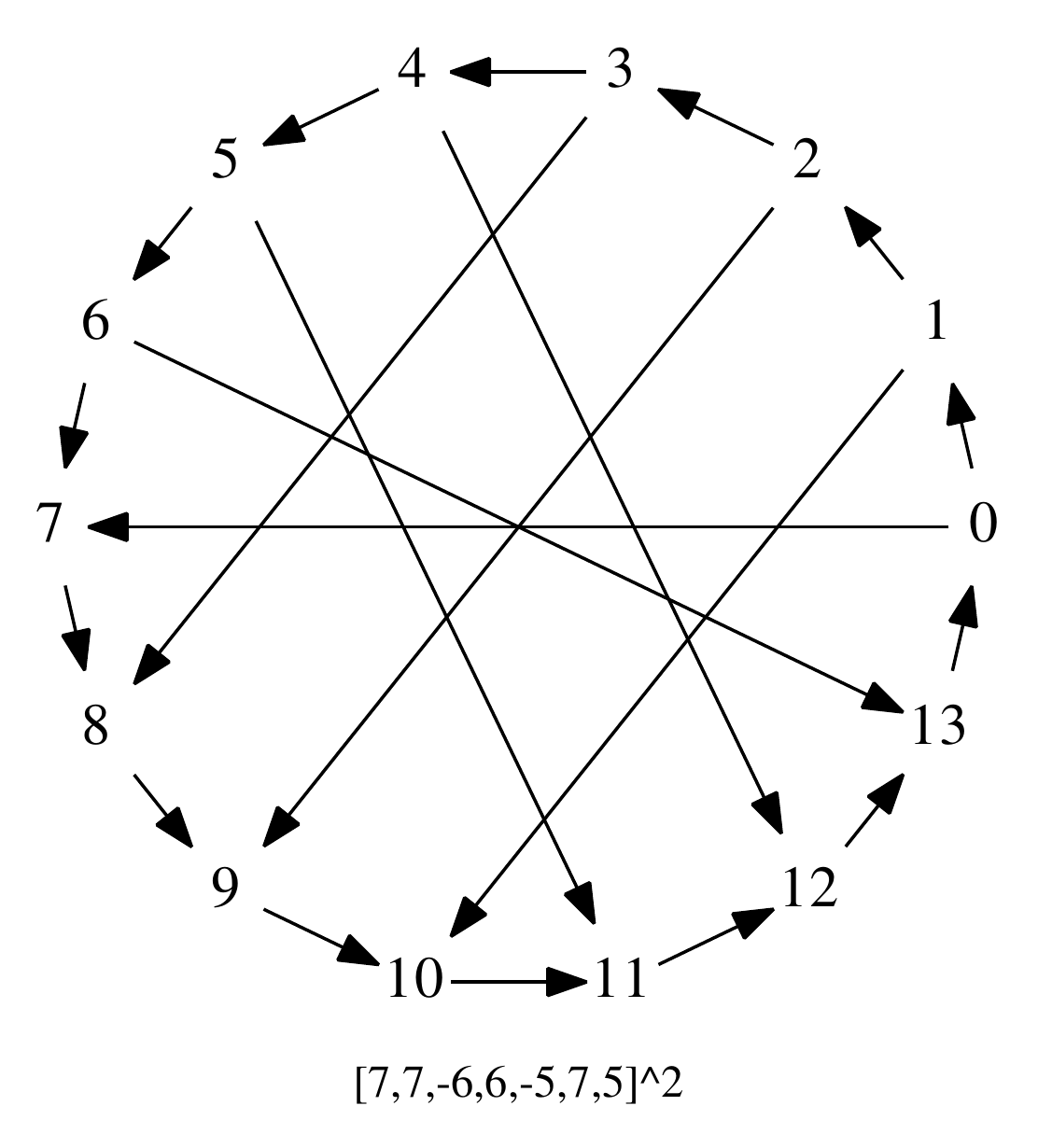}
\includegraphics[scale=0.45]{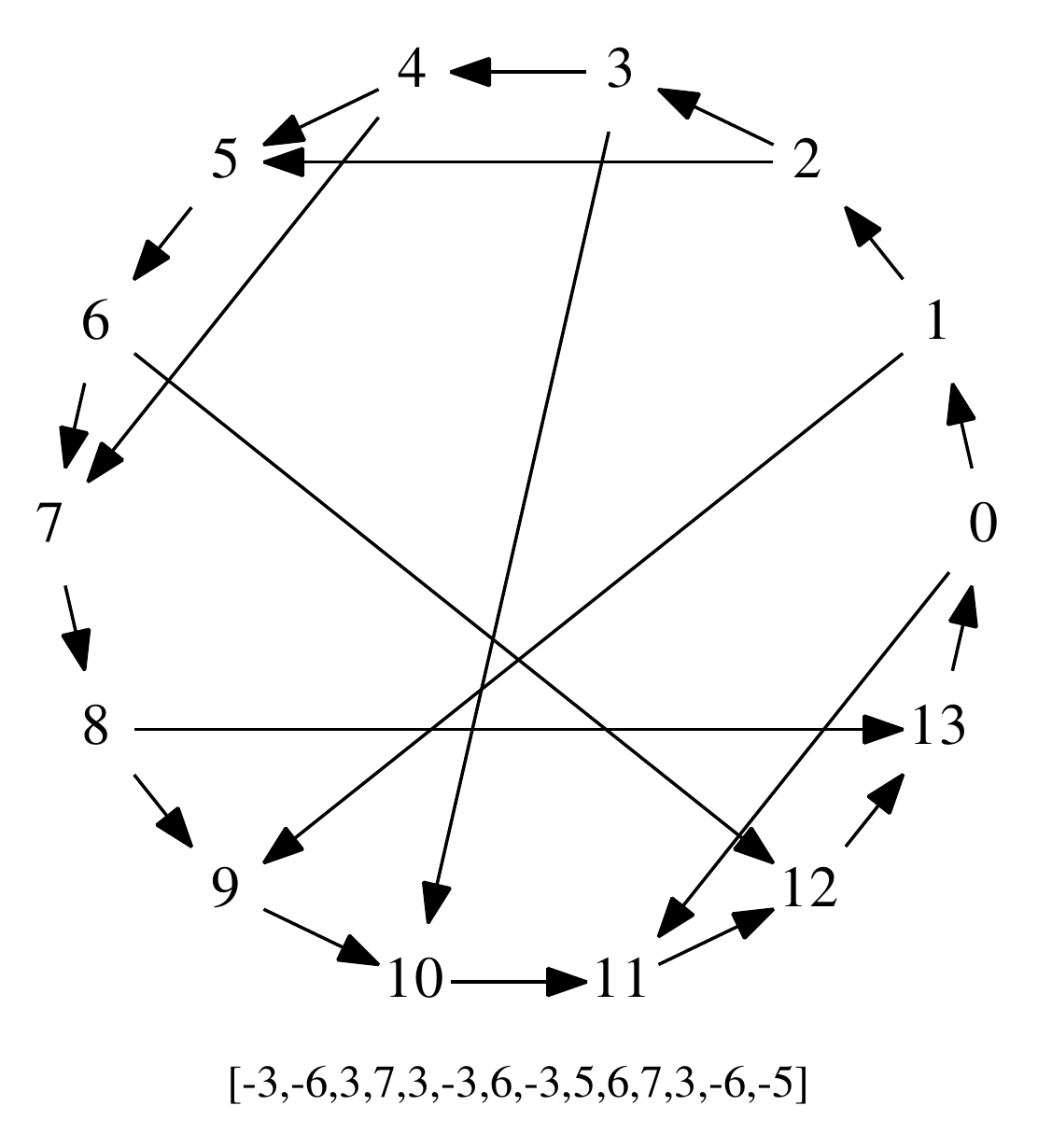}
\includegraphics[scale=0.45]{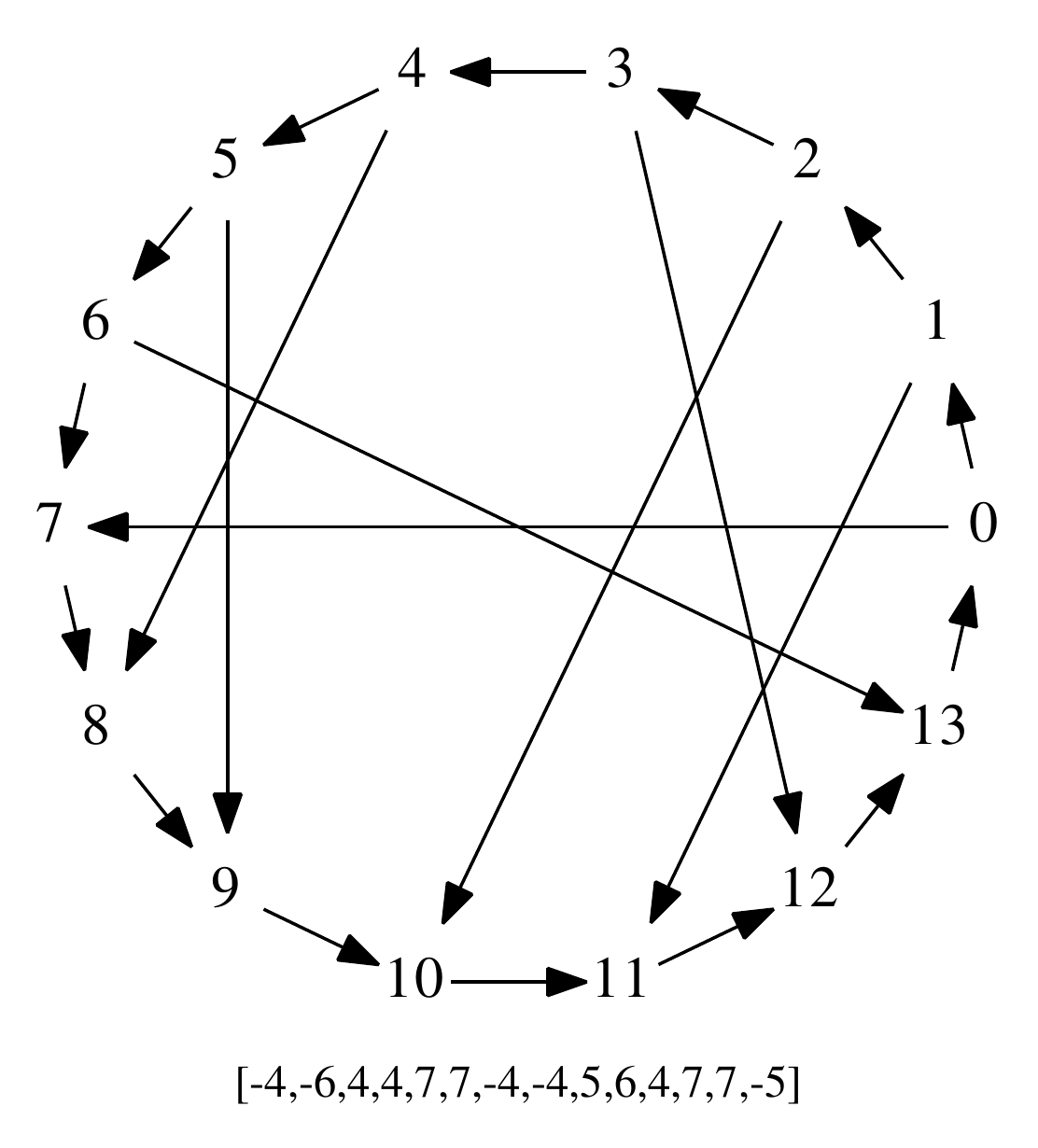}
\includegraphics[scale=0.45]{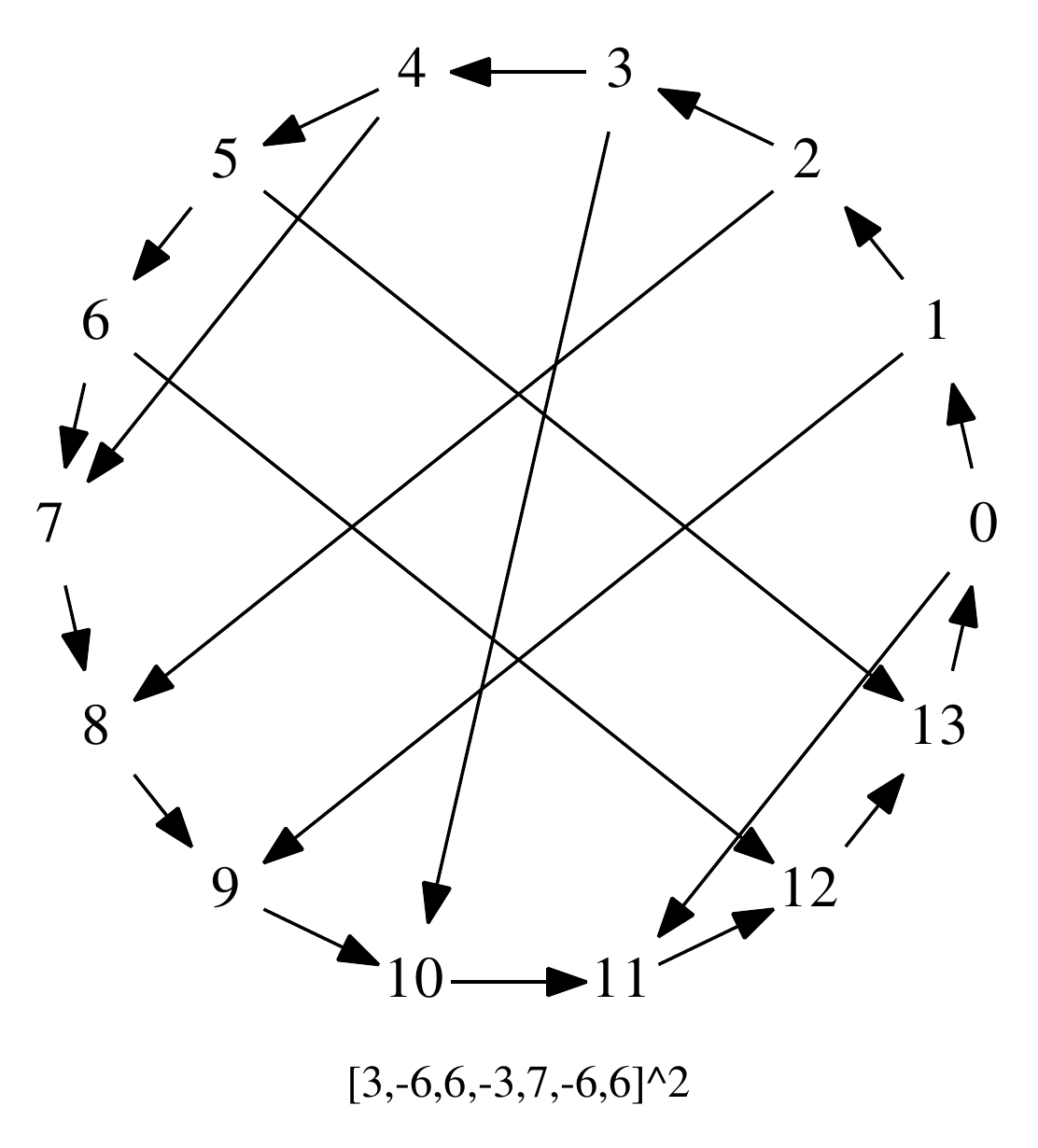}
\caption{Graphs on $n=14$ vertices which are irreducible (start).
}
\label{fig.14ns}
\end{figure}

\begin{figure}
\includegraphics[scale=0.45]{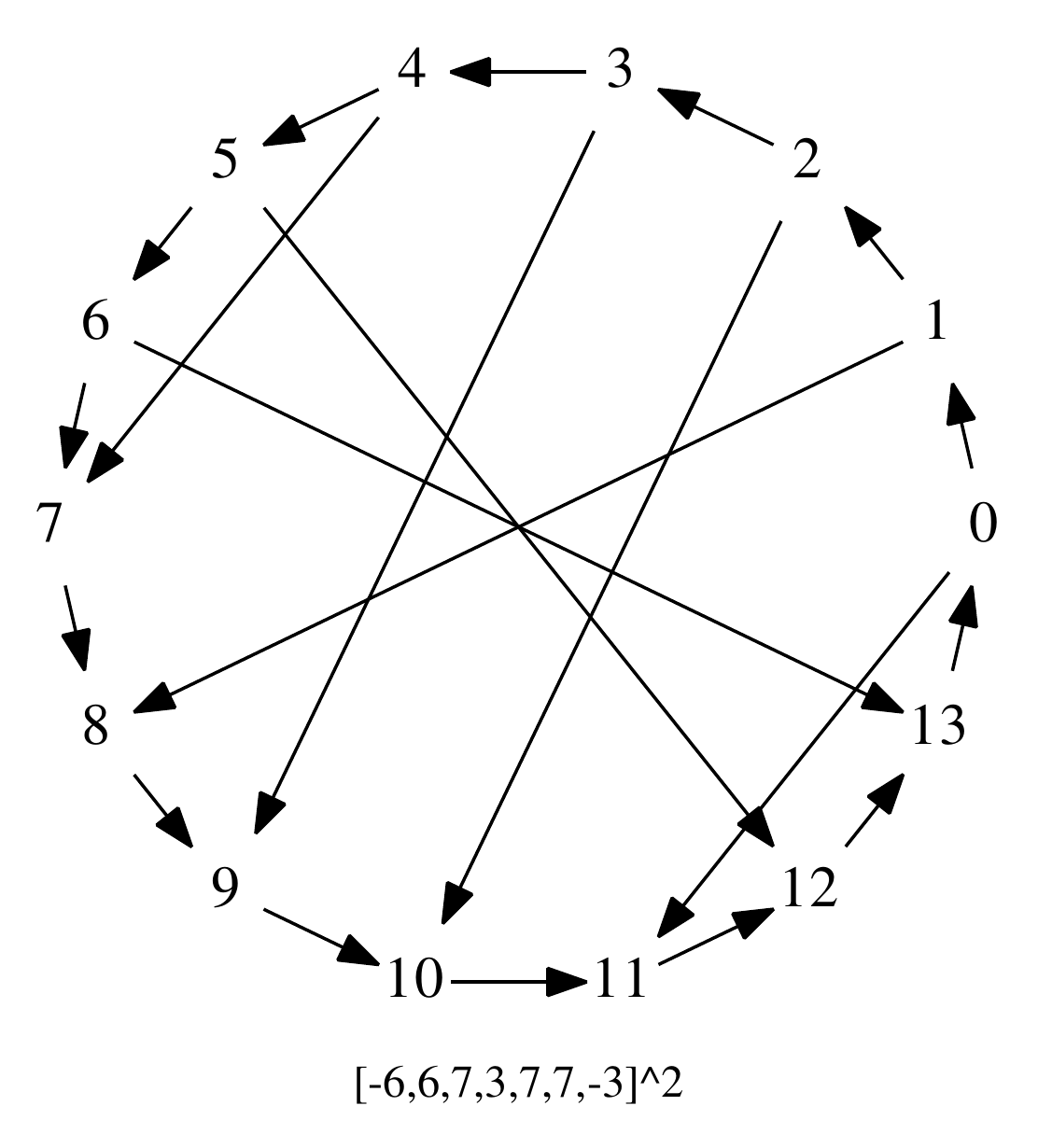}
\includegraphics[scale=0.45]{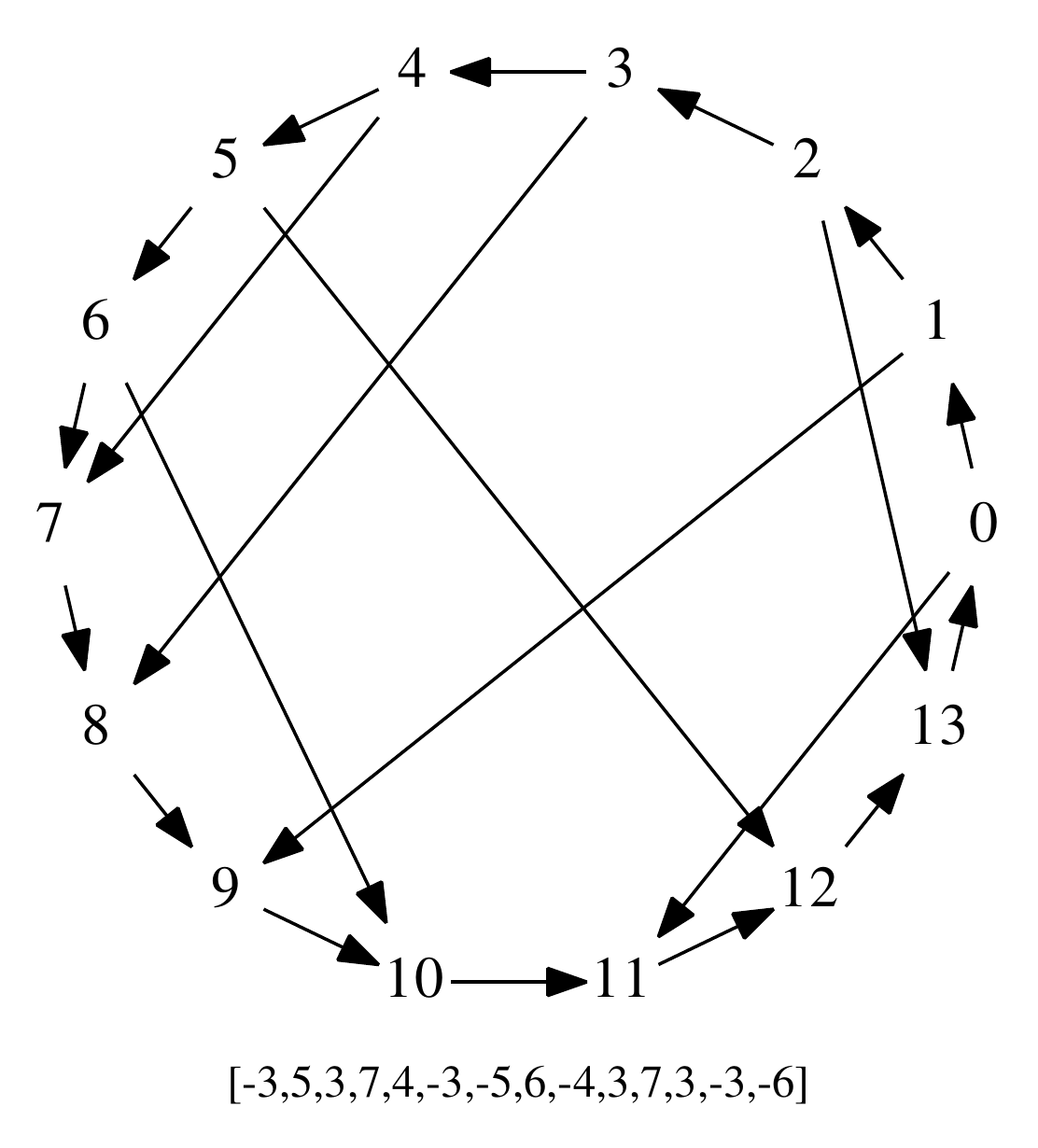}
\includegraphics[scale=0.45]{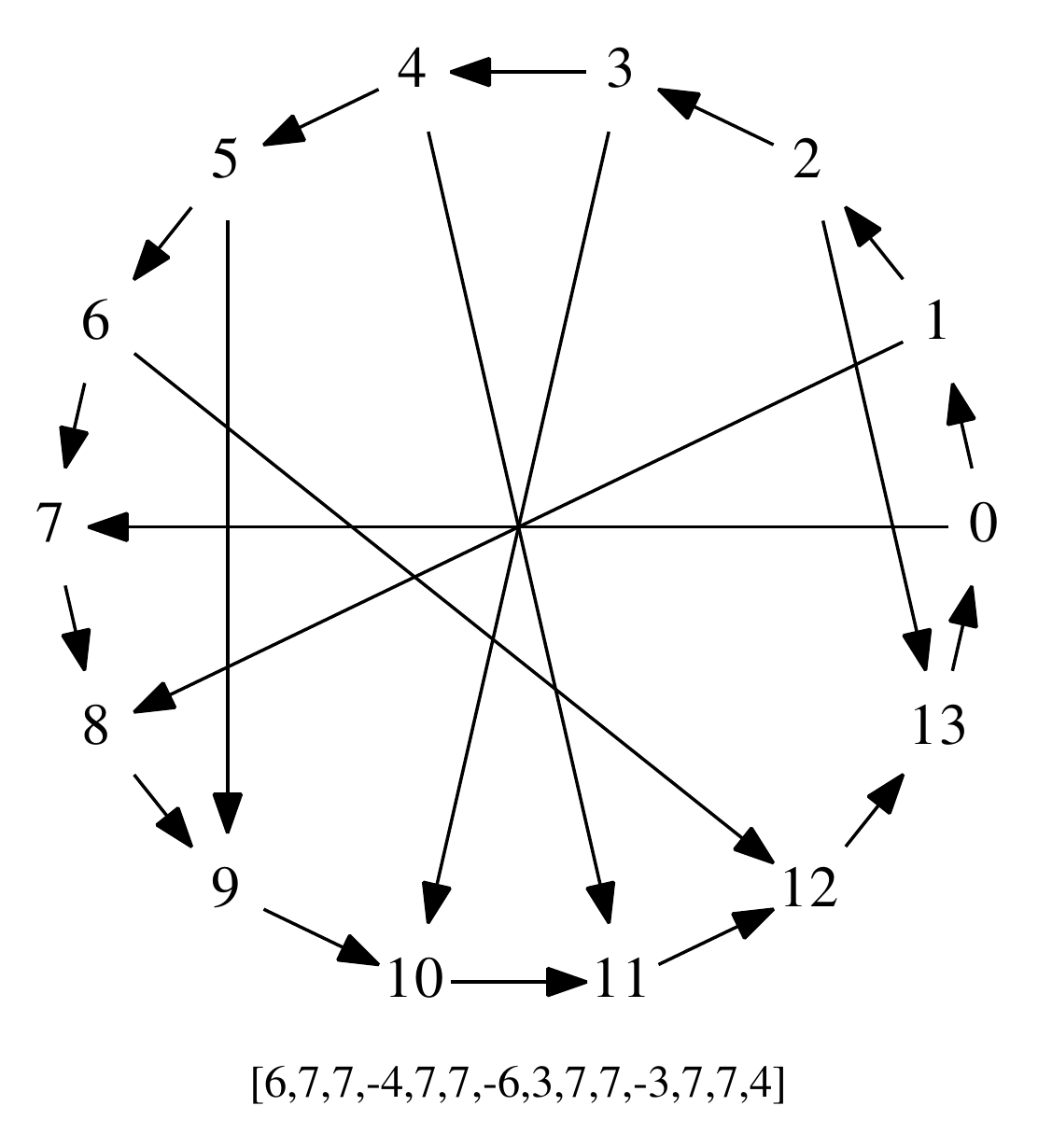}
\includegraphics[scale=0.45]{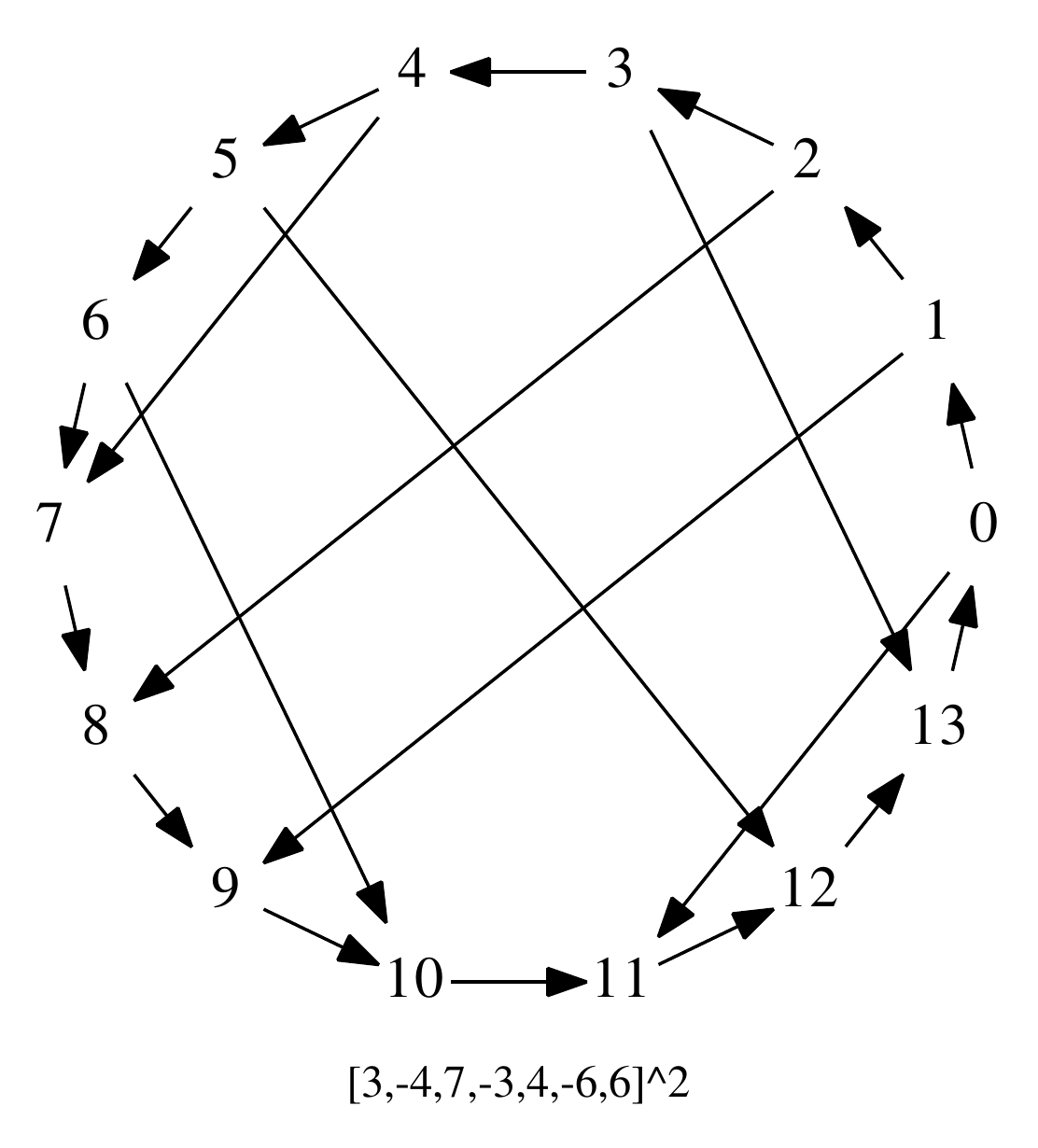}
\includegraphics[scale=0.45]{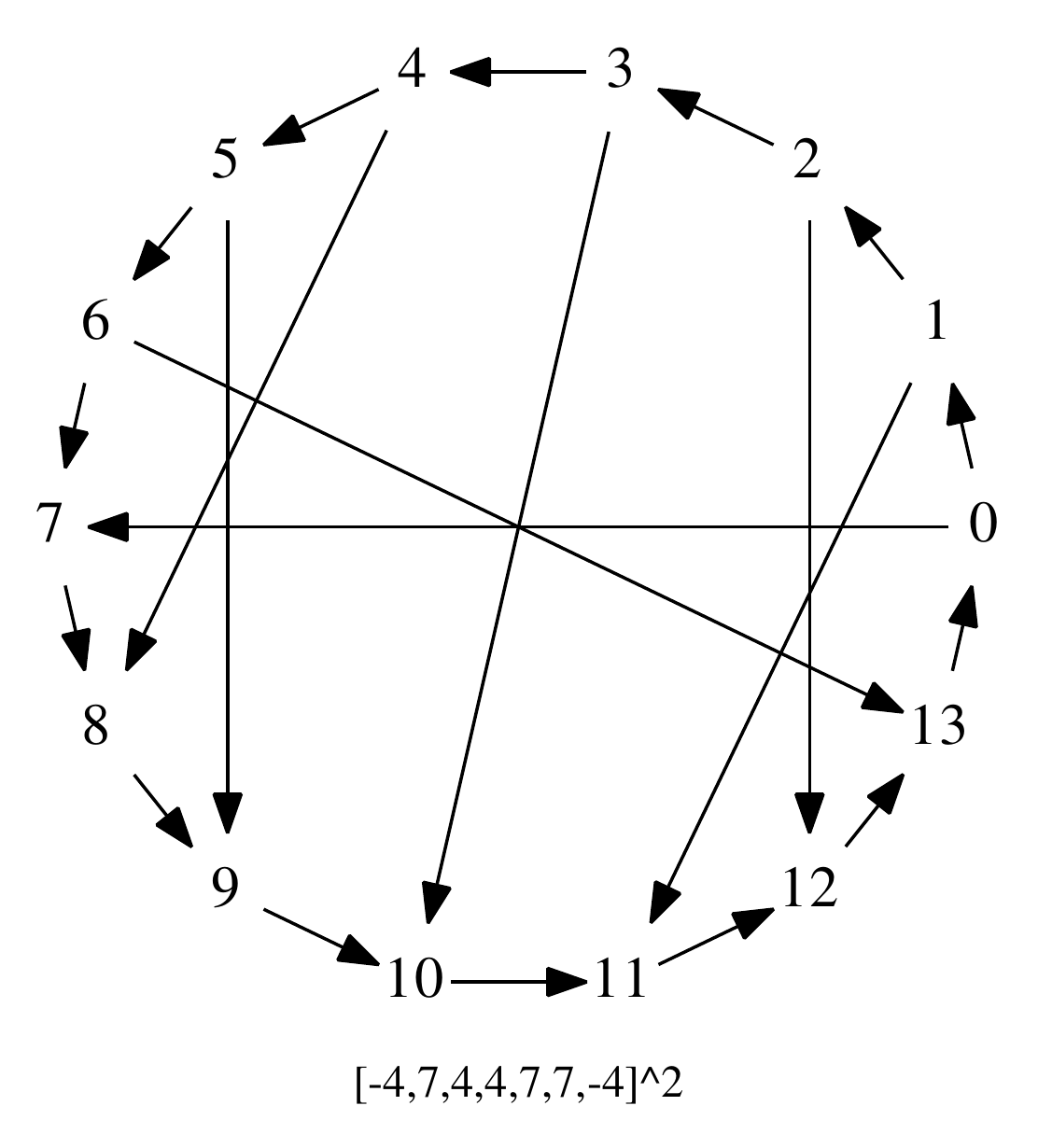}
\includegraphics[scale=0.45]{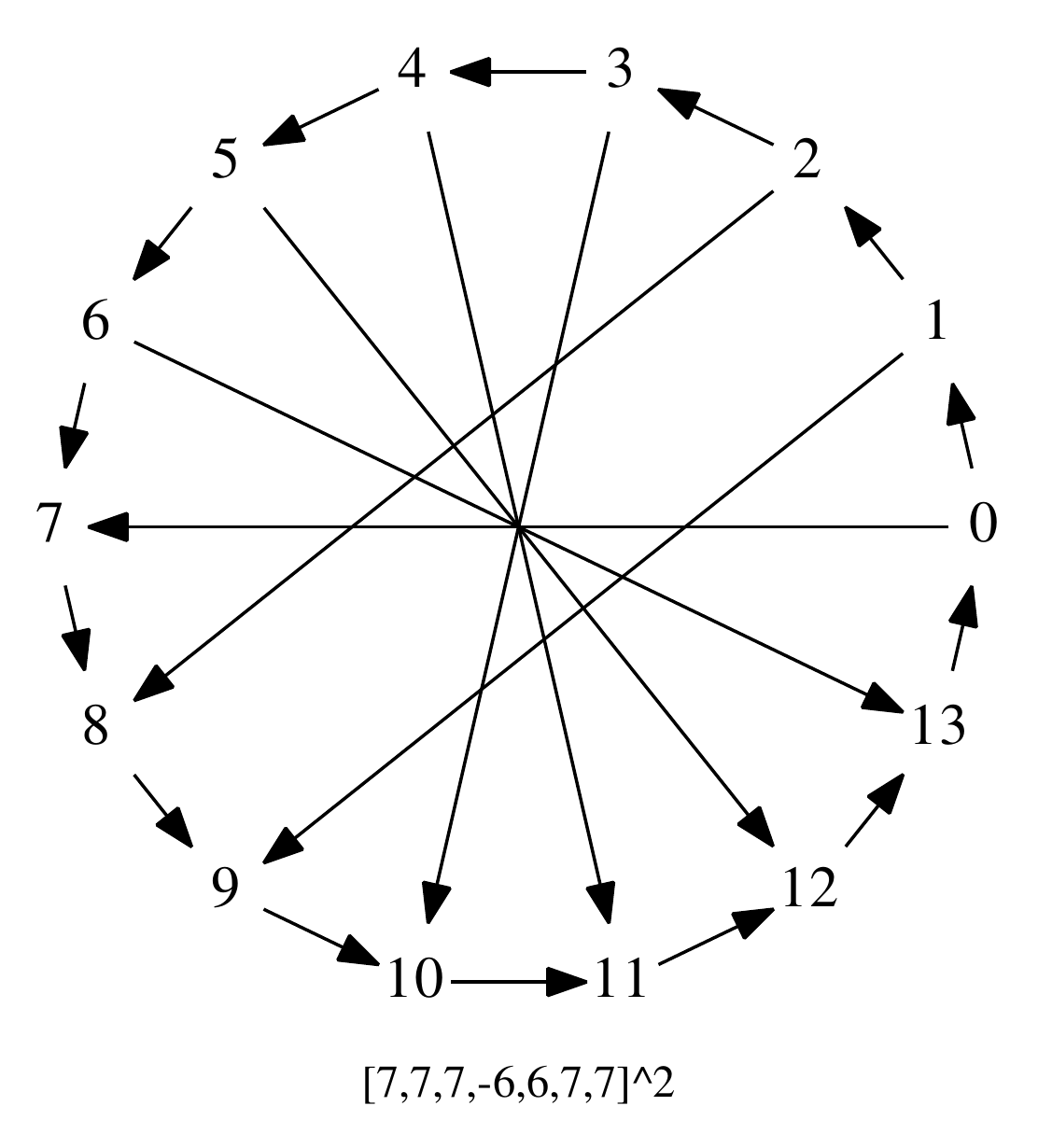}
\includegraphics[scale=0.45]{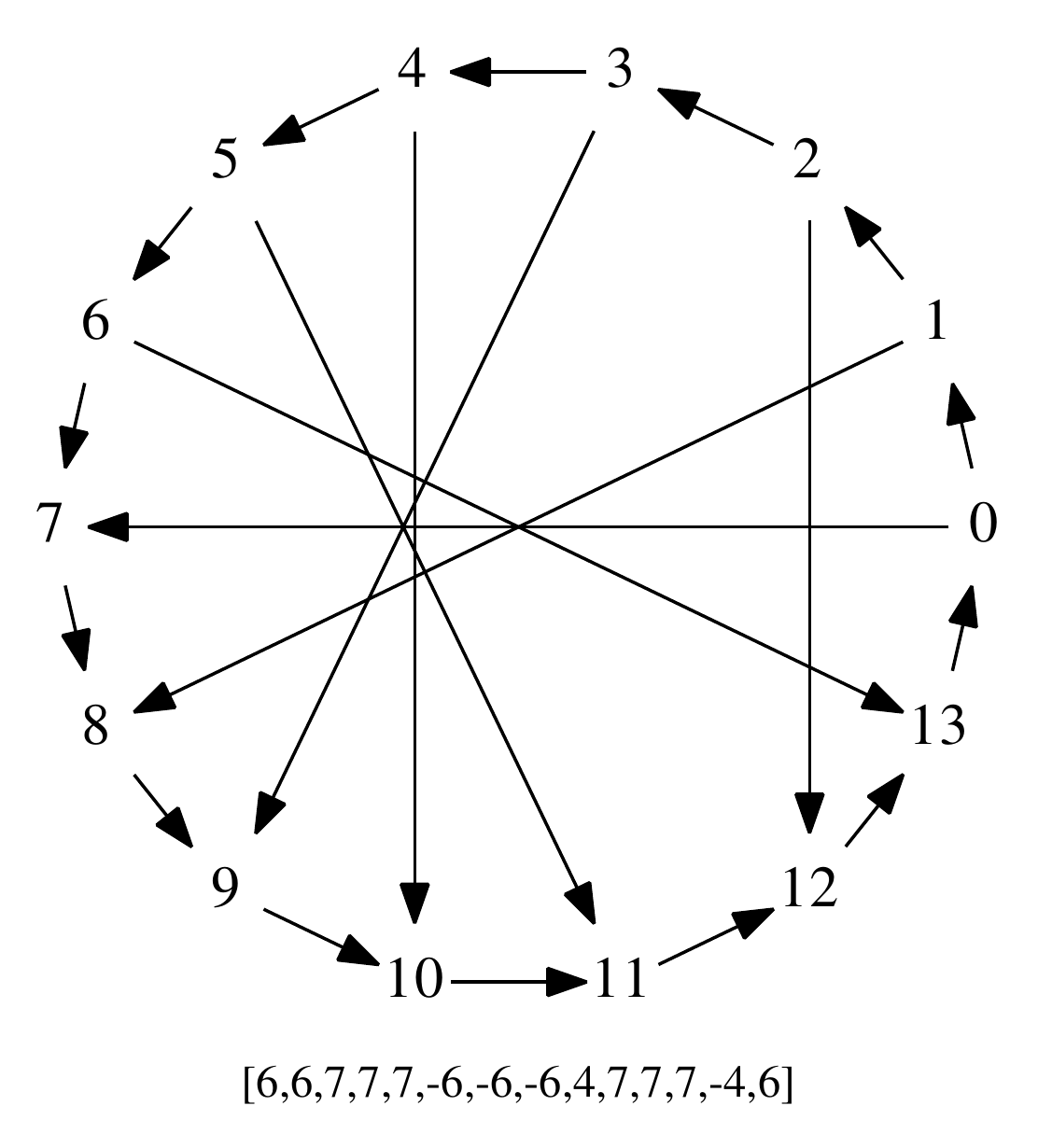}
\includegraphics[scale=0.45]{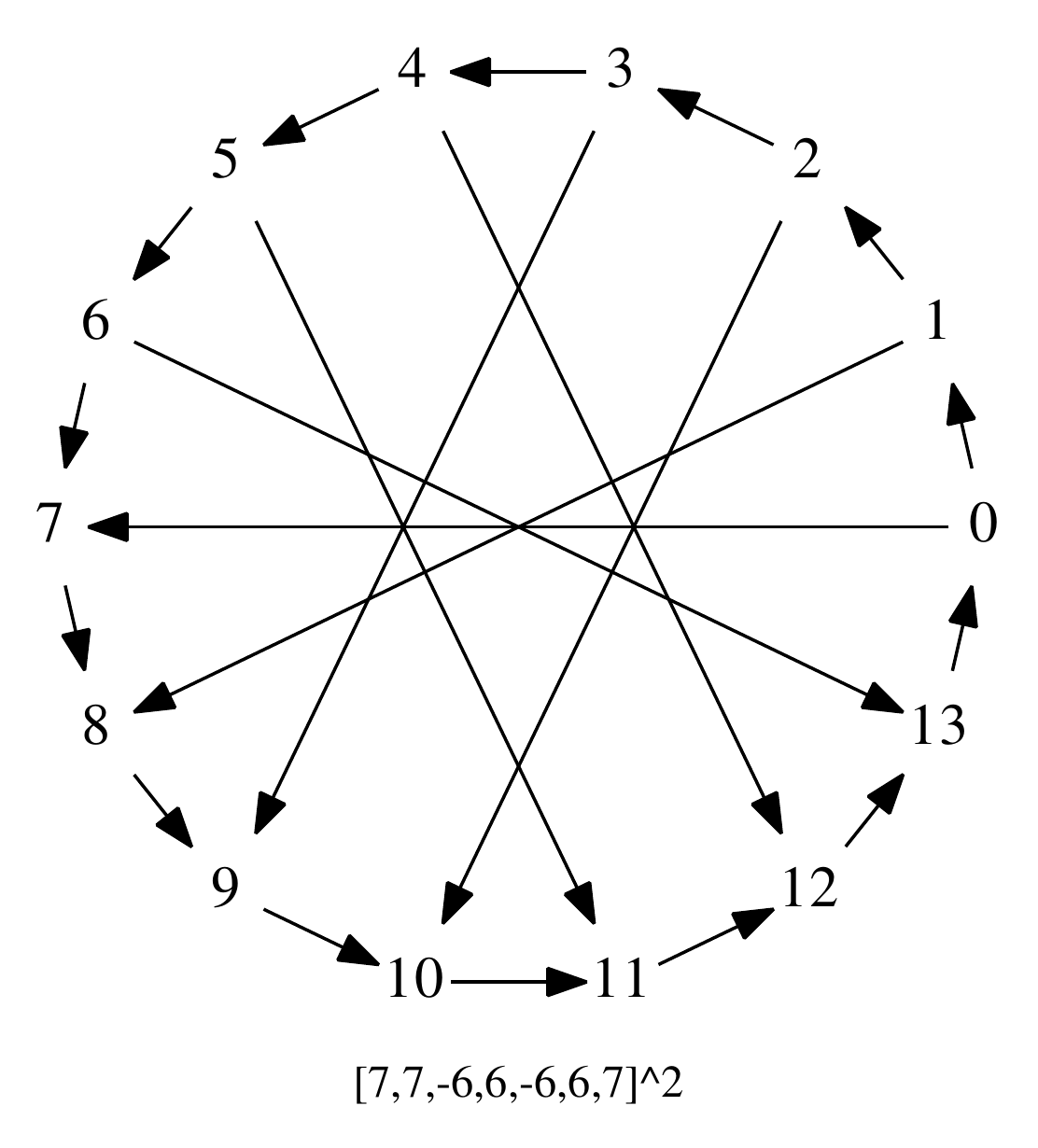}
\includegraphics[scale=0.45]{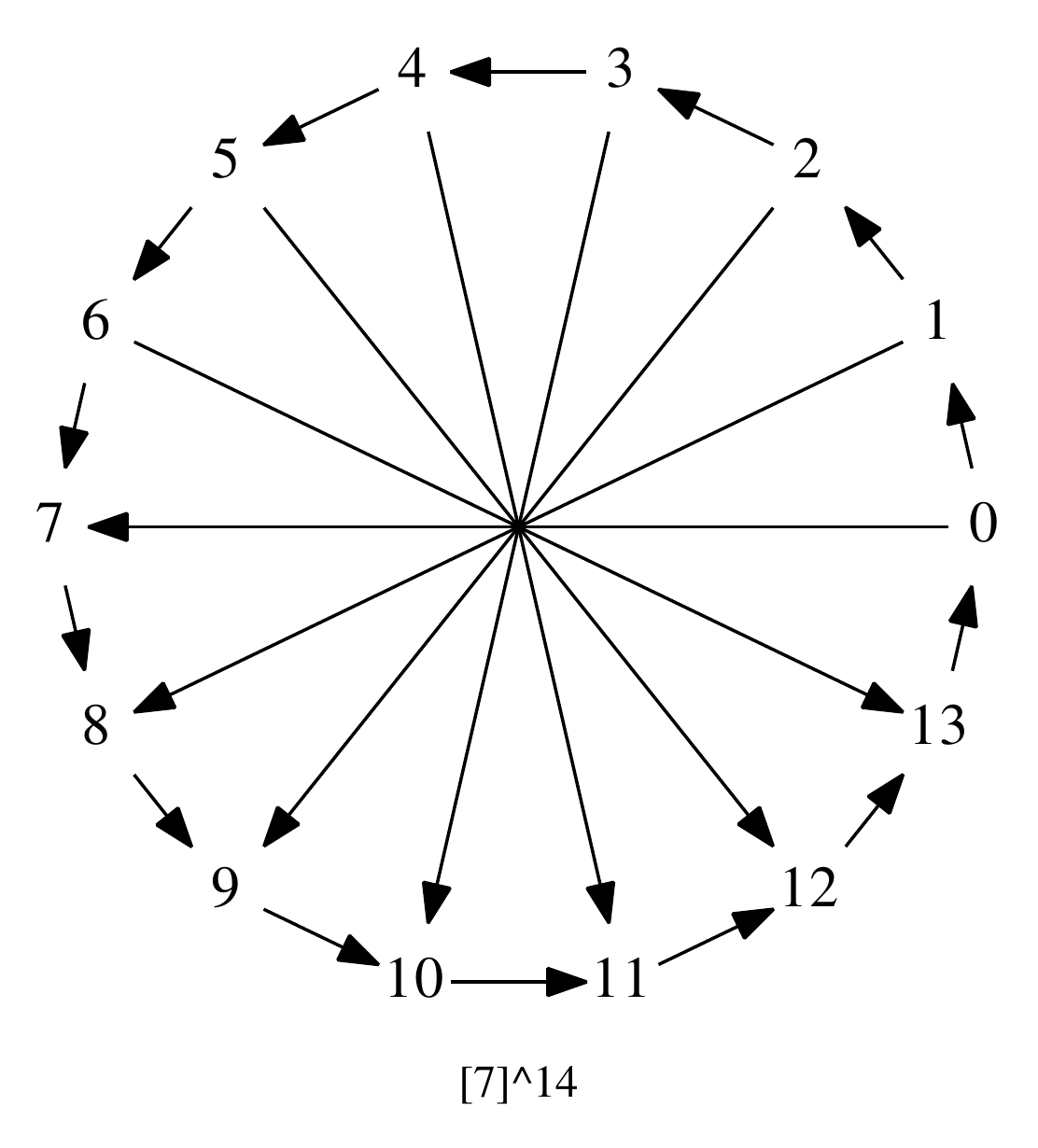}
\includegraphics[scale=0.45]{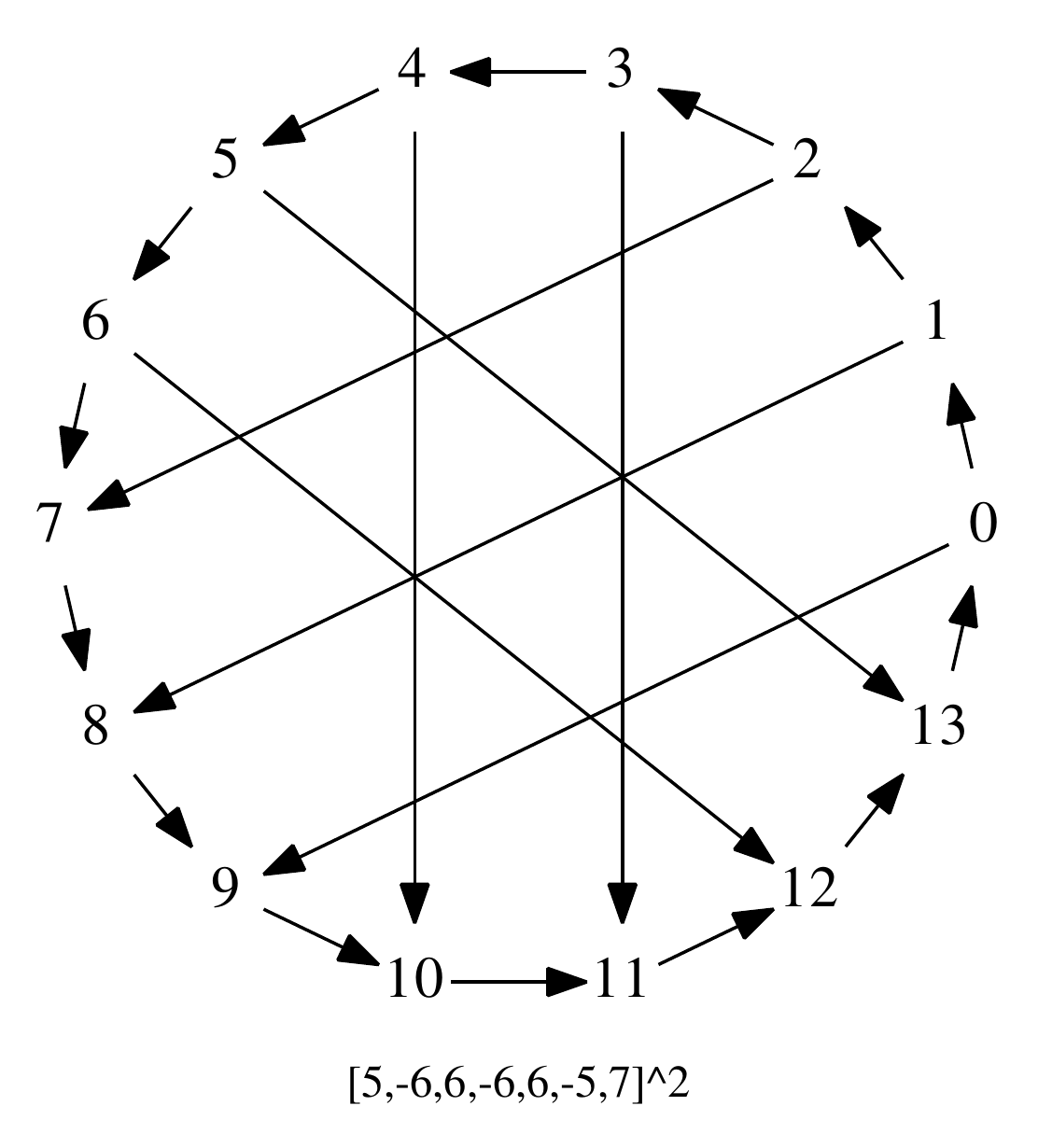}
\includegraphics[scale=0.45]{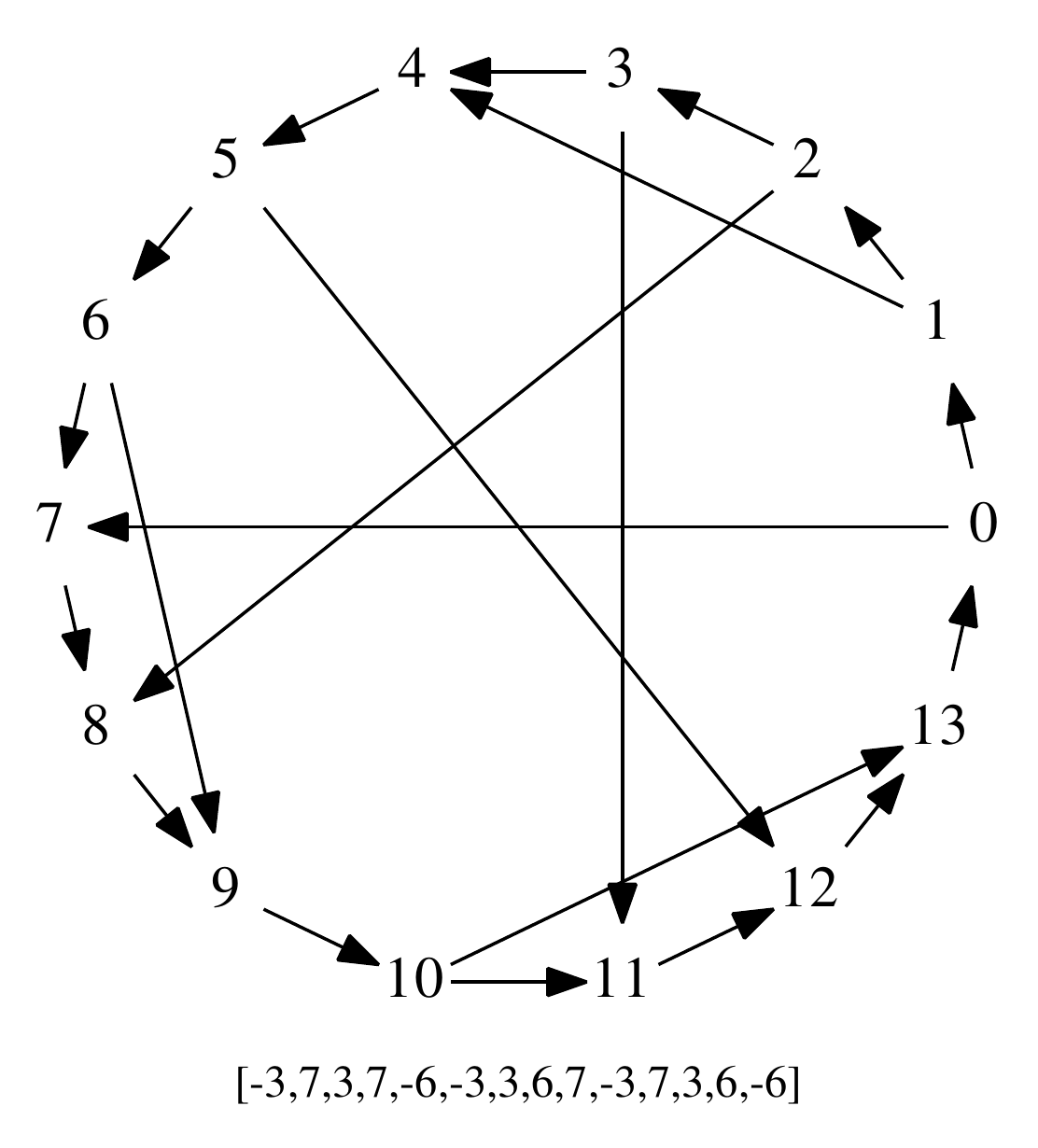}
\includegraphics[scale=0.45]{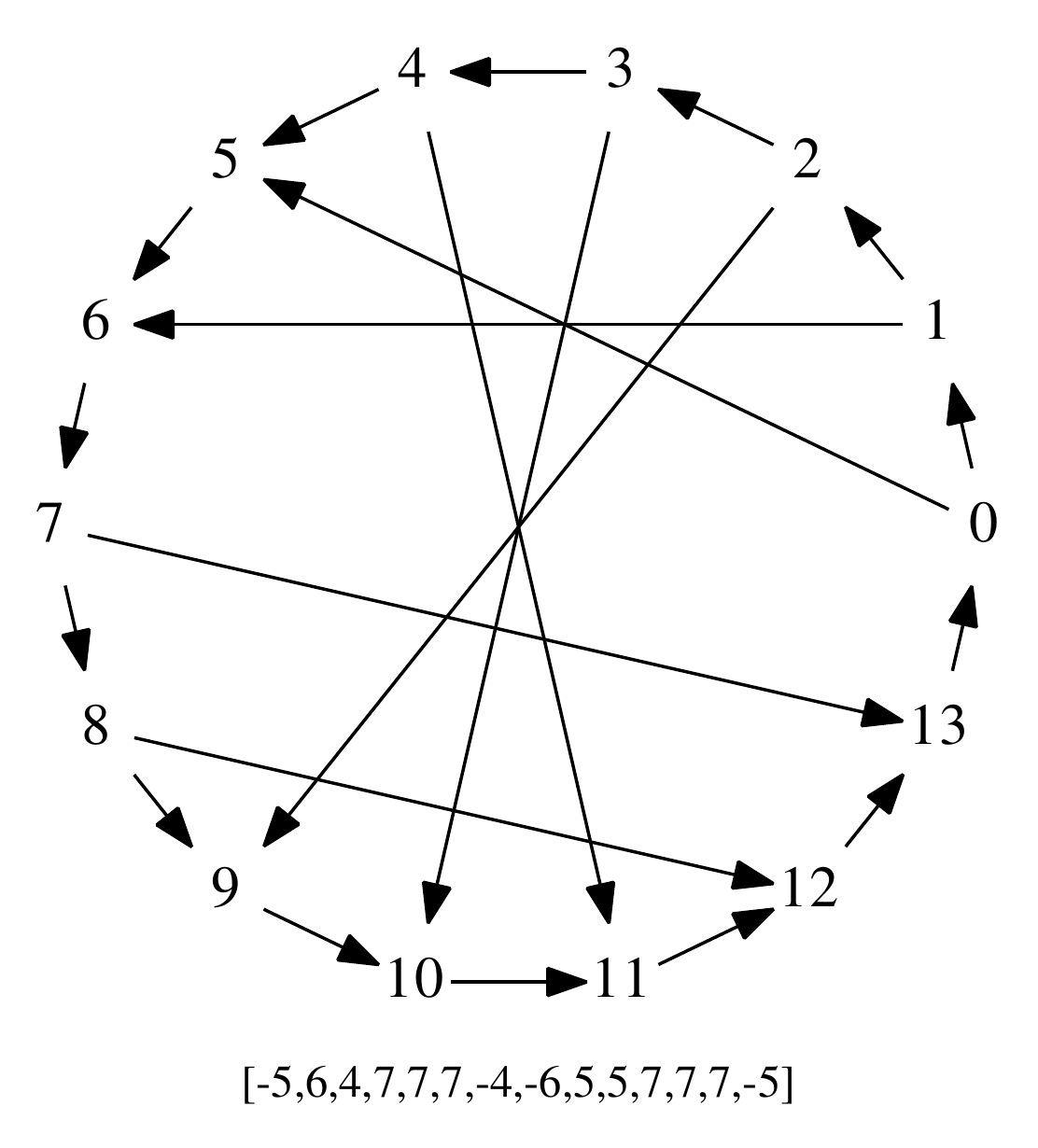}
\caption{Graphs on $n=14$ vertices which are irreducible (continued).
}
\label{fig.14n2}
\end{figure}

\begin{figure}
\includegraphics[scale=0.45]{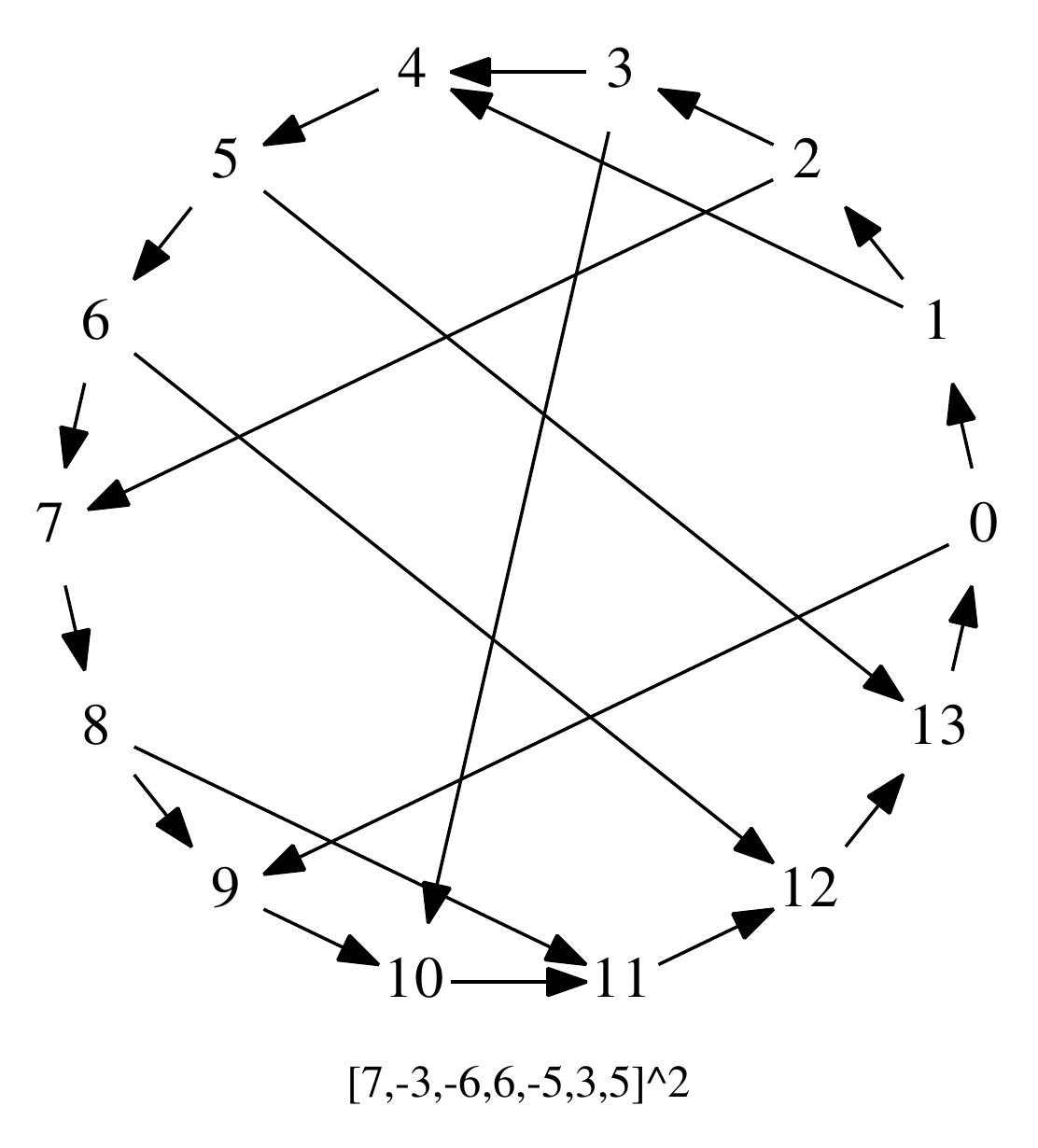}
\includegraphics[scale=0.45]{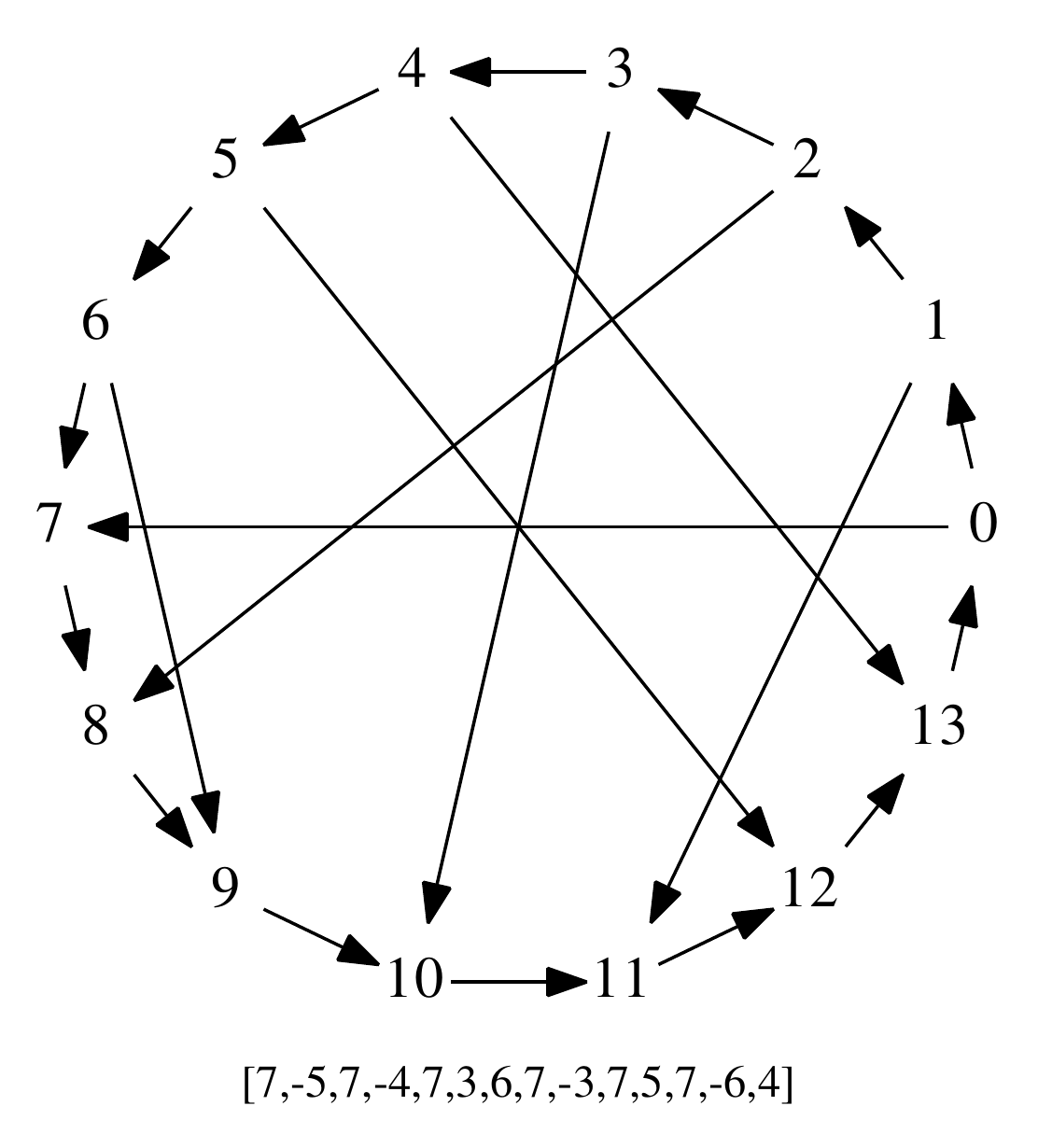}
\includegraphics[scale=0.45]{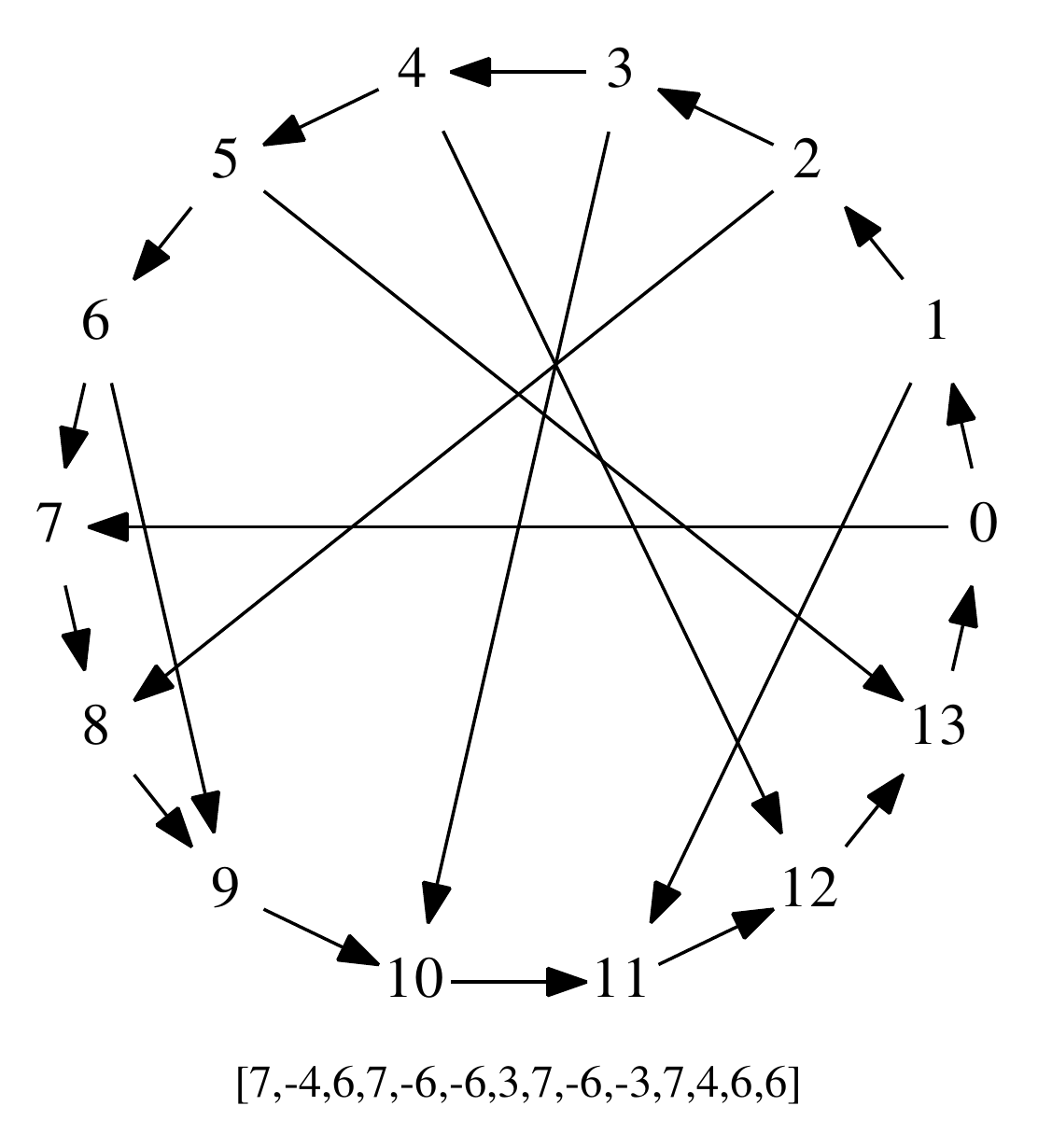}
\includegraphics[scale=0.45]{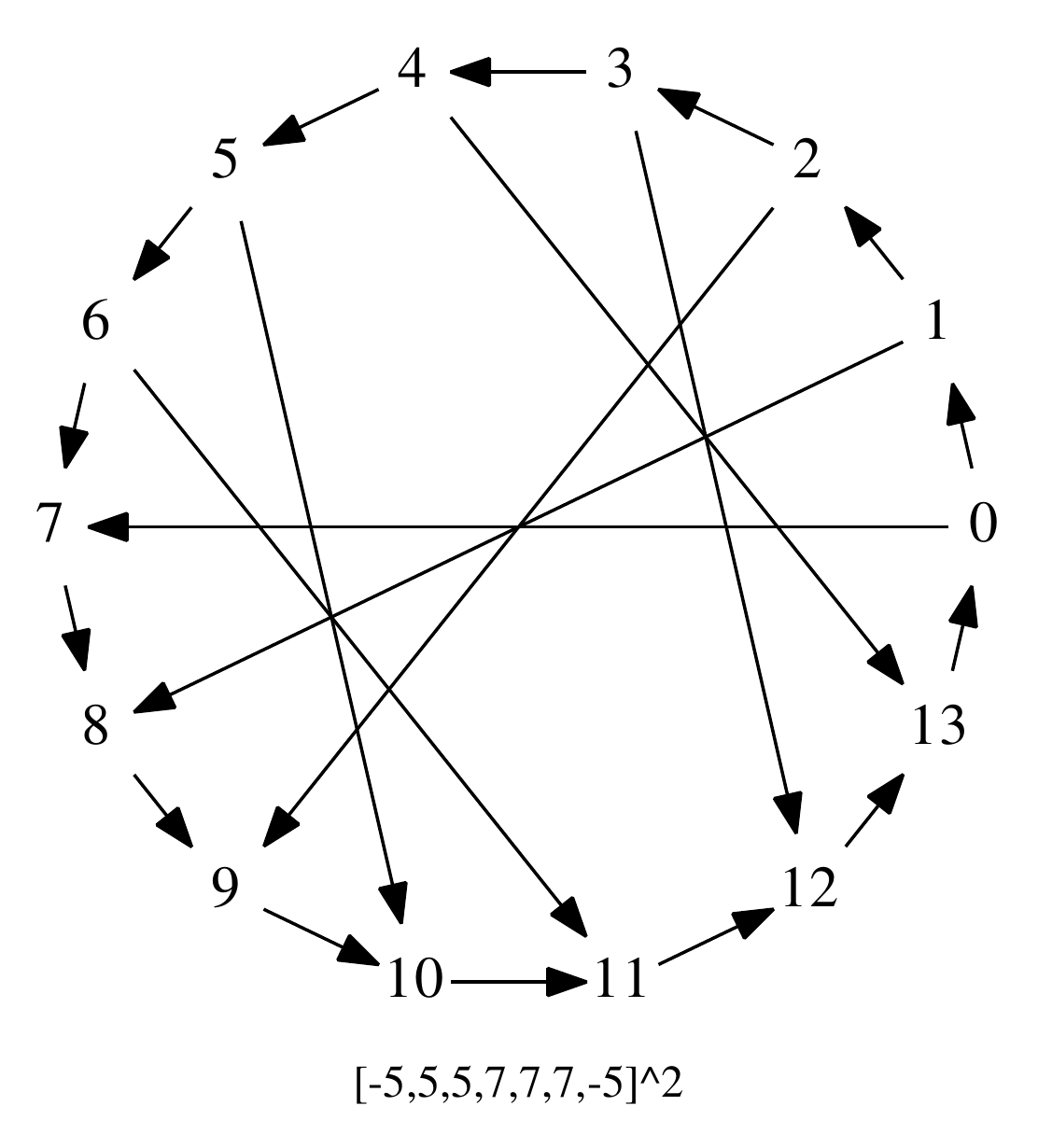}
\includegraphics[scale=0.45]{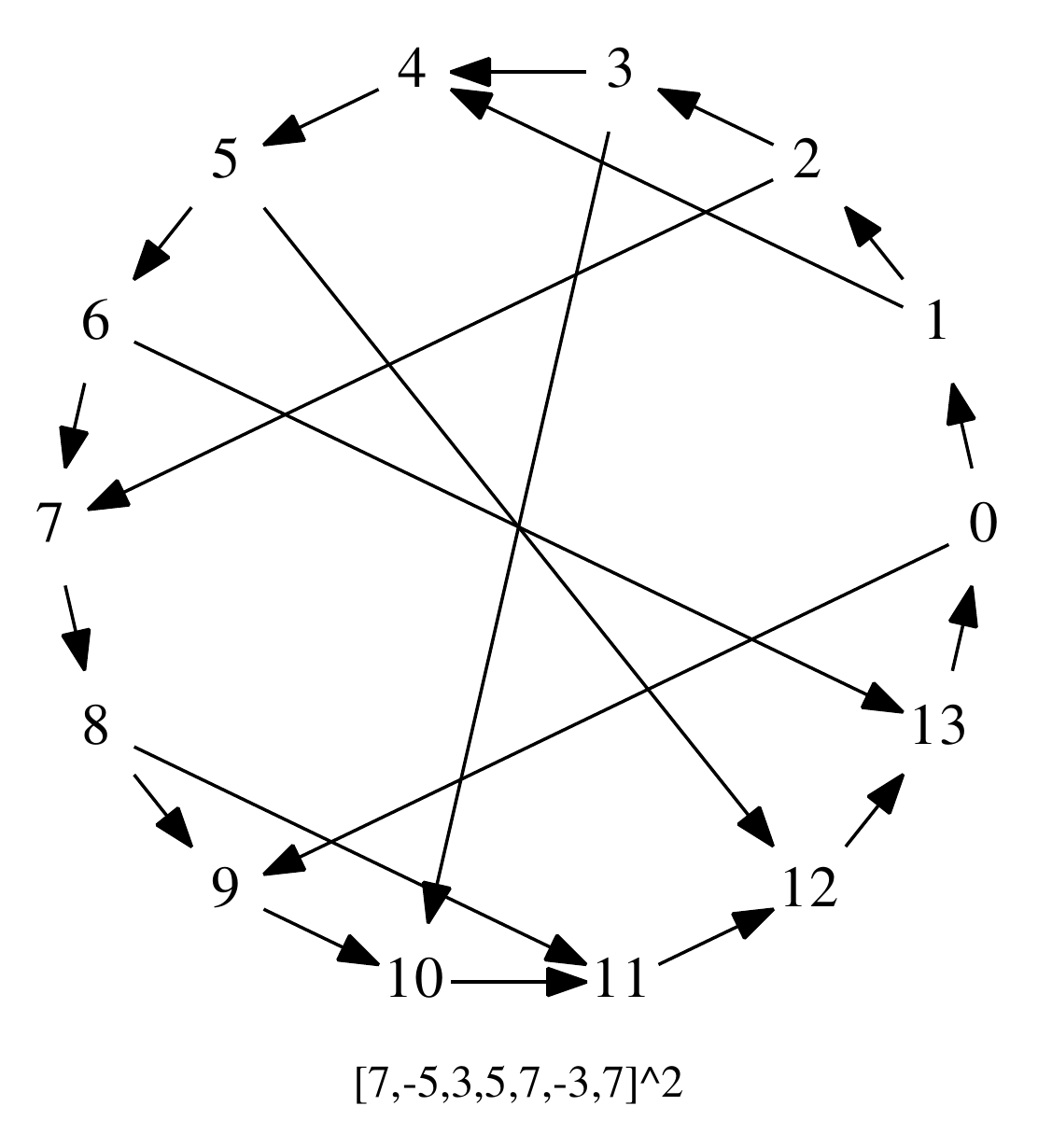}
\includegraphics[scale=0.45]{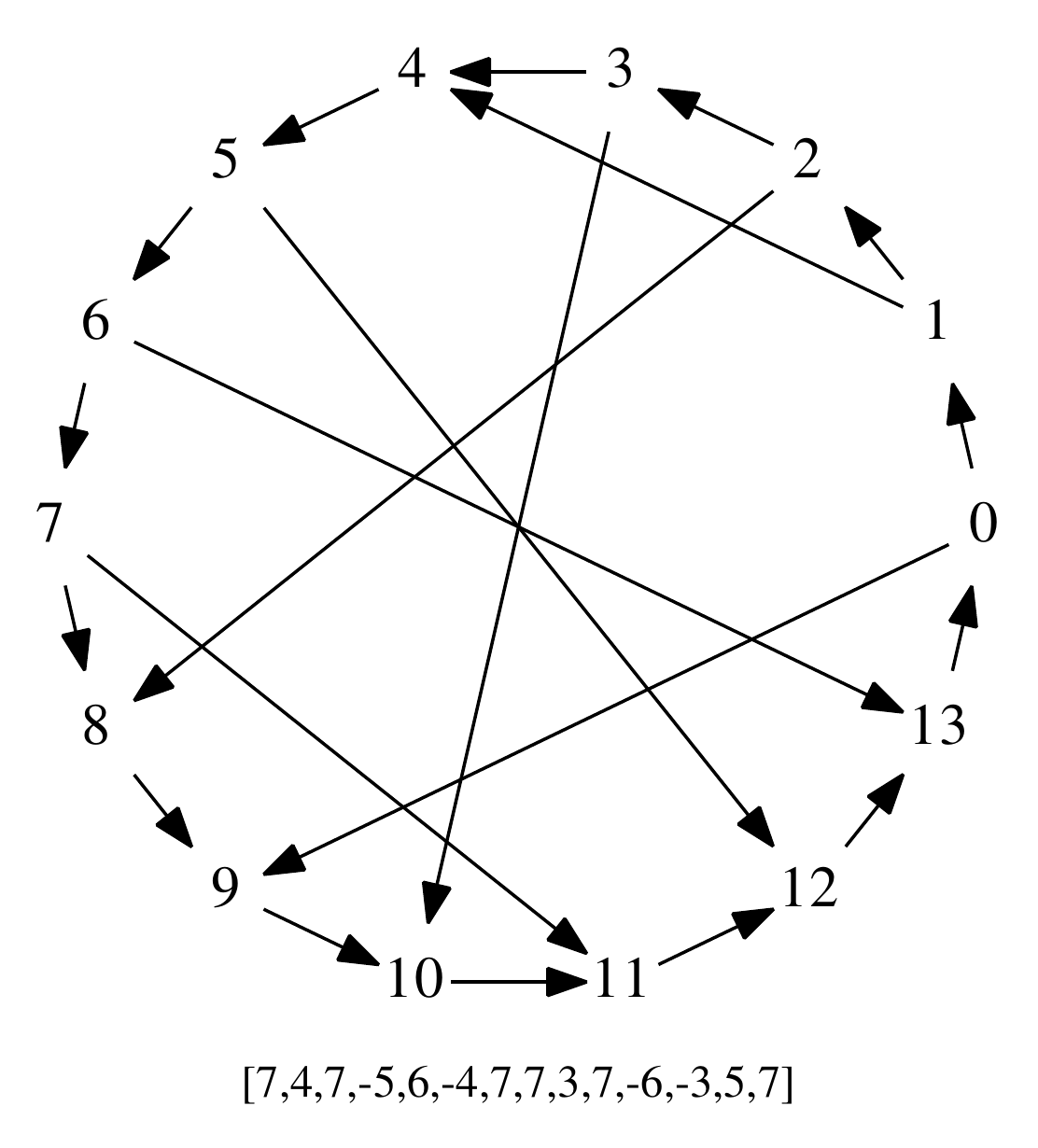}
\includegraphics[scale=0.45]{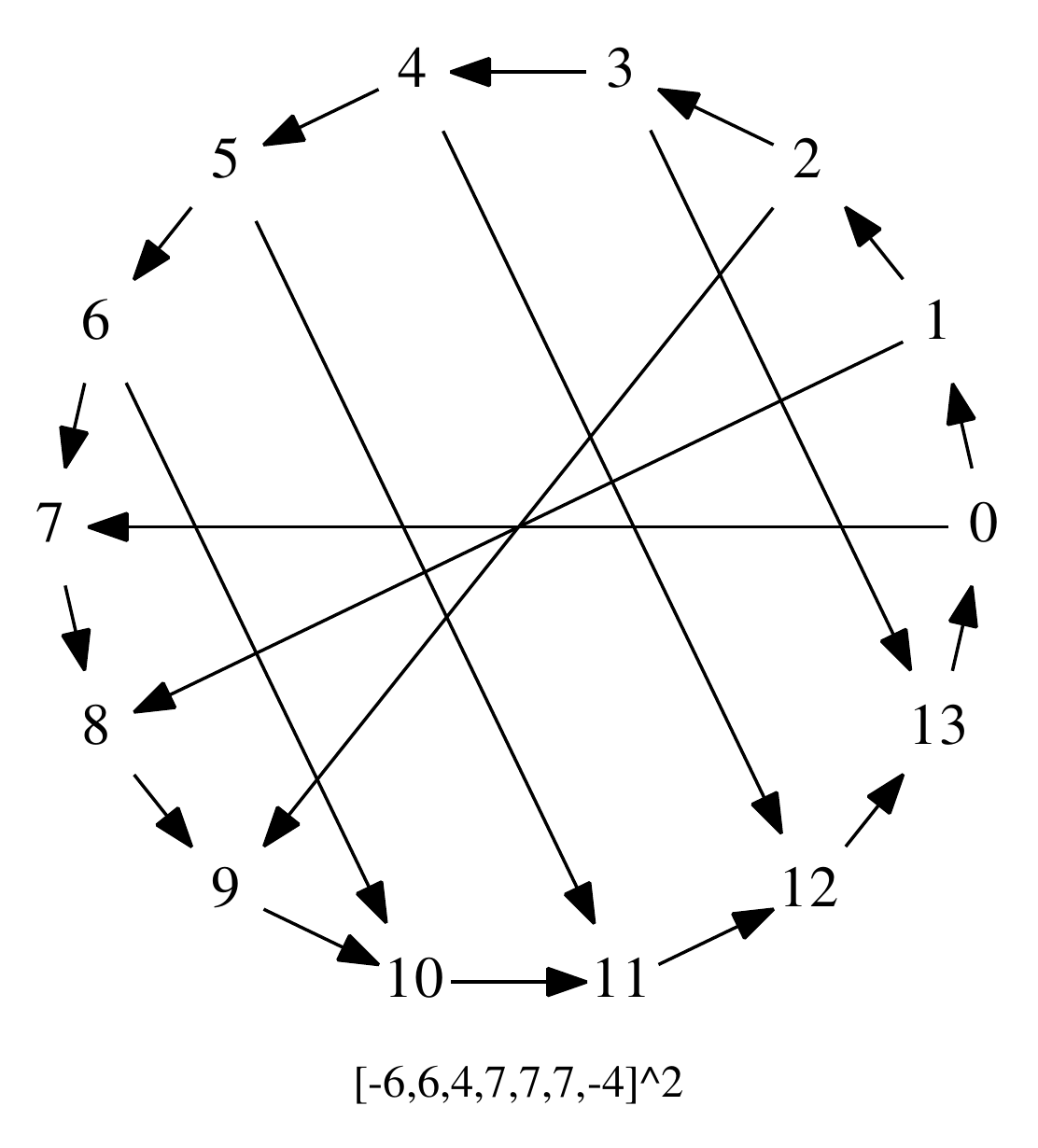}
\includegraphics[scale=0.45]{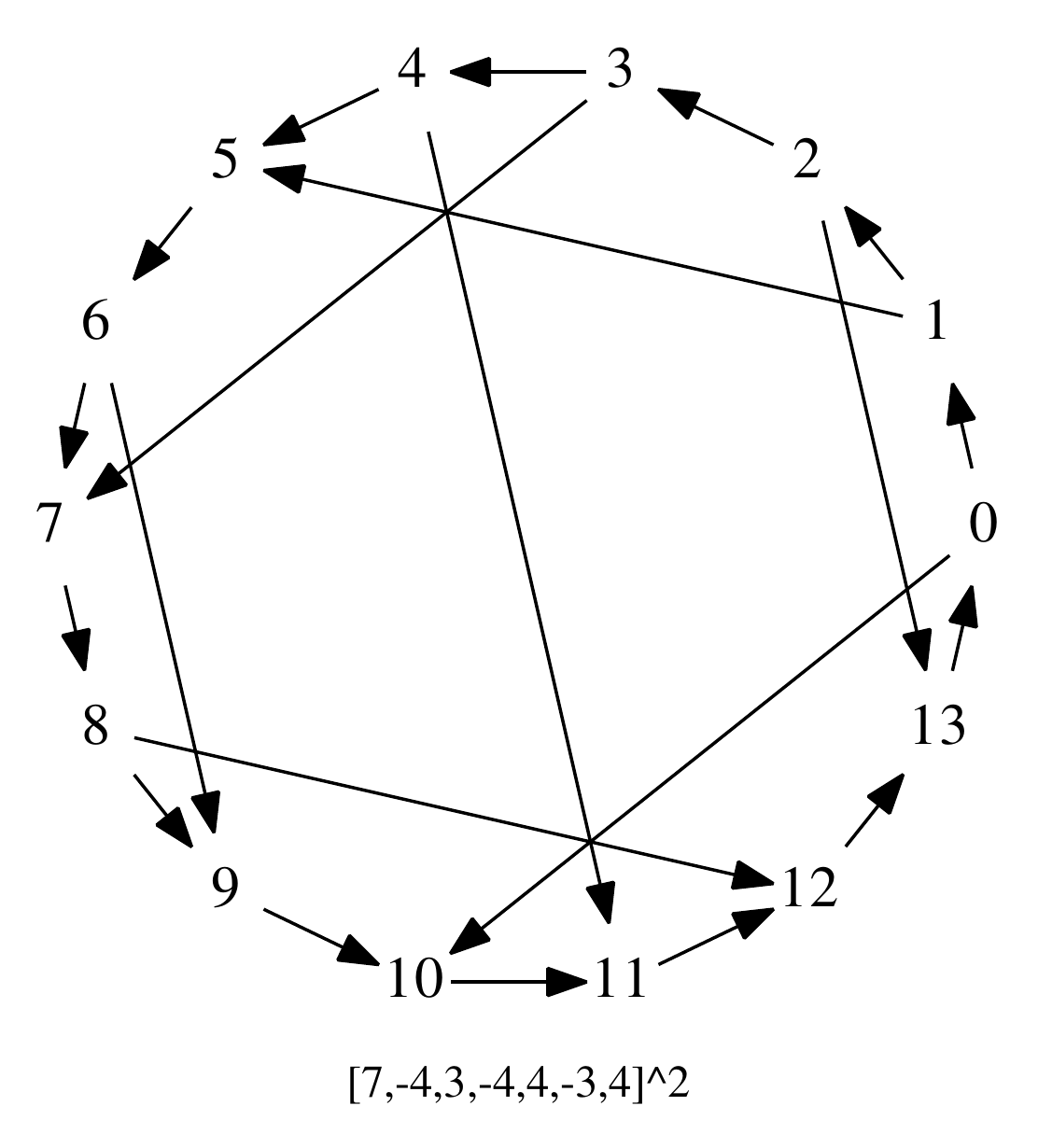}
\includegraphics[scale=0.45]{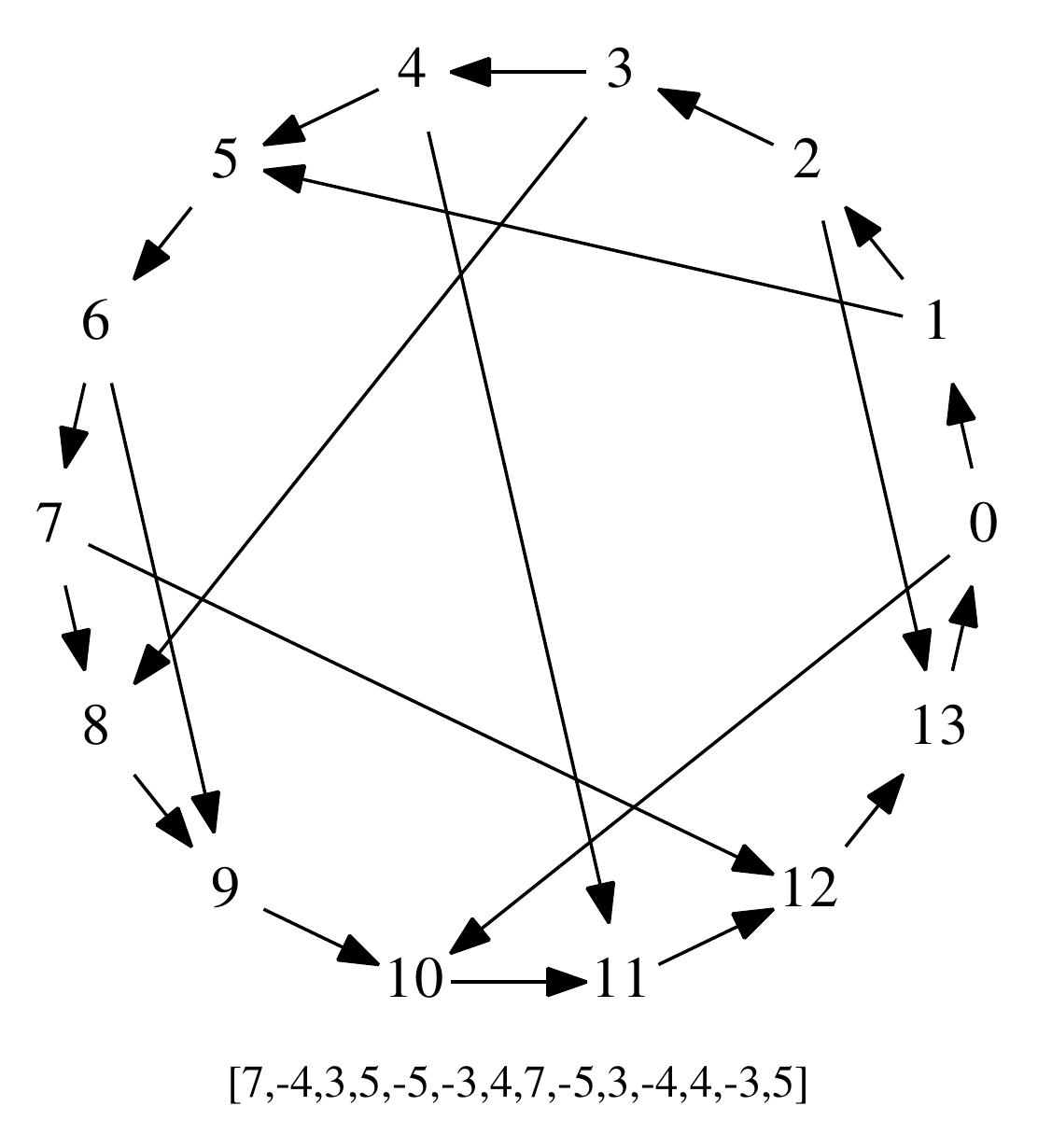}
\includegraphics[scale=0.45]{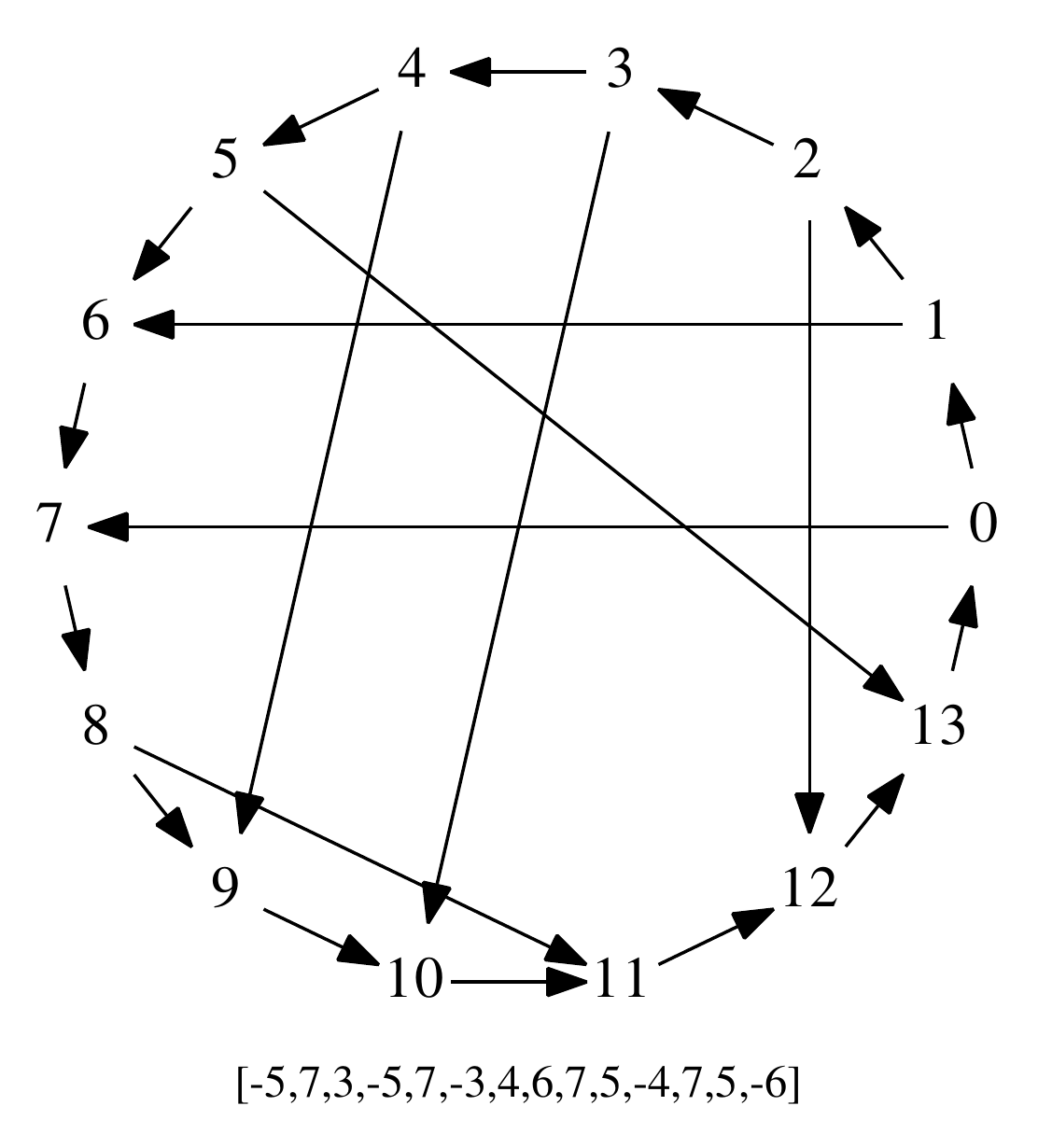}
\includegraphics[scale=0.45]{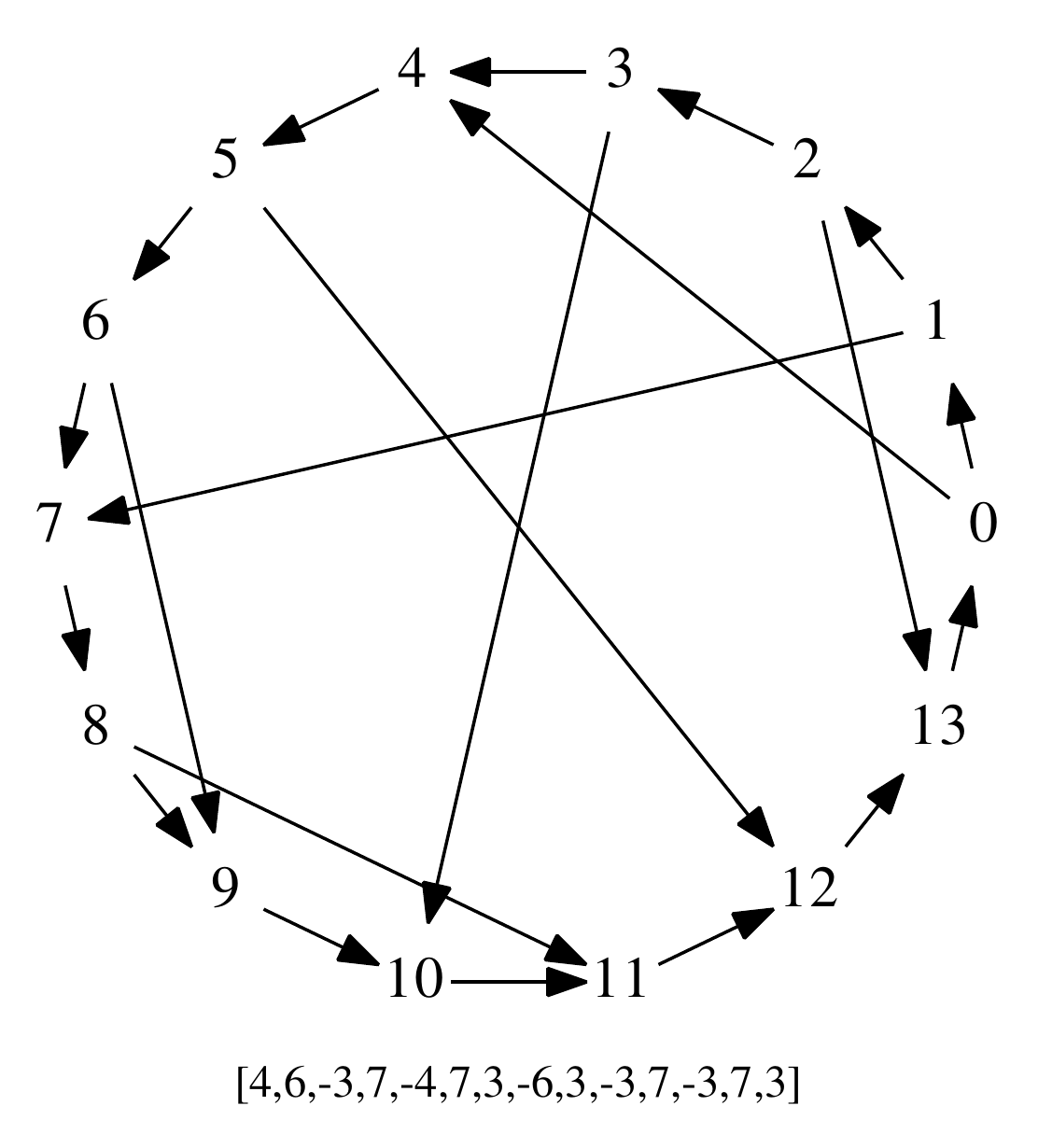}
\includegraphics[scale=0.45]{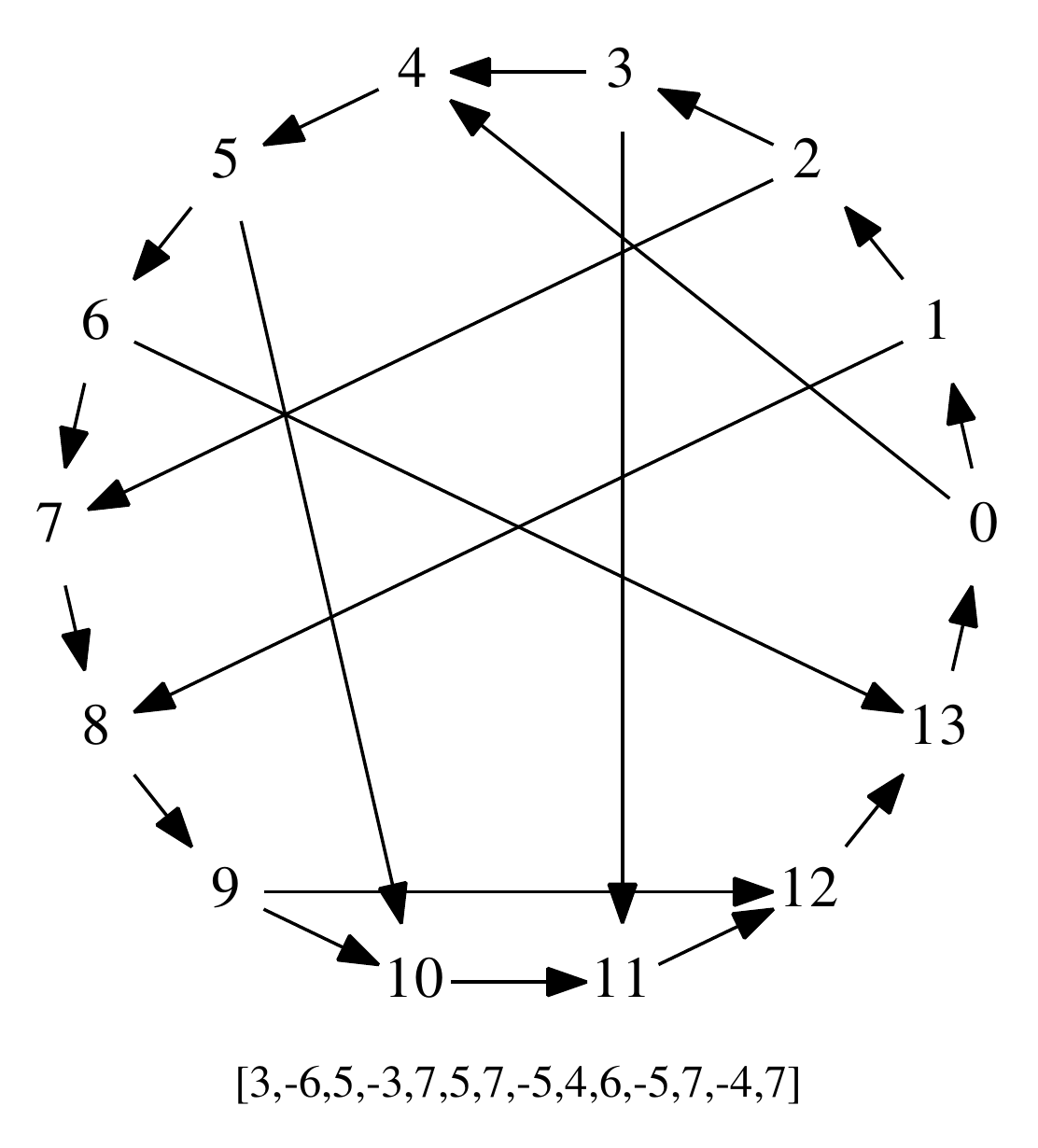}
\caption{Graphs on $n=14$ vertices which are irreducible (continued).
}
\label{fig.14n3}
\end{figure}

\begin{figure}
\includegraphics[scale=0.45]{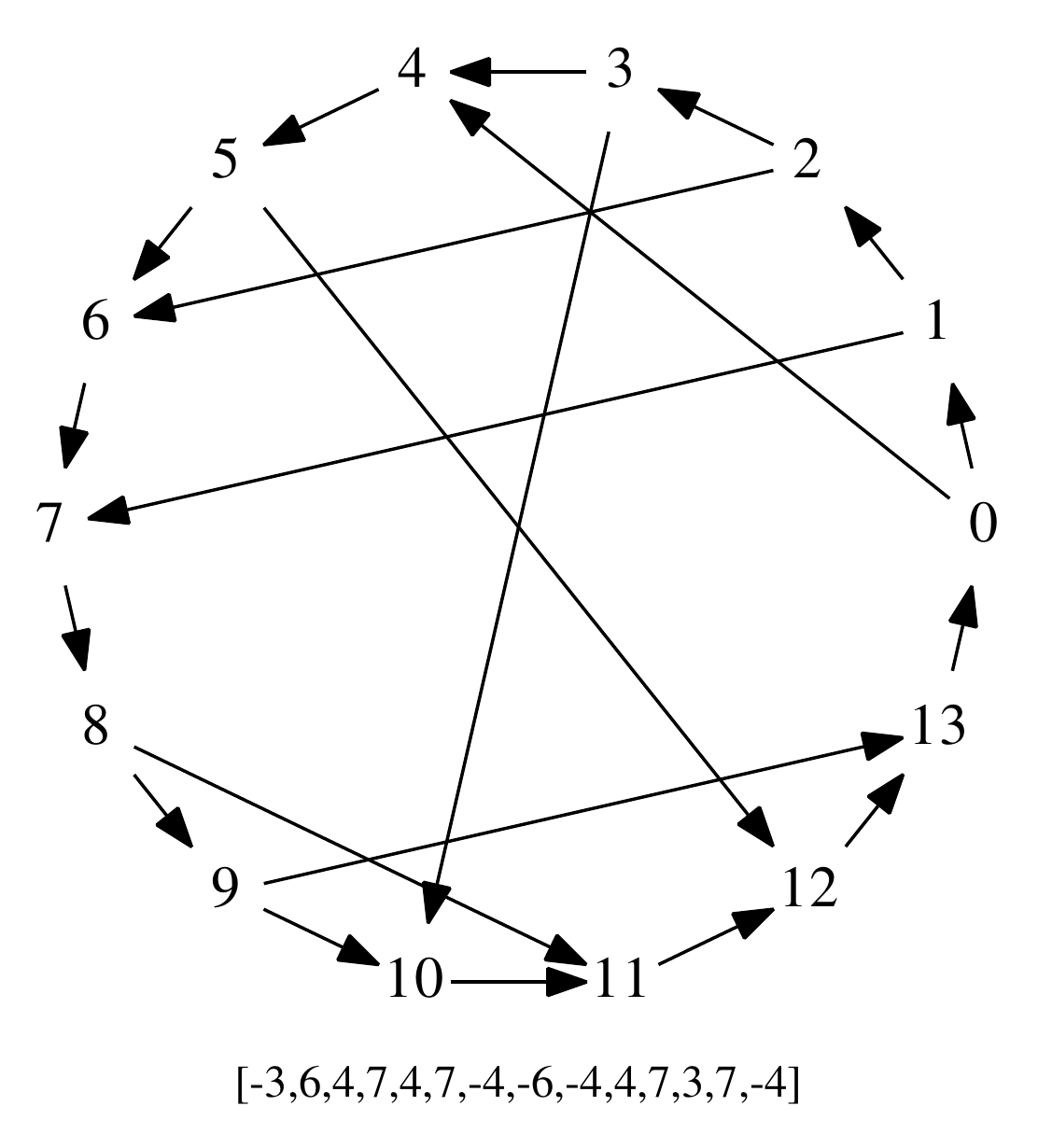}
\includegraphics[scale=0.45]{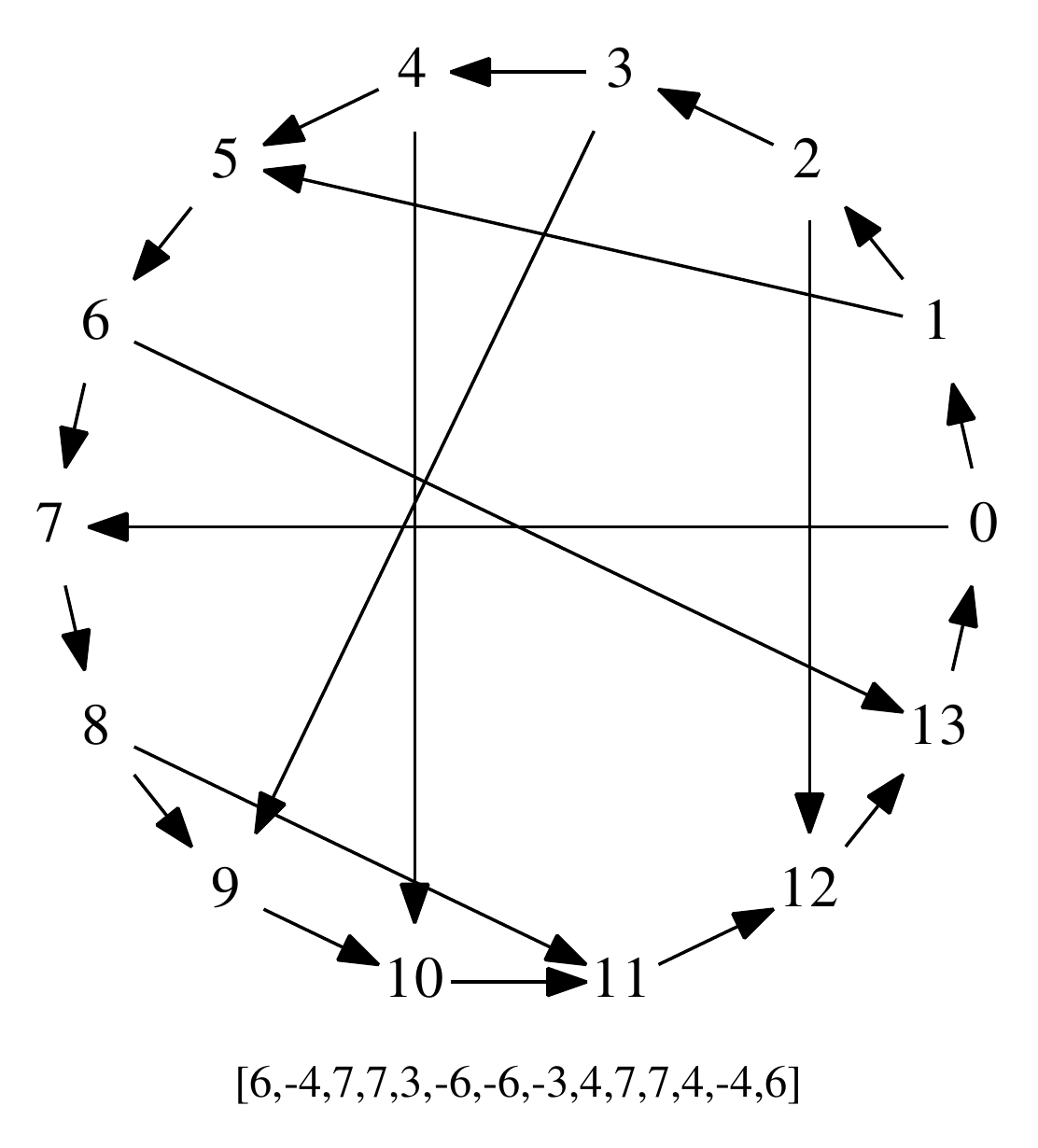}
\includegraphics[scale=0.45]{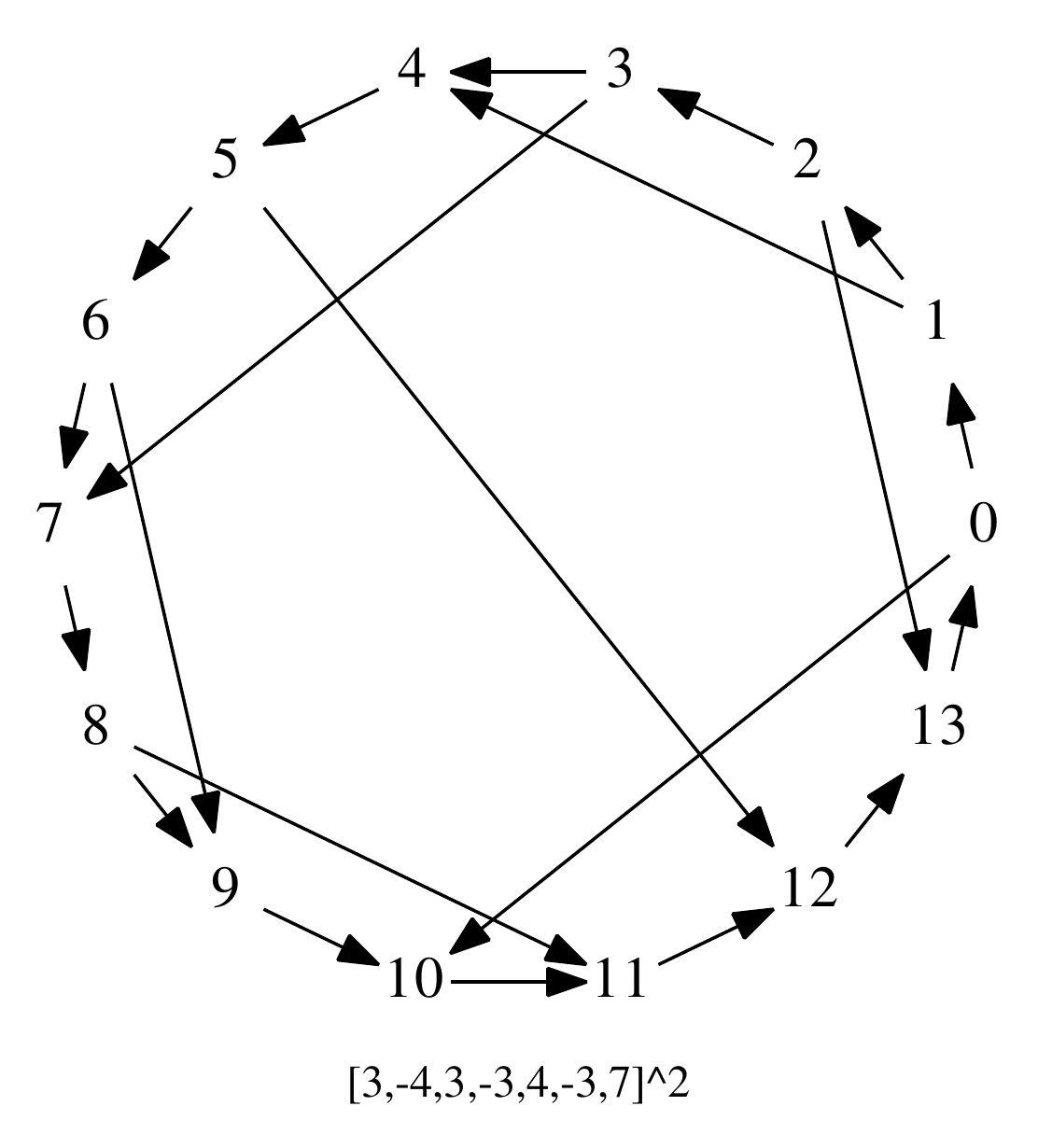}
\includegraphics[scale=0.45]{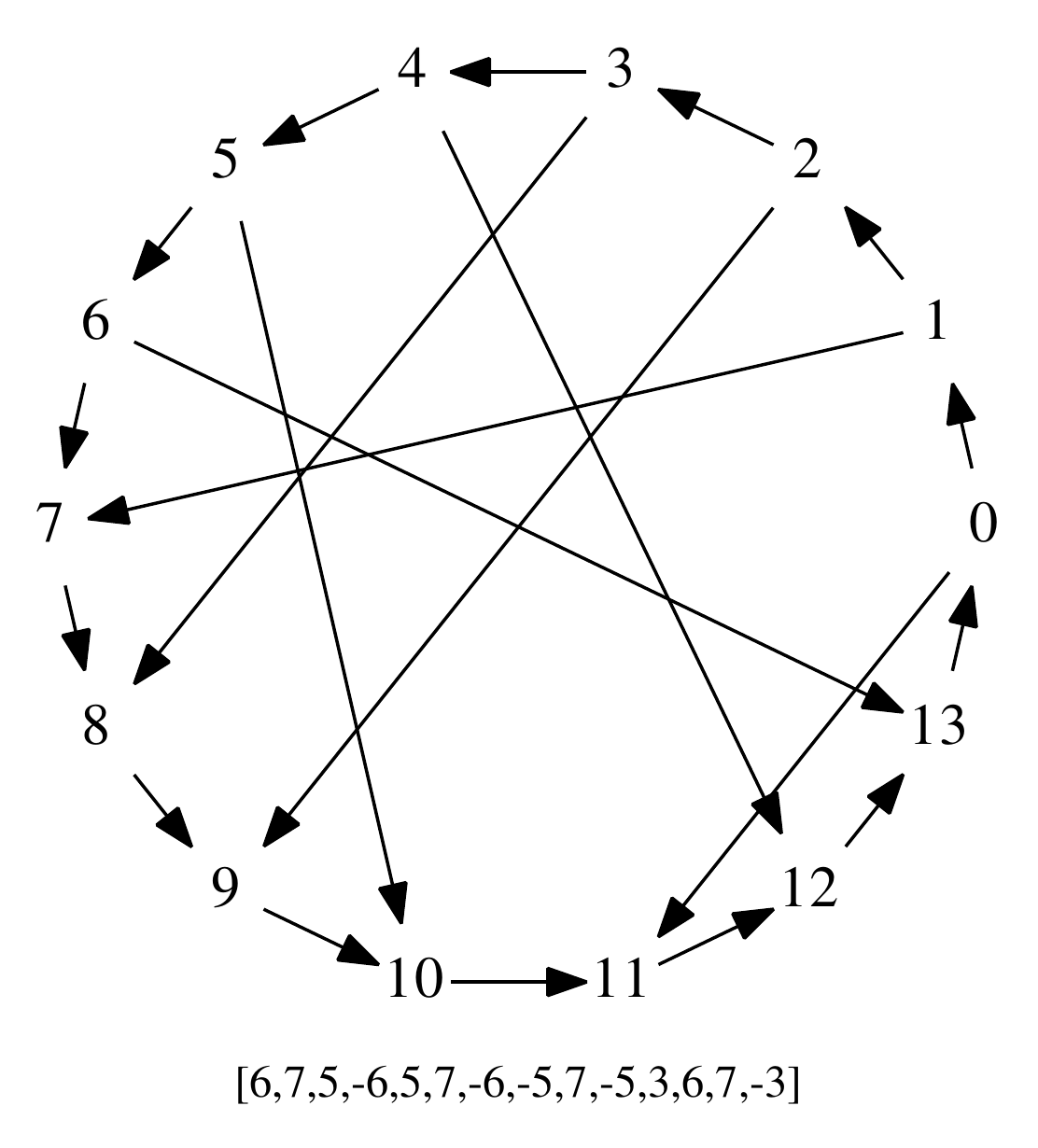}
\includegraphics[scale=0.45]{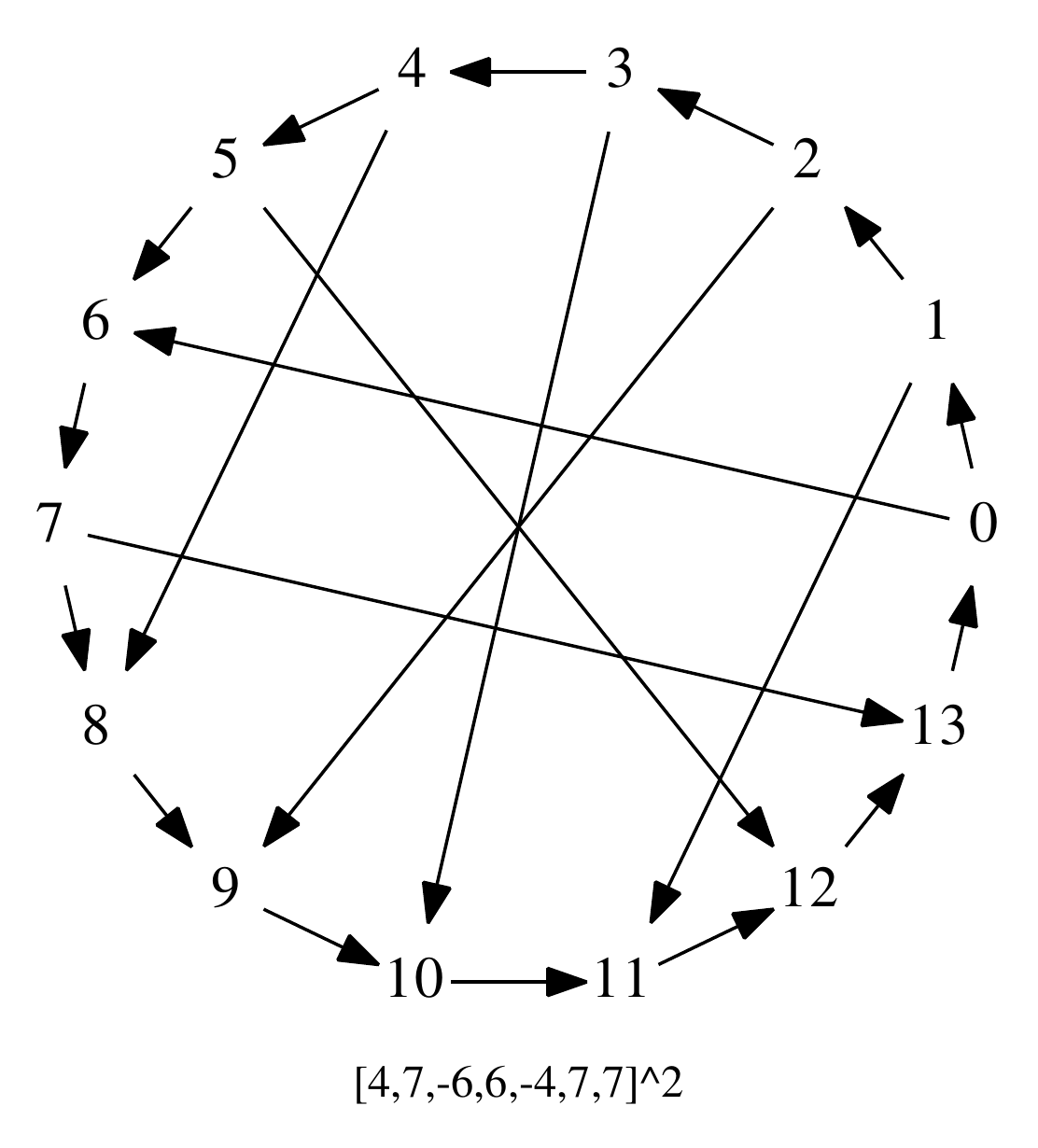}
\includegraphics[scale=0.45]{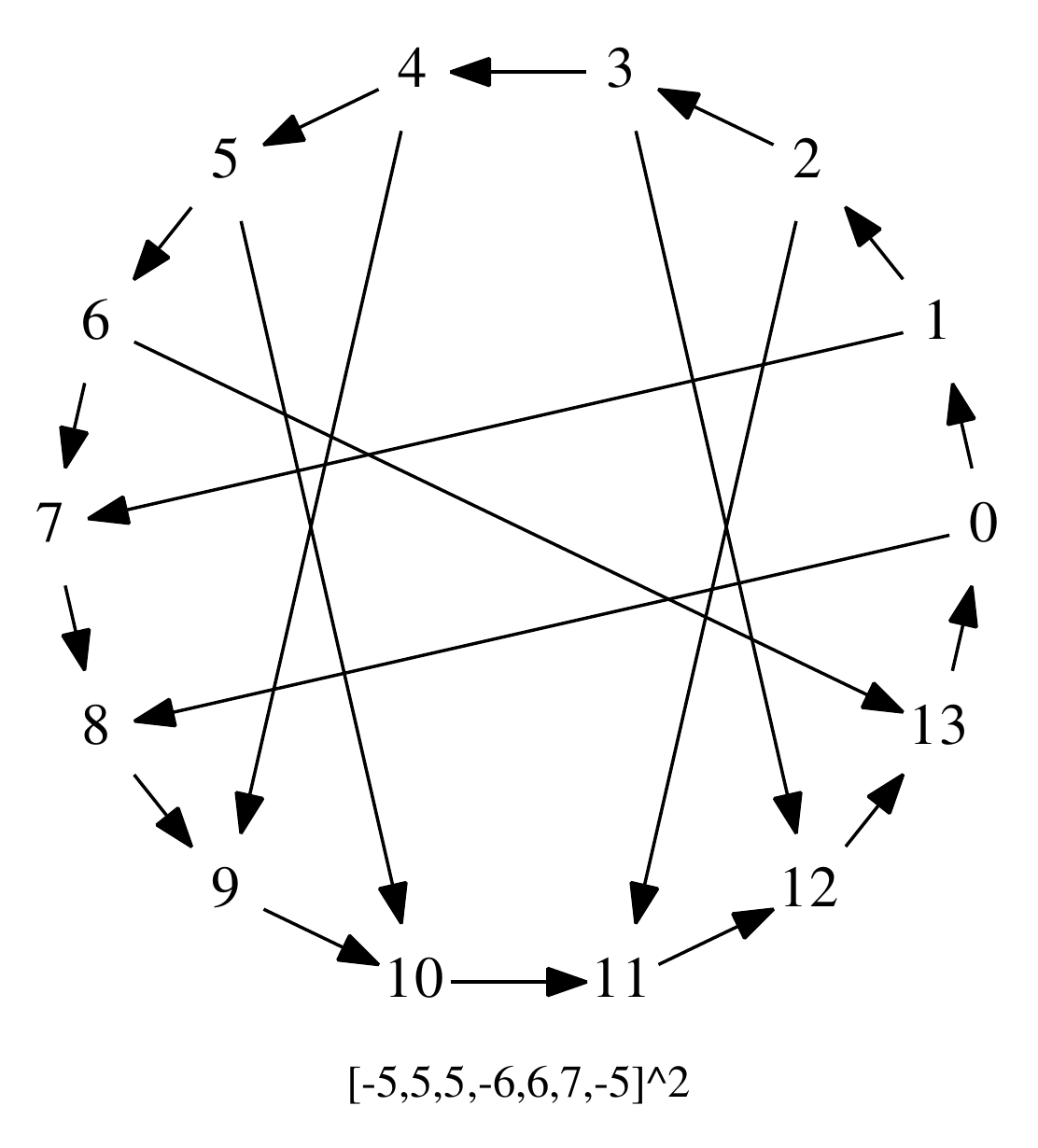}
\includegraphics[scale=0.45]{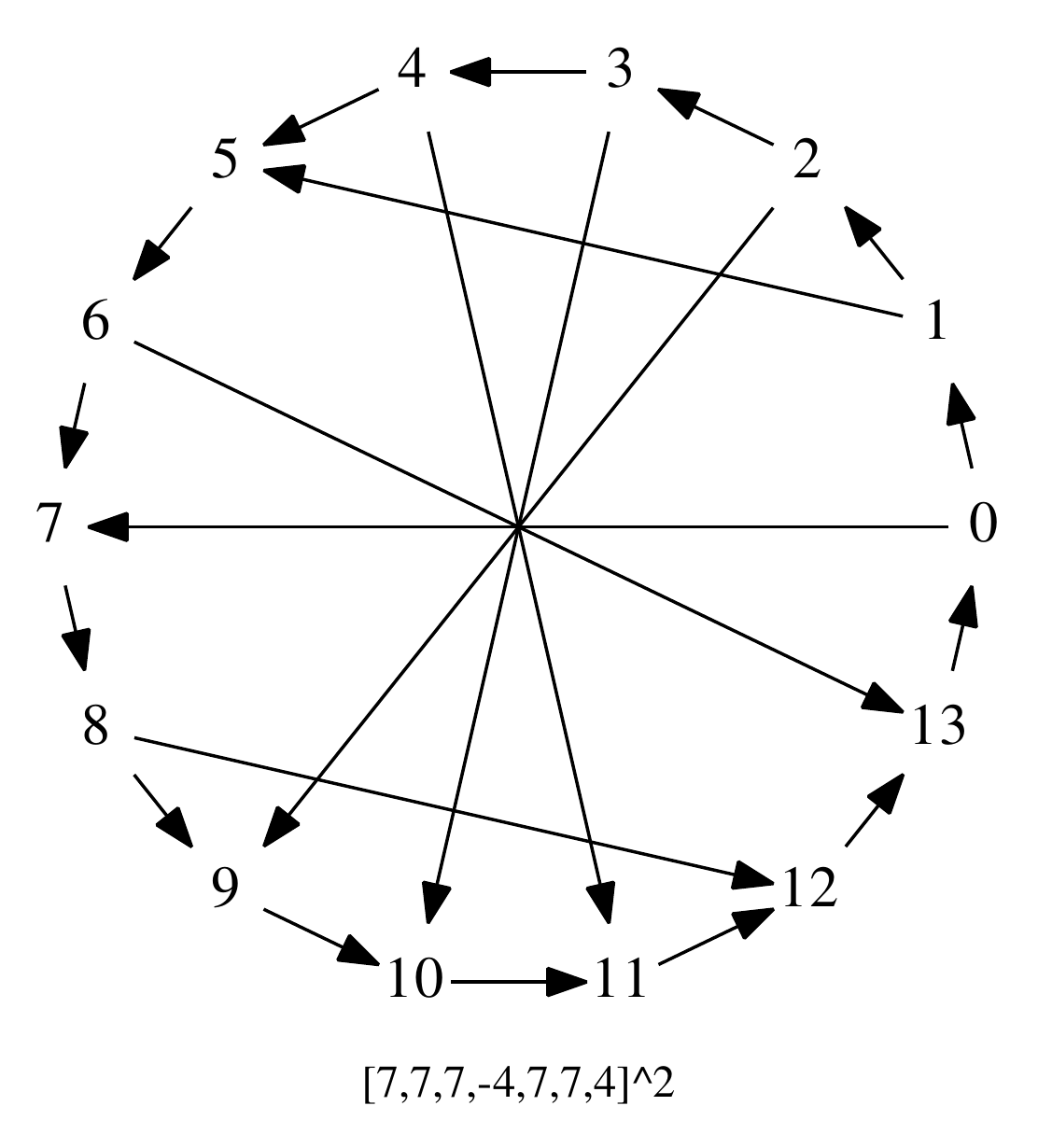}
\includegraphics[scale=0.45]{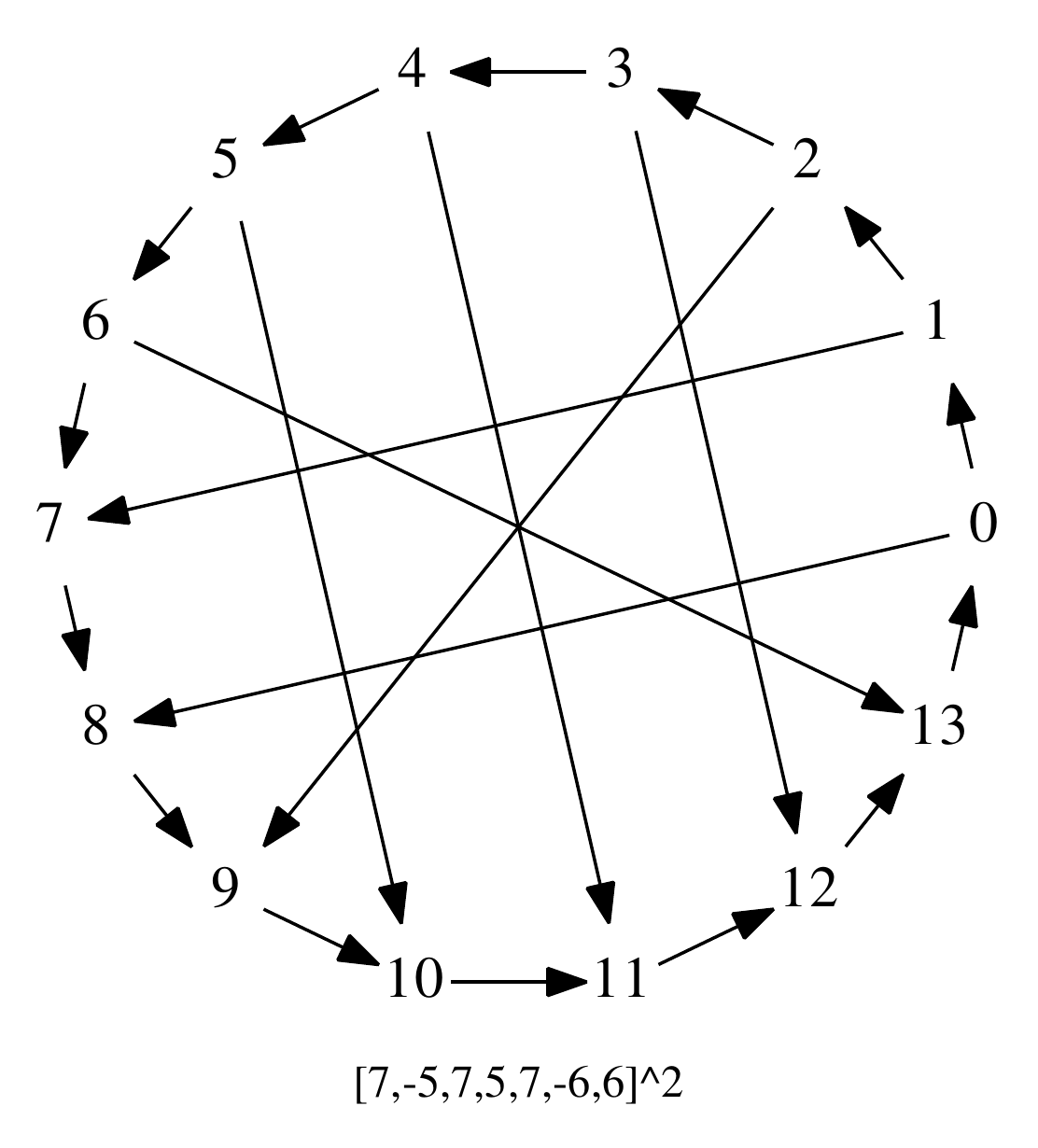}
\includegraphics[scale=0.45]{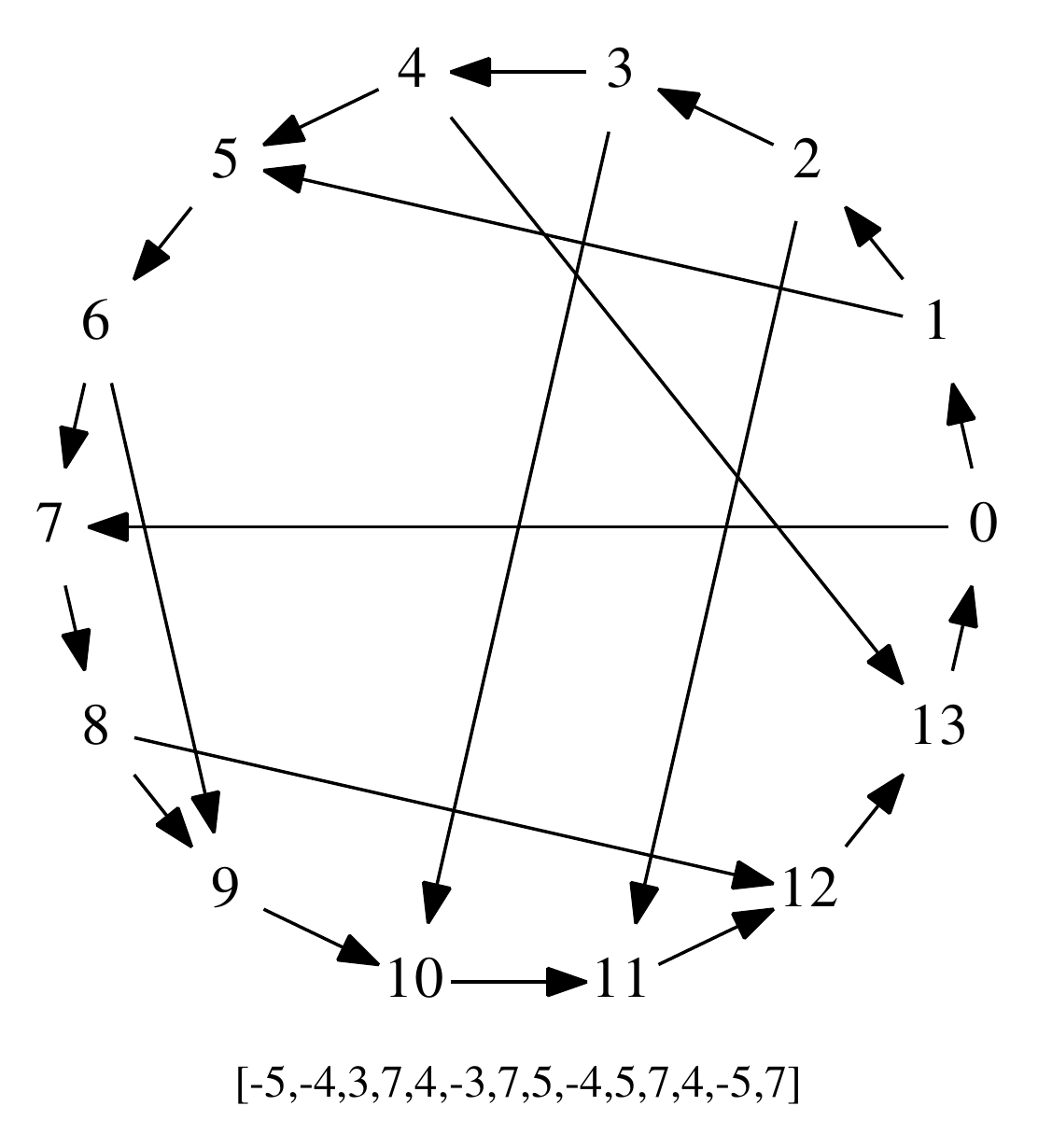}
\includegraphics[scale=0.45]{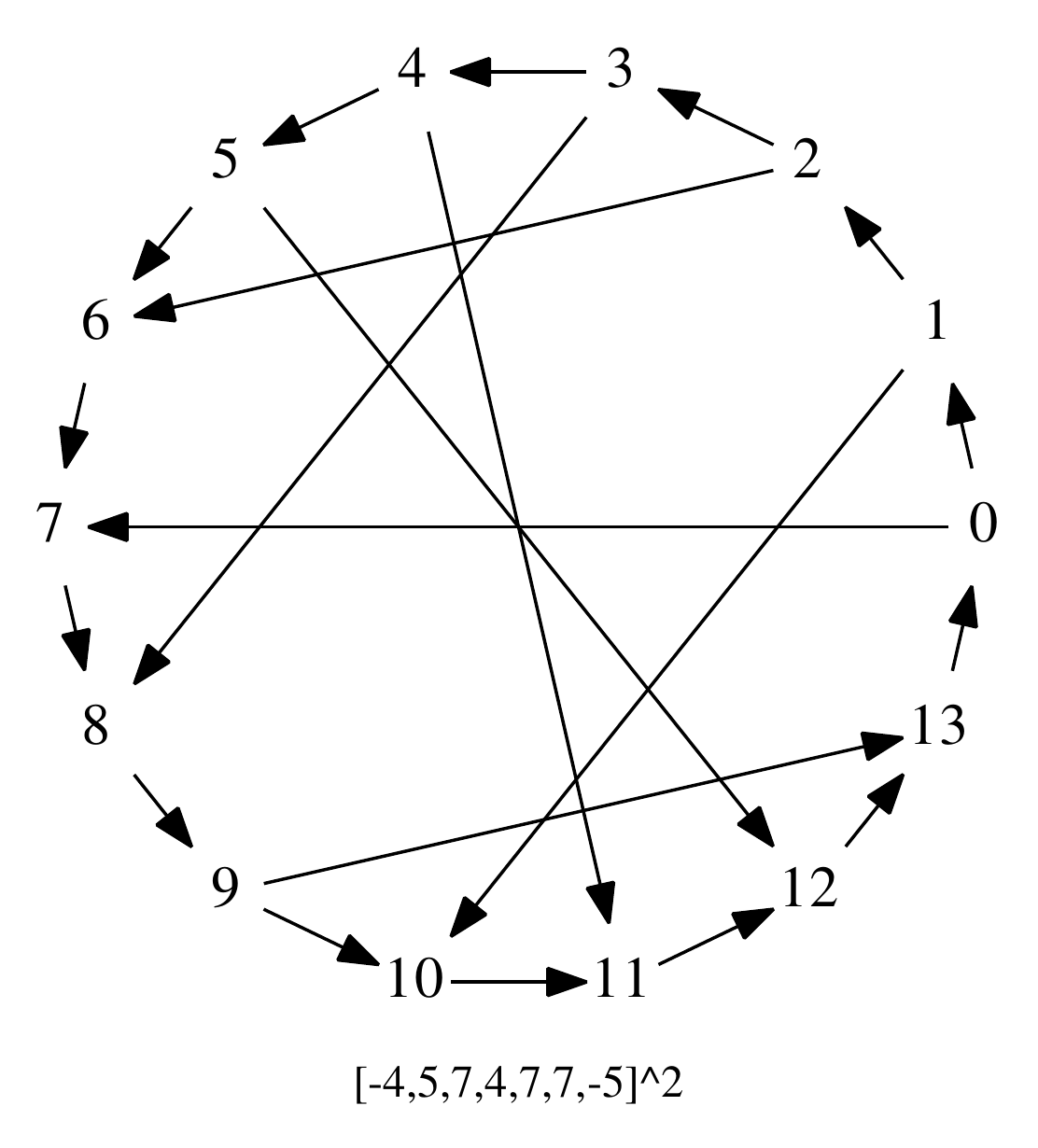}
\includegraphics[scale=0.45]{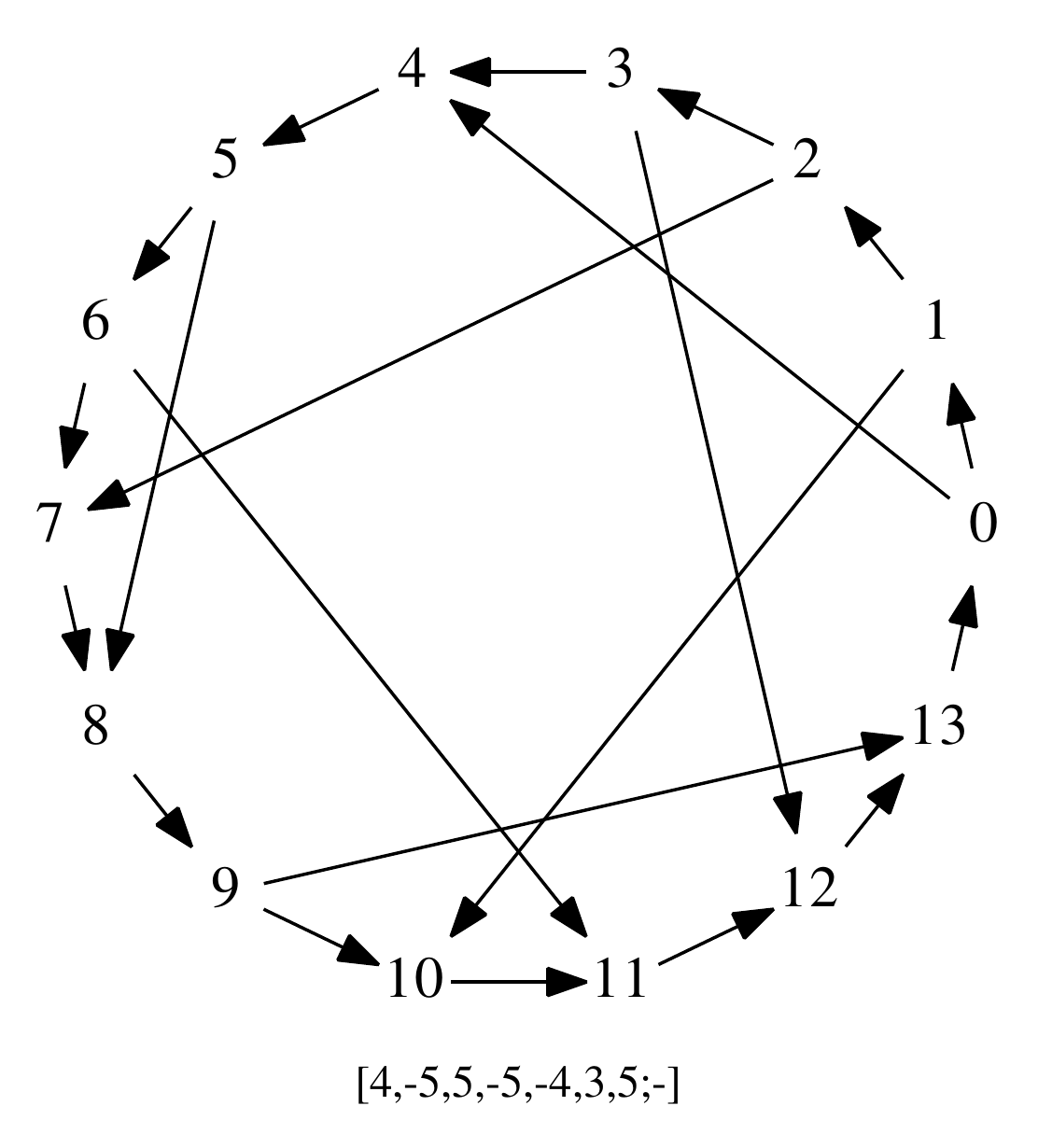}
\includegraphics[scale=0.45]{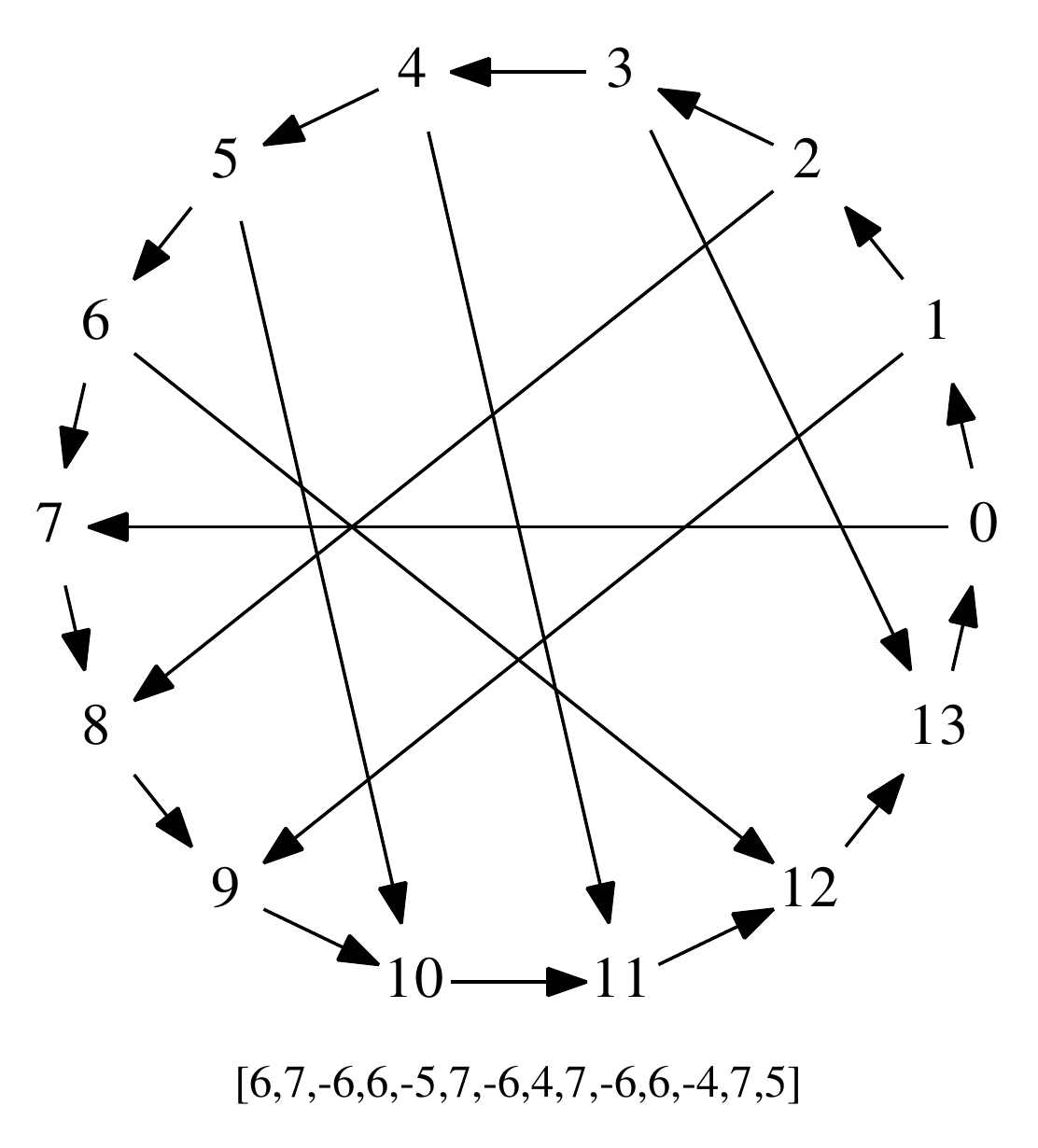}
\caption{Graphs on $n=14$ vertices which are irreducible (continued).
}
\label{fig.14n4}
\end{figure}

\begin{figure}
\includegraphics[scale=0.45]{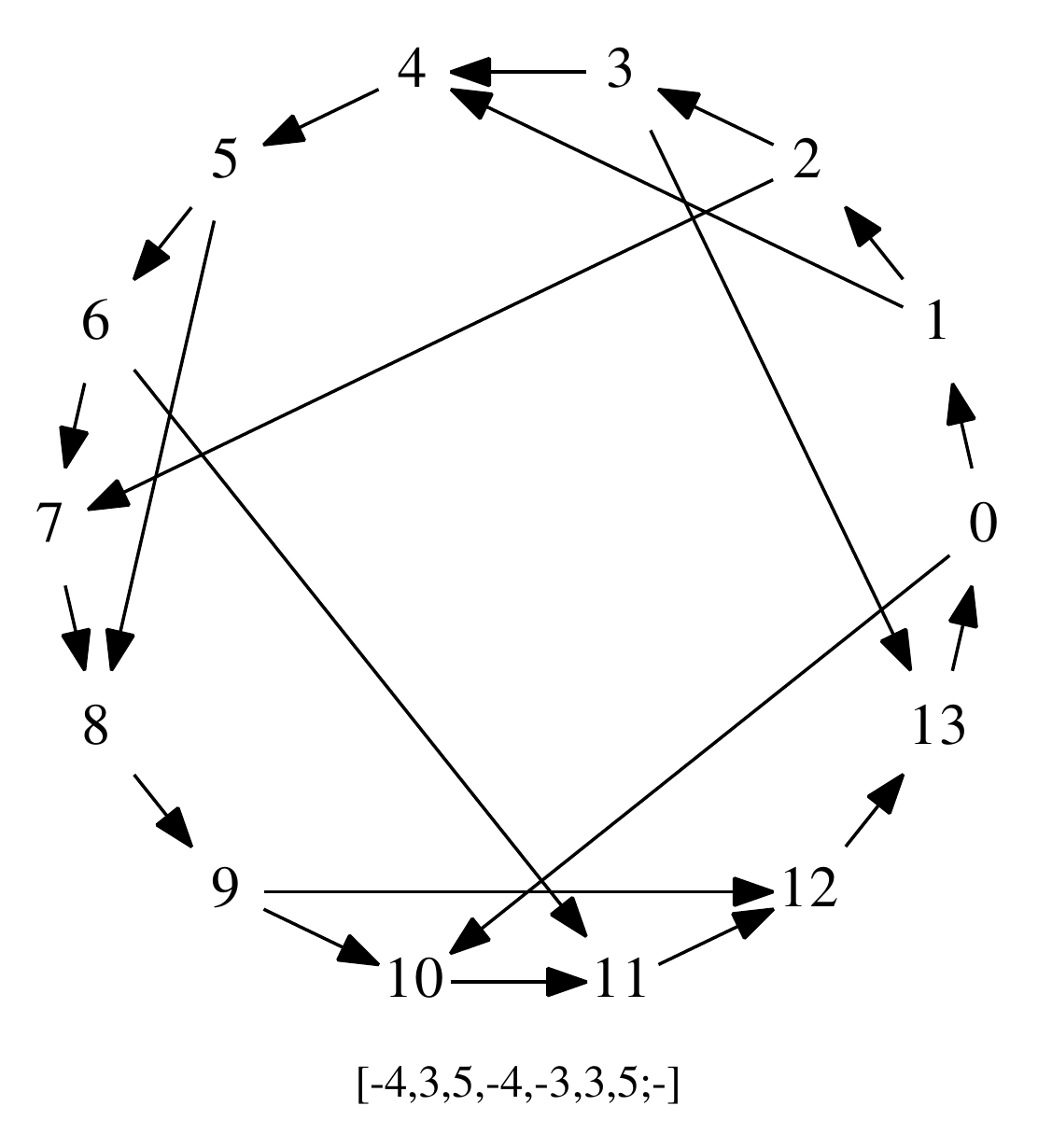}
\includegraphics[scale=0.45]{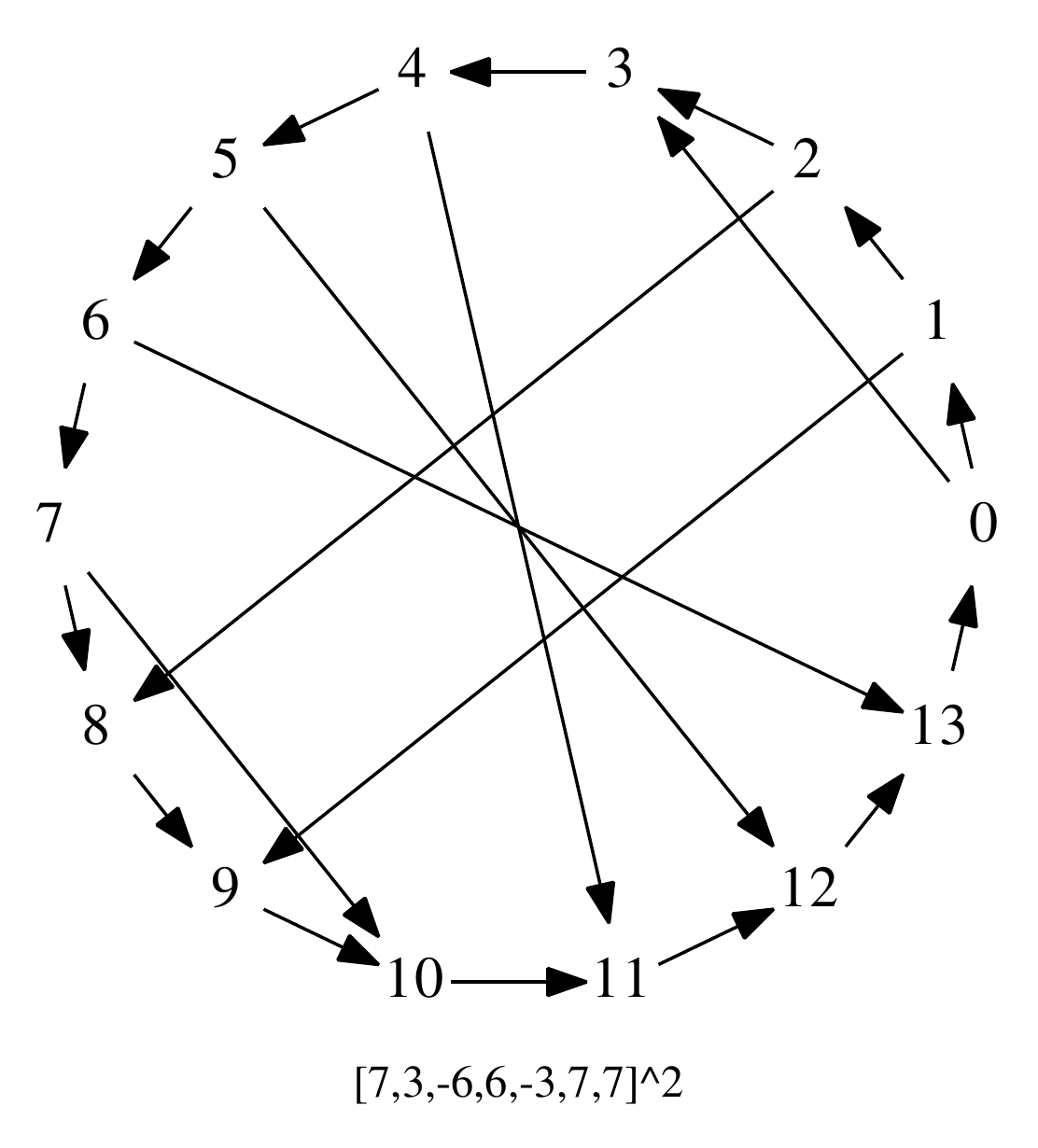}
\includegraphics[scale=0.45]{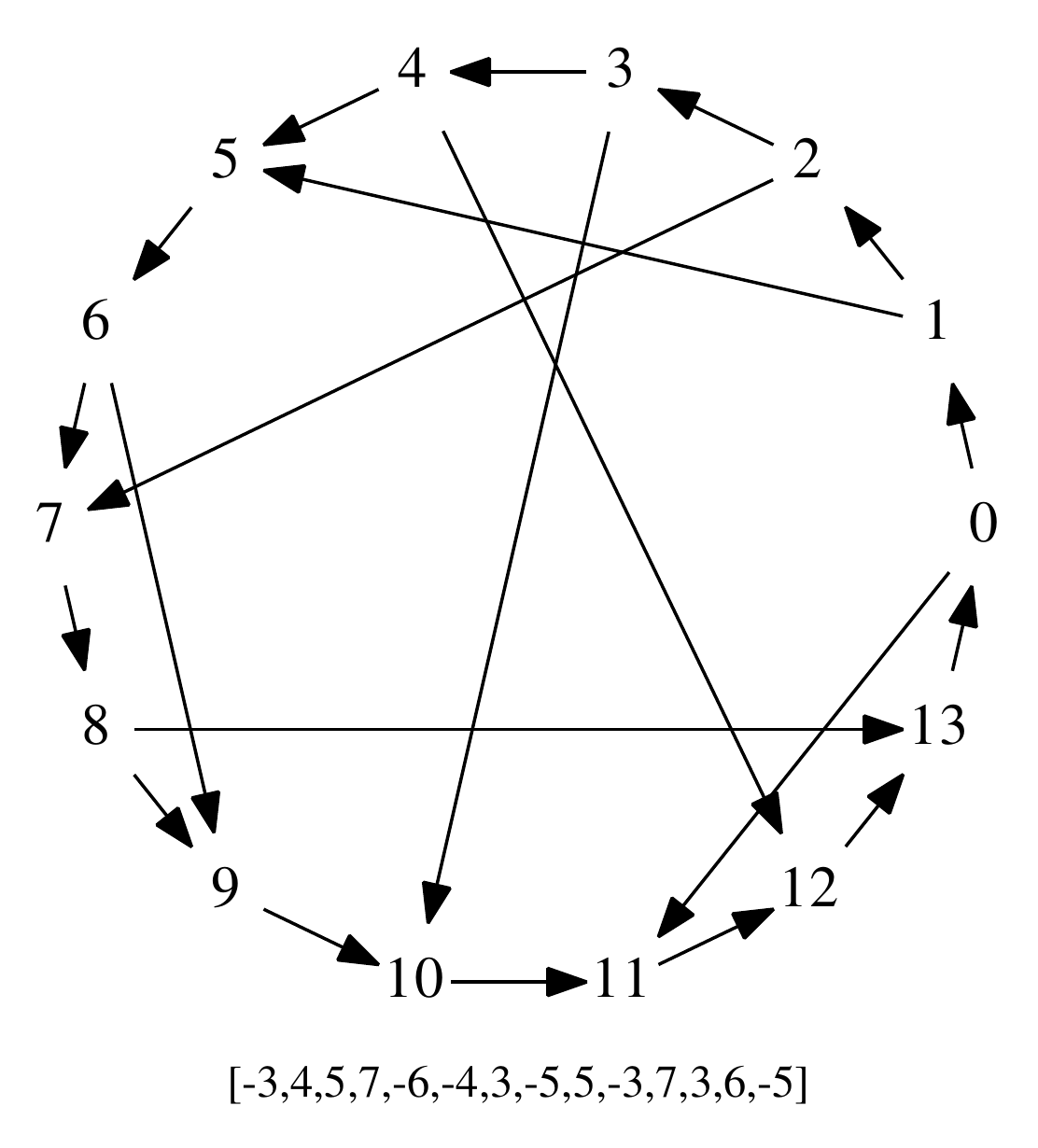}
\includegraphics[scale=0.45]{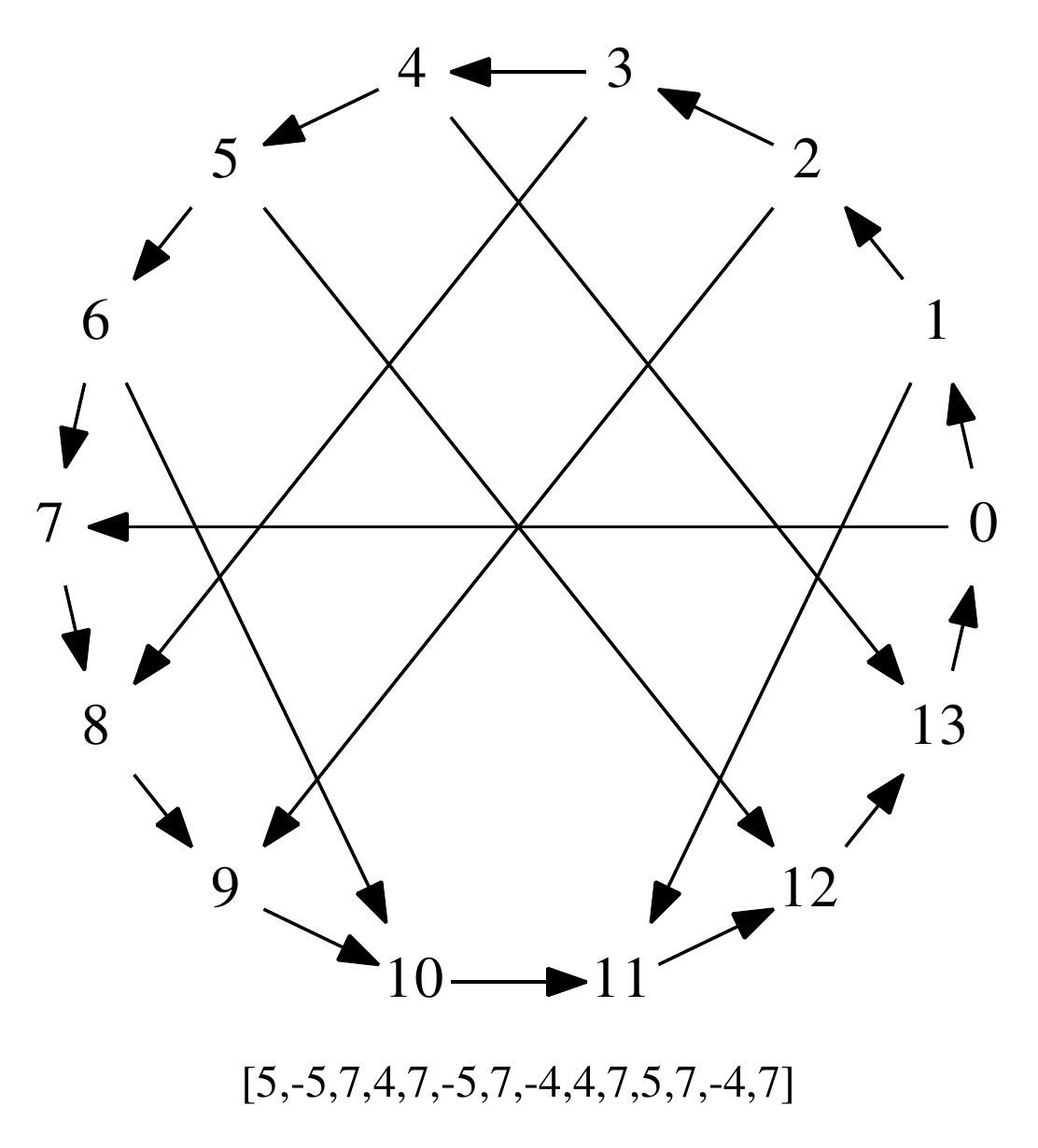}
\includegraphics[scale=0.45]{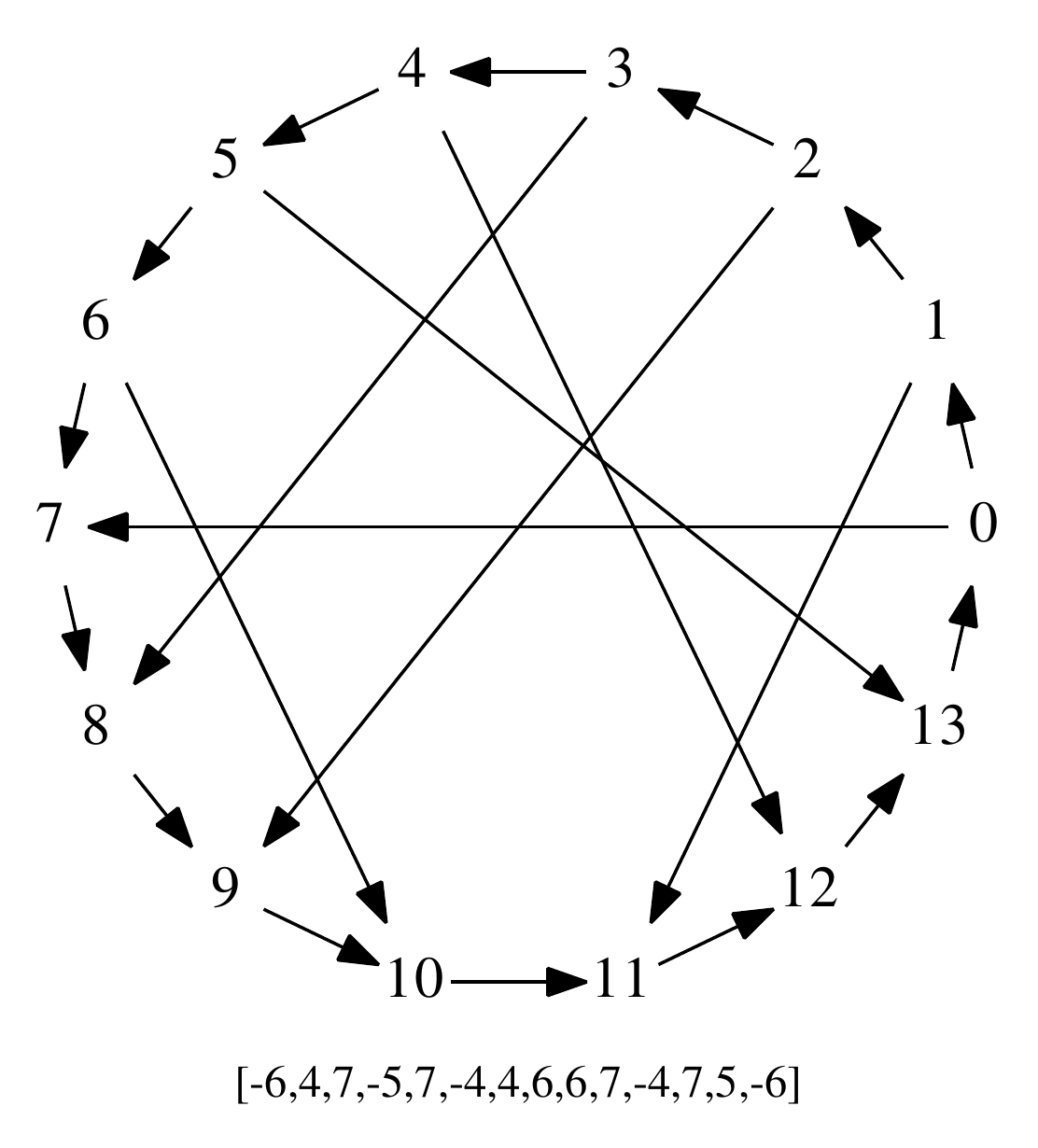}
\includegraphics[scale=0.45]{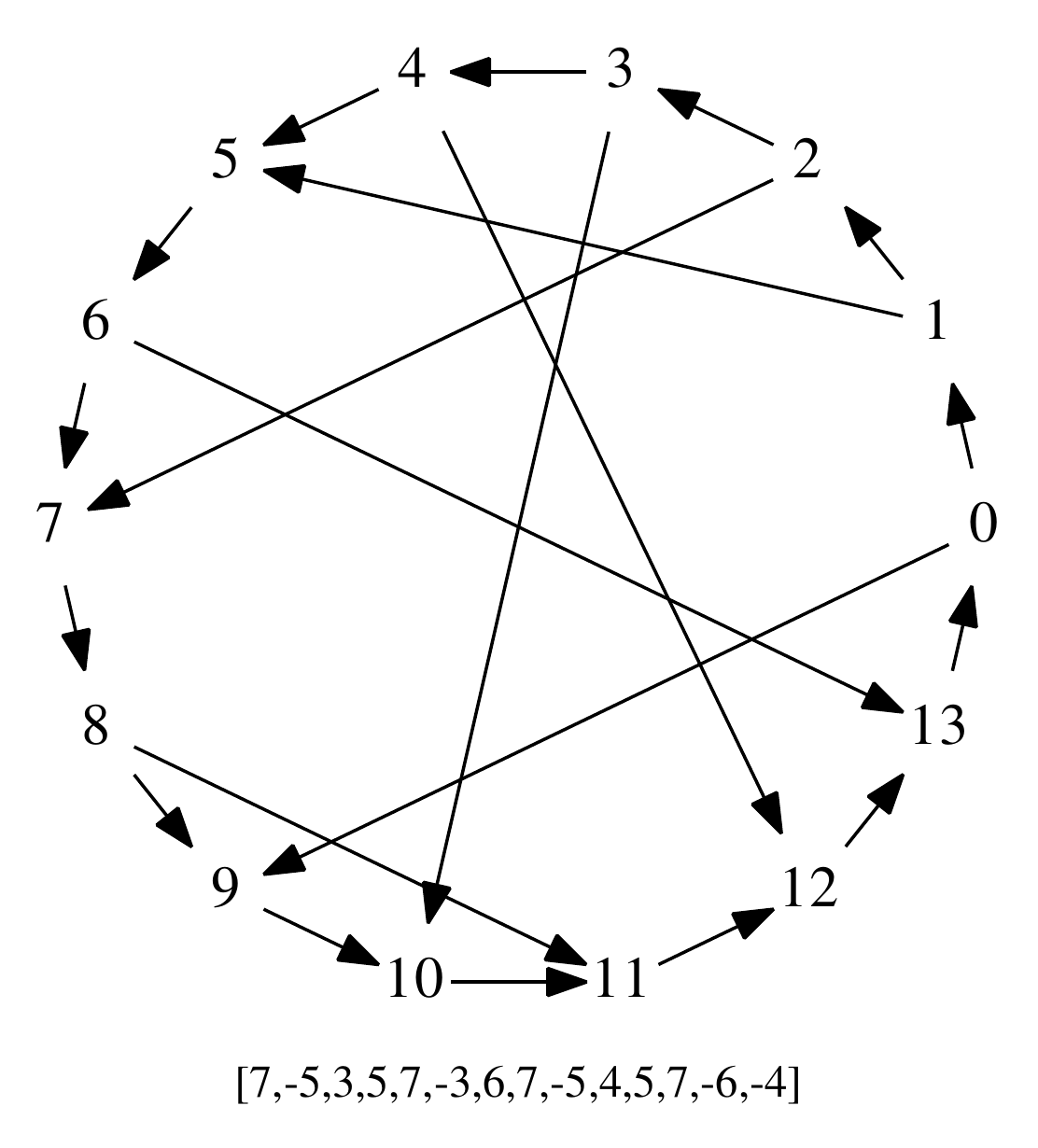}
\includegraphics[scale=0.45]{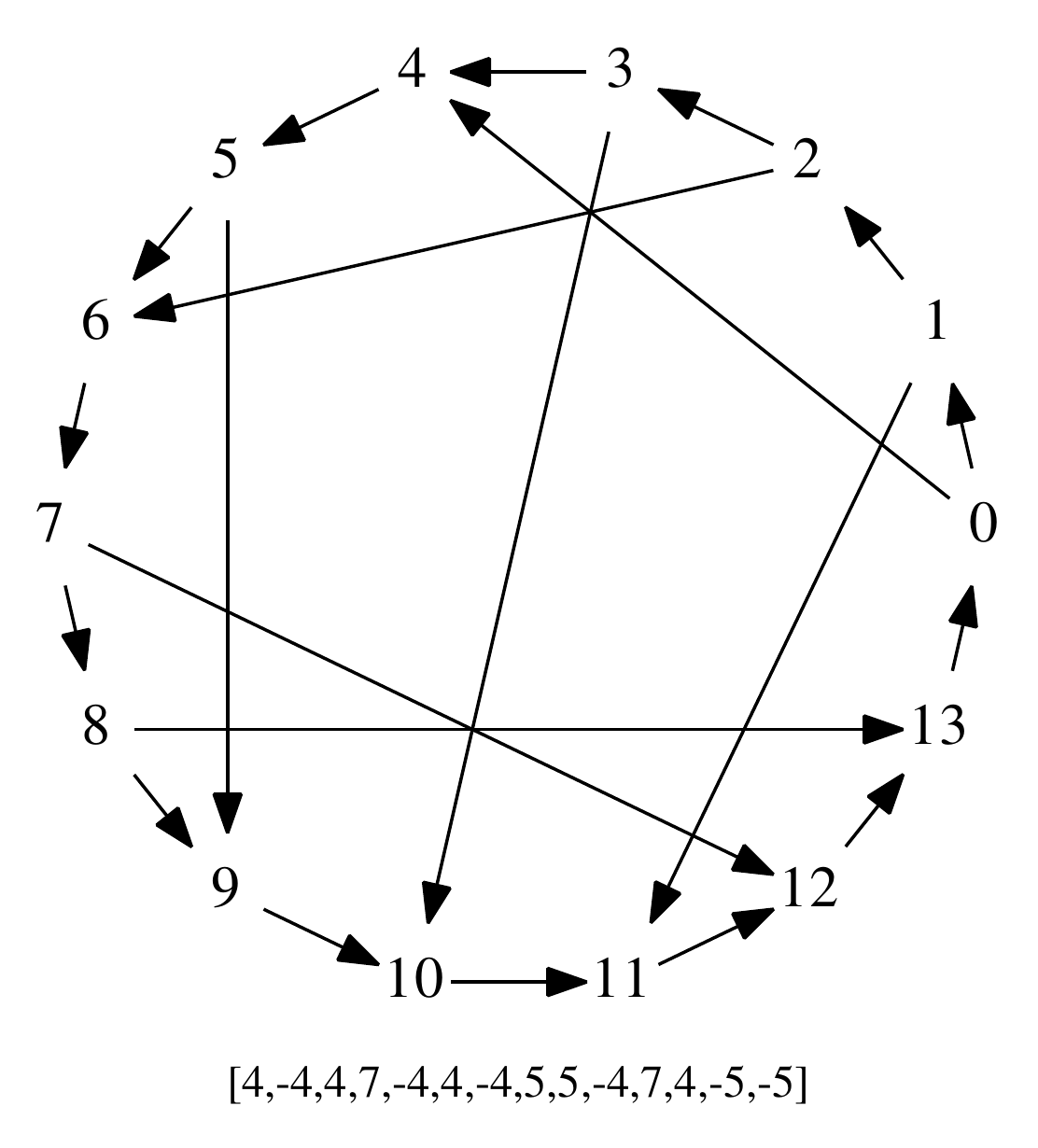}
\includegraphics[scale=0.45]{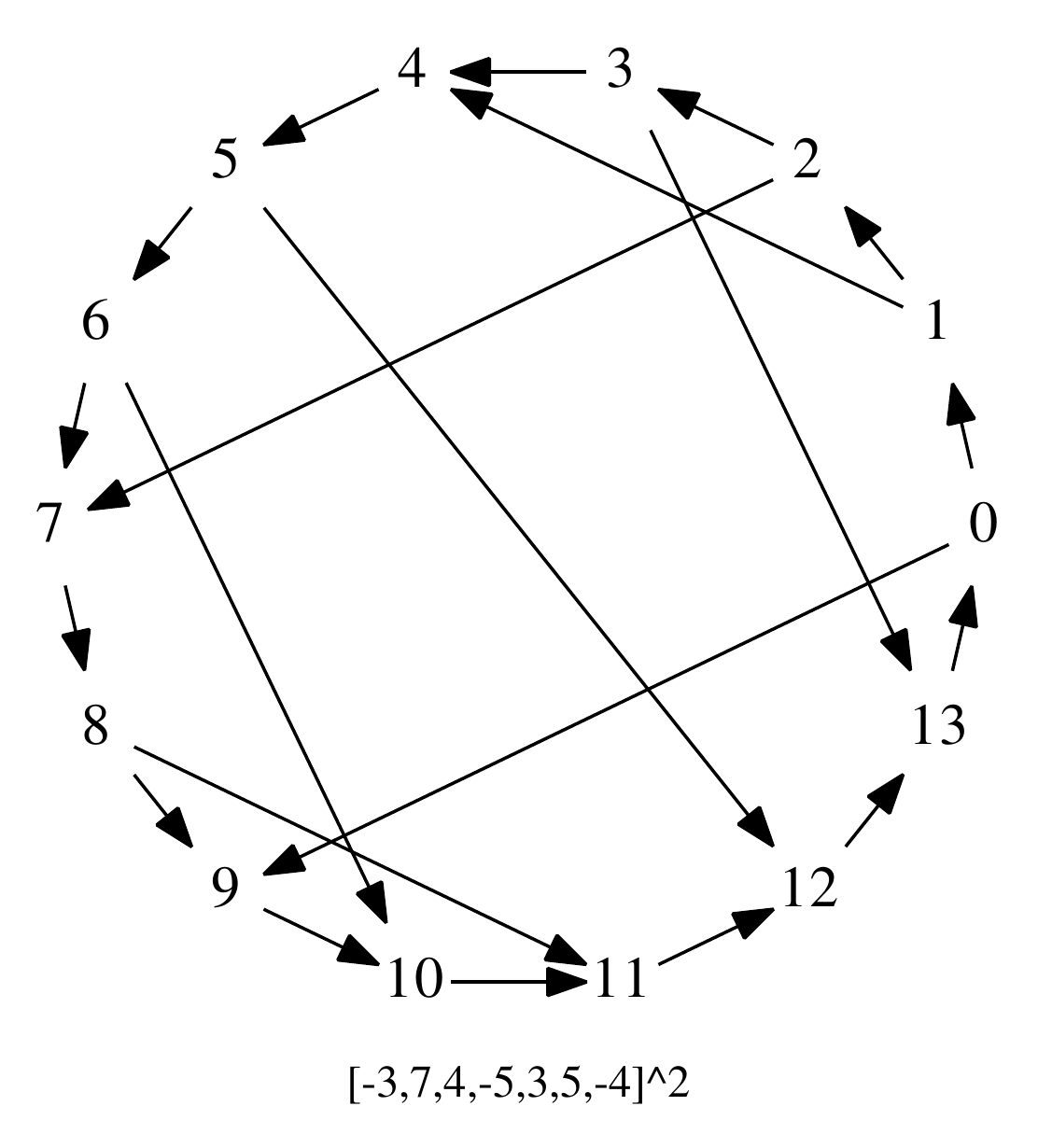}
\includegraphics[scale=0.45]{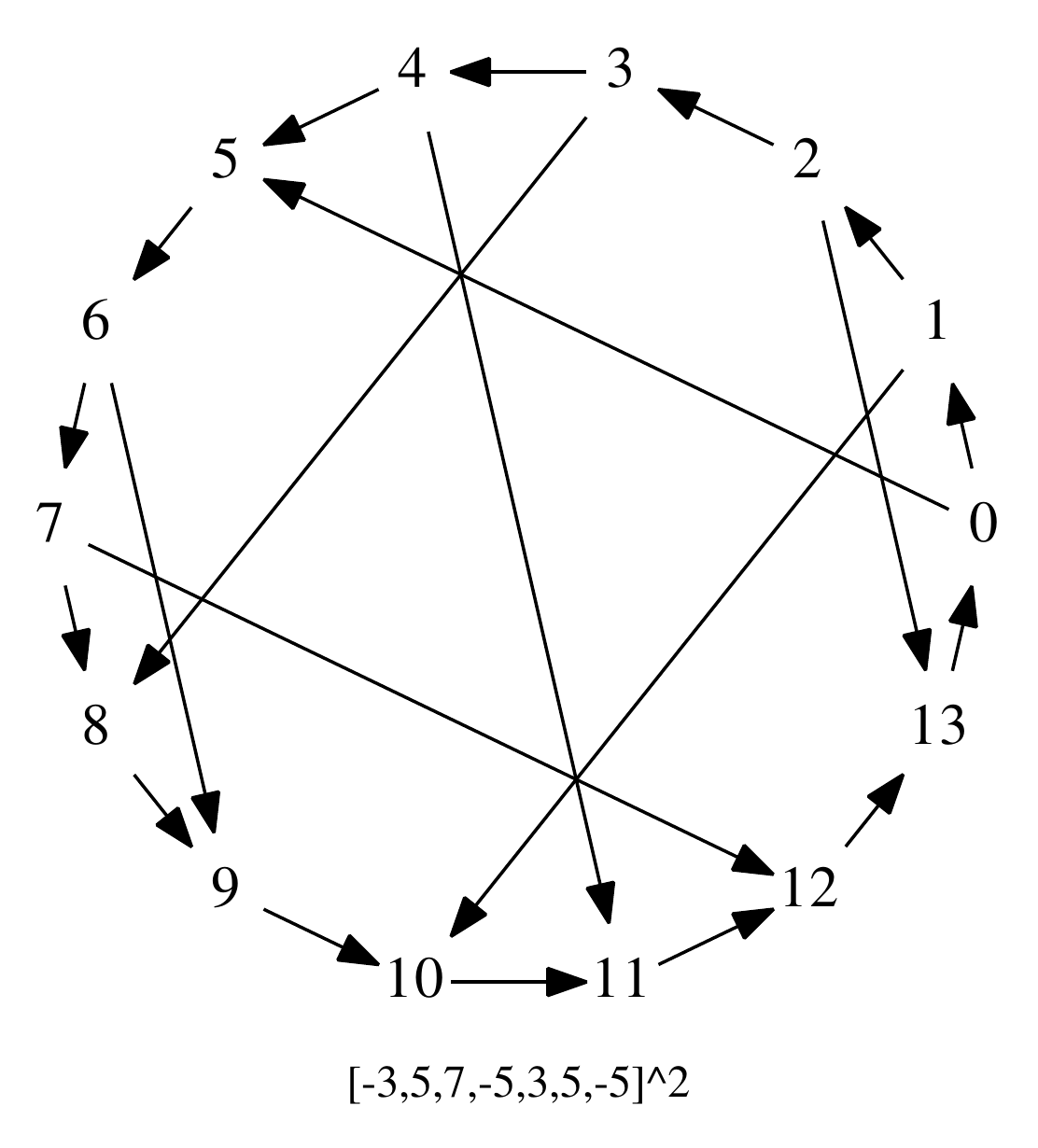}
\includegraphics[scale=0.45]{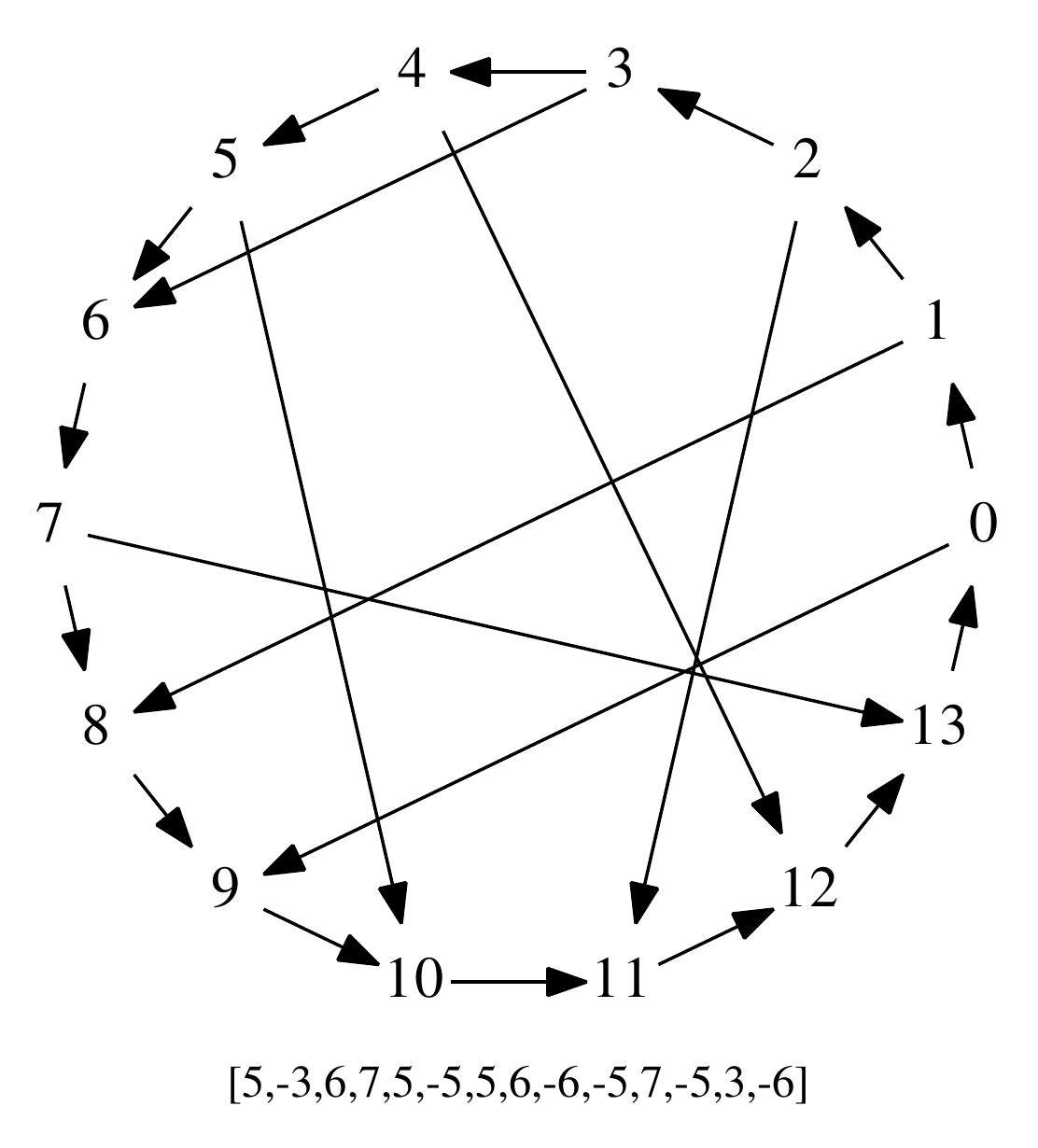}
\includegraphics[scale=0.45]{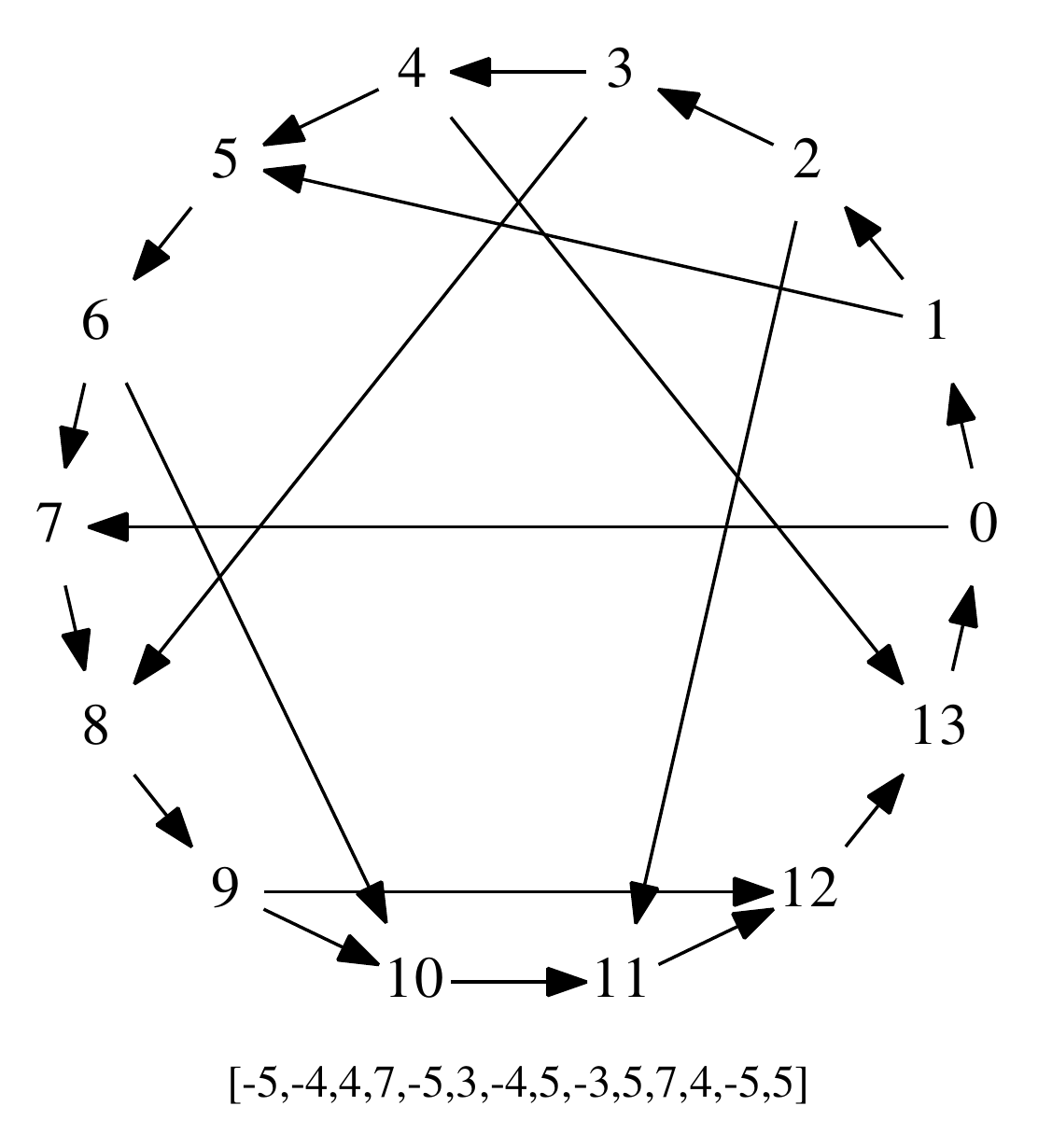}
\includegraphics[scale=0.45]{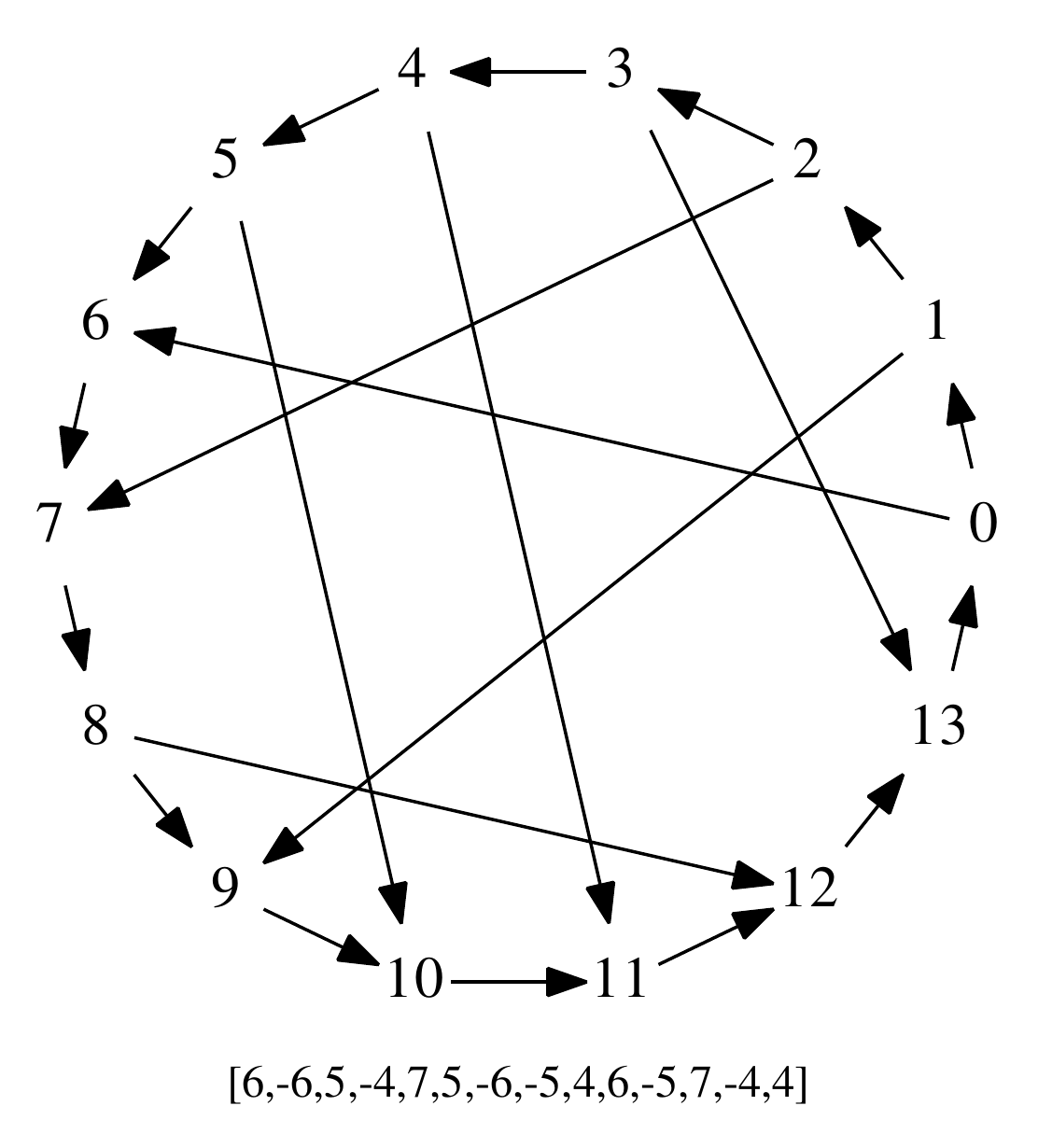}
\caption{Graphs on $n=14$ vertices which are irreducible (continued).
}
\label{fig.14n5}
\end{figure}

\begin{figure}
\includegraphics[scale=0.45]{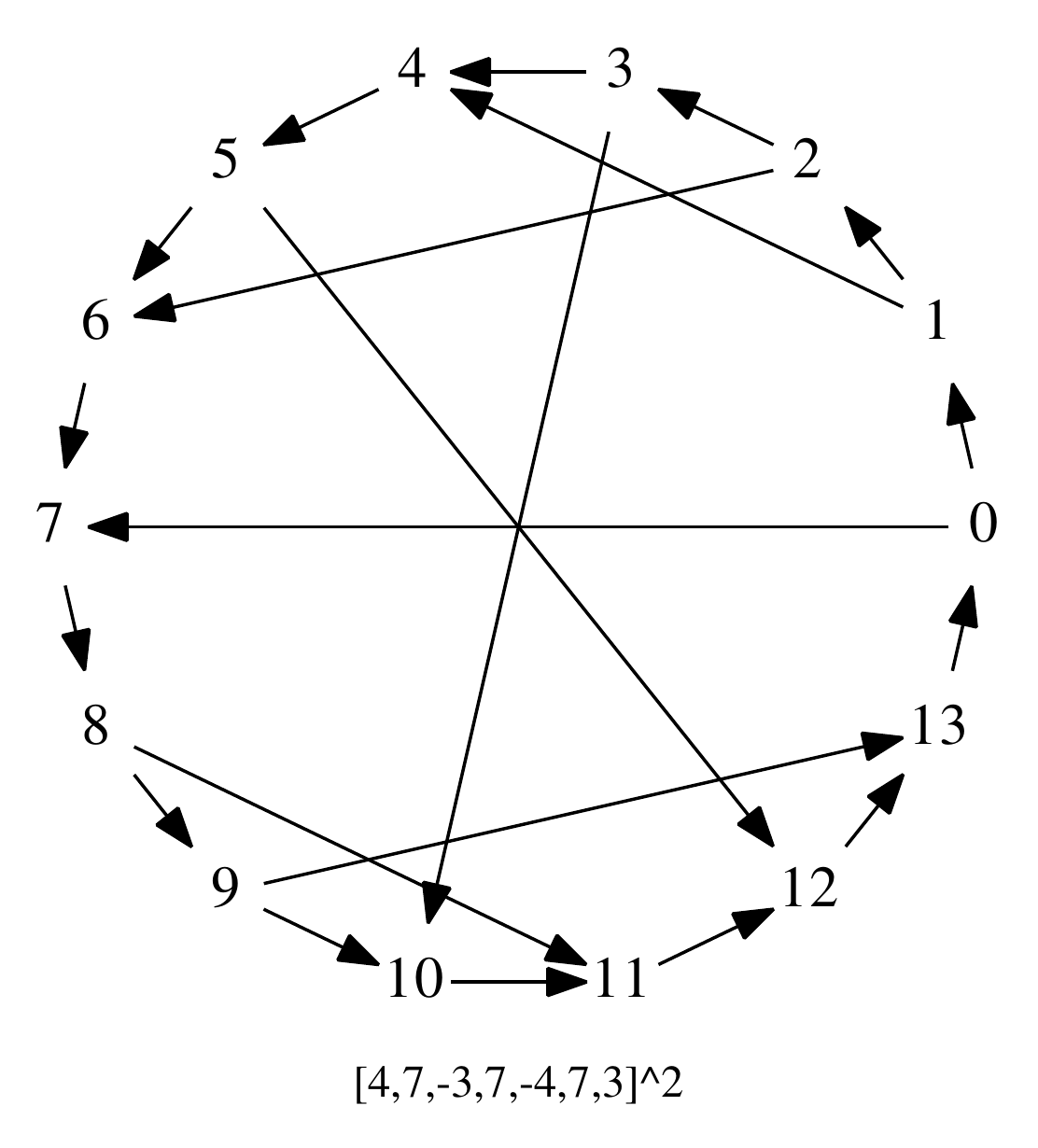}
\includegraphics[scale=0.45]{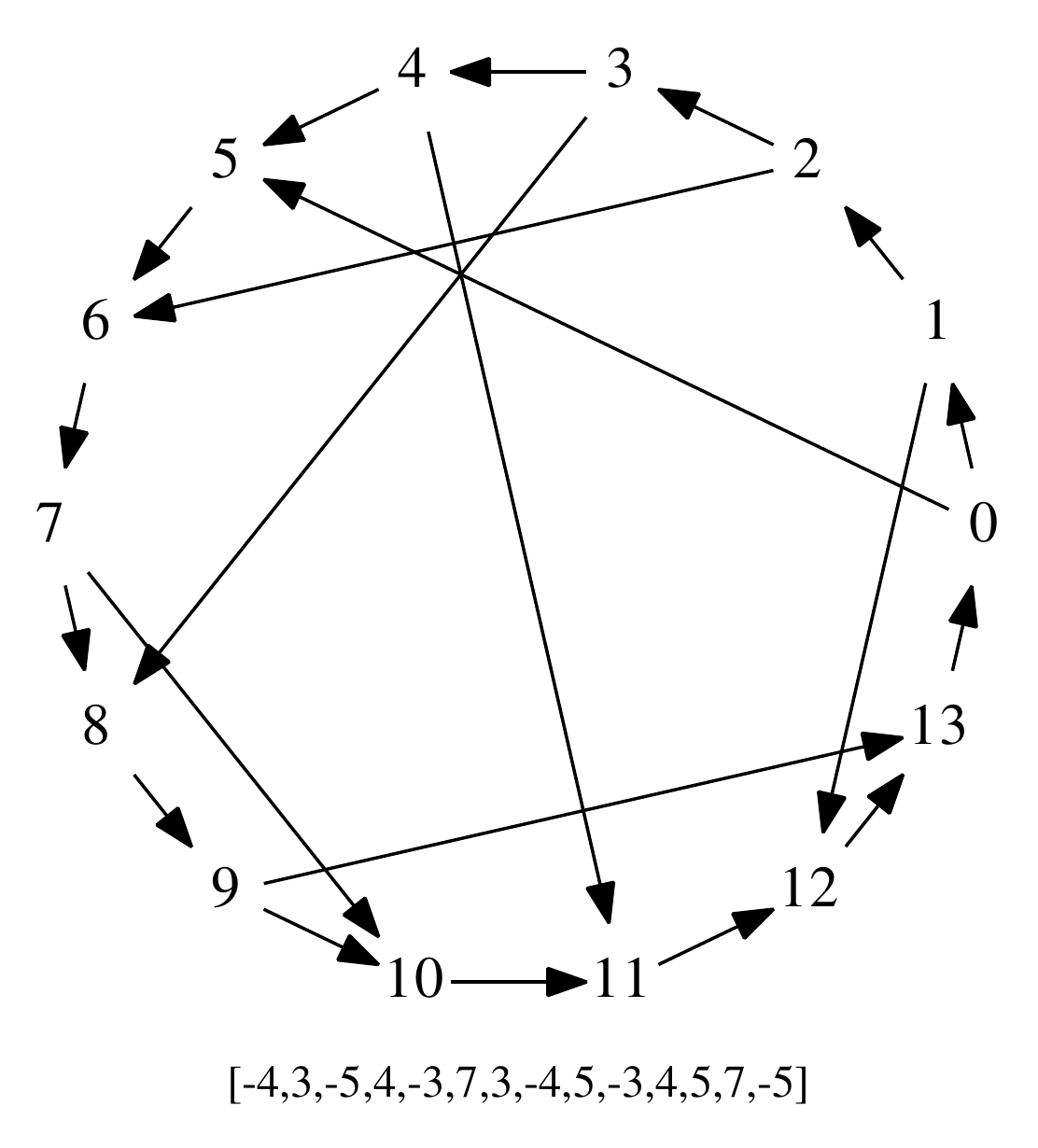}
\includegraphics[scale=0.45]{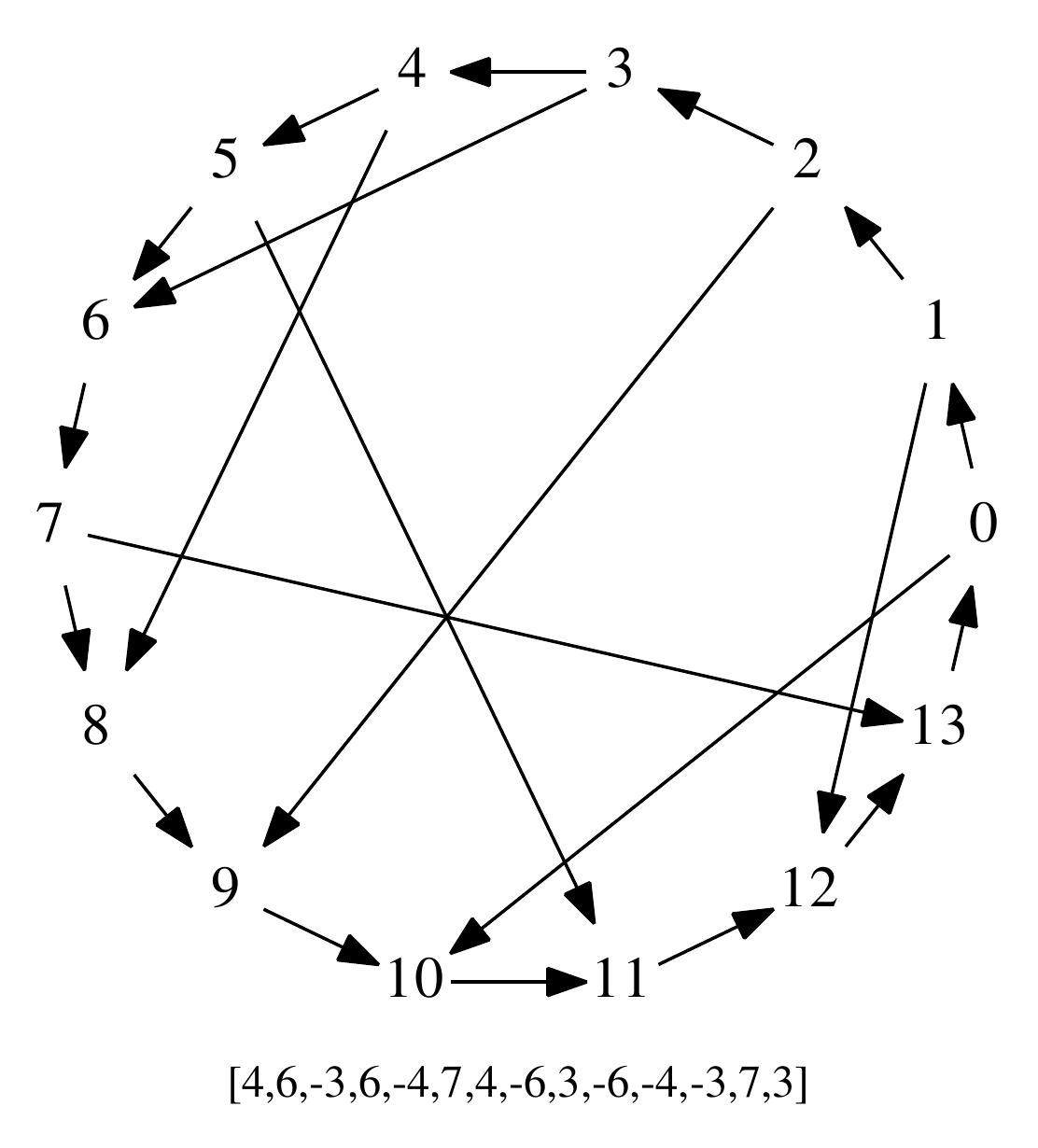}
\includegraphics[scale=0.45]{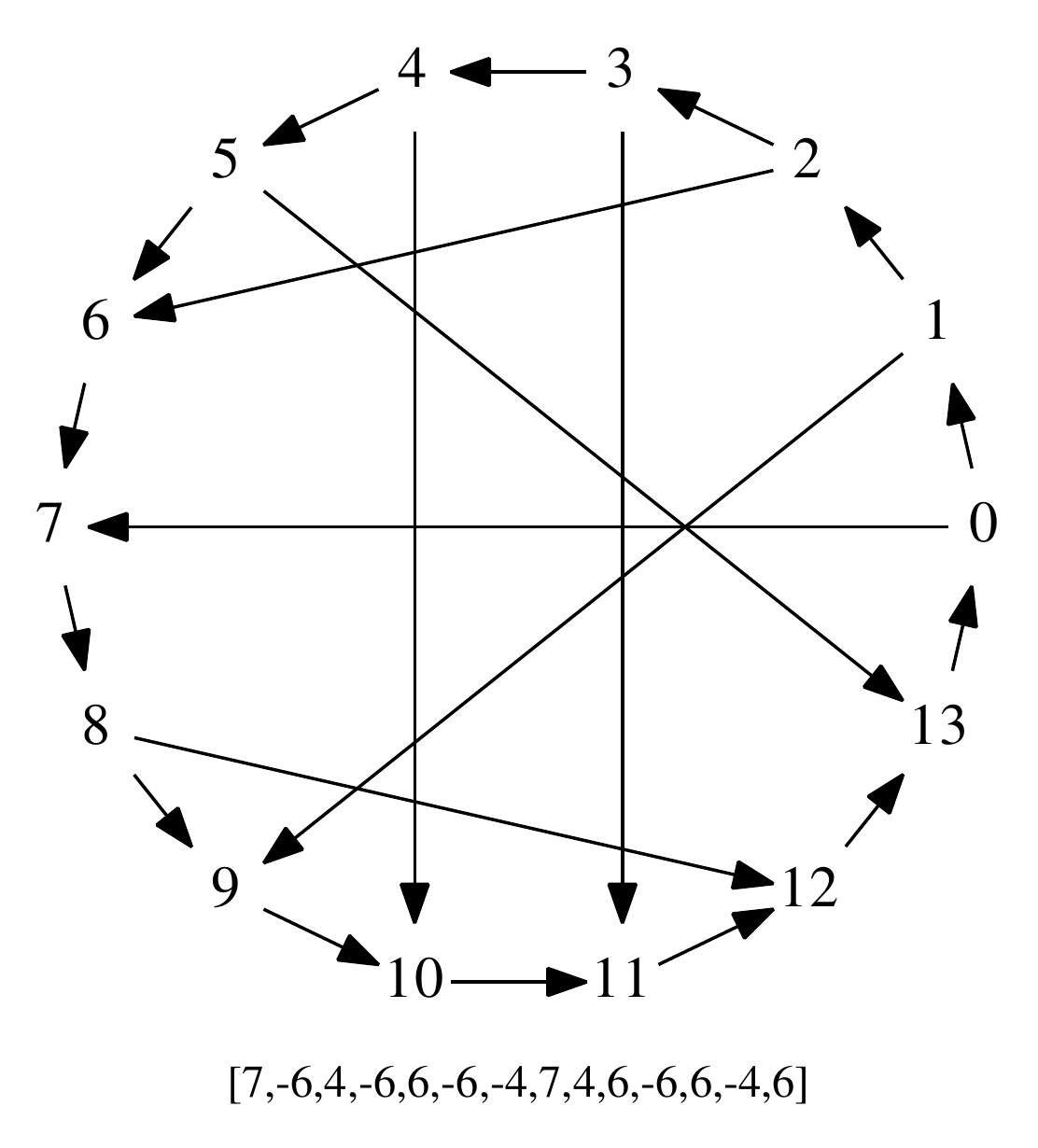}
\includegraphics[scale=0.45]{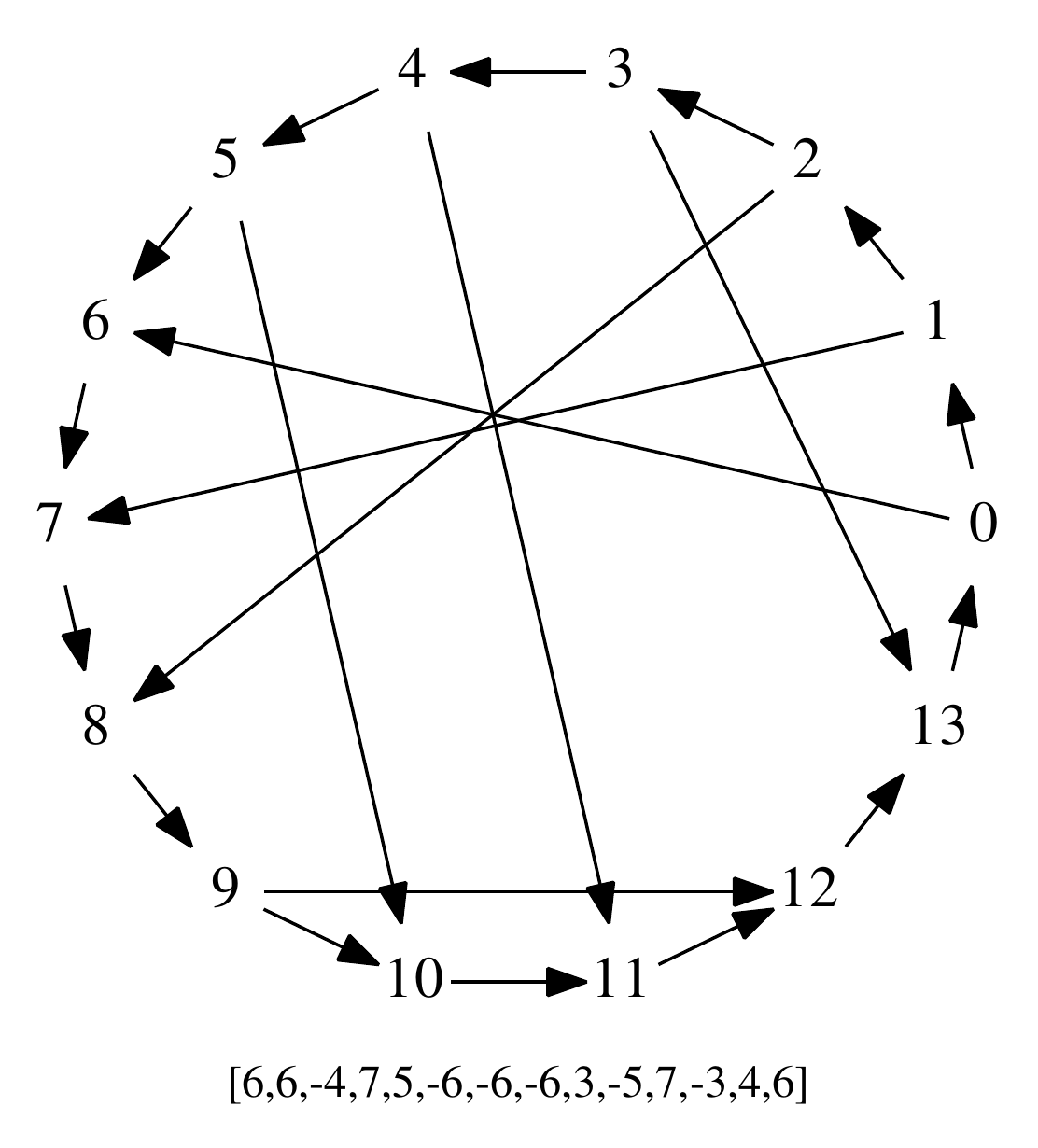}
\includegraphics[scale=0.45]{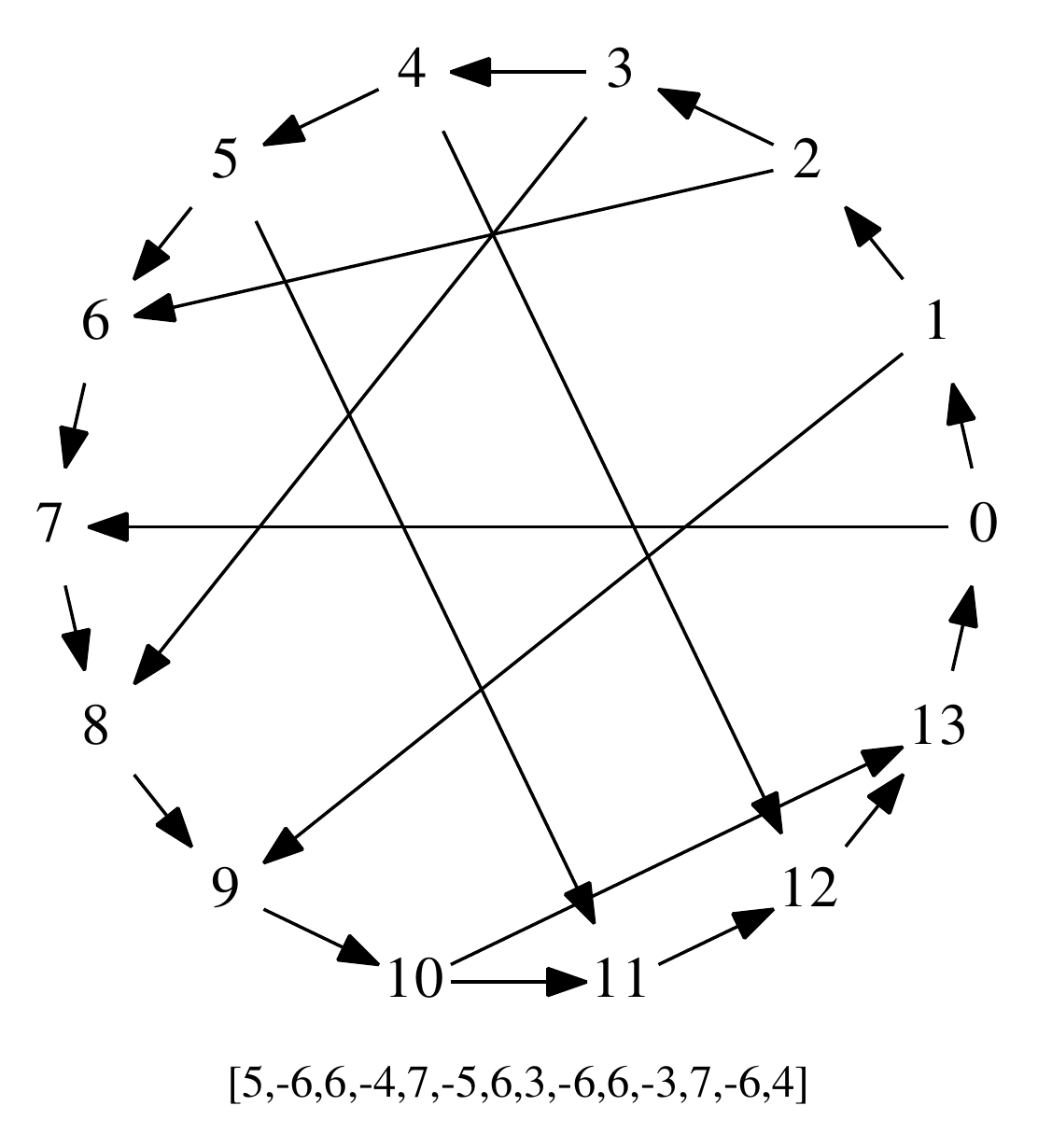}
\includegraphics[scale=0.45]{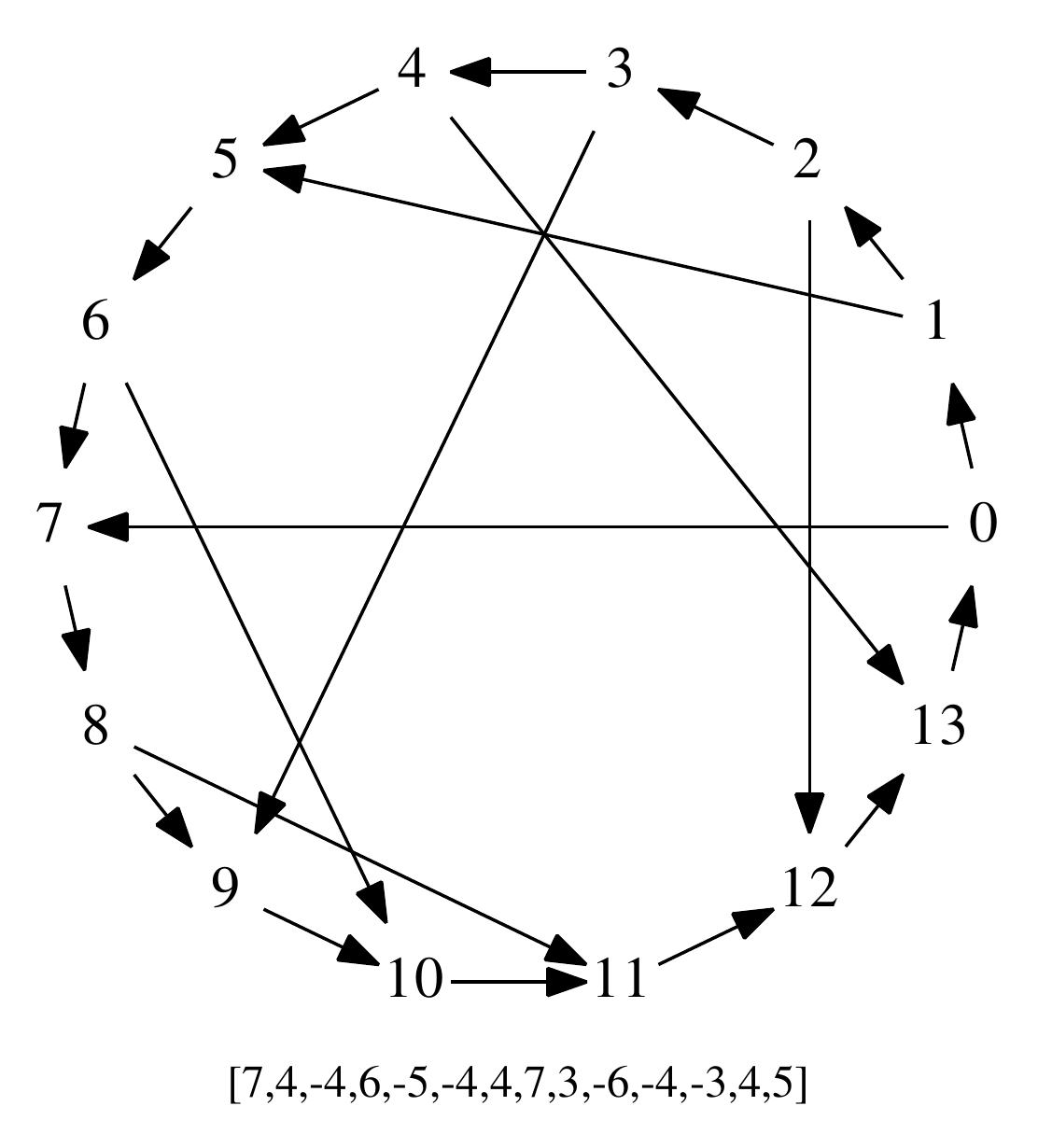}
\includegraphics[scale=0.45]{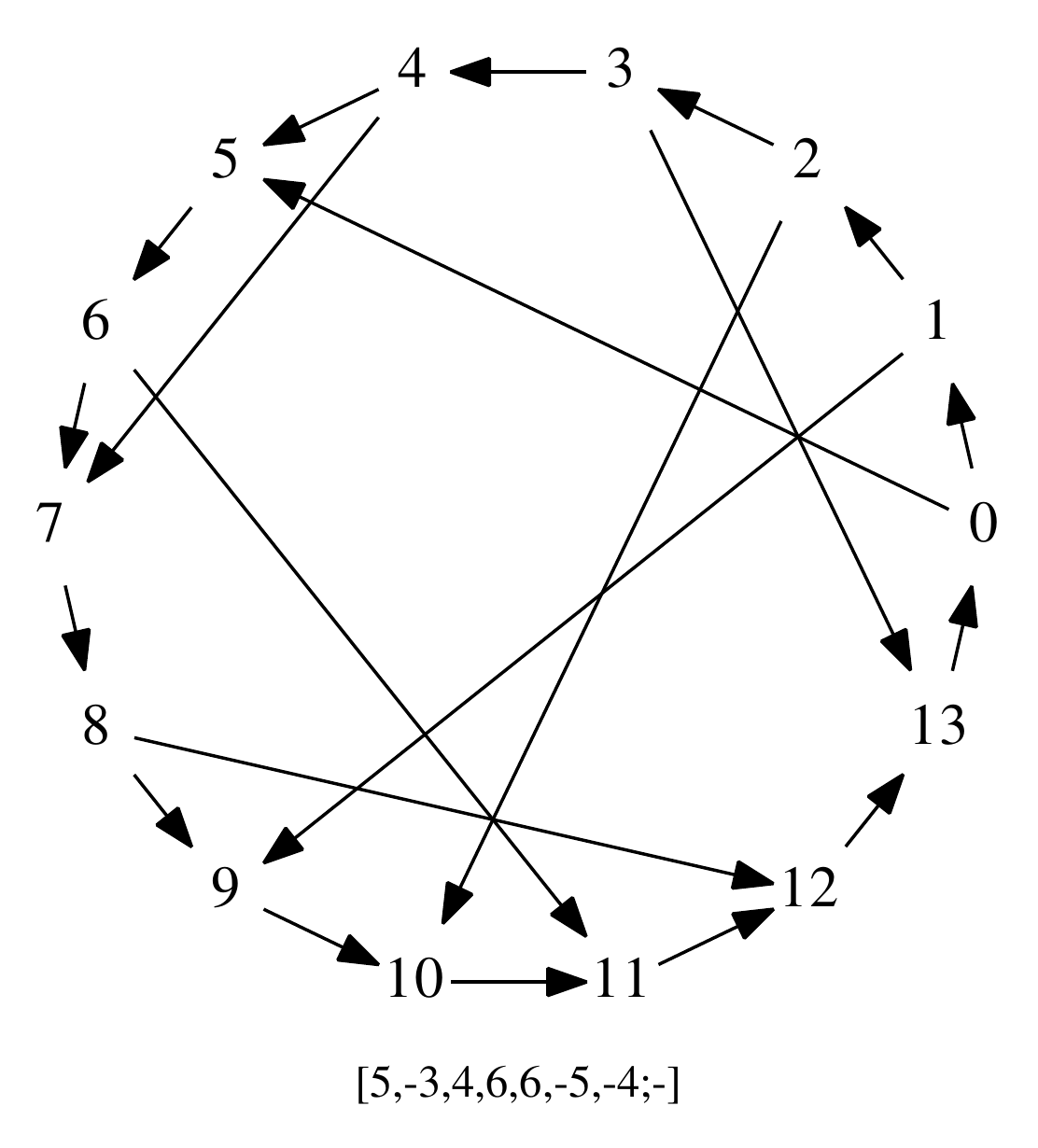}
\includegraphics[scale=0.45]{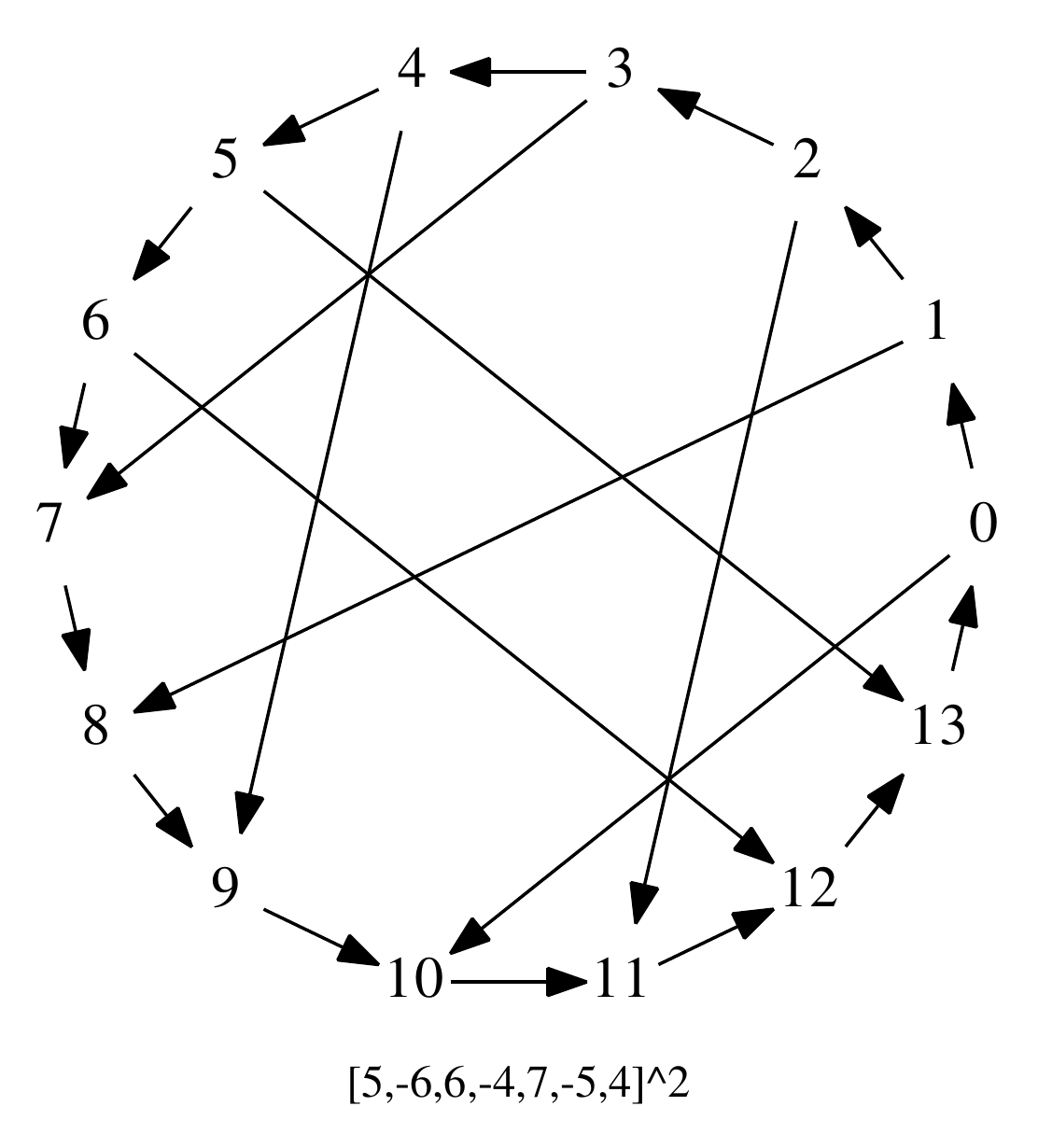}
\includegraphics[scale=0.45]{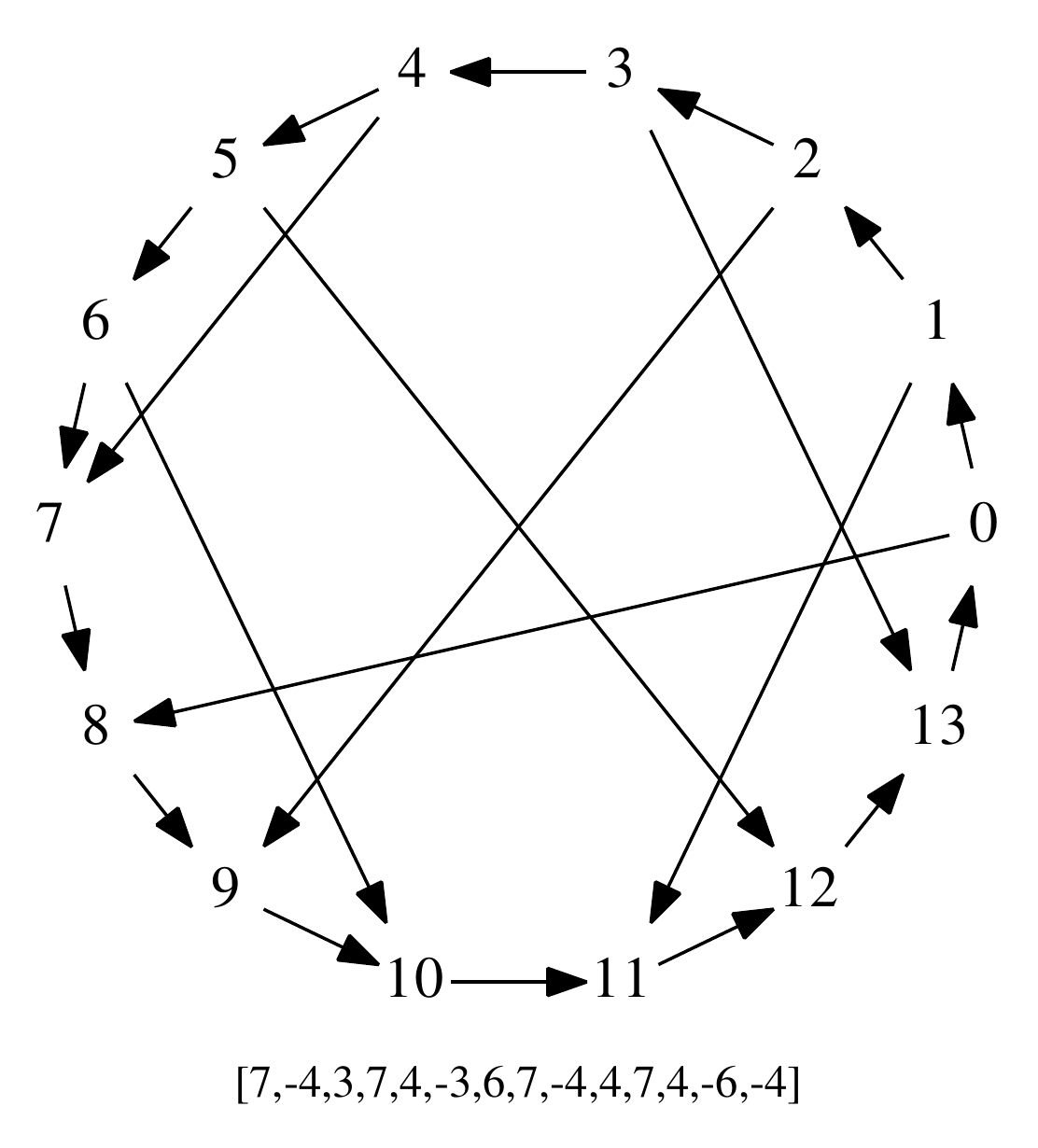}
\includegraphics[scale=0.45]{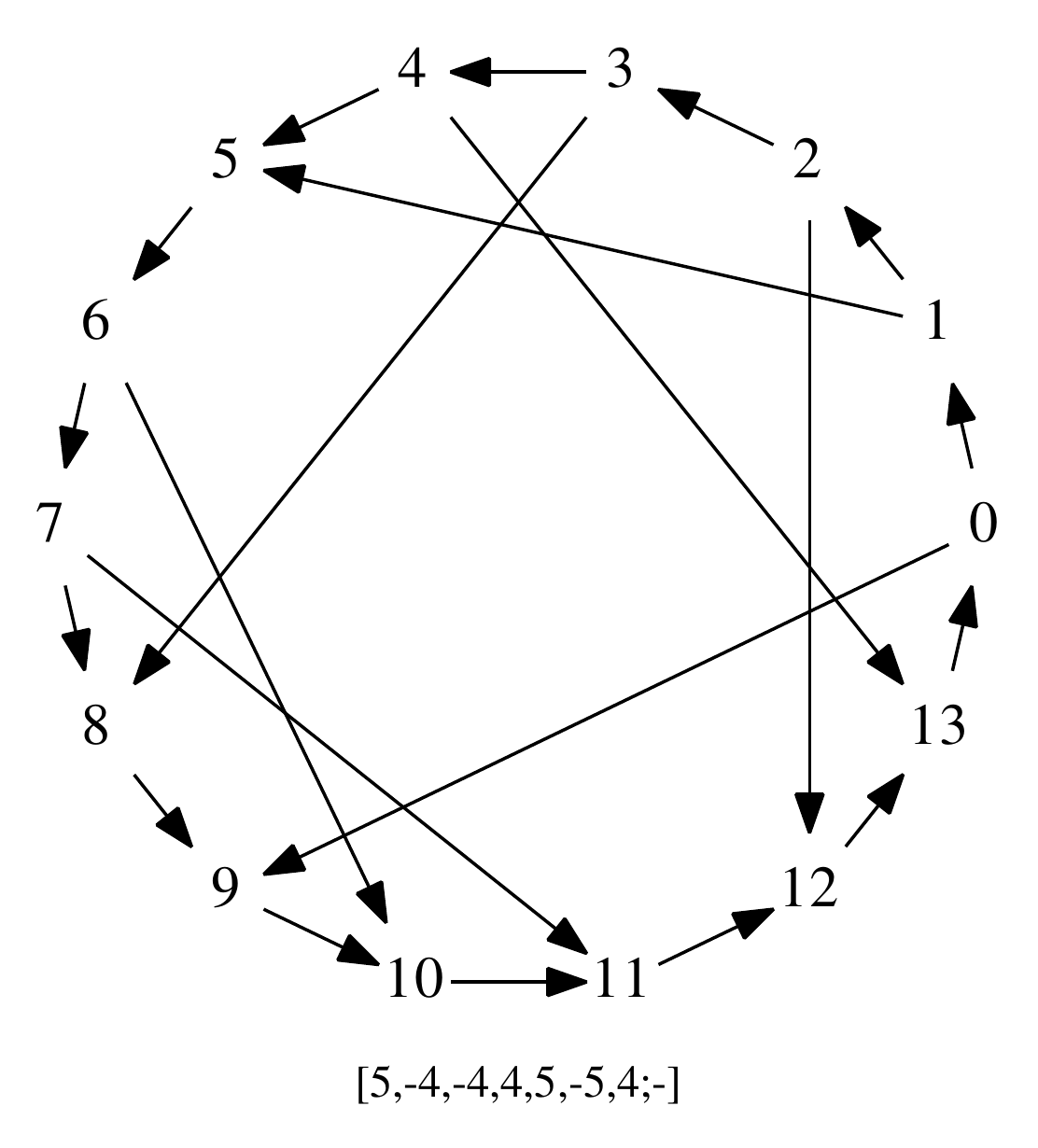}
\includegraphics[scale=0.45]{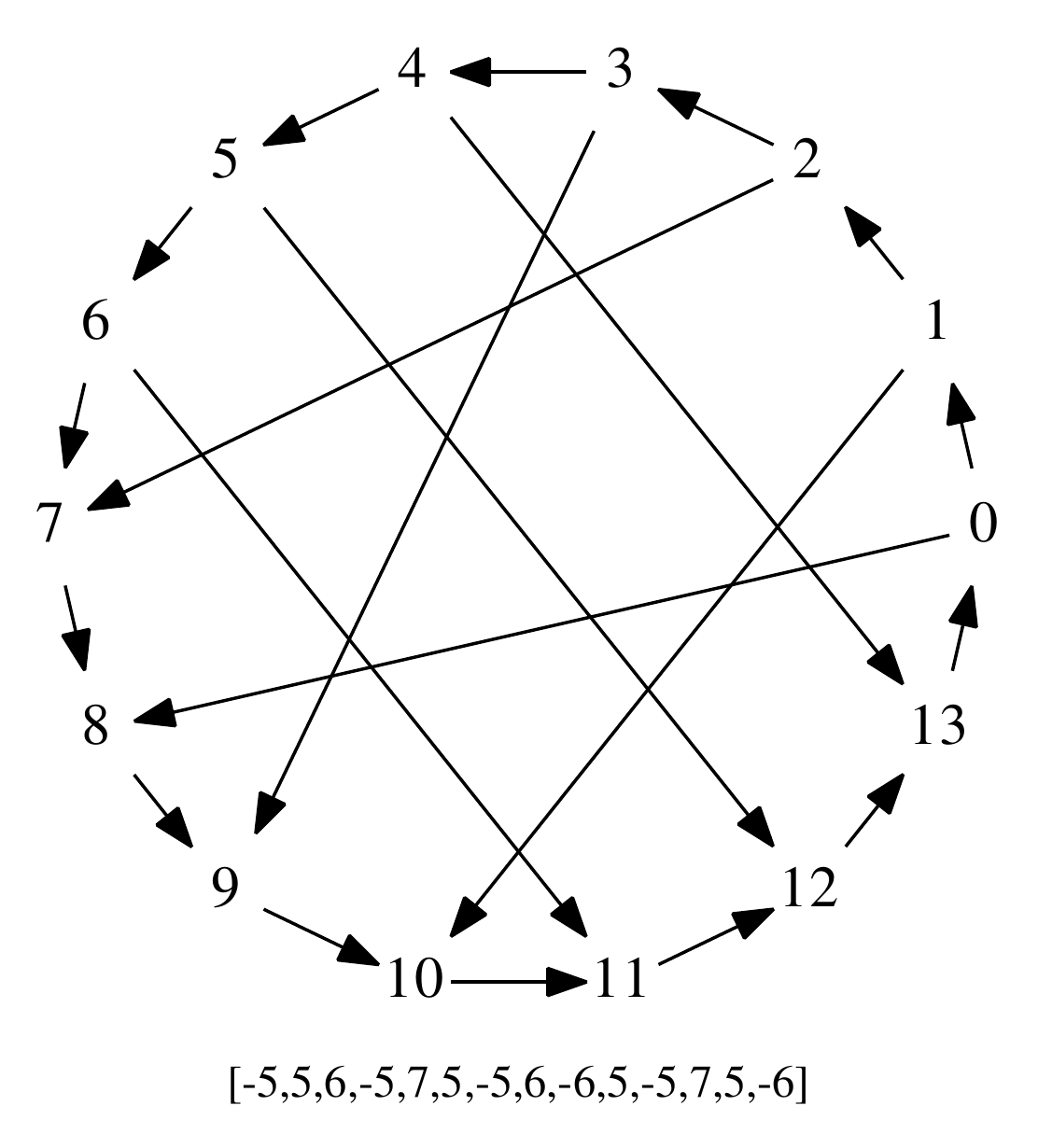}
\caption{Graphs on $n=14$ vertices which are irreducible (continued).
}
\label{fig.14n6}
\end{figure}

\begin{figure}
\includegraphics[scale=0.45]{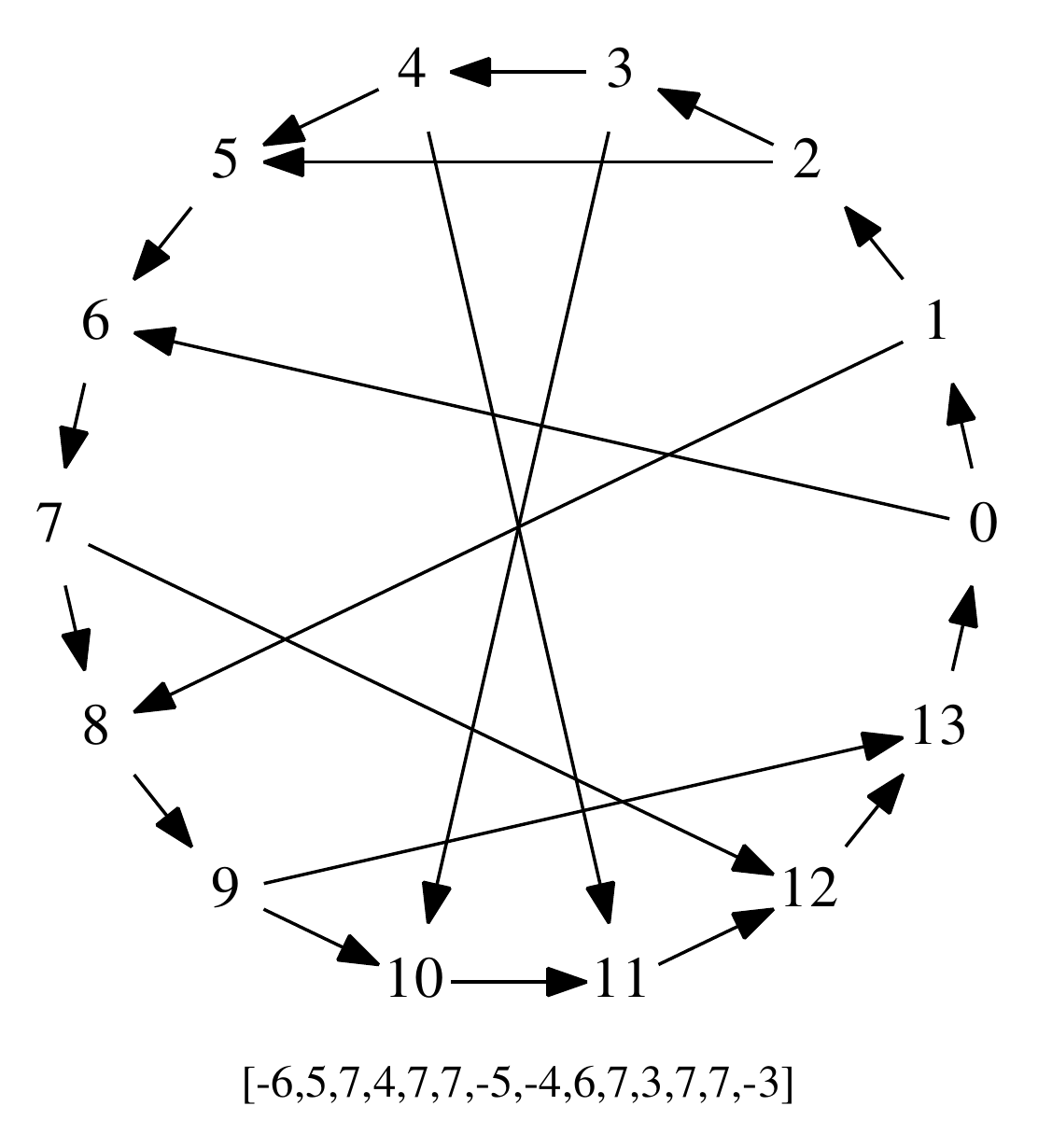}
\includegraphics[scale=0.45]{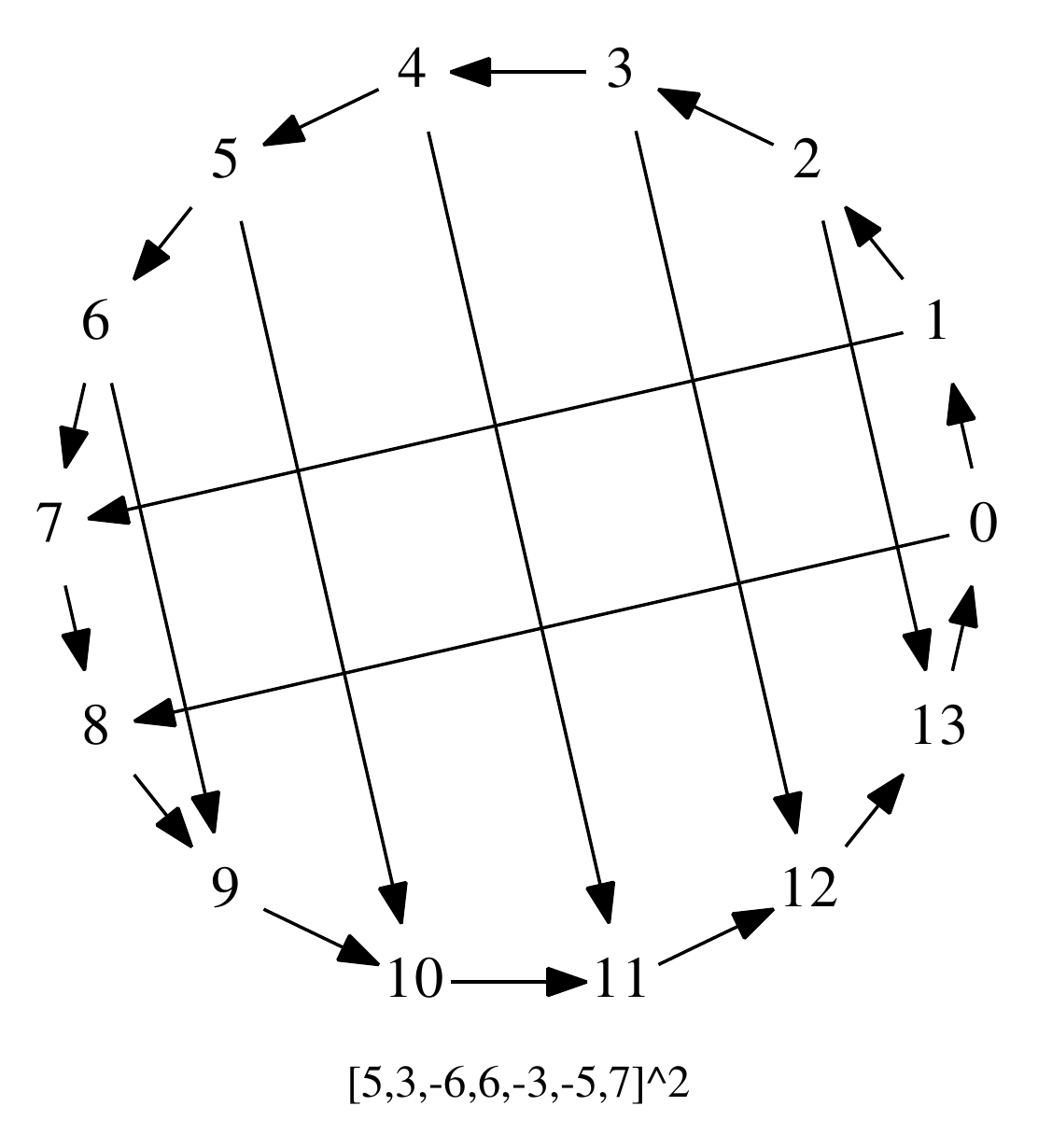}
\includegraphics[scale=0.45]{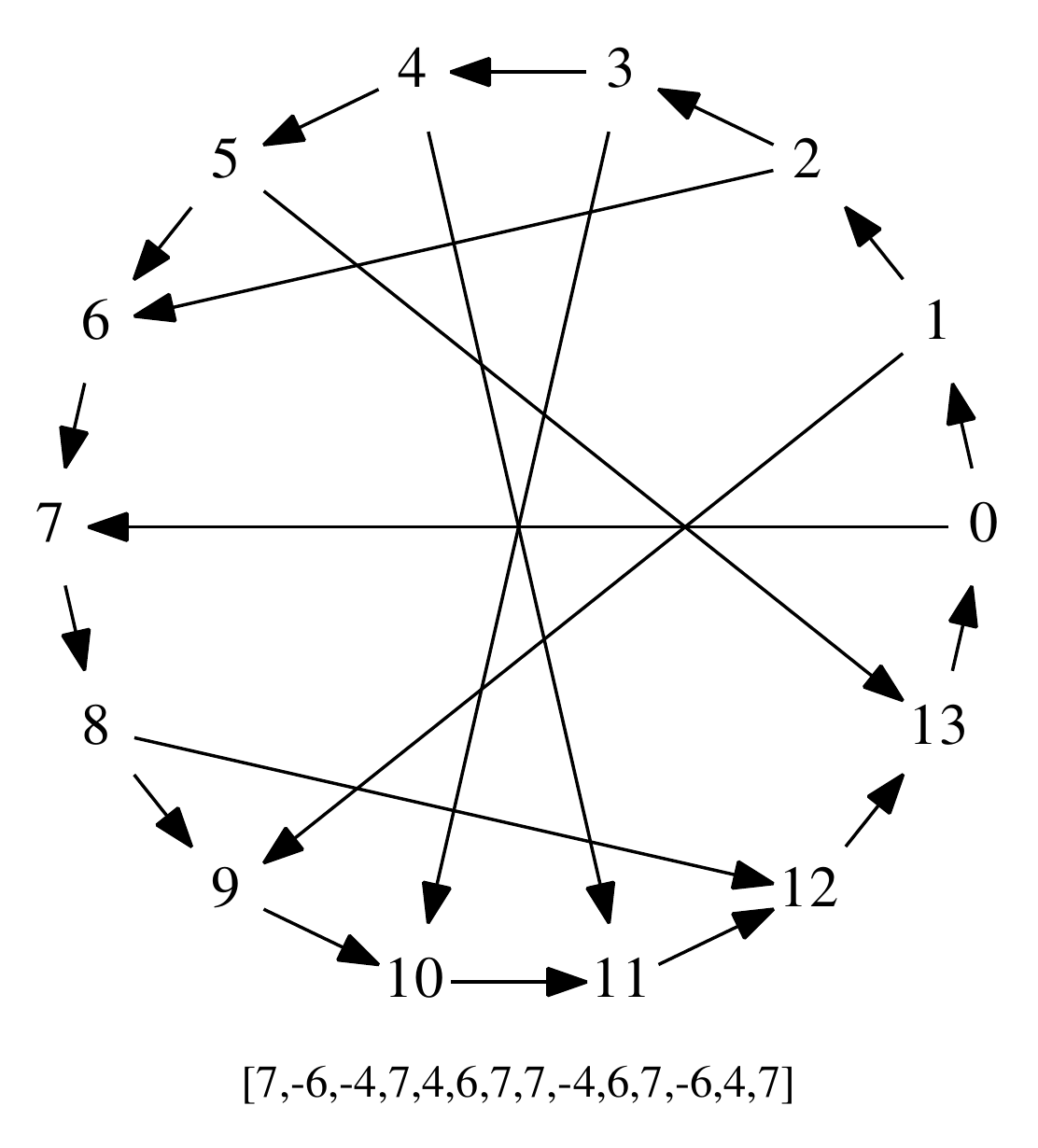}
\includegraphics[scale=0.45]{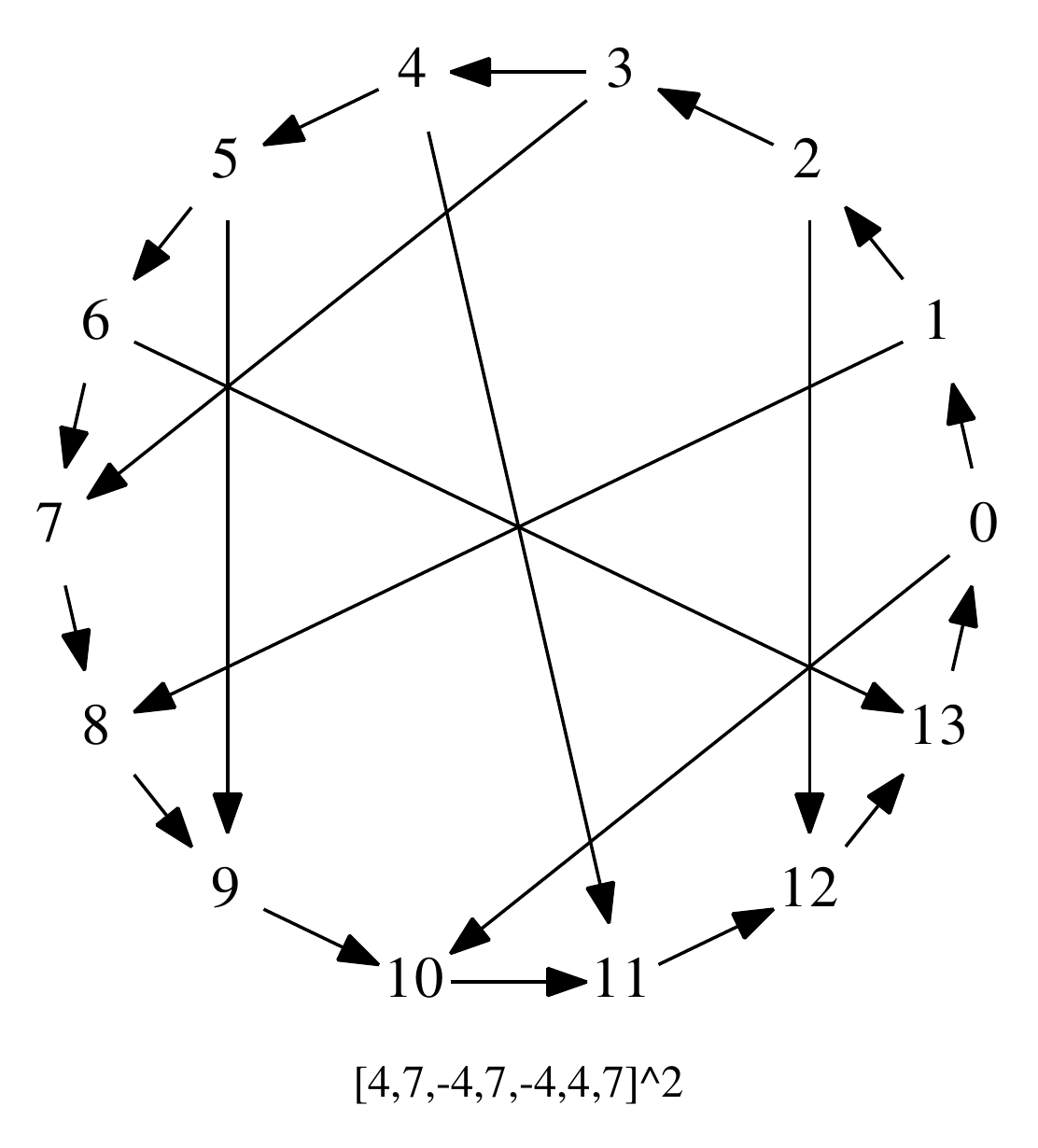}
\includegraphics[scale=0.45]{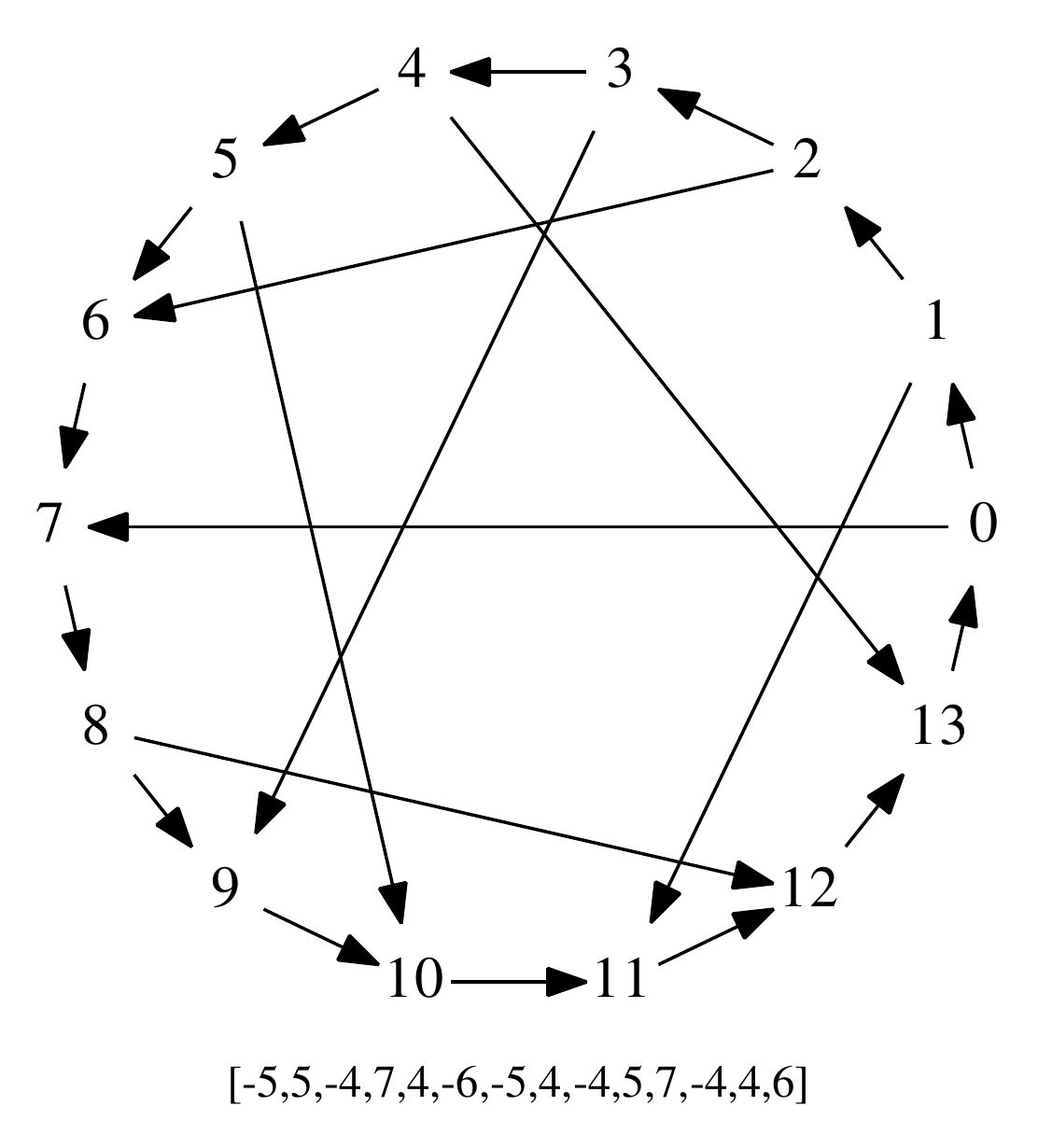}
\includegraphics[scale=0.45]{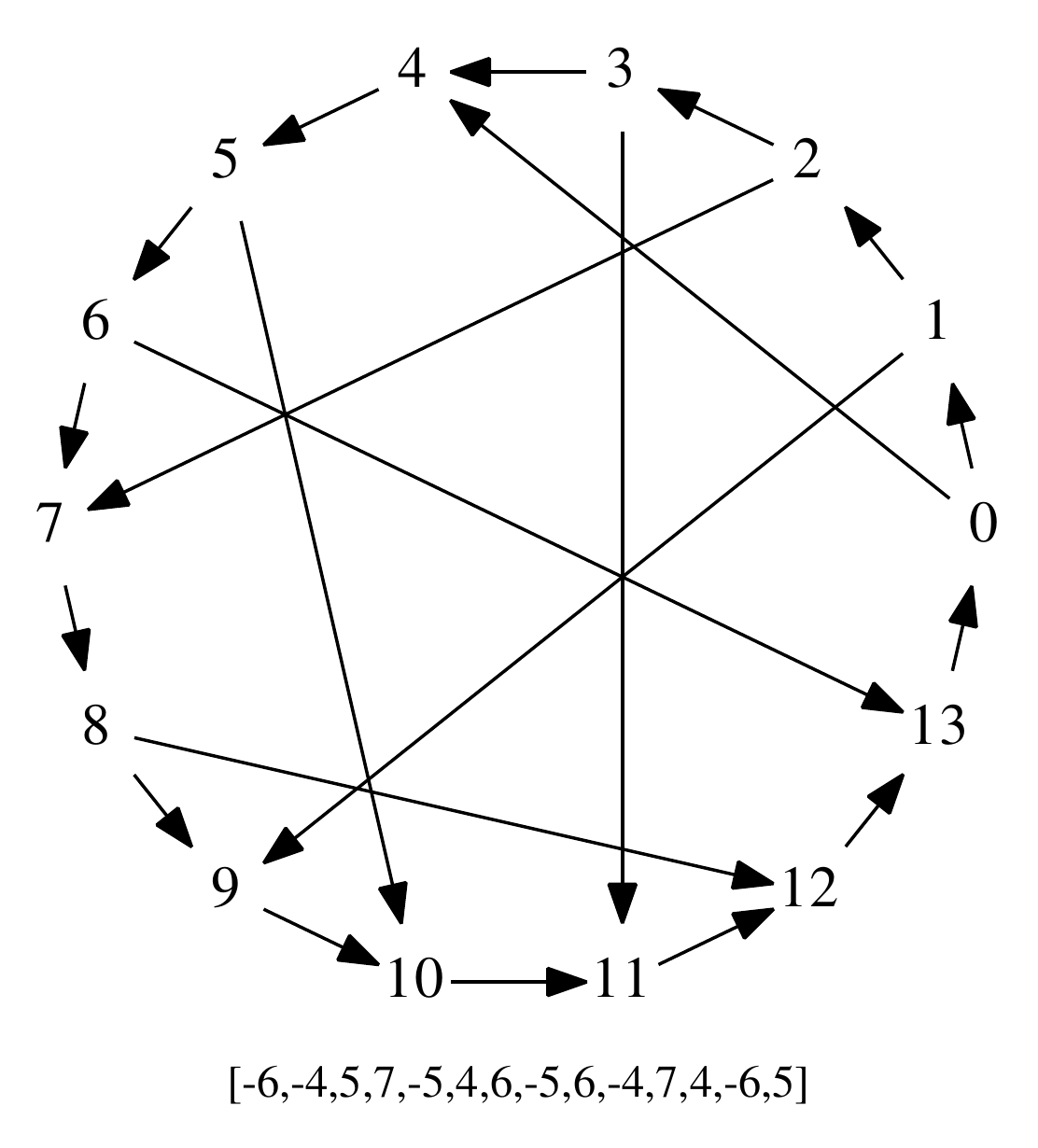}
\includegraphics[scale=0.45]{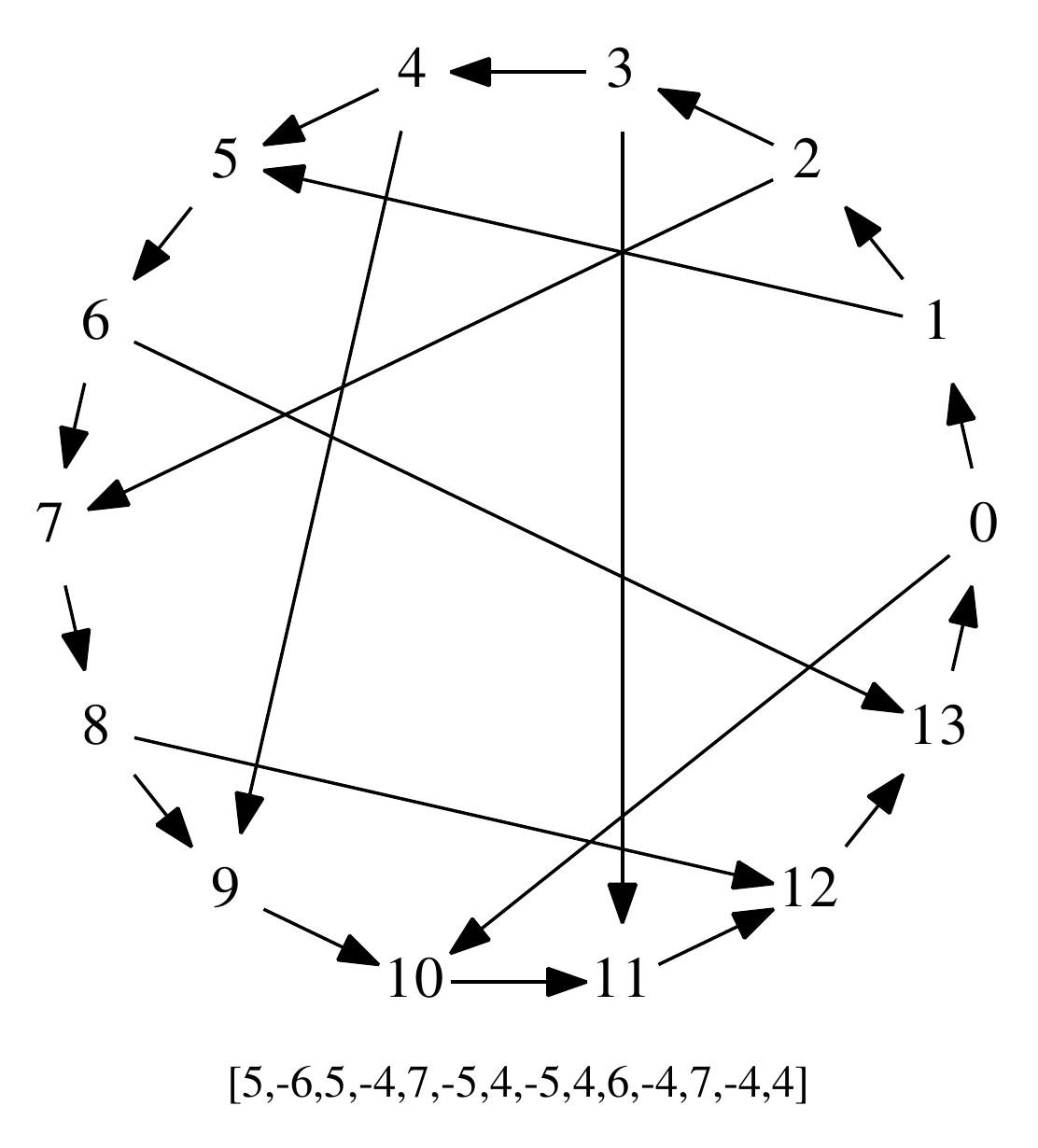}
\includegraphics[scale=0.45]{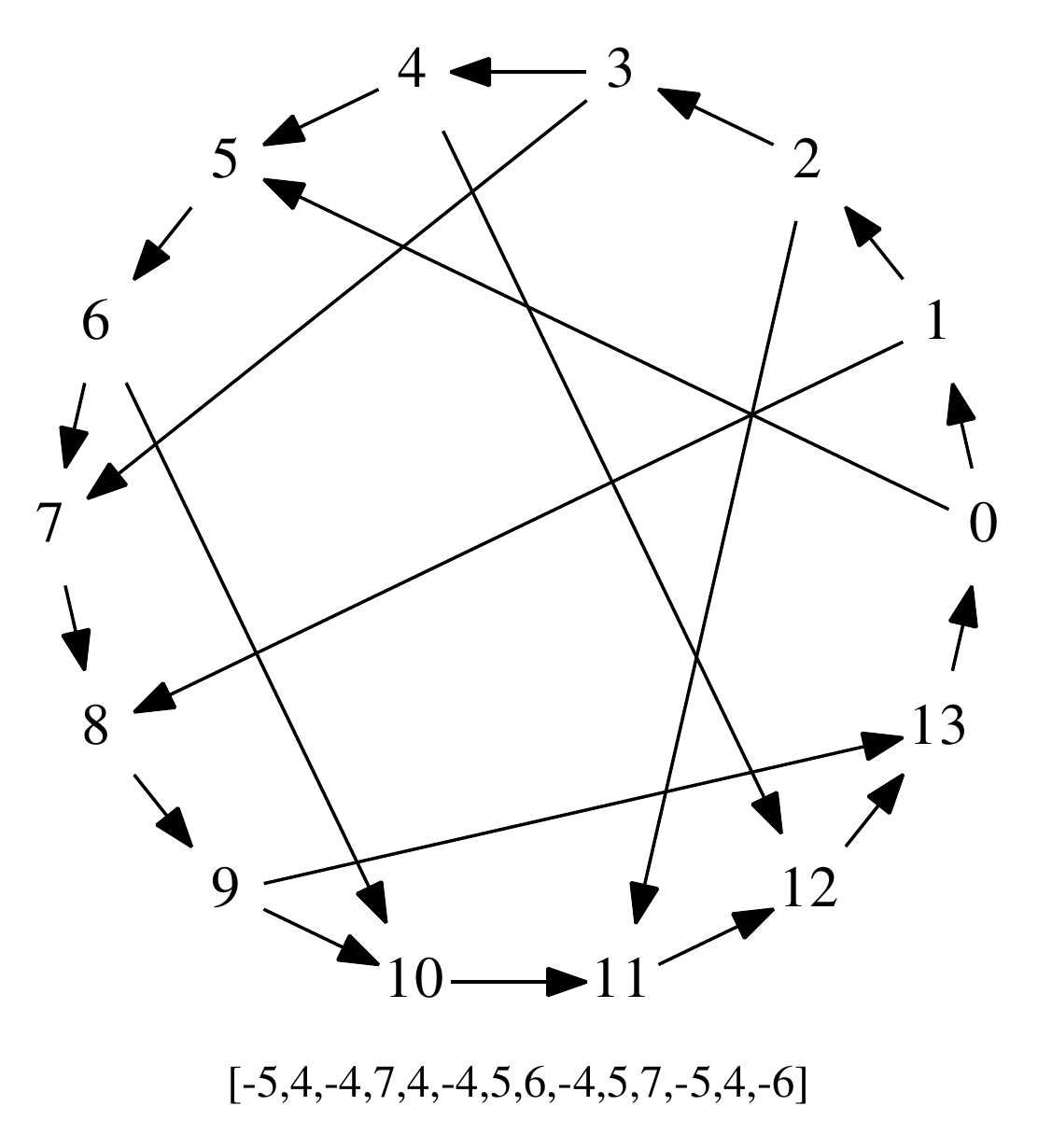}
\includegraphics[scale=0.45]{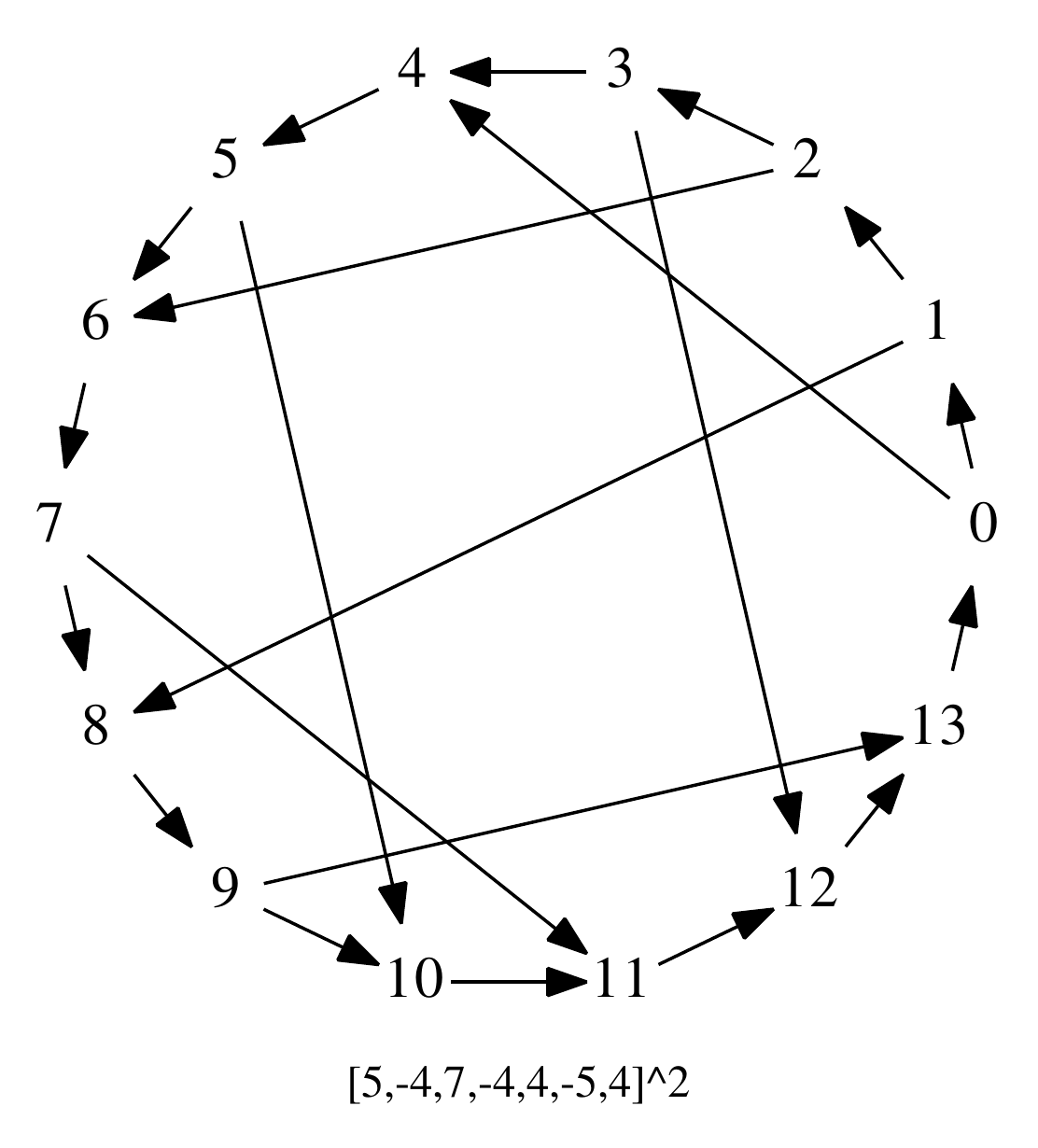}
\includegraphics[scale=0.45]{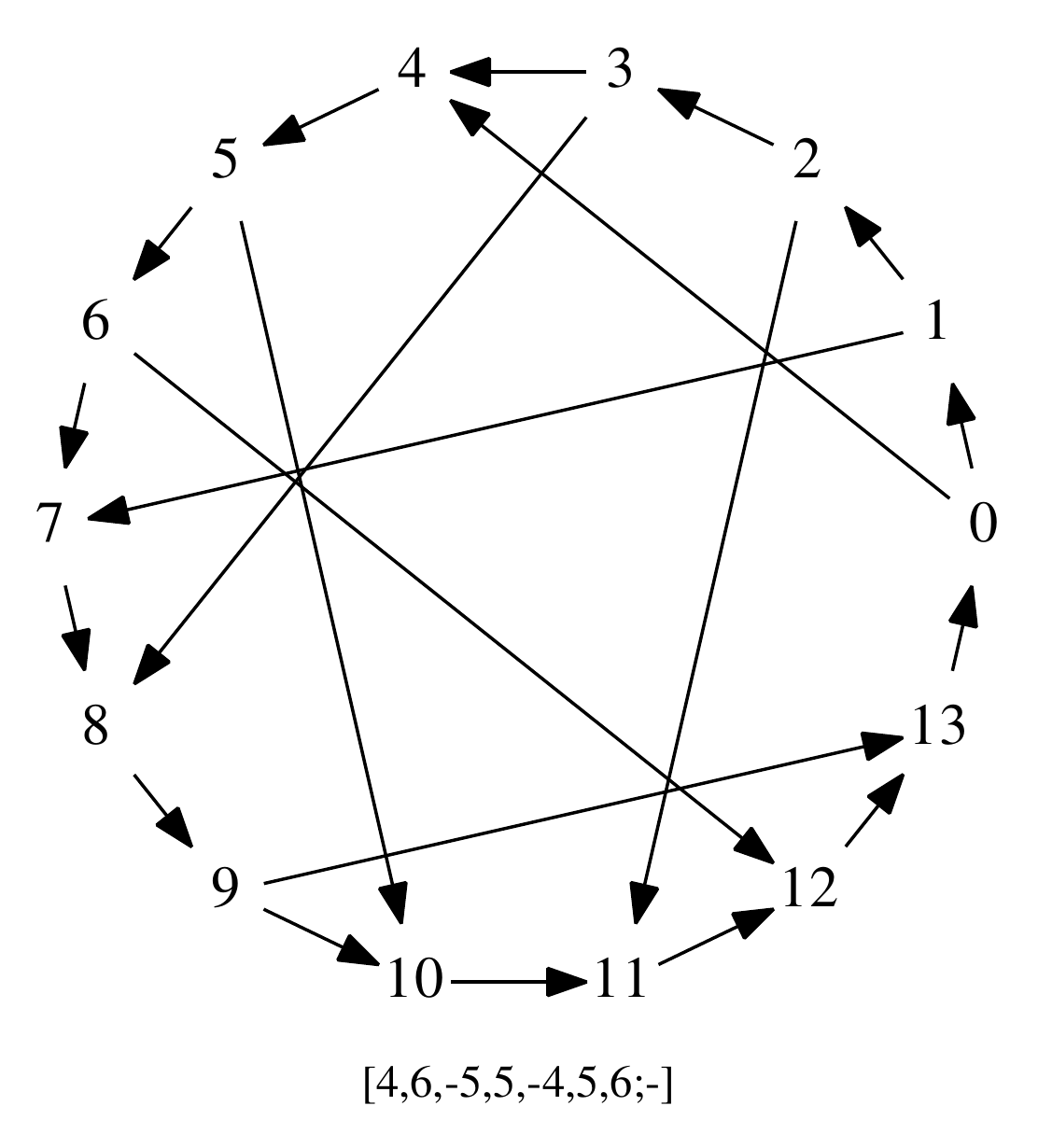}
\includegraphics[scale=0.45]{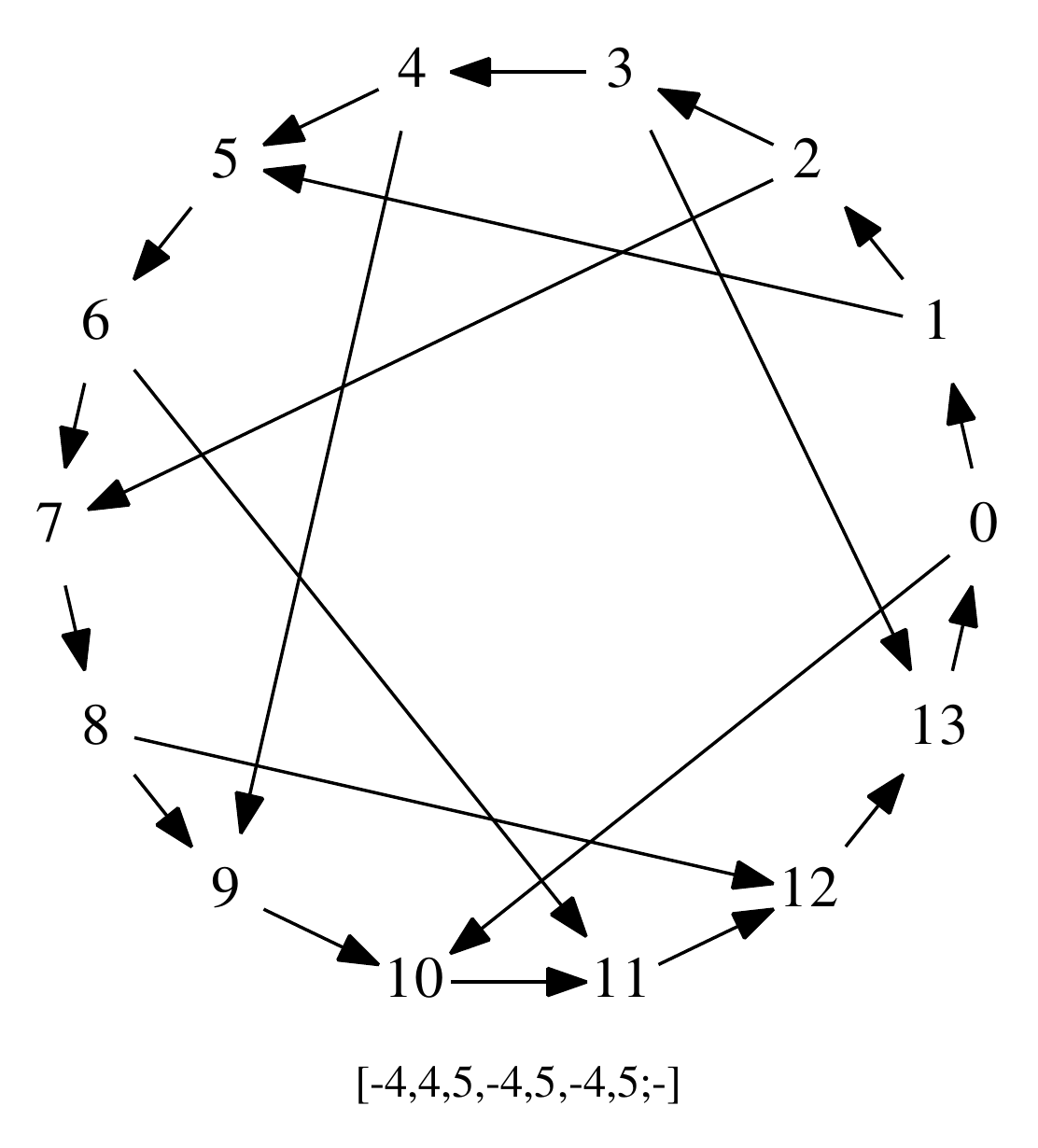}
\includegraphics[scale=0.45]{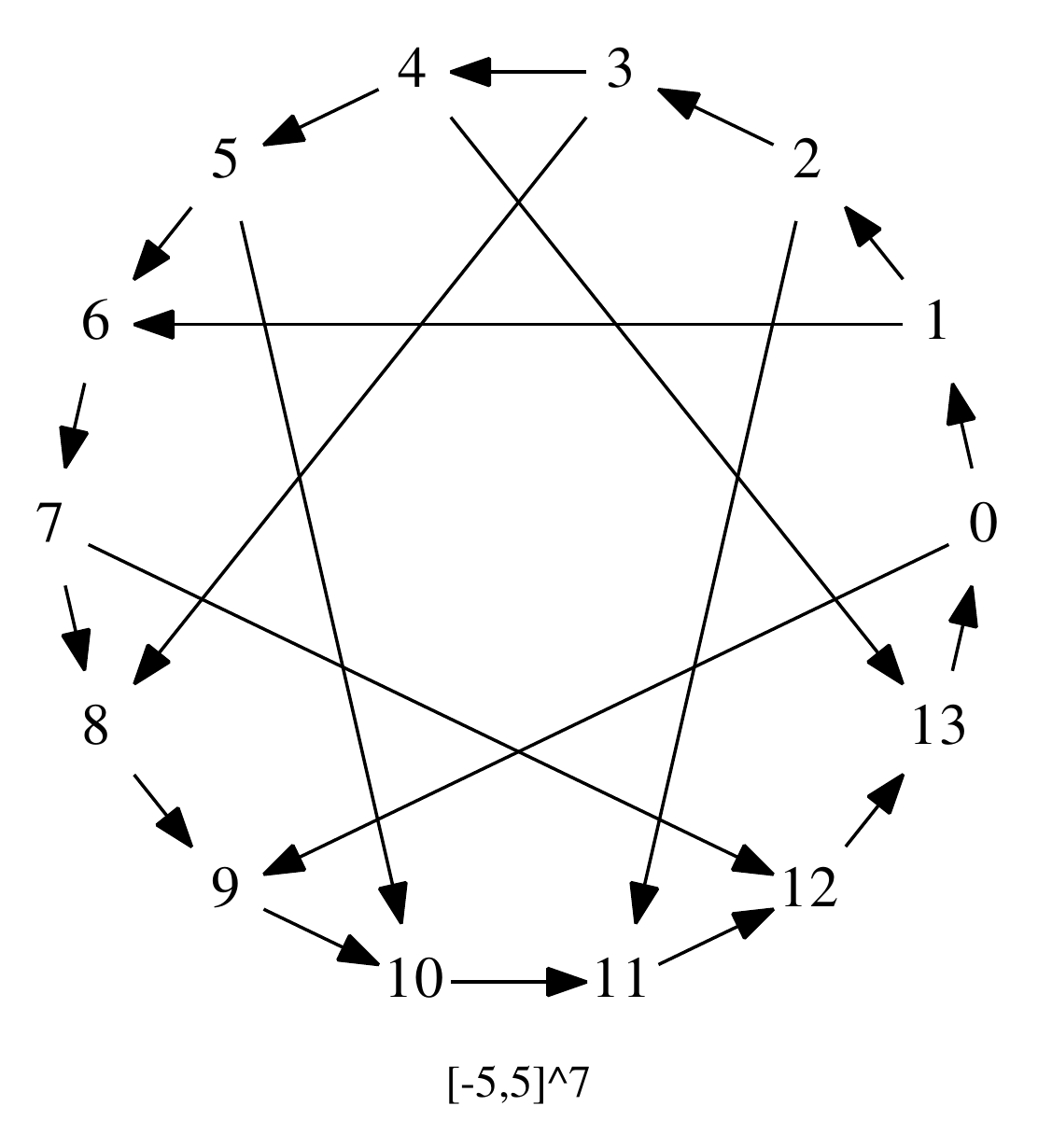}
\caption{Graphs on $n=14$ vertices which are irreducible (end).
}
\label{fig.14ne}
\end{figure}
\clearpage

\end{widetext}
\section{Summary}
We have plotted the non-isomorphic simple cubic graphs up to 12 vertices ($18j$-symbols)
plus the subset on 14 vertices that defines classes of $21j$-symbols.
Hamiltonian cycles have been identified. The associated LCF notation introduces
a convenient ordering representation which combats the bewildering
variety of planar graphical representations as the number of edges
becomes large.

\acknowledgments
The graphs were generated with Meringer's program \texttt{genreg} \cite{MeringerJGT30}
and have been plotted with the
\texttt{neato}
program of the \texttt{graphviz}
package.

\bibliographystyle{apsrmp}
\bibliography{all}

\begin{thebibliography}{45}
\expandafter\ifx\csname natexlab\endcsname\relax\def\natexlab#1{#1}\fi
\expandafter\ifx\csname bibnamefont\endcsname\relax
  \def\bibnamefont#1{#1}\fi
\expandafter\ifx\csname bibfnamefont\endcsname\relax
  \def\bibfnamefont#1{#1}\fi
\expandafter\ifx\csname citenamefont\endcsname\relax
  \def\citenamefont#1{#1}\fi
\expandafter\ifx\csname url\endcsname\relax
  \def\url#1{\texttt{#1}}\fi
\expandafter\ifx\csname urlprefix\endcsname\relax\def\urlprefix{URL }\fi
\providecommand{\bibinfo}[2]{#2}
\providecommand{\eprint}[2][]{\url{#2}}

\bibitem[{\citenamefont{Aldred} \emph{et~al.}(2009)\citenamefont{Aldred, {Van
  Dyck}, Brinkmann, Fack, and McKay}}]{AldredDAM157}
\bibinfo{author}{\bibnamefont{Aldred}, \bibfnamefont{R.~E.~L.}},
  \bibinfo{author}{\bibfnamefont{D.}~\bibnamefont{{Van Dyck}}},
  \bibinfo{author}{\bibfnamefont{G.}~\bibnamefont{Brinkmann}},
  \bibinfo{author}{\bibfnamefont{V.}~\bibnamefont{Fack}}, and
  \bibinfo{author}{\bibfnamefont{B.~D.} \bibnamefont{McKay}},
  \bibinfo{year}{2009}, \bibinfo{journal}{Discrete Math.}
  \textbf{\bibinfo{volume}{157}}(\bibinfo{number}{2}), \bibinfo{pages}{377}.

\bibitem[{\citenamefont{Ali\v{s}auskas}(2002)}]{AlisauskasJMP43}
\bibinfo{author}{\bibnamefont{Ali\v{s}auskas}, \bibfnamefont{S.}},
  \bibinfo{year}{2002}, \bibinfo{journal}{J. Math. Phys.}
  \textbf{\bibinfo{volume}{43}}(\bibinfo{number}{3}), \bibinfo{pages}{1547}.

\bibitem[{\citenamefont{Anderson} \emph{et~al.}(2009)\citenamefont{Anderson,
  Aquilanti, and Marzuoli}}]{AndersonJPCA113}
\bibinfo{author}{\bibnamefont{Anderson}, \bibfnamefont{R.~W.}},
  \bibinfo{author}{\bibfnamefont{V.}~\bibnamefont{Aquilanti}}, and
  \bibinfo{author}{\bibfnamefont{A.}~\bibnamefont{Marzuoli}},
  \bibinfo{year}{2009}, \bibinfo{journal}{J. Phys. Chem. A}
  \textbf{\bibinfo{volume}{113}}(\bibinfo{number}{52}), \bibinfo{pages}{15106}.

\bibitem[{\citenamefont{Aquilanti} \emph{et~al.}(2009)\citenamefont{Aquilanti,
  Bitencourt, {da S. Ferreira}, Marzuoli, and Ragni}}]{AquilantiTCA123}
\bibinfo{author}{\bibnamefont{Aquilanti}, \bibfnamefont{V.}},
  \bibinfo{author}{\bibfnamefont{A.~C.~P.} \bibnamefont{Bitencourt}},
  \bibinfo{author}{\bibfnamefont{C.}~\bibnamefont{{da S. Ferreira}}},
  \bibinfo{author}{\bibfnamefont{A.}~\bibnamefont{Marzuoli}}, and
  \bibinfo{author}{\bibfnamefont{M.}~\bibnamefont{Ragni}},
  \bibinfo{year}{2009}, \bibinfo{journal}{Theor. Chem. Acc.}
  \textbf{\bibinfo{volume}{123}}(\bibinfo{number}{3--4}), \bibinfo{pages}{237}.

\bibitem[{\citenamefont{Balcar and Lovesey}(2009)}]{BalcarSTMP234}
\bibinfo{author}{\bibnamefont{Balcar}, \bibfnamefont{E.}}, and
  \bibinfo{author}{\bibfnamefont{S.~W.} \bibnamefont{Lovesey}},
  \bibinfo{year}{2009}, \emph{\bibinfo{title}{Introduction to the graphical
  theory of angular momentum}}, volume \bibinfo{volume}{234} of
  \emph{\bibinfo{series}{Springer Tracts in Modern Physics}}
  (\bibinfo{publisher}{Springer}).

\bibitem[{\citenamefont{Bar-Shalom and Klapisch}(1988)}]{BarCPC50}
\bibinfo{author}{\bibnamefont{Bar-Shalom}, \bibfnamefont{A.}}, and
  \bibinfo{author}{\bibfnamefont{M.}~\bibnamefont{Klapisch}},
  \bibinfo{year}{1988}, \bibinfo{journal}{Comp. Phys. Commun.}
  \textbf{\bibinfo{volume}{50}}(\bibinfo{number}{3}), \bibinfo{pages}{375}.

\bibitem[{\citenamefont{Bau}(1990)}]{BauAJC2}
\bibinfo{author}{\bibnamefont{Bau}, \bibfnamefont{S.}}, \bibinfo{year}{1990},
  \bibinfo{journal}{Australas. J. Comb.} \textbf{\bibinfo{volume}{2}},
  \bibinfo{pages}{57}.

\bibitem[{\citenamefont{Brinkmann}(1996)}]{BrinkmannJGT23}
\bibinfo{author}{\bibnamefont{Brinkmann}, \bibfnamefont{G.}},
  \bibinfo{year}{1996}, \bibinfo{journal}{J. Graph Theory}
  \textbf{\bibinfo{volume}{23}}(\bibinfo{number}{2}), \bibinfo{pages}{139}.

\bibitem[{\citenamefont{Brinkmann} \emph{et~al.}(2011)\citenamefont{Brinkmann,
  Goedgebeur, and McKay}}]{BrinkmannDMTCS13}
\bibinfo{author}{\bibnamefont{Brinkmann}, \bibfnamefont{G.}},
  \bibinfo{author}{\bibfnamefont{J.}~\bibnamefont{Goedgebeur}}, and
  \bibinfo{author}{\bibfnamefont{B.~D.} \bibnamefont{McKay}},
  \bibinfo{year}{2011}, \bibinfo{journal}{Disc. Math. Theor. Comp. Sci.}
  \textbf{\bibinfo{volume}{13}}(\bibinfo{number}{2}), \bibinfo{pages}{69}.

\bibitem[{\citenamefont{Brouder and Brinkmann}(1997)}]{BrouderJEESR86}
\bibinfo{author}{\bibnamefont{Brouder}, \bibfnamefont{C.}}, and
  \bibinfo{author}{\bibfnamefont{G.}~\bibnamefont{Brinkmann}},
  \bibinfo{year}{1997}, \bibinfo{journal}{J. Electron. Spectrosc.}
  \textbf{\bibinfo{volume}{86}}(\bibinfo{number}{1--3}), \bibinfo{pages}{127}.

\bibitem[{\citenamefont{Clark and Entringer}(1983)}]{ClarkPMH14}
\bibinfo{author}{\bibnamefont{Clark}, \bibfnamefont{L.}}, and
  \bibinfo{author}{\bibfnamefont{R.}~\bibnamefont{Entringer}},
  \bibinfo{year}{1983}, \bibinfo{journal}{Period. Math. Hung.}
  \textbf{\bibinfo{volume}{14}}(\bibinfo{number}{1}), \bibinfo{pages}{57}.

\bibitem[{\citenamefont{Coxeter} \emph{et~al.}(1981)\citenamefont{Coxeter,
  Frucht, and Powers}}]{Coxeter}
\bibinfo{author}{\bibnamefont{Coxeter}, \bibfnamefont{H.~S.~M.}},
  \bibinfo{author}{\bibfnamefont{R.}~\bibnamefont{Frucht}}, and
  \bibinfo{author}{\bibfnamefont{D.~L.} \bibnamefont{Powers}},
  \bibinfo{year}{1981}, \emph{\bibinfo{title}{Zero-symmetry graphs: trivalent
  graphical regular representations of groups}} (\bibinfo{publisher}{Academic
  Press}), ISBN \bibinfo{isbn}{0-12-19458-04}.

\bibitem[{\citenamefont{Dalby} \emph{et~al.}(1992)\citenamefont{Dalby, Nourse,
  Hounshell, Gushurst, Grier, Leland, and Laufer}}]{DalbyJCIC32}
\bibinfo{author}{\bibnamefont{Dalby}, \bibfnamefont{A.}},
  \bibinfo{author}{\bibfnamefont{J.~G.} \bibnamefont{Nourse}},
  \bibinfo{author}{\bibfnamefont{W.~D.} \bibnamefont{Hounshell}},
  \bibinfo{author}{\bibfnamefont{A.~K.~I.} \bibnamefont{Gushurst}},
  \bibinfo{author}{\bibfnamefont{D.~L.} \bibnamefont{Grier}},
  \bibinfo{author}{\bibfnamefont{B.~A.} \bibnamefont{Leland}}, and
  \bibinfo{author}{\bibfnamefont{J.}~\bibnamefont{Laufer}},
  \bibinfo{year}{1992}, \bibinfo{journal}{J. Chem. Inf. Comput. Sci.}
  \textbf{\bibinfo{volume}{32}}(\bibinfo{number}{3}), \bibinfo{pages}{244}.

\bibitem[{\citenamefont{Danos and Fano}(1998)}]{DanosPR304}
\bibinfo{author}{\bibnamefont{Danos}, \bibfnamefont{M.}}, and
  \bibinfo{author}{\bibfnamefont{U.}~\bibnamefont{Fano}}, \bibinfo{year}{1998},
  \bibinfo{journal}{Phys. Rep.}
  \textbf{\bibinfo{volume}{304}}(\bibinfo{number}{4}), \bibinfo{pages}{155}.

\bibitem[{\citenamefont{{Di Leva} and Ponzano}(1967)}]{LevaNCIM51}
\bibinfo{author}{\bibnamefont{{Di Leva}}, \bibfnamefont{A.}}, and
  \bibinfo{author}{\bibfnamefont{G.}~\bibnamefont{Ponzano}},
  \bibinfo{year}{1967}, \bibinfo{journal}{Nuovo Cimento}
  \textbf{\bibinfo{volume}{51A}}(\bibinfo{number}{4}), \bibinfo{pages}{1107}.

\bibitem[{\citenamefont{D\"urr and Wagner}(1968)}]{DurrNC53}
\bibinfo{author}{\bibnamefont{D\"urr}, \bibfnamefont{H.~P.}}, and
  \bibinfo{author}{\bibfnamefont{F.}~\bibnamefont{Wagner}},
  \bibinfo{year}{1968}, \bibinfo{journal}{Nuovo Cimento}
  \textbf{\bibinfo{volume}{53A}}(\bibinfo{number}{1}), \bibinfo{pages}{255}.

\bibitem[{\citenamefont{Edmonds}(1957)}]{Edmonds}
\bibinfo{author}{\bibnamefont{Edmonds}, \bibfnamefont{A.~R.}},
  \bibinfo{year}{1957}, \emph{\bibinfo{title}{Angular momentum in quantum
  mechanics}} (\bibinfo{publisher}{Princeton University Press}),
  \bibinfo{note}{{E:} the factor $\sqrt{2j_2+1}$ in the denominator of (3.7.3)
  ought read $\sqrt{2j_3+1}$.}

\bibitem[{\citenamefont{El-Batanoni}
  \emph{et~al.}(1966)\citenamefont{El-Batanoni, El-Nadi, and
  Vysotsky}}]{ElbatanoniNP82}
\bibinfo{author}{\bibnamefont{El-Batanoni}, \bibfnamefont{F.}},
  \bibinfo{author}{\bibfnamefont{M.}~\bibnamefont{El-Nadi}}, and
  \bibinfo{author}{\bibfnamefont{G.~L.} \bibnamefont{Vysotsky}},
  \bibinfo{year}{1966}, \bibinfo{journal}{Nucl. Phys.}
  \textbf{\bibinfo{volume}{82}}(\bibinfo{number}{2}), \bibinfo{pages}{407}.

\bibitem[{\citenamefont{Fack} \emph{et~al.}(1997)\citenamefont{Fack, Pitre, and
  {Van der Jeugt}}}]{FackCPC101}
\bibinfo{author}{\bibnamefont{Fack}, \bibfnamefont{V.}},
  \bibinfo{author}{\bibfnamefont{S.~N.} \bibnamefont{Pitre}}, and
  \bibinfo{author}{\bibfnamefont{J.}~\bibnamefont{{Van der Jeugt}}},
  \bibinfo{year}{1997}, \bibinfo{journal}{Comput. Phys. Commun.}
  \textbf{\bibinfo{volume}{101}}(\bibinfo{number}{1--2}), \bibinfo{pages}{155}.

\bibitem[{\citenamefont{Frucht}(1976)}]{FruchtJGT1}
\bibinfo{author}{\bibnamefont{Frucht}, \bibfnamefont{R.}},
  \bibinfo{year}{1976}, \bibinfo{journal}{J. Graph Theory}
  \textbf{\bibinfo{volume}{1}}(\bibinfo{number}{1}), \bibinfo{pages}{45}.

\bibitem[{\citenamefont{Granovskii and Zhednov}(1993)}]{GranovskiiJPA26}
\bibinfo{author}{\bibnamefont{Granovskii}, \bibfnamefont{Y.~I.}}, and
  \bibinfo{author}{\bibfnamefont{A.~S.} \bibnamefont{Zhednov}},
  \bibinfo{year}{1993}, \bibinfo{journal}{J. Phys. A.: Math. Gen.}
  \textbf{\bibinfo{volume}{26}}(\bibinfo{number}{17}), \bibinfo{pages}{4339}.

\bibitem[{\citenamefont{Graovac and Pisanski}(1991)}]{GraovacJMC8}
\bibinfo{author}{\bibnamefont{Graovac}, \bibfnamefont{A.}}, and
  \bibinfo{author}{\bibfnamefont{T.}~\bibnamefont{Pisanski}},
  \bibinfo{year}{1991}, \bibinfo{journal}{J. Math. Chem.}
  \textbf{\bibinfo{volume}{8}}(\bibinfo{number}{1}), \bibinfo{pages}{53}.

\bibitem[{\citenamefont{Gutman and Graovac}(2007)}]{GutmanCPL436}
\bibinfo{author}{\bibnamefont{Gutman}, \bibfnamefont{I.}}, and
  \bibinfo{author}{\bibfnamefont{A.}~\bibnamefont{Graovac}},
  \bibinfo{year}{2007}, \bibinfo{journal}{Chem. Phys. Lett.}
  \textbf{\bibinfo{volume}{436}}(\bibinfo{number}{1--4}), \bibinfo{pages}{294}.

\bibitem[{\citenamefont{Imrich}(1971)}]{ImrichAE6}
\bibinfo{author}{\bibnamefont{Imrich}, \bibfnamefont{W.}},
  \bibinfo{year}{1971}, \bibinfo{journal}{Aequationes mathematicae}
  \textbf{\bibinfo{volume}{6}}(\bibinfo{number}{1}), \bibinfo{pages}{6}.

\bibitem[{\citenamefont{Jahn and Hope}(1954)}]{JahnPR93}
\bibinfo{author}{\bibnamefont{Jahn}, \bibfnamefont{H.~A.}}, and
  \bibinfo{author}{\bibfnamefont{J.}~\bibnamefont{Hope}}, \bibinfo{year}{1954},
  \bibinfo{journal}{Phys. Rev.}
  \textbf{\bibinfo{volume}{93}}(\bibinfo{number}{2}), \bibinfo{pages}{318}.

\bibitem[{\citenamefont{Kent and Schlesinger}(1989)}]{KentPRA40}
\bibinfo{author}{\bibnamefont{Kent}, \bibfnamefont{R.~D.}}, and
  \bibinfo{author}{\bibfnamefont{M.}~\bibnamefont{Schlesinger}},
  \bibinfo{year}{1989}, \bibinfo{journal}{Phys. Rev. A}
  \textbf{\bibinfo{volume}{40}}(\bibinfo{number}{2}), \bibinfo{pages}{536}.

\bibitem[{\citenamefont{Lievens and {Van der Jeugt}}(2002)}]{LievensJMP43}
\bibinfo{author}{\bibnamefont{Lievens}, \bibfnamefont{S.}}, and
  \bibinfo{author}{\bibfnamefont{J.}~\bibnamefont{{Van der Jeugt}}},
  \bibinfo{year}{2002}, \bibinfo{journal}{J. Math. Phys.}
  \textbf{\bibinfo{volume}{43}}(\bibinfo{number}{7}), \bibinfo{pages}{3824}.

\bibitem[{\citenamefont{Lima}(1991)}]{LimaCPC66}
\bibinfo{author}{\bibnamefont{Lima}, \bibfnamefont{P.~M.}},
  \bibinfo{year}{1991}, \bibinfo{journal}{Comput. Phys. Commun.}
  \textbf{\bibinfo{volume}{66}}(\bibinfo{number}{1}), \bibinfo{pages}{89}.

\bibitem[{\citenamefont{Louck}(2008)}]{Louck}
\bibinfo{author}{\bibnamefont{Louck}, \bibfnamefont{J.~D.}},
  \bibinfo{year}{2008}, \emph{\bibinfo{title}{Unitary symmetry and
  combinatorics}} (\bibinfo{publisher}{World Scientific},
  \bibinfo{address}{Singapore}), ISBN \bibinfo{isbn}{981-281-472-8}.

\bibitem[{\citenamefont{Martello}(1983)}]{MartelloTOMS9}
\bibinfo{author}{\bibnamefont{Martello}, \bibfnamefont{S.}},
  \bibinfo{year}{1983}, \bibinfo{journal}{ACM Trans. Math. Software}
  \textbf{\bibinfo{volume}{9}}(\bibinfo{number}{1}), \bibinfo{pages}{131}.

\bibitem[{\citenamefont{Massot} \emph{et~al.}(1967)\citenamefont{Massot,
  El-Baz, and Lafoucri\'ere}}]{MassotRMP39}
\bibinfo{author}{\bibnamefont{Massot}, \bibfnamefont{J.-N.}},
  \bibinfo{author}{\bibfnamefont{E.}~\bibnamefont{El-Baz}}, and
  \bibinfo{author}{\bibfnamefont{J.}~\bibnamefont{Lafoucri\'ere}},
  \bibinfo{year}{1967}, \bibinfo{journal}{Rev. Mod. Phys.}
  \textbf{\bibinfo{volume}{39}}(\bibinfo{number}{2}), \bibinfo{pages}{288}.

\bibitem[{\citenamefont{Mattis and Braaten}(1989)}]{MattisPRD39}
\bibinfo{author}{\bibnamefont{Mattis}, \bibfnamefont{M.~P.}}, and
  \bibinfo{author}{\bibfnamefont{E.}~\bibnamefont{Braaten}},
  \bibinfo{year}{1989}, \bibinfo{journal}{Phys. Rev. D.}
  \textbf{\bibinfo{volume}{39}}(\bibinfo{number}{9}), \bibinfo{pages}{2737}.

\bibitem[{\citenamefont{Meringer}(1999)}]{MeringerJGT30}
\bibinfo{author}{\bibnamefont{Meringer}, \bibfnamefont{M.}},
  \bibinfo{year}{1999}, \bibinfo{journal}{J. Graph Theory}
  \textbf{\bibinfo{volume}{30}}(\bibinfo{number}{2}), \bibinfo{pages}{137}.

\bibitem[{\citenamefont{Newman and Wallis}(1976)}]{NewmanJPA9}
\bibinfo{author}{\bibnamefont{Newman}, \bibfnamefont{D.~J.}}, and
  \bibinfo{author}{\bibfnamefont{J.}~\bibnamefont{Wallis}},
  \bibinfo{year}{1976}, \bibinfo{journal}{J. Phys. A: Math. Gen.}
  \textbf{\bibinfo{volume}{9}}(\bibinfo{number}{12}), \bibinfo{pages}{2021}.

\bibitem[{\citenamefont{Ord-Smith}(1954)}]{OrdPR94}
\bibinfo{author}{\bibnamefont{Ord-Smith}, \bibfnamefont{R.~J.}},
  \bibinfo{year}{1954}, \bibinfo{journal}{Phys. Rev.}
  \textbf{\bibinfo{volume}{94}}(\bibinfo{number}{5}), \bibinfo{pages}{1227}.

\bibitem[{\citenamefont{Ponzano}(1965)}]{PonzanoNC36}
\bibinfo{author}{\bibnamefont{Ponzano}, \bibfnamefont{G.}},
  \bibinfo{year}{1965}, \bibinfo{journal}{Nuovo Cimento}
  \textbf{\bibinfo{volume}{36}}(\bibinfo{number}{2}), \bibinfo{pages}{385}.

\bibitem[{\citenamefont{Schwenk}(1989)}]{SchwenkJCTB47}
\bibinfo{author}{\bibnamefont{Schwenk}, \bibfnamefont{A.~J.}},
  \bibinfo{year}{1989}, \bibinfo{journal}{J. Comb. Theory B}
  \textbf{\bibinfo{volume}{47}}, \bibinfo{pages}{53}.

\bibitem[{\citenamefont{Sloane}(2003)}]{EIS}
\bibinfo{author}{\bibnamefont{Sloane}, \bibfnamefont{N.~J.~A.}},
  \bibinfo{year}{2003}, \bibinfo{journal}{Notices Am.\ Math.\ Soc.}
  \textbf{\bibinfo{volume}{50}}(\bibinfo{number}{8}), \bibinfo{pages}{912},
  \bibinfo{note}{http://oeis.org/}, \urlprefix\url{http://oeis.org/}.

\bibitem[{\citenamefont{Stone}(1956)}]{StoneMPCR52}
\bibinfo{author}{\bibnamefont{Stone}, \bibfnamefont{A.~P.}},
  \bibinfo{year}{1956}, \bibinfo{journal}{Math. Proc. Cambr. Phil. Soc.}
  \textbf{\bibinfo{volume}{52}}, \bibinfo{pages}{424}.

\bibitem[{\citenamefont{Stone}(1957)}]{StonePPS70}
\bibinfo{author}{\bibnamefont{Stone}, \bibfnamefont{A.~P.}},
  \bibinfo{year}{1957}, \bibinfo{journal}{Proc. Phys. Soc. A}
  \textbf{\bibinfo{volume}{70}}(\bibinfo{number}{12}), \bibinfo{pages}{908}.

\bibitem[{\citenamefont{{van Dyck}} \emph{et~al.}(2005)\citenamefont{{van
  Dyck}, Brinkmann, Fack, and McKay}}]{vanDyckCPC173}
\bibinfo{author}{\bibnamefont{{van Dyck}}, \bibfnamefont{D.}},
  \bibinfo{author}{\bibfnamefont{G.}~\bibnamefont{Brinkmann}},
  \bibinfo{author}{\bibfnamefont{V.}~\bibnamefont{Fack}}, and
  \bibinfo{author}{\bibfnamefont{B.~D.} \bibnamefont{McKay}},
  \bibinfo{year}{2005}, \bibinfo{journal}{Comput. Phys. Commun.}
  \textbf{\bibinfo{volume}{173}}(\bibinfo{number}{1--2}), \bibinfo{pages}{61}.

\bibitem[{\citenamefont{{van Dyck} and Fack}(2007)}]{vanDyckDM307}
\bibinfo{author}{\bibnamefont{{van Dyck}}, \bibfnamefont{D.}}, and
  \bibinfo{author}{\bibfnamefont{V.}~\bibnamefont{Fack}}, \bibinfo{year}{2007},
  \bibinfo{journal}{Disc. Math.} \textbf{\bibinfo{volume}{307}},
  \bibinfo{pages}{1506}.

\bibitem[{\citenamefont{Weininger}(1988)}]{WeiningerJCIC28}
\bibinfo{author}{\bibnamefont{Weininger}, \bibfnamefont{D.}},
  \bibinfo{year}{1988}, \bibinfo{journal}{J. Chem. Inf. Comput. Sci.}
  \textbf{\bibinfo{volume}{28}}(\bibinfo{number}{1}), \bibinfo{pages}{31}.

\bibitem[{\citenamefont{Wormer and Paldus}(2006)}]{WormerAQC51}
\bibinfo{author}{\bibnamefont{Wormer}, \bibfnamefont{P.~E.~S.}}, and
  \bibinfo{author}{\bibfnamefont{J.}~\bibnamefont{Paldus}},
  \bibinfo{year}{2006}, \bibinfo{journal}{Adv. Quant. Chem.}
  \textbf{\bibinfo{volume}{51}}, \bibinfo{pages}{59}.

\bibitem[{\citenamefont{Yutsis} \emph{et~al.}(1962)\citenamefont{Yutsis,
  Levinson, Vanagas, and Sen}}]{Yutsis}
\bibinfo{author}{\bibnamefont{Yutsis}, \bibfnamefont{A.~P.}},
  \bibinfo{author}{\bibfnamefont{I.~B.} \bibnamefont{Levinson}},
  \bibinfo{author}{\bibfnamefont{V.~V.} \bibnamefont{Vanagas}}, and
  \bibinfo{author}{\bibfnamefont{A.}~\bibnamefont{Sen}}, \bibinfo{year}{1962},
  \emph{\bibinfo{title}{Mathematical Apparatus of the Theory of Angular
  Momentum}} (\bibinfo{publisher}{Israel program for scientific translations}),
  \bibinfo{note}{{E}: Eq. (20.6) in \cite{ElbatanoniNP82}, Fig.\ 17.1 in
  \cite{MattisPRD39}}.

\end{thebibliography}

\end{document}